\documentclass[phd, copyrightpage]{mqthesis} 
\newcommand{\bra}[1]{\langle#1|}
\newcommand{\ket}[1]{|#1\rangle}
\newcommand{\braket}[2]{\langle{#1}|{#2}\rangle}
\newcommand{\ketbra}[2]{|{#1}\rangle\langle{#2}|}

\newcommand{\overlap}[2]{\langle{#1}|{#2}\rangle}

\newcommand{\BS}{{\scshape BosonSampling\ }}
\newcommand{\BStwo}{{\scshape BosonSampling}}
\newcommand{\expec}[1]{\langle #1\rangle}

\newcommand{\smallfrac}[2]{\mbox{$\frac{#1}{#2}$}}
\newcommand{\gdisp}[1]{ \smallfrac{ g {#1} }{ \sqrt{2} } }

\usepackage{hyperref}
\hypersetup{
colorlinks = true, 
citecolor = [rgb]{0.3568627450980392, 0.4745098039215686, 
 0.10196078431372549}, 
urlcolor = [rgb]{0.00392156862745098, 0.24705882352941178, 
 0.8666666666666667}, 
linkcolor = [rgb]{0.6078431372549019, 0.2980392156862745, 
 0.8705882352941177} 
}

\usepackage{subfig}
\usepackage{mathtools}

\usepackage{natbib}
\newcommand\imageWidth{.4}
\newcommand\imageWidthTwo{.7}
\newcommand\imageWidthThree{.55}

\begin{document}

\frontmatter
 
\title{Routes Towards Optical Quantum Technology --- New Architectures and Applications}
\author{Keith R. Motes}
\department{Physics}  

\titlepage
\chapter{Abstract}

As this century unfolds we will witness the rise of the quantum computer. A quantum computer is a device that utilizes the laws of one of the most fundamental theories of physics --- quantum mechanics. By utilizing properties of quantum mechanics a quantum computer is able to perform certain information processing tasks much more efficiently than an ordinary classical computer. The last century of modern society was revolutionized by some of the simplest ideas in quantum mechanics like energy quantization and quantum tunnelling, which led to technologies like the laser and the transistor. We can expect to see additional technological revolutions occur in this century since we are beginning to build technologies with some of the more complex properties of quantum mechanics such as quantum entanglement and quantum superposition. These quantum effects make more radical technologies like the quantum computer possible.

This thesis is based upon the work I have done during my PhD candidature at Macquarie University. In this work we develop quantum technologies that are directed towards realising a quantum computer. Specifically, we have made many theoretical advancements in a type of quantum information processing protocol called \BStwo. This device efficiently simulates the interaction of quantum particles called bosons, which no classical computer can efficiently simulate. In this thesis we explore quantum random walks, which are the basis of how the bosons in \BS interfere with each other. We explore implementing \BS using the most readily available photon source technology. We invented a completely new architecture which can implement \BS in time rather than space and has since been used to make the worlds largest \BS experiment ever performed. We look at variations to the traditional \BS architecture by considering other quantum states of light. We show a worlds first application inspired by \BS in quantum metrology where measurements may be made more accurately than with any classical method. Lastly, dealing with \BStwo, we look at reformulating the formalism of \BS using a quantum optics approach. In addition, but not related to \BStwo, we show a protocol for efficiently generating large-photon Fock states, which are a type of quantum state of light, that are useful for quantum computation. Also, we show a method for generating a specific quantum state of light that is useful for quantum error correction --- an essential component of realising a quantum computer --- by coupling together light and atoms. 
\chapter{List of Publications}

\begin{enumerate}

\item{\textit{Encoding qubits into oscillators with atomic ensembles and squeezed light.} \textbf{Keith R. Motes}, Ben Q. Baragiola, Alexei Gilchrist, Nicolas C. Menicucci. \href{https://arxiv.org/abs/1703.02107}{arXiv:1703.02107 (2017)}}

\item{\textit{Linear optical quantum metrology with single photons --- Experimental errors, resource counting, and quantum Cram\'{e}r-Rao bounds.} Jonathan P. Olson, \textbf{Keith R. Motes}, Patrick M. Birchall, Nick M. Studer, Margarite LaBorde, Todd Moulder, Peter P. Rohde, Jonathan P. Dowling. \href{https://arxiv.org/abs/1610.07128}{arXiv:1610.07128 (2016)}}

\item{\textit{Adaptive phase estimation with two-mode squeezed-vacuum and parity measurement.} Zixin Huang, \textbf{Keith R. Motes}, Petr M. Anisimov, Jonathan P. Dowling, Dominic W. Berry. \href{http://arxiv.org/abs/1609.04689}{arXiv:1609.04689 (2016)}}

\item{\textit{A Quantum Optics Argument for the \#P-hardness of a Class of Multidimensional Integrals.} Peter P. Rohde, Dominic W. Berry, \textbf{Keith R. Motes}, Jonathan P. Dowling. \href{http://arxiv.org/abs/1607.04960}{arXiv:1607.04960 (2016)}}

\item{\textit{Efficient recycling strategies for preparing large Fock states from single-photon sources --- applications to quantum metrology.} \textbf{Keith R. Motes}, Ryan L. Mann, Jonathan P. Olson, Nicholas M. Studer, E. Annelise Bergeron, Alexei Gilchrist, Jonathan P. Dowling, Dominic W. Berry, Peter P. Rohde. \href{http://journals.aps.org/pra/abstract/10.1103/PhysRevA.94.012344}{Phys. Rev A \textbf{94}, 012344 (2016)}}

\item{\textit{Quantum random walks on congested lattices and the effect of dephasing.} \textbf{Keith R. Motes}, Alexei Gilchrist, Peter P. Rohde. \href{http://www.nature.com/articles/srep19864}{Scientific Reports 6, 19864 (2016)}}

\item{\textit{Implementing Scalable Boson Sampling with Time-Bin Encoding: Analysis of Loss, Mode Mismatch, and Time Jitter.} \textbf{Keith R. Motes}, Jonathan P. Dowling, Alexei Gilchrist, Peter P. Rohde. \href{http://journals.aps.org/pra/abstract/10.1103/PhysRevA.92.052319}{Phys. Rev. A 92, 052319 (2015)}}

\item{\textit{Linear Optical Quantum Metrology with Single Photons: Exploiting Spontaneously Generated Entanglement to Beat the Shot-Noise Limit.} \textbf{Keith R. Motes}, Jonathan P. Olson, Evan J. Rabeaux, Jonathan P. Dowling, S. Jay Olson, Peter P. Rohde. \href{http://journals.aps.org/prl/abstract/10.1103/PhysRevLett.114.170802}{Phys. Rev. Lett. 114, 170802 (2015)}}

\item{\textit{Sampling arbitrary photon-added or photon-subtracted squeezed states is in the same complexity class as boson sampling.} Jonathan P. Olson, Kaushik P. Seshadreesan, \textbf{Keith R. Motes}, Peter P. Rohde, Jonathan P. Dowling. \href{http://journals.aps.org/pra/abstract/10.1103/PhysRevA.91.022317}{Phys. Rev. A 91, 022317 (2015)}}

\item{\textit{An introduction to boson-sampling.} Bryan T. Gard, \textbf{Keith R. Motes}, Jonathan P. Olson, Peter P. Rohde, Jonathan P. Dowling. Chapter 8 of From Atomic to Mesoscale: The Role of Quantum Coherence in Systems of Various Complexities (2015). \href{http://arxiv.org/abs/1406.6767}{Free Link}}

\item{\textit{Boson sampling with displaced single-photon Fock states versus single-photon-added coherent states---The quantum-classical divide and computational-complexity transitions in linear optics.} Kaushik P. Seshadreesan, Jonathan P. Olson, \textbf{Keith R. Motes}, Peter P. Rohde, Jonathan P. Dowling. \href{https://journals.aps.org/pra/abstract/10.1103/PhysRevA.91.022334}{Phys. Rev. A 91, 022334 (2015)}}

\item{\textit{Evidence for the conjecture that sampling generalized cat states with linear optics is hard.} Peter P. Rohde, \textbf{Keith R. Motes}, Paul Knott, Joseph Fitzsimons, William Munro, Jonathan P. Dowling. \href{http://journals.aps.org/pra/abstract/10.1103/PhysRevA.91.012342}{Phys. Rev. A 91, 012342 (2015)}}

\item{\textit{Scalable boson-sampling with time-bin encoding using a loop-based architecture.} \textbf{Keith R. Motes}, Alexei Gilchrist, Jonathan P. Dowling, Peter P. Rohde. \href{http://journals.aps.org/prl/abstract/10.1103/PhysRevLett.113.120501}{Phys. Rev. Lett. 113, 120501 (2014)}}

\item{\textit{Will boson-sampling ever disprove the Extended Church-Turing thesis?} Peter P. Rohde, \textbf{Keith R. Motes}, Paul A. Knott, William J. Munro. \href{http://arxiv.org/abs/1401.2199}{arXiv:1401.2199 (2014)}}

\item{\textit{Spontaneous parametric down-conversion photon sources are scalable in the asymptotic limit for boson-sampling.} Keith R. Motes, Jonathan P. Dowling, Peter P. Rohde. \href{http://journals.aps.org/pra/abstract/10.1103/PhysRevA.88.063822}{Phys. Rev. A 88, 063822 (2013)}}

\item{\textit{Phase estimation with two-mode squeezed-vacuum and parity detection restricted to finite resources.} \textbf{Keith R. Motes}, Petr M. Anisimov, Jonathan P. Dowling. \href{http://arxiv.org/abs/1110.1587}{arXiv:1110.1587 (2013)}}

\end{enumerate}

\chapter{Acknowledgements}

I would like to thank Peter P. Rohde and Alexei Gilchrist for taking me in as their PhD student. They were always supportive, gave excellent advice, and taught me many new concepts in physics. A special thanks to Peter who was interacting with me on a daily basis for the first half or so of my candidature discussing with me topics ranging from physics to life in general. Without their support none of what I have accomplished would be possible. 

I would like to thank Jonathan P. Dowling who originally took me in as an undergraduate researcher at Louisiana State University and is continuing to have such a huge impact on my life. He recommended me for this PhD position at Macquarie University and has visited Macquarie University once or twice every year throughout my PhD continuing to support my career in physics. Without him none of what I have accomplished would be possible.   

I am extremely grateful for all of the wonderful places that my adventures during my PhD has led me to including Sydney, Brisbane, Melbourne, Gold Coast, Sunshine Coast, Byron Bay, Cairns, Great Ocean Road, Japan, China, United Kingdom, Netherlands, Germany, New Zealand, Bali, Utah, Louisiana, and Baltimore. 

I would like to thank in no particular order all of my co-authers who have played a critical role in my development as a physicist: Peter P. Rohde, Alexei Gilchrist, Jonathan P. Dowling, Nicolas C. Menicucci, Dominic W. Berry, William Munro, Jonathan P. Olson, Ben Q. Baragiola, Bryan T. Gard, Ryan L. Mann, Kaushik P. Shshadreesan, Paul Knott, Joseph Fitzsimons, Evan J. Rabeaux, S. Jay Olson, Nicholas M. Studer, and E. Annelise Bergeron. 

I would like to thank in no particular order people who I've had helpful discussions with: Matthew Broome, Mark Wilde, Andrew White, Gerard Milburn, Giulia Ferrini, Gavin Brennen, Michael Steel, Geoff Pryde, Michael Bremner, Hwang Lee, Dave Wecker, Dave Aldred, and many others. 

For funding I would like to acknowledge my scholarship from Macquarie University titled the Macquarie Research Excellence Scholarship (MQRES) as well as support from the the Australian Research Council Centre of Excellence for Engineered Quantum Systems (Project number CE110001013).
{
\hypersetup{
colorlinks = true, 
linkcolor = [rgb]{0.00392156862745098, 0.24705882352941178, 
 0.8666666666666667} 
}

\tableofcontents
}

\mainmatter

\begin{savequote}[45mm]
The boundaries are imaginary, the rules are made up, the limits do not exist. 
\qauthor{William Paul Wiebers}
\end{savequote}

\chapter{Introduction to Linear Optical Quantum Computing} \label{ch:LOQC}

\section{Synopsis}

In the 20th century the first quantum revolution occurred. Due to most basic quantum effects, namely wave-particle duality and quantisation, new technologies emerged that revolutionised the world as we know it. Some of these technologies are lasers, transistors (which allow for computers), magnetic resonance imaging, and the electron microscope. Due to these quantum based technologies human civilisation has experienced the most rapid cultural changes over the last century than ever before. These technologies however only harness the most basic aspects of the quantum world. There are much more subtle quantum effects like quantum entanglement that promise a whole new era of human civilisation and perhaps even larger changes as we advance through the 21st century. One of the most promising of these new technologies is the quantum computer as it promises significant advantages over ordinary classical computing. There are many architectures for building a quantum computer but they are all technically very challenging. A useful approach is to build a quantum computer optically as it has many advantages and is the focus of this thesis.

In section \ref{sec:LOQCMotivation} we will motivate a particular type of quantum computer called a linear optical quantum computer (LOQC) and provide a brief history of the field of quantum computing. In section \ref{sec:LOQCLOQC} we will review quantum computing in the context of linear optics, discussing the requirements of quantum computing. In section \ref{sec:LOQCSolvableProblems} we discuss what classes of problems quantum computers are known to show advantages over classical computers. In section \ref{sec:LOQCcomplexity} we review some basic computational complexity classes in computer science that are important to this thesis. Lastly, in section \ref{sec:LOQChard}, we explain why linear optical quantum computing remains challenging.

\section{Motivation} \label{sec:LOQCMotivation}

Quantum Computers are information processing devices that utilise the laws of quantum mechanics to process information. It was the famous physicist Richard Feynman who first postulated the idea of a quantum computer. Since then quantum computing has inspired a huge field of research commonly referred to as quantum information processing \cite{bib:NielsenChuang00}. For a quantum computer to utilise the laws of quantum mechanics the quantum computer requires the use of various quantum objects like atoms and photons. These objects are some of nature's smallest and most fundamental building blocks and so they are extremely sensitive to being disturbed by their environment which is a greatly simplified reason of why realising a quantum computer is extremely challenging.

Many different physical architectures for building a quantum computer have been proposed. Some of these proposals use atom and ion traps \cite{bib:cirac95, bib:kielpinski02}, superconducting qubits \cite{bib:wallraff04}, nuclear magnetic resonance \cite{bib:cory98}, quantum dots \cite{bib:Loss98}, nuclear spin \cite{bib:kane98}, and linear optical interferometers \cite{bib:LOQC}. Which of these models is likely to yield the first quantum computer? 
It is likely not just one, but a composite of these technologies that will be used. Whatever technology \emph{is} used, we will need something that is efficient, by which we mean that the resources used by the quantum computer scale polynomially with the size of the computation. Likewise, if the quantum computer is inefficient we mean that the resources required scale exponentially with the size of the computation. Inefficient devices by definition  will not be scalable and will be too costly to realise.  

Optical quantum information processing is one promising candidate. It has a particularly useful advantage over many other proposals since it harnesses light which is highly resistant to certain forms of decoherence. Looking into the history of linear optics for quantum computing, we find that it was not always believed to be a viable method for building a quantum computer. The physics community has investigated linear optical interferometers to process quantum information for quite some time. Before 2001 it was believed that a linear optical interferometer alone could not be engineered to make a universal quantum computer. For example, in 1993 {\u C}ern{\' y} proposed that a linear interferometer could be used to solve \textbf{NP}-complete problems in polynomial time, but he found that the scheme suffered from an exponential overhead in energy \cite{cerny}, which is inefficient. In 1996 Clauser \& Dowling showed that a linear optics Talbot interferometer could factor integers in polynomial time, which is efficient, but with either an exponential overhead in energy or physical size \cite{clauser}, which is inefficient. Again in 1996, Cerf, Adami \& Kwiat demonstrated a programmable linear optical interferometer that could implement any universal logic gate (a requirement for universal quantum computing that we talk about in section \ref{sec:LOQCLOQC}) with single photon inputs but this suffered an exponential overhead in the spatial dimension which makes the scheme inefficient. In 2002, Bartlett \emph{et al.} showed that any interferometer with Gaussian states at the input and with Gaussian measurements at the output can be efficiently simulated classically even in the presence of quadratic nonlinearities which created a continuous variable analog of the Gottesman-Knill theorem for discrete variables in the ordinary circuit quantum computation model \cite{bib:Bartlett02b}. 

These examples, among others, led to the widespread belief that photonic linear interferometry alone could not be used to build a universal quantum computer. As a corollary it was believed that linear optical interferometers were efficiently simulateable on an ordinary classical computer. 
However, in 2001, Knill, Laflamme \& Milburn (KLM) \cite{bib:LOQC} showed that efficient universal quantum computation can be implemented using only photon sources, beam splitters, phase shifters, photo-detectors, and feedback from the photo-detector's proving that linear optical interferometers are a viable architecture. This is known as linear optical quantum computing (LOQC).

There are many models for processing quantum information. These models include cluster (or graph) states \cite{bib:Raussendorf01, bib:Raussendorf03}, topological \cite{bib:freedman03}, adiabatic \cite{bib:farhi}, quantum random walks \cite{bib:ADZ}, quantum Turing machines \cite{bib:bern}, permutational \cite{bib:jordan2},  the one-clean qubit model \cite{bib:moussa} and the gate model \cite{bib:NielsenChuang00}. Some of these models are \emph{universal}, meaning they can efficiently implement any quantum algorithm, whereas others are \emph{restricted}, meaning they implement a specific subset of quantum algorithms. There are always errors in these schemes, such as decoherence, so the scheme needs to also be made fault-tolerant, which means that a protocol for quantum error correction is implementable. The gate model is perhaps the most intuitive and most familiar model as it resembles classical circuit models so we will use this model to describe LOQC. In chapters \ref{Ch:BSIntro}, \ref{Ch:SPDC}, \ref{Ch:FiberLoop}, \ref{Ch:sampOther}, and \ref{Ch:MORDOR} we discuss the results we have on \BStwo, which is a special purpose implementation of LOQC and the main focus of this thesis. 

\section{Linear Optical Quantum Computing} \label{sec:LOQCLOQC}

In this section we introduce the LOQC approach to universal quantum computing of KLM \cite{bib:LOQC, bib:KokLovett11}. As mentioned above it is a promising route forward for realising a universal quantum computer and many researchers around the world are constantly improving the associated technologies of linear optics. These technologies also promise a simple implementation of \BStwo, which is the primary focus of this work and facilitates development of key technologies that will ultimately become building blocks for universal LOQC. 

There are three main results in the original work of KLM \cite{bib:LOQC}. Firstly, they showed that linear optical elements are sufficient for efficient non-deterministic universal quantum computation with photons. Secondly, the success probability of implementing the quantum gates may be made asymptotically close to unity using a certain encoding technique. Thirdly, the resources for obtaining accurately encoded qubits scale efficiently using quantum coding. In addition they show that by iterating their method LOQC can be made to be fault-tolerant \cite{bib:Aharonov96, bib:kitaev97, bib:knill98, bib:preskill98}. 

In this section we review some of the fundamental requirements for quantum computing including the photonic qubit, some of the optical elements such as the beam splitter, phase shifter, and photo-detectors, logic gates, the nonlinear sign-flip gate, and quatum gate teleportation. In addition we discuss qubits on the Bloch sphere. 

\subsection{Photonic Qubit}
The fundamental unit of information of a quantum computer is the qubit or \textit{quantum bit}. This is analogous to the classical bit but is instead quantum in nature and can be in a superposition of the logical zero state $\ket{0}$ and the logical one state $\ket{1}$ as well as have arbitrary phase relationships. A qubit can also be entangled with other qubits. These properties of qubits, superposition and entanglement, give quantum computers significant advantages over classical computers in solving many algorithms. A popular encoding of a qubit in LOQC is one photon in two optical modes, known as \emph{dual rail} encoding. Another technique is polarization encoding where the qubit is encoded in the horizontal and vertical polarizations of light. Mathematically, a qubit is represented as a unit vector in the complex two-dimensional vector space $\mathbb{C}^2$. The vector space is spanned by the basis
\begin{eqnarray}
\ket{0} &=&  \left[ \begin{array}{l}
1 \\
0 \\
\end{array} \right] \\
\ket{1} &=& \left[ \begin{array}{l}
0 \\
1 \\
\end{array} \right].
\end{eqnarray}
A general quantum state may be then written as a normalised complex sum of these basis states, i.e.
\begin{equation}
\ket{\psi} = \cos\left(\frac{\theta}{2}\right) \ket{0} + e^{i \phi} \sin\left(\frac{\theta}{2}\right)\ket{1}.
\end{equation}
By setting $\theta =\frac{\pi}{2}$ and $\phi =0$ a particular qubit is obtained which has the form
\begin{equation}
\ket{\psi}= \frac{1}{\sqrt{2}}\left( \ket{0} +\ket{1} \right).
\end{equation}
This superposition state has a 50\% probability of being in logical state zero $\ket{0}$ and a 50\% probability of being in logical state one $\ket{1}$ once a measurement is made on the system in the $\ket{0}/\ket{1}$ logical basis. The probability of measuring a particular logical state is obtained by taking the absolute square of the corresponding amplitude.  

\subsection{Qubit Visualisation on a Bloch Sphere}
\begin{figure}[!htb]
\centering
 \includegraphics[width=0.35\columnwidth]{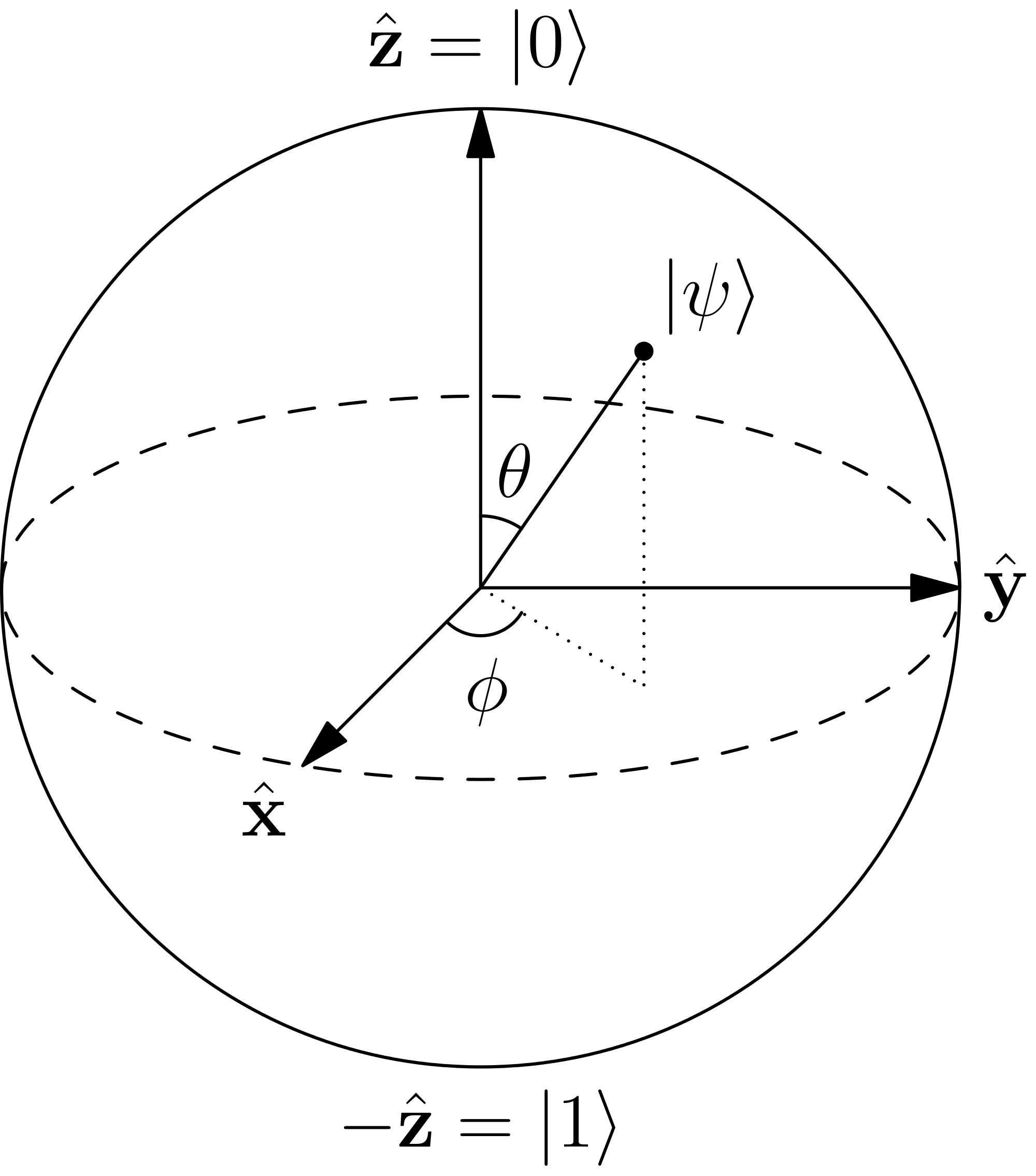}
 \caption{The Bloch sphere is a convenient way to visualize qubits. The qubit $\ket{\psi}$ is generally in a complex superposition of the $\ket{0}$ and $\ket{1}$ logical states. The sphere is also a convenient tool to visualize how the Pauli matrices rotate the qubit state. Pure states lie on the surface of the sphere while mixed states are contained within the sphere.}
 \label{fig:Bloch}
\end{figure}

A elegant way to visualize superposition states is to draw them on the so called Bloch sphere as shown in Fig.~\ref{fig:Bloch}. The $+z$ component of the Bloch sphere represents the $\ket{0}$ state while the $-z$ component represents the $\ket{1}$ state. A qubit $\ket{\psi}$ drawn on the Bloch sphere is in a superposition of the $\ket{0}$ and $\ket{1}$ if it is pointing anywhere except along the $\pm z$ axes, where the state is a $\ket{0}$ or $\ket{1}$ respectively. Pure states lie on the surface of the sphere while mixed states are contained within the sphere. Once the qubit is measured in the logical (i.e $\ket{0}/\ket{1}$) basis the quantum state will probabilistically collapse to either the $\ket{0}$ or $\ket{1}$. 

\subsection{Optical Elements for Quantum Computing}
There are certain optical elements necessary to implement linear optical quantum computing including: beamsplitters, phase-shifters, photodetectors, and feedforward from the photodetector outputs. With these elements, it has been shown that \cite{bib:LOQC},
\begin{enumerate}
\item{Universal quantum computation can be implemented non-deterministically.} 
\item{The probability of implementing non-deterministic quantum gates can be made asymptotically close to unity using an encoding technique with efficient overhead \cite{bib:lemr}.}
\item{Error correcting codes can be implemented, enabling fault-tolerant quantum computation.} 
\item{Quantum computation can be efficiently implemented.}
\end{enumerate}

In general, to build a quantum computer three essential components are required: a way to prepare quantum states, a way to implement any operation from a universal gate set on the qubits, and to measure quantum states. Using linear optical approaches we prepare a single photon in the Fock basis. There are several technologies that exist such as spontaneous parametric down conversion \cite{bib:OuLu99} and quantum dot sources \cite{bib:Santori01} to prepare single photons, which can be described by adding a photon to the vacuum state $\hat{a}^{\dag}\ket{0}=\ket{1}$. This is non-deterministic with all existing methods of generating single photon states but even with this non-determinism it may be used for quantum computing. As technologies improve single-photon sources are becoming more and more deterministic.

The two core optical elements that are used to implement optical gates on our prepared quantum states are phase-shifters and beamsplitters, which are both unitary transformations on a qubit. 

\subsubsection{Phase-Shifter}
The unitary for a phase-shifter acting on a single mode is
\begin{equation}
\hat{\Phi}_{\phi}=e^{i \hat{n}\phi},
\end{equation} 
where $\hat{n}$ is the number operator defined as $\hat{n}=\hat{a}^{\dag}\hat{a}$. 

\subsubsection{Beamsplitter and Hong-Ou-Mandel effect}

The unitary matrix for a beamsplitter may be represented by
\begin{eqnarray}
\hat{B}(\theta,\phi)=
\left(
\begin{array}{cc}
\cos{\theta}& -e^{i \phi} \sin{\theta} \\
e^{-i \phi} \sin{\theta}& \cos{\theta} 
\end{array}
\right),
\end{eqnarray}
in the basis of optical modes, where $\phi$ gives relative phase between the modes and $\theta$ is related to the reflectivity of the beamsplitter, which is $r=\cos^2{\theta}$. One of the most important quantum interference effects in quantum optics is the Hong-Ou-Mandel (HOM) effect \cite{bib:HOM87}. This effect is can be seen when identical single photons are incident on both of the input modes of a 50:50 beamsplitter at the same time (i.e. $\ket{\psi_{\mathrm{in}}}= \hat{a}^{\dag}\hat{b}^{\dag}\ket{0}\ket{0}$). At the output you see the HOM effect where the photons always tend to bunch. In other words both photons will always come out of the same output mode but you do not know which one it will be. This effect can be described mathematically as,
\begin{eqnarray}
\ket{\psi_{\mathrm{out}}} &=& \hat{B}\left(\frac{\pi}{4},0\right) \ket{\psi_{\mathrm{in}}} \nonumber \\
&=& \hat{B}\left(\frac{\pi}{4},0\right) \hat{a}^{\dag}\hat{b}^{\dag}\ket{0}\ket{0} \nonumber \\
&=& \frac{1}{2}\left((\hat{a}^{\dag}-\hat{b}^{\dag})(\hat{a}^{\dag}+\hat{b}^{\dag}) \right)\ket{0}\ket{0} \nonumber \\
&=&  \frac{1}{2}\left(\hat{a}^{\dag 2} + \hat{a}^{\dag}\hat{b}^{\dag} - \hat{a}^{\dag}\hat{b}^{\dag} - \hat{b}^{\dag 2} \right) \ket{0}\ket{0} \nonumber \\
&=& \frac{1}{2}\left(\hat{a}^{\dag 2} - \hat{b}^{\dag 2} \right) \ket{0}\ket{0} \nonumber \\
&=& \frac{1}{\sqrt{2}}\left(\ket{2}\ket{0} - \ket{0}\ket{2} \right).
\end{eqnarray}
Note the quantum interference, where the middle terms from line four to line five cancel, causing the HOM effect. 

\subsubsection{Photon Measurement}

In order to measure the state, photodetectors are used. They destructively determine the number of photons in a mode. A bucket detector is the simplest kind of detector which only measures if a mode contains zero photons or more than zero photons. For states with more than one photon a photon counting detector may be used to distinguish numbers but it is harder to realise than a bucket detector. An effective photon-number-counting detector referred to as a multiplexed photodetector \cite{bib:RohdeWebb07, bib:Fitch03, bib:Achilles04, bib:LPOR201400027, bib:ma2011experimental} may be realised with bucket detectors by using a series of beamsplitters to evenly spread out the photons over $N$ modes and then use bucket detectors at the output of those modes. $N$ should be large enough that the photons are sufficiently spread out so the probability of more than one photon going into the same bucket detector is negligible. The probability of under-counting given that the photon number is $n$ is at most $n(n-1)/(2N)$. When measuring a photon-number state projective measurements may be used with measurement operators given by \cite{bib:NielsenChuang00} 
\begin{equation}
\hat{\Pi}(n) = \ket{n}\bra{n},
\end{equation}
where $n$ is the photon number being measured. 
\subsection{Logic Gates}
There is a set of logic gates that are needed to manipulate qubits. We think of these gates as performing operations on the qubit \cite{bib:Nielsen04}. Some of the most common gates are: 
\begin{eqnarray}
\textrm{Controlled-NOT (CNOT):} && \left[ 
\begin{array}{cccc}
1 & 0 & 0 & 0 \\
0 & 1 & 0 & 0 \\
0 & 0 & 0 & 1 \\
0 & 0 & 1 & 0 \\
\end{array} \right] \\ \nonumber
\textrm{Hadamard (H):} &&  \frac{1}{\sqrt{2}}\left[ 
\begin{array}{cc}
1&1 \\
1&-1 \\
\end{array} \right] \\ \nonumber
\textrm{Pauli-X ($\sigma_x$):} && \left[
\begin{array}{cc}
0&1\\
1&0\\
\end{array} \right] \\ \nonumber
\textrm{Pauli-Y ($\sigma_y$):} && \left[
\begin{array}{cc}
0&-i\\
i&0\\
\end{array} \right] \\ \nonumber
\textrm{Pauli-Z ($\sigma_z$):} && \left[
\begin{array}{cc}
1&0\\
0&-1\\
\end{array} \right] \\ \nonumber
\textrm{Phase:} && \left[
\begin{array}{cc}
1&0\\
0&i\\
\end{array} \right] \\
\frac{\pi}{8}: &\;& \left[
\begin{array}{cc}
1&0\\
0&e^{i \pi/4} \\
\end{array} \right]. \nonumber
\end{eqnarray}
These gates form a universal gate set. 
To be universal requires that any unitary operation can be expressed as a finite sequence of gates from the set, thus there are many universal gate sets that can be chosen. A convenient way to visualise how some of these gates act on our quantum state $\ket{\psi}$ is to use the Bloch sphere as shown in Fig. \ref{fig:Bloch}. 

The first in this list of logic gates, the CNOT gate, is a maximally entangling gate, which is the quantum equivalent of the classical XOR gate. The latter gates are single qubit gates, which implement rotations on the Bloch sphere. The Hadamard gate $H$ will take a logical $\ket{0}$ or $\ket{1}$ and rotate them to a 50:50 superposition of $\ket{0}$ and $\ket{1}$ with a relative phase difference. Specifically, $H\ket{0}= (\ket{0}+\ket{1})/\sqrt{2}$ and $H\ket{1}= (\ket{0}-\ket{1})/\sqrt{2}$. The Pauli-X gate $\sigma_x$, Pauli-Y gate $\sigma_y$, and Pauli-Z gate $\sigma_z$ will rotate the state around the $x$, $y$, and $z$ axes by $\pi$ radians of the Bloch sphere respectively. The phase gate leaves the $\ket{0}$ state alone and maps $\ket{1}\to i\ket{1}$. Similar to the phase gate, the $\pi/8$ gate applies a phase to the $\ket{1}$ state. An advantage of using linear optics is that waveplates easily implement all of these single qubit gates with polarisation encoded qubits \cite{bib:humphreys13} but a disadvantage is that implementing the CNOT gate requires an effective Kerr non-linearity which is quite challenging in optics. 

\subsection{Nonlinear Sign-Flip Gate}
One way to realise a CNOT gate is to use a nonlinear sign-flip (NS) gate \cite{bib:LOQC} which implements the transformation 
\begin{equation}
\mathrm{NS}: \alpha_0 \ket{0} + \alpha_1 \ket{1} +\alpha_2 \ket{2} \rightarrow \alpha_0 \ket{0} + \alpha_1 \ket{1} - \alpha_2 \ket{2},
\end{equation}
and is used as a building block to implement the CNOT gate. The required universal gate set to perform quantum computing is given by the two qubit CNOT gate along with the previously discussed single qubit gates. With a set of one- and two-qubit universal gates multi-qubit gates may be constructed. With photons, linear optical elements, and photodetection, the NS gate may be implemented but is non-deterministic. 

\subsection{Quantum Gate Teleportation}
With multiple non-deterministic gates in a quantum computing circuit, the success probability of performing the computation drops exponentially with the number of gates. Overcoming this exponential drop may be achieved by using quantum gate teleportation which increases the probability of successfully implementing non-deterministic gates \cite{bib:LOQC, bib:GottesmanChuang99}. To do this, two Bell pairs (which are maximally entangled two-qubit states) act as a resource to teleport the action of a gate onto two qubits. This teleportation trick is also non-deterministic but still increases the success probability of the non-deterministic gate close to unity which enables efficient quantum computation since it can be concatenated \cite{bib:LOQC}. 

\section{Classes of Solvable Problems on a Quantum Computer} \label{sec:LOQCSolvableProblems}
The reason people are interested in building a quantum computer is because it promises to solve certain problems much more efficiently than a classical computer can. So far, there are three provable classes of problems in which a quantum computer outperforms classical computers. These are the quantum Fourier transform (which forms the basis for various other quantum algorithms), quantum search algorithms, and quantum simulation.

\subsection{Quantum Fourier Transform}


The first class is the quantum Fourier transform which is employed in Shor's factoring algorithm and discrete logarithms \cite{bib:Shor97}. As an example of the speedup that quantum Fourier transforms give us versus classical Fourier transforms we consider a problem with $ N=2^n$ numbers. A classical fast Fourier transform requires $ N \log{N} \approx 2^n n$ steps but a quantum computer does this in $\log^2{N} \approx n^2$ steps \cite{bib:Nielsen04}, which is exponentially faster and thus makes fast Fourier transforms efficient. Shor's factoring algorithm is an interesting example in terms of the impact it may have. Many security protocols that secure records such as financial transactions and perhaps government databases around the world rely on the conviction that factoring a large number cannot be efficiently performed on a classical computer. Shor's factoring algorithm however \emph{can} factor large numbers efficiently and may put many security protocols at risk. Luckily, there are quantum security protocols that rely on the laws of quantum mechanics to secure information that even quantum computers can not hack. This is known as quantum cryptography \cite{bib:BennetBrassard84}.

\subsection{Quantum Search Algorithms}
The second class, quantum search algorithms, make use of superposition to decrease the time it takes to search an unstructured database. The most famous example of such an algorithm was discovered by Grover \cite{bib:grover}. In his work he stated the problem to be a search of an unstructured database of $N$ elements with the goal of finding an element that satisfies a specific property. On a classical computer this search would require $O(N)$ operations, whilst a quantum search could accomplish this in $O(\sqrt{N})$ operations using Grover's algorithm. 


\subsection{Quantum Simulation}

The third class is quantum simulation, where one simulates complex interacting quantum systems with a quantum computer. The ramifications of a practical quantum simulator are huge and would have direct applications in condensed-matter physics, high-energy physics, atomic physics, quantum chemistry, and cosmology among others \cite{bib:georgescu14}. For example, in condensed matter physics, where one could study quantum phase transitions, quantum magnetism, and high temperature superconductivity (one of the most sought after open problems in condensed matter physics). It intuitively makes sense that quantum computers would be required to simulate quantum phenomena since quantum systems are described with an exponentially increasing Hilbert space and the number of available states in a quantum computer also increases as exponentially, specifically as $2^n$, where $n$ is the number of qubits. Classical computers are good at simulating most classical phenomena such as how air plane wings should be optimised for certain parameters such as lift and reducing drag and so too should quantum computers be used to simulate quantum phenomena. In general a classical computer requires $\mathrm{exp}(n)$ resources to simulate a typical quantum system that has $n$ distinct components while a quantum computer requires $\mathrm{poly}(n)$ qubits and time. 

\section{Computational Complexity} \label{sec:LOQCcomplexity}

Computer scientists have devised a way to analyse how difficult a problem is to solve. It is based on the fact that algorithms are used to solve certain problems and depending on the resources required to solve a problem it is classified in a particular way. There are three relevant classes of problems important to this thesis: decision problems, counting problems, and sampling problems. Decision problems ask what is the answer to a particular problem. Counting problems ask how many solutions there are to a problem. Sampling problems ask what are the samples drawn from some distribution. More specifically the classification that a problem falls into may be characterized by a complexity class where similar problems will fall into the same complexity class. Here we summarize just a few important complexity classes important for this thesis:

\begin{itemize}
\item \textbf{P}: Decision problems solvable on a deterministic Turing machine in time that scales polynomially with the size of the system.
\item \textbf{NP}: Decision problems with potentially multiple solutions where any particular solution is verifiable in time that scales polynomially with the size of the system.
\item \textbf{NP-Hard}: The set of problems that can be reduced to any \textbf{NP} problem with at most polynomial resource overhead. 
\item \textbf{NP-Complete}: The set of all problems that are in \textbf{NP} and \textbf{NP-Hard}. These contain the most difficult problems in \textbf{NP}.
\item \textbf{$\#$P}: The set of problems that count the number of solutions to a deterministic, polynomial time problem.
\item \textbf{$\#$P-Hard}: The set of problems that can be reduced to any \textbf{$\#$P} problem with polynomial resource overhead. 
\item \textbf{$\#$P-Complete}: The set of all problems that are in \textbf{$\#$P} and \textbf{$\#$P-Hard}. These contain the most difficult problems in \textbf{$\#$P}.
\item \textbf{PostBPP}: The class of problems that may be solved on a probabilisitic classical computer with postselection in time that scales polynomially with the size of the system and with a given bounded error of $1/3$.
\item \textbf{BQP}: The class of problems that may be solved by a universal quantum computer in time that scales polynomially with the size of the system.
\item \textbf{PostBQP}: The class of problems that may be solved on a universal quantum computer with postselection in time that scales polynomially with the size of the system and with a given bounded error of $2/3$.
\item \textbf{\BStwo{P}}: The class of sampling problems that can be described by the formalism of \BStwo, which is the main focus of this thesis, and is summarized in Ch. \ref{Ch:BSIntro}.  
\end{itemize}

\section{Why is Linear Optical Quantum 
Computing Hard?} \label{sec:LOQChard}

So far we have made this linear optical implementation of quantum computing seem quite simple, so how come we have not built a universal linear optical quantum computer yet? There are a myriad of technicalities spread through all of the different components and stages of implementing the linear optical quantum computer. Some of these include generating indistinguishable single photons, synchronising the photons, mode-matching, fast controllable delay lines, fast-feedforward, tuneable beamsplitters and phase-shifters, and accurate, fast, single-photon detectors. Much of these technicalities are realisable with current technologies but the success probability of a quantum computation reduces exponentially with the number of photons at the output; therefore, efficiencies need to be very high. Another major problem is inefficiency, such as loss, which can happen anywhere within the circuit. Also, with current efficiencies of photodetectors the implementation of teleportation and the more complex two qubit gate operations are challenging. 


Even though quantum computing remains challenging and seems like a distant dream there remain many interesting problems to investigate with more frugal resource requirements. One such problem is \BS which is the topic of Chapters \ref{Ch:BSIntro}, \ref{Ch:SPDC}, \ref{Ch:FiberLoop}, \ref{Ch:sampOther}, \ref{Ch:MORDOR}, and \ref{ch:sharpPhard}. \BS does away with the requirement of fast-feedforward, teleportation, and number-resolved photodetectors yet still implements an interesting problem since no classical computer can efficiently simulate \BStwo.

In the next chapter, however, we will review our work on quantum random walks which may be implemented with optical techniques. Quantum random walks are a route towards implementing quantum information processing tasks \cite{bib:ADZ, bib:AAKV, bib:Kempe08, bib:Salvador12} and are also an important aspect of understanding \BS as the bosons in \BS are undergoing a quantum random walk as they evolve. In fact \BS is equivalent to a quantum random walk, albeit a very complex one \cite{bib:increasing13}.

\begin{savequote}[45mm]
It is the mark of an educated mind to be able to entertain a thought without accepting it. 
\qauthor{Aristotle}
\end{savequote}

\chapter{Quantum Random Walks on Congested Lattices and the Effect of Dephasing} \label{ch:QRW}

\section{Synopsis}
Quantum random walks are an important aspect of quantum computing. They form the basis of several quantum algorithms that give speedups over their classical counterparts such as in some oracular problems \cite{bib:childs03}, Grover's search algorithm \cite{bib:grover}, and the element distinctness problem \cite{bib:ambainis07}. In this work we consider quantum random walks on congested lattices and contrast them to classical random walks. Congestion is modelled on lattices that contain static defects which reverse the walker's direction. We implement a dephasing process after each step which allows us to smoothly interpolate between classical and quantum random walks as well as study the effect of dephasing on the quantum walk. Our key results show that a quantum walker escapes a finite boundary dramatically faster than a classical walker and that this advantage remains in the presence of heavily congested lattices. It follows that quantum walks on congested lattices remain advantageous over classical random walks.

In section \ref{ch:QRWintro} we provide some motivation for our work and give an overview of what we have done. In section \ref{ch:QRWQRW} we introduce quantum random walks, write out a mathematical formalism, explain their evolution, and introduce the metrics we use to characterise our quantum random walk simulations which are variance and escape probability. In section \ref{ch:QRWcongestion} we introduce lattice congestion in the random walks and show our metrics for the walker on a congested lattice in both the classical and quantum case. In section \ref{ch:QRWdephasing} we introduce a model of dephasing into the quantum random walk and show how, with full dephasing, the quantum walker behaves like a classical walker. In section \ref{ch:QRWcombined} we show how our metrics behave in the presence of both congestion and dephasing for the quantum random walk.

\section{Motivation} \label{ch:QRWintro}


One route to implementing quantum information processing tasks with quantum computing is via quantum random walks \cite{bib:ADZ, bib:AAKV, bib:Kempe08, bib:Salvador12} whereby a particle, such as a photon, `hops' between the vertices in a lattice. The effects of a congested, or obstructed, lattice on a quantum random walk (QRW) are studied and compared to a classical random walk (CRW). Congestion may also be thought of as traffic and the walker is like a car trying to avoid the traffic. The quantum walkers also suffer a dephasing process as they propagate. This study provides insight into how random errors in the lattice and dephasing affect the dynamics of random walks and the robustness of certain quantum features. In our model, congestion refers to where the lattice through which the walker propagates has defects, which are like blocked streets that the walker encounters and has to back out of during the next step. These defects are stationary during the evolution of the random walk, though we average over many such random lattices. Dephasing occurs when the state decoheres and is implemented via a dephasing channel acting after each step. In the limit of full dephasing the QRW becomes a CRW, so that dephasing also allows us to interpolate between the classical and quantum regimes. For an experimental demonstration of dephasing in a QRW see Broome \emph{et al.} \cite{bib:Broome10}, and for related theoretical work on QRWs with phase damping see Lockhart \emph{et al.} \cite{bib:lockhart13}.

For characterising the resulting probability distributions for QRWs and CRWs we use variance and `escape probability', that is the probability that the walker escapes a finite region of the lattice, or more picturesquely, the probability that the walker `beats the traffic'.

\section{Quantum Random Walks} \label{ch:QRWQRW}

A QRW describes the evolution of a quantum particle through a given topological structure. A common choice of structure is a $d$ dimensional lattice. In a CRW, the walker probabilistically follows edges through a lattice to step to an adjacent vertex. In a QRW on the other hand, the walker spreads as a superposition of different paths through the graph. Physically, the walker can be a wide range of quantum particles, though of particular interest is the photon as photons are readily produced, manipulated and measured using off-the-shelf components in the laboratory. Photons have found widespread use in quantum information processing, most notably linear optical quantum computing (LOQC) \cite{bib:LOQC}. The technologies required in LOQC provide the topological structure for implementing a QRW. They also allow for multi-photon QRWs \cite{bib:increasing13}, which increases the dimensionality of the walk. For a further review on QRWs see Refs. \cite{bib:ADZ, bib:AAKV, bib:Kempe08, bib:Salvador12}, and see Refs. \cite{bib:Hagai08, bib:Schreiber10, bib:Broome10, bib:Peruzzo10, bib:Schreiber11b, bib:Matthews11, bib:Owens11, bib:Schreiber12, bib:Sansoni12} for the numerous optical demonstrations of elementary QRWs that have been performed.
  
\subsection{Quantum Random Walk Formalism}
To illustrate our QRW formalism we present the details for a one-dimensional discrete QRW on an unbounded lattice without any defects. The state of a one-dimensional QRW at any given time has the form,
\begin{equation} \label{eq:State}
\ket{\Psi(t)}=\sum_{x, c} \gamma_{x,c} \ket{x, c},
\end{equation}
where $x \in [-x_\mathrm{max},x_\mathrm{max}]$ represents the position of the particle; $x_\mathrm{max}$ represents the size of the lattice; $c \in \{-1,1\}$ is the coin value, unique for each time step, that tells the walker whether to evolve to the left ($c=-1$) or right ($c=1$); and $|\gamma_{x,c}|^{2}$ is the probability amplitude for a given position and coin value. The dimension of the lattice is $2 x_\mathrm{max}+1$. Since there are two coin values for each position, the probability that the walker is at position $x$ is given by,
\begin{equation} \label{eq:probX}
P(x)=|\gamma_{x,-1}|^{2}+|\gamma_{x,1}|^{2}.
\end{equation}

The one-dimensional walker begins at some specified input state $\ket{\Psi(0)}=\ket{x_{0},c_{0}}$ before it begins to evolve at time $t=0$, where $x_{0}$ and $c_{0}$ are the starting position and coin values respectively. The state then evolves for a finite number of time steps. The evolution is described by two operators: the coin $\hat{C}$ and step $\hat{S}$ operators, 
\begin{eqnarray} 
\hat{C} \ket{x, \pm 1}&=&(\ket{x,1}\pm \ket{x,-1})/\sqrt{2} \label{eq:coin} \\ \nonumber
\hat{S}\ket{x, c}&=&\ket{x+c,c} \label{eq:step}.
\end{eqnarray}
The coin operator takes a state and maps it to a superposition of new states using the Hadamard coin,
\begin{equation} \label{eq:H}
\hat{H}= \frac{1}{\sqrt{2}} \begin{pmatrix}
1 & 1 \\
1 & -1\\
\end{pmatrix},\\
\end{equation}
exploiting both possible degrees of freedom in the coin while maintaining the same position. Next, the step operator $\hat{S}$ moves the walker to an adjacent position according to the value of $c$. $\hat{C}$ and $\hat{S}$ act on the state at every time step and thus the evolution of the system after $t$ steps is given by,
\begin{equation}
\ket{\Psi(t)} = (\hat{S}\cdot \hat{C})^{t}\ket{\Psi(0)}.
\end{equation}
If the walker begins at the origin or on an even lattice position then, as the walker evolves, it lies on odd positions for odd time steps and on even positions for even time steps. Thus, as the walker evolves, the allowed locations for the walker oscillate between even and odd sites.

It is straightforward to generalise Eq. (\ref{eq:State}) to multiple dimensions by expanding the Hilbert space. For example, a two-dimensional walk would have the form,
\begin{equation}
\ket{\Psi^{(2)}(t)}=\sum_{x,y,c_{x},c_{y}} \gamma_{x,y,c_{x},c_{y}} \ket{x,y, c_{x},c_{y}},
\end{equation}
where $x\in [-x_\mathrm{max},x_\mathrm{max}]$ and $y\in [-y_\mathrm{max},y_\mathrm{max}]$ denote the two spatial dimensions, $c_{x} \in \{-1,1\}$ indicates for the walker to move left or right, $c_{y} \in \{-1,1\}$ indicates for the walker to move down or up, and the superscript represents the dimension. The dimension of the two-dimensional system is $(2 t_{\mathrm{max}}+1)^2$, when the lattice is both dimensions consist of the same number of sites in each direction from the origin as total time steps $t_{\mathrm{max}}$, which is the case for this work. The coin and step operator can be generalised by taking a tensor product for each respective dimension, or alternately a coin could be employed which entangles the two dimensions. In the case of a spatially separable two-dimensional coin one obtains $\hat{C}^{(2)}=\hat{C}_{x}\otimes \hat{C}_{y}$ and $\hat{S}^{(2)}=\hat{S}_{x}\otimes \hat{S}_{y}$. Likewise, the Hadamard coin for two dimensions becomes $H \otimes H$.

After the system evolves, a measurement is made on either the position or the coin degree of freedom yielding the  output probability distribution. With this probability distribution various metrics can be defined to characterise the evolution of the system, which we define next.

\subsection{Random Walk Metrics}

The two common metrics that we use to quantify a QRW are the variance $\sigma^{2}$ and the escape probability $\mathcal{P}_\mathrm{esc}$. All simulations done in this work have the initial condition that the walker begins at the origin $\ket{\Psi(0)}=\ket{0,0,1,1}$. Also, all statistics are averaged over one hundred simulations unless the walk was deterministic (i.e. there was no congestion or dephasing introduced) in which case only one simulation was needed. The sample space is exponential in size, and so averaging over an exponential number of simulations is not tractable; however, one hundred simulations for the size of systems we consider is sufficient for our work because it produces stable statistics that converge to fixed values and it smooths out the oscillations between data points. 

\subsubsection{Variance}

The variance $\sigma^{2}$ is a measure of how much the walker has spread out during its evolution. It is defined as,
\begin{equation}
\sigma^2=\sum_{i=1}^{n}{p_{i}(i-\mu)^{2}},
\end{equation}
where $p_i$ is the position probability distribution of the walker in the spatial degree of freedom, $n=2\,t_\mathrm{max}+1$ is the number of lattice sites, and $\mu=\sum_{i=1}^{n}{p_{i} i}$ is the mean of the distribution. For calculating the variance in two-dimensions we take the variance of the marginal probability distribution where the probability distribution becomes $p_i=\sum_{j=1}^{}p_{i,j}$ and $p_{i,j}$ is the two-dimensional probability distribution. Fig.~\ref{fig:PescVsTime}a illustrates the variance versus time for both a QRW and a CRW on a two-dimensional square lattice of size $t_\mathrm{max}=20$. The QRW demonstrates a quadratic rate of spreading across the lattice while the CRW demonstrates a linear rate of spreading. This quadratic spreading is one of the distinguishing features of a QRW compared to the CRW. It forms the basis of some QRW algorithms such as the QRW search algorithm, which is quadratically faster than the best corresponding classical algorithm. For simulations of the variance we do not impose boundary conditions because the walker never reaches the boundary. 

\subsubsection{Escape Probability}

The escape probability $\mathcal{P}_\mathrm{esc}$ is a measure of how much of the walker's amplitude leaks outside of a certain region in the lattice. To calculate $\mathcal{P}_\mathrm{esc}$ a boundary must first be defined which depends on the size of the lattice. For the square two-dimensional lattice we let the walker begin in the state $\ket{\Psi(0)}=\ket{0,0,1,1}$ and let the escape boundary be two vertical lines at $x=\pm x_{\mathrm{b}}$, where $t_\mathrm{b}$ is the distance the escape boundary is from the origin ($x=y=0$). To calculate the escape probability on this square lattice we use,
\begin{equation} \label{eq:Pesc}
\mathcal{P}_{\mathrm{esc}}=\sum_{|x|>x_{\mathrm{b}}} \sum_{y}P^{(2)}(x,y),
\end{equation}
where $P^{(2)}(x,y)=|\gamma_{x,y,1,1}|^{2}+|\gamma_{x,y,1,-1}|^{2}+|\gamma_{x,y,-1,1}|^{2}+|\gamma_{x,y,-1,-1}|^{2}$ is the two-dimensional version of Eq. (\ref{eq:probX}).
\begin{figure}[h] 
\centering
\includegraphics[width=0.45\columnwidth]{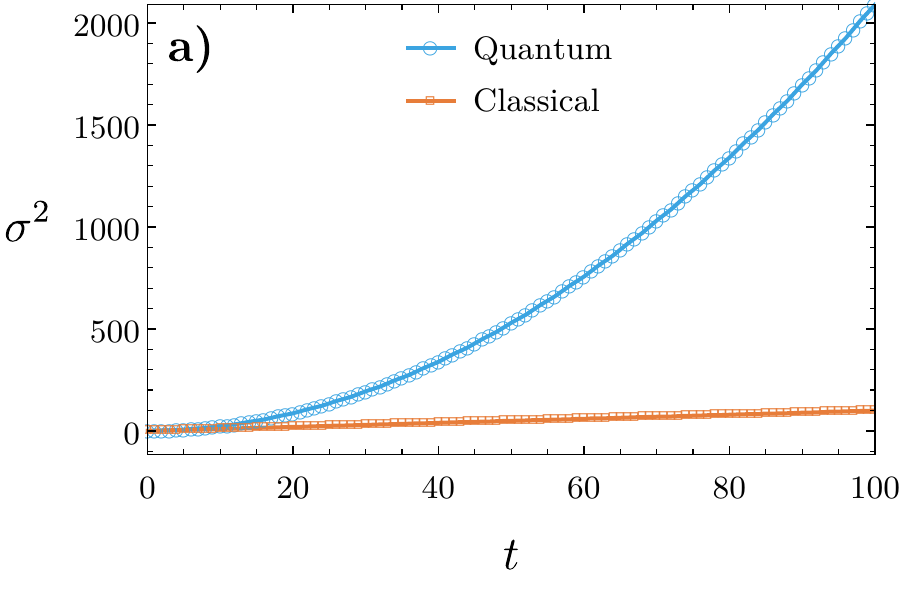}
\includegraphics[width=0.45\columnwidth]{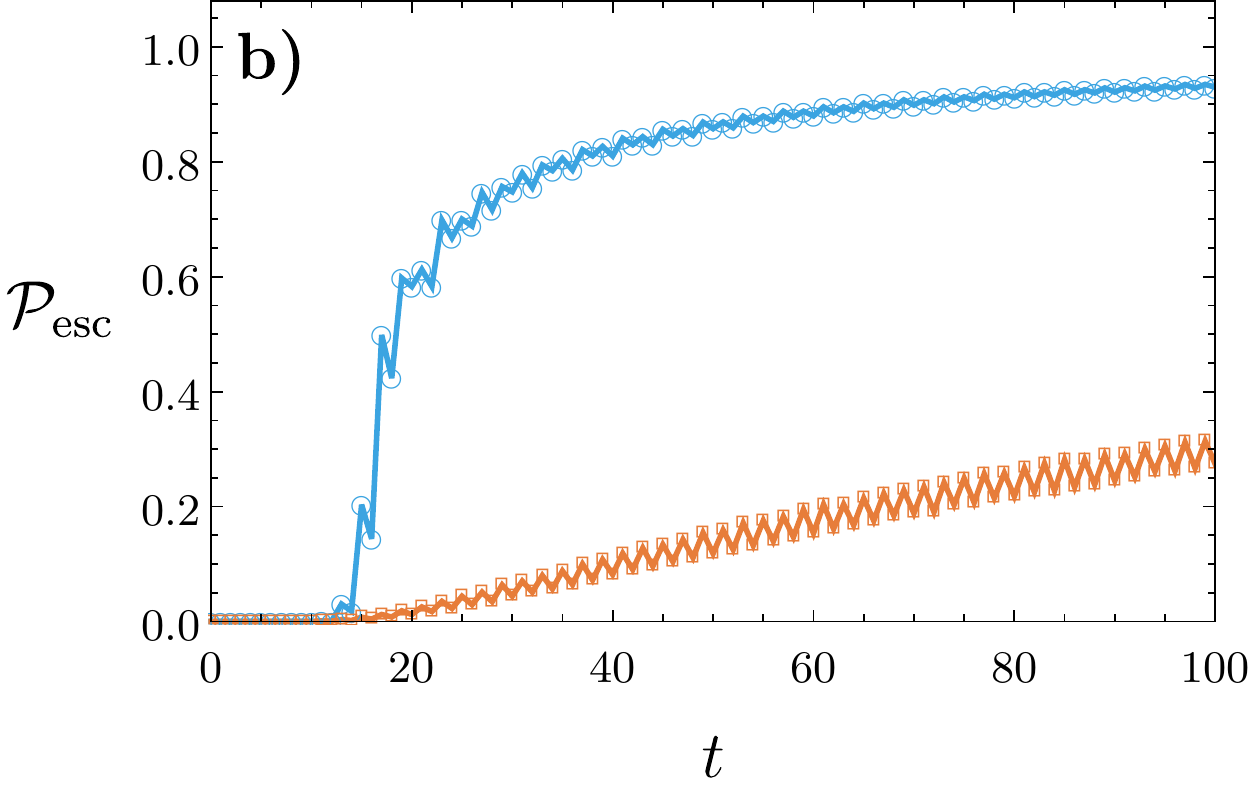}
\caption{\label{fig:PescVsTime} (a) The variance $\sigma^{2}$ versus time $t$ for the CRW and QRW on a two-dimensional square lattice defined by $t_\mathrm{max}=100$. The rate of spreading is quadratic for the QRW and linear for the CRW. (b) The escape probability $P_\mathrm{esc}$ against time for the CRW and QRW on a two-dimensional square lattice defined by $t_\mathrm{max}=100$ with a boundary defined by $t_\mathrm{b}=10$. In the quantum case, the probability of escape is significantly larger for any given time after escaping than in the CRW.}
\end{figure}

Fig.~\ref{fig:PescVsTime}b illustrates $\mathcal{P}_\mathrm{esc}$ versus $t$ for both a QRW and a CRW on a square lattice of size $t_\mathrm{max}=100$ with a boundary given by $t_\mathrm{b}=10$. Here the QRW exhibits a dramatic jump in escape probability compared to the CRW. This is due to both the faster rate of spreading of the QRW, and to the QRW having larger amplitudes at the tails of its distribution. This dramatic jump is a key feature pointed out in this work that demonstrates an advantage that QRWs have over CRWs.  

For all escape probability simulations the walker is allowed to walk back into the unescaped region which subtracts from the probability that the walker has escaped. This, in conjunction with the fact that the walker occupies alternating even and odd positions as the walker evolves, explains the oscillatory nature of the escape probability.

The two metrics, $\sigma^{2}$ and $\mathcal{P}_{\mathrm{esc}}$, are closely related. If the walker has a large spread in its distribution then the walker also has a better chance to fall outside of the escape boundary, since it is centered around the origin. At any given time step $t$ during the evolution we can determine the probability distribution over the lattice with Eq. (\ref{eq:probX}) and then calculate these various metrics to be used for quantifying a random walk. 

Any non-deterministic distribution obtained in this work was obtained using a Monte-Carlo averaging technique. Since the sample space we are averaging over grows quadratically we are limited to about $t_{\mathrm{max}}\leq 100$ time steps. Next, we demonstrate how to add spatial defects, which cause congestion, into the walkers' lattice and explore how the variance and escape probability are affected by this lattice congestion. 

\section{Lattice Congestion} \label{ch:QRWcongestion}

Lattice congestion is a model of defects in a medium. For the QRW and CRW the medium is the walkers' lattice and the defects are modelled as blocked pathways where the walker has to enter the pathway to realise it is blocked and then reverse out on the next step. This model is closely related to percolation theory \cite{bib:broadbent57} which models defects as missing lattice nodes. For a detailed introduction on percolation theory see \cite{bib:Shante71, bib:blanc86}. Percolation is generally modeled on a $d$ dimensional lattice with a given geometry such as a square, triangle or honeycomb. Regardless of geometry, the lattice consists of two components: $\textit{sites}$ and $\textit{bonds}$. A site is a point on the lattice and a bond is the connection between the sites. These components give two strategies for introducing the random fluctuations that define percolation theory: \emph{site percolations} and \emph{bond percolations}. In site percolation the lattice sites exist with probability $p\in [0,1]$ and when a site does not exist it is a defect in the lattice. In bond percolation the positions in a lattice are fixed while the bonds between the positions exist with probability $p$. The model in this work is a variant of site percolation whereby the walker can occupy any site, but with probability $1-p$ will find an obstruction and reverse direction upon hitting the respective site. 

Percolation theory has an associated scaling hypothesis that predicts critical values, such as percolation thresholds \cite{bib:grimmett99}, which we do not reproduce in this work due to our small lattice sizes. Instead we observe the behavior of QRWs on congested lattices and compare them to CRWs. However, we expect the same percolation characteristics such as percolation thresholds to exist in the underlying lattice that the walkers are exploring. For a two-dimensional square lattice with site percolations that most closely resemble the lattice used in this work, the percolation threshold is $p_c\approx 0.6$ \cite{bib:yonezawa89}. Values of $p$ higher than this threshold produce long-range connectedness in the lattice. We make the comparison to percolation in this work because our spatial defects are equivalent to the defects in percolation theory; however, we do not observe the critical values that percolation theory predicts so we call it congestion to avoid confusion.

To generate a lattice with spatial defects a matrix of coin operators is constructed. The matrix is the same size as the lattice and each position in the matrix corresponds to a spatial position on the lattice. The coin operator corresponding to a given position then determines the behaviour of the walker. The coin operators are defined as either a Hadamard coin, Eq. (\ref{eq:H}), if the site is present, or a bit-flip coin,
\begin{equation} \label{eq:bitflip}
X= \begin{pmatrix}
0 & 1\\
1 & 0\\
\end{pmatrix},
\end{equation}
if the site contains a defect. For the two-dimensional case the bit-flip coin becomes $X\otimes X$.

As the quantum or classical walker evolves it will walk into these defects that signify congested points on the lattice. Upon reaching a defect the walker reverses direction, thus slowing the walker's rate of spread. In this work we define $p$ as the probability that the site is not a defect; therefore, the probability that a site is a defect is $1-p$.  

\subsection{Classical Random Walk on a Congested Lattice}

The lattices we are considering contain randomly distributed defects, or points of congestion that impede the walker's progress. Questions such as what is the probability that there is an open path from one side of the lattice to the other, are answered by \emph{percolation theory}. There are many known applications for percolation theory \cite{bib:sahimi94}. A common example is asking whether a liquid can flow through a porous material. If enough pores (or sites) exist then the liquid can make it through. Another example is whether or not an electric current can flow through some medium where conductive sites are spread throughout some insulator. If enough conductive sites are present then a path will exist through the medium in the asymptotic limit.

\begin{figure}[h] 
\centering
\includegraphics[width=0.45\columnwidth]{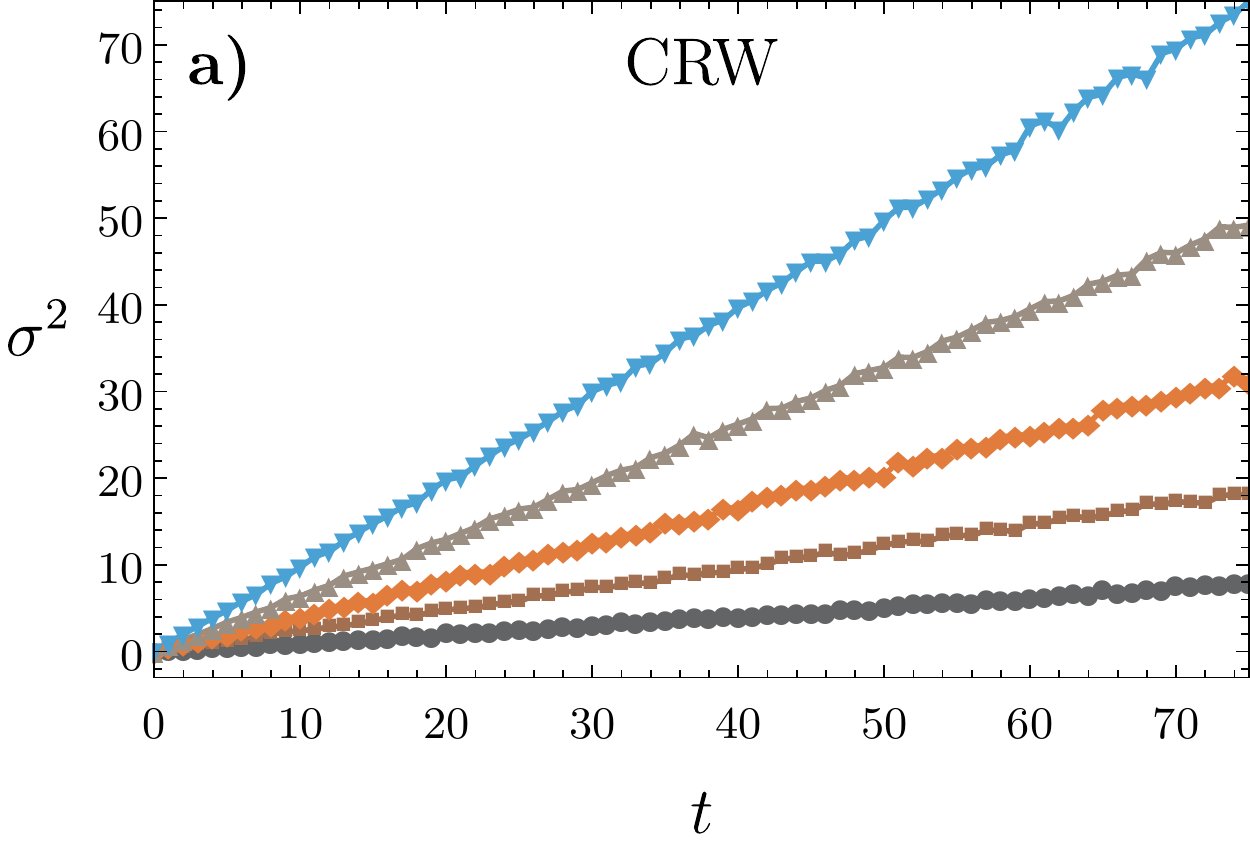}
\includegraphics[width=0.45\columnwidth]{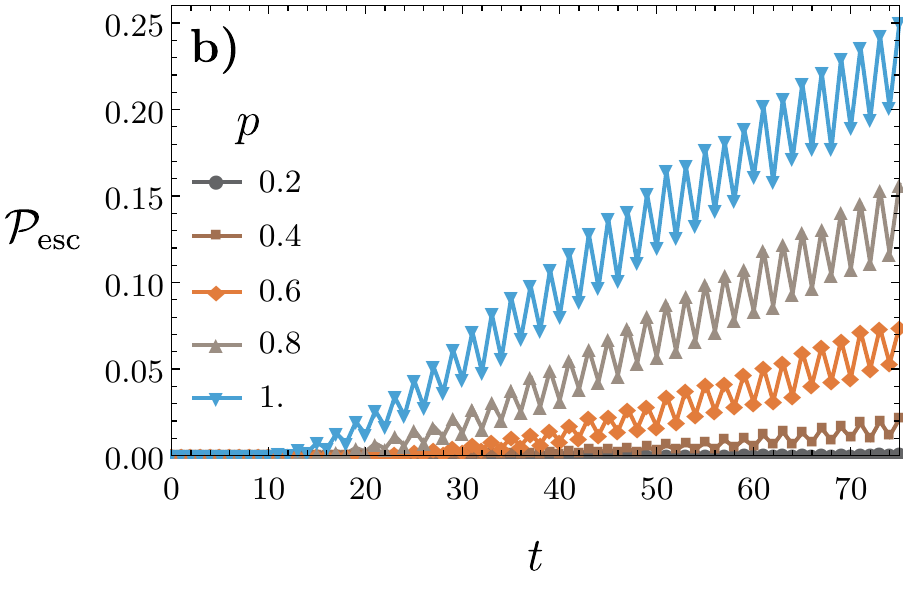}
\caption{The variance $\sigma^2$ and the escape probability $\mathcal{P}_{\mathrm{esc}}$ for a CRW plotted as a function of time $t$ for varying congestion probabilities $1-p$ on a two-dimensional square lattice of size $t_\mathrm{max}=75$. (a) Reduced spreading is observed as congestion increases but the linear dependence remains. (b) The escape probability $\mathcal{P}_{\mathrm{esc}}$ decreases as shown with an escape boundary of $t_\mathrm{b}=10$. The walker tends to escape linearly with time.} \label{fig:ClassVar}
\end{figure}

Within the congested lattice we examine the spread of walkers. Defects have the effect of reducing the rate of spread of the walker, or stopping it entirely if the lattice is so congested that there is no escape possible from the region the walker finds itself in. Fig. \ref{fig:ClassVar}a shows the variance $\sigma^2$ of a CRW versus time $t$ in the presence of varying values of congestion $1-p$ on a lattice of size $t_{\mathrm{max}}=75$. As the congestion increases the classical walker becomes trapped. In each case the variance preserves the linear dependence as is expected in a CRW. 

In Fig. \ref{fig:ClassVar}b the escape probability $\mathcal{P}_{\mathrm{esc}}$ of a CRW is shown versus time $t$ in the presence of varying values of congestion $1-p$ on a lattice of size $t_{\mathrm{max}}=75$ and escape boundary $t_b=10$. $\mathcal{P}_{\mathrm{esc}}$ decreases as congestion is increased but remains linear modulo the oscillations being averaged out. Again there is a threshold where in terms of $\mathcal{P}_{\mathrm{esc}}$ the walker stops escaping the boundary and the lattice becomes insulating.

\subsection{Quantum Random Walk on a Congested Lattice}

Classically, the state can only move in one direction at a time while quantum mechanically the state spreads in a superposition of every direction simultaneously. As with a classical walker, the quantum walker escapes the bounded region more often if there are less defects. The significance of the quantum walker is both the quadratic spreading behaviour and the resulting probability distribution having more weight in the tails. For a review of work done on QRWs with percolation see \cite{bib:kollar12} for asymptotic results and analytic solutions. See \cite{bib:kendon10} for quantum tunneling effects on a one-dimensional QRW and, for a two-dimensional lattice, average distance measures and the order of quadratic scaling. This work is unique from these two for several reasons. First, properties of a QRW on congested lattices with the $\sigma^2$ and $\mathcal{P}_{\mathrm{esc}}$ metrics were not studied previously. Second, we compare QRWs to CRWs and observe whether QRWs maintain their advantages over CRWs on congested lattices. Third, we tune the random walks on congested lattices between being fully quantum and fully classical using a dephasing process, described later in this chapter, which acts as an error model that describes coupling of the walker to the outside environment. 

\begin{figure}[h]
\centering
\includegraphics[width=0.45\columnwidth]{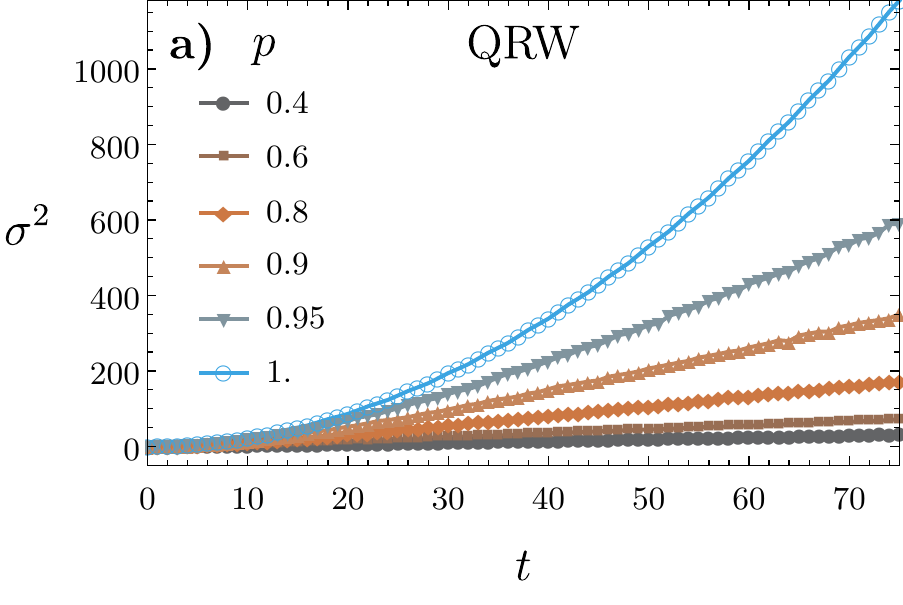}
\includegraphics[width=0.45\columnwidth]{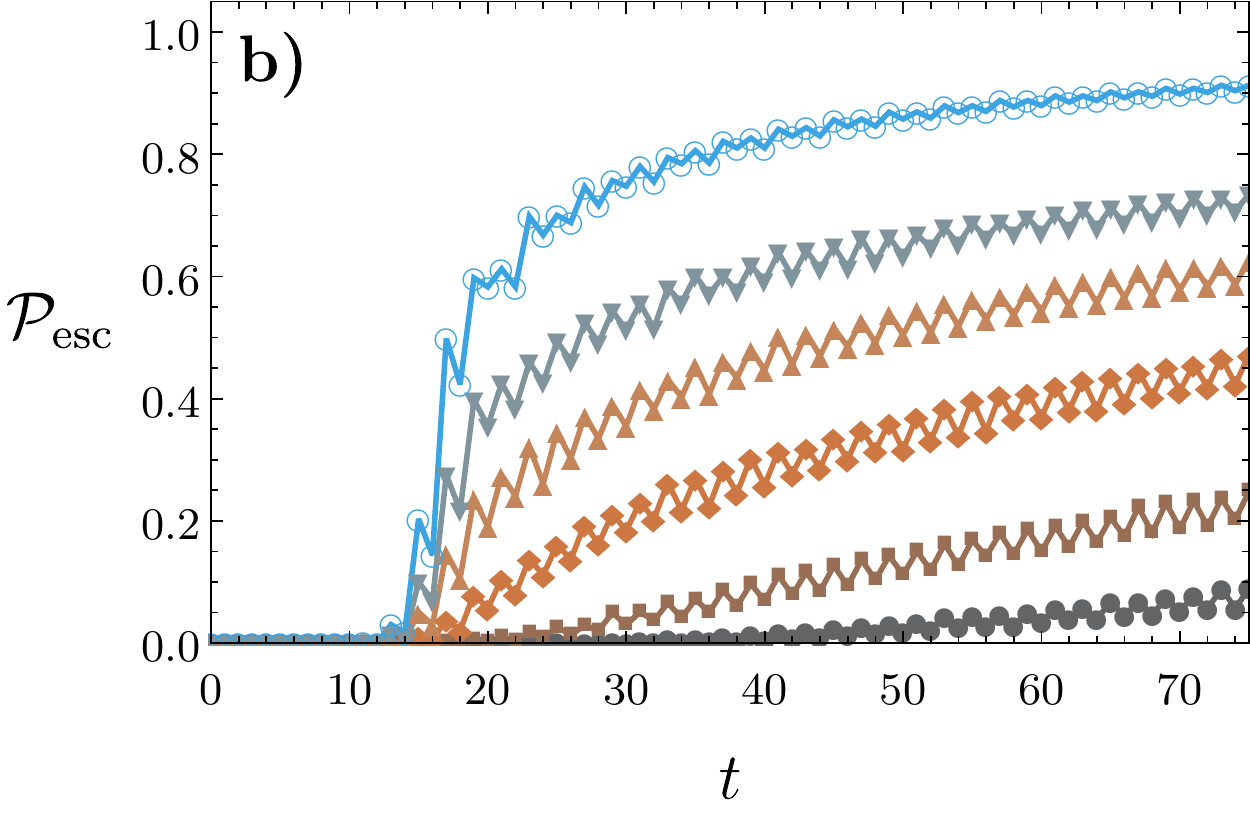}
\caption{The variance $\sigma^2$ and the escape probability $\mathcal{P}_{\mathrm{esc}}$ for a QRW plotted as a function of time $t$ for varying congestion probabilities $1-p$ on a two-dimensional square lattice of size $t_\mathrm{max}=75$. (a) Reduced spreading is observed as congestion increases but the QRW maintains its advantages over CRWs with congestion. (b) The walker quickly escapes the boundary as compared to the classical walker. As $p$ decreases the jump in $\mathcal{P}_{\mathrm{esc}}$ becomes less prominent as shown for an escape boundary of $t_\mathrm{b}=10$.} \label{fig:PescVsTVaryingP}
\end{figure}

Fig. \ref{fig:PescVsTVaryingP}a shows the variance $\sigma^2$ versus time $t$ for a QRW with varying values of congestion $1-p$ for $t_{\mathrm{max}}=75$. As congestion increases the variance of the walker decreases; however, it retains its quadratic (i.e. ballistic) spreading albeit with a different quadratic coefficient. This property shows that QRWs remain advantageous over CRWs, since the quantum walker spreads faster, even in the presence of lattice defects.

Fig.~\ref{fig:PescVsTVaryingP}b shows the escape probability $\mathcal{P}_{\mathrm{esc}}$ versus time $t$ for varying values of congestion probability $1-p$ on a lattice of size $t_\mathrm{max}=75$ and boundary $t_b=10$. For $p=1$ there is no congestion present and the $\mathcal{P}_{\mathrm{esc}}$ metric experiences a sudden jump from $t=10$ to $t=11$. This is because the QRW has most of its amplitude in its tails as it evolves. When $p$ decreases and the lattice becomes more and more congested the sudden jump is still present at the same value of $t$ but with a much smaller amplitude. This shows that QRWs retain their advantage over CRWs in the presence of heavy congestion. Note that the percolation threshold is around $p\approx 0.6$, below which we expect that on average there is no clear route across the graph. 

\subsection{Varying Escape Boundary}

\begin{figure}[h] 
\centering
\includegraphics[width=0.45\columnwidth]{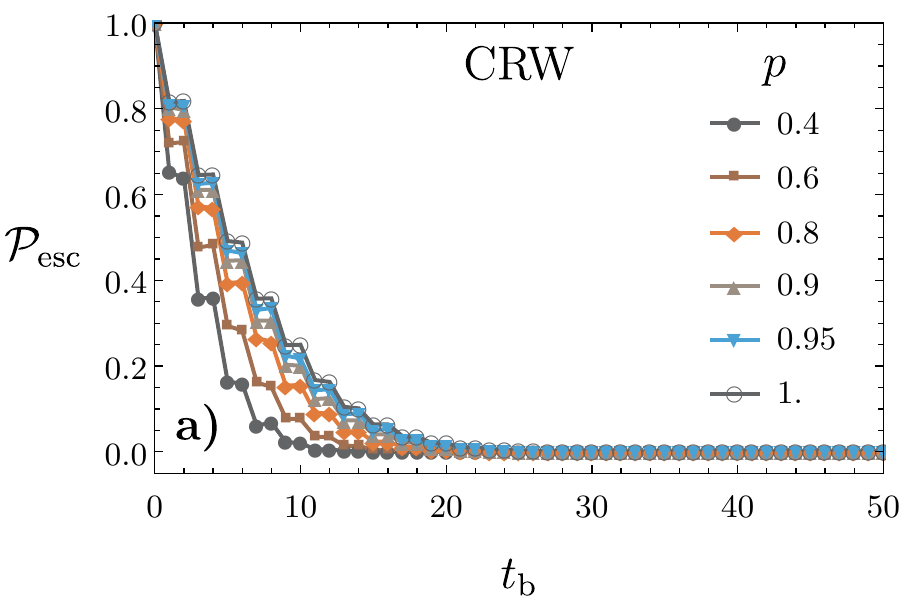}
\includegraphics[width=0.45\columnwidth]{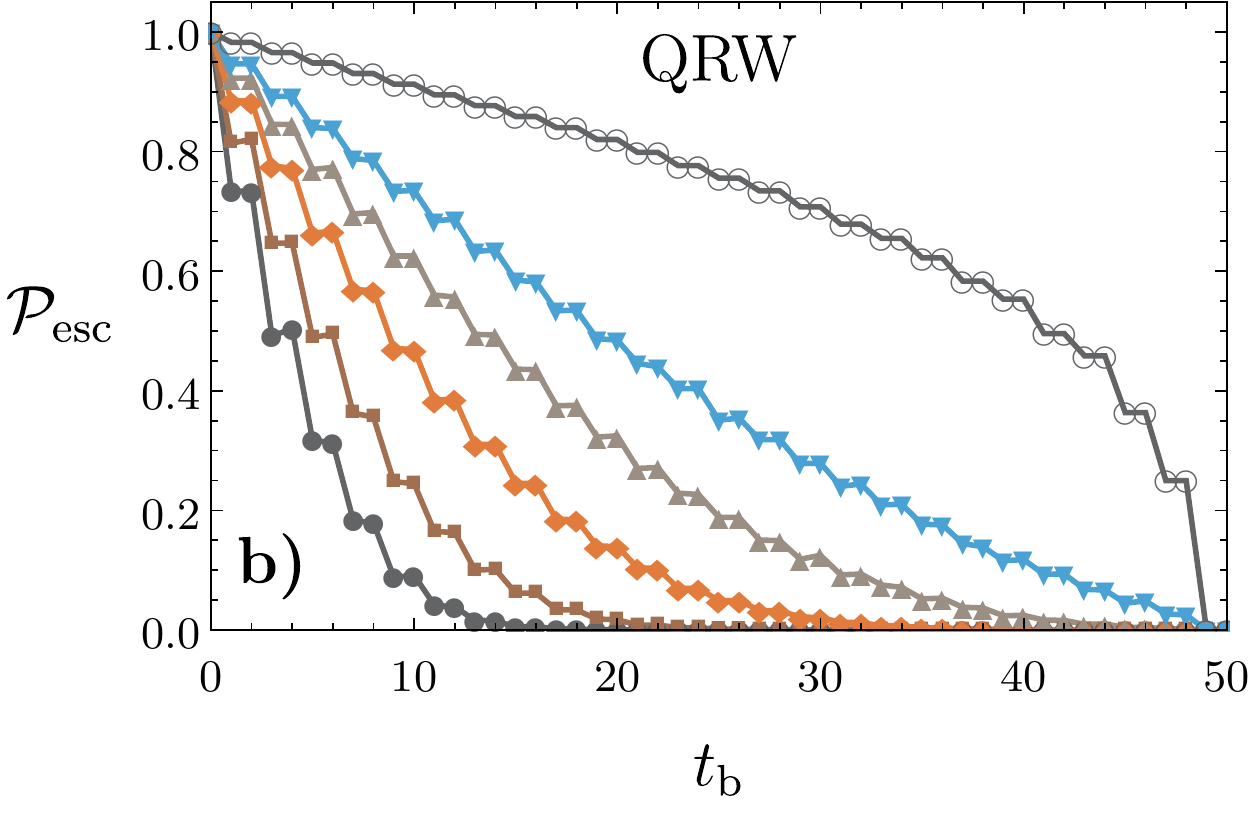}
\caption{The escape probability $\mathcal{P}_{\mathrm{esc}}$ versus a varying escape boundary $t_\mathrm{b}$ for several values of congestion $1-p$ for the CRW (a) and the QRW (b). The walker evolves for $t_{\mathrm{max}}=50$ time steps. The QRW maintains a significantly larger $\mathcal{P}_{\mathrm{esc}}$ as the boundary moves away from the initial starting position than the CRW. In both cases $\mathcal{P}_{\mathrm{esc}}$ goes to zero as the boundary approaches the end of the lattice.} \label{fig:VarTb}
\end{figure}

In the previous simulations involving escape probability the escape boundary was set to be near the initialised position of the walker. The next topic we consider on a congested lattice is how the escape probability on a congested lattice changes as $t_\mathrm{b}$ varies. Consider Fig. \ref{fig:VarTb} which shows $\mathcal{P}_{\mathrm{esc}}$ as a function of $t_\mathrm{b}$ with varying values of congestion $p$ for the CRW (a) and the QRW (b). Both walkers evolve for $t_{\mathrm{max}}=50$ steps and $\mathcal{P}_{\mathrm{esc}}$ is calculated at $t=t_{\mathrm{max}}$. In both the CRW and QRW $\mathcal{P}_{\mathrm{esc}}$ reduces with increased congestion and when $t_\mathrm{b}$ is farther from the walker's initial position. What is interesting is that the QRW maintains a significantly larger $\mathcal{P}_{\mathrm{esc}}$ than the CRW as the escape boundary moves away. 

\section{Dephasing} \label{ch:QRWdephasing}

Next, we consider what happens to a QRW subject to dephasing. Dephasing represents decoherence caused by unwanted interaction with the environment which can be related to measurement errors caused by thermal fluctuations, white noise, photons interfering with the quantum walker, etc. To explore this we first introduce a model of dephasing and characterise it with our two metrics: variance and escape probability. 

Consider a QRW where after each step, each basis state has probability $p_{\mathrm{d}}$ of acquiring a $\pi$ phase flip. We can model this process as choosing to apply one of a set $\{F_j\}$ of unitary matrices covering all the combinations of $\pm 1$ on the diagonal. If $F_j$ has $s$ -1's on the diagonal we choose it with probability $p_{\mathrm{d}}^s(1-p_{\mathrm{d}})^{m-s}$. 

The probability of a particular sequence of $F_j$'s will be the product of the probabilities of the $F_j$ appearing since they are independently chosen at each step. If $\rho_\mathrm{seq}$ is the final pure density matrix appearing with probability $p_\mathrm{seq}$, then in general the final state of the system is described by,
\begin{equation}
    \rho = \sum_\mathrm{seq} p_\mathrm{seq} \rho_\mathrm{seq}.
\end{equation}
For any POVM element $E$ we have,
\begin{equation}
    \sum_\mathrm{seq} p_\mathrm{seq} \mathrm{Tr}\{E \rho_\mathrm{seq}\} = \mathrm{Tr}\{E \rho\}.
\end{equation}

We algorithmically implement dephasing by randomly flipping the signs of individual basis states in the walker's state with probability $p_{\mathrm{d}}$, and average measurement results over a large number of independent trials. This in effect samples from the  distribution represented by $\rho$ and is automatically weighted by the probability of a given sequence. 

That this whole process represents dephasing is not immediately obvious. To see it, we first rewrite $\rho$ as the vector $\ket{\rho}$ using the \emph{vec} operation which simply stacks its columns on top of each other \cite{bib:magnus1995matrix}. Using the identity $\ket{ABC}=C^T\!\!\otimes\!A\,\ket{B}$ for any three square matrices $A$, $B$, and $C$; then grouping the terms that turn up, we can write,
\begin{equation} \label{eq:vecrho}
    \ket{\rho} = \ldots \left(\sum_k p_k D_k U\right)\left(\sum_j p_j D_j U\right)\left(\sum_i p_i D_i U\right)\ket{\rho_0},
\end{equation}
where $D_j = F_j^*\!\otimes\!F_j=F_j^{\otimes 2}$, $U$ represents the step and coin operations, and $\ket{\rho_0}$ is the vectorised initial density matrix. This shows that after each step we apply the process described by the dynamical matrix,
\begin{equation}
    D = \sum_j p_j F_j^{\otimes 2}.
\end{equation}

The matrices $F_j$ are diagonal so we write the diagonal as a vector denoted by $\ket{f}_j$, so that the diagonal of $F_j^{\otimes 2}$ is $\ket{f}_j\ket{f}_j$. Since $\ket{f}_j$ has only real entries we can rearrange it into the matrix $\ket{f}_j\bra{f}$. We can do a similar arrangement with $D$ so that,
\begin{equation}
    \ket{d}\!\bra{d} =  \sum_j p_j \ket{f}_j\bra{f}.
\end{equation}
It is worthwhile pausing and noting what this matrix represents. From Eq. (\ref{eq:vecrho}) we can see that the diagonal of $D$ multiplies the elements of the vectorised $\ket{\rho}$. Hence when we arrange the values into a matrix, the entries of $\ket{d}\!\bra{d}$ multiply the corresponding entries in $\rho$.

The first thing to note is that this matrix is symmetric. We will denote the entries of $\ket{f}_j$ by $f_k$ and drop the reference $j$ for clarity. The diagonals of $\ket{f}_j\bra{f}$ are of the form $f_k^2=1$ and since $\sum_j p_j = 1$ the diagonal of $\ket{d}\bra{d}$ is unity and the process does not change the amplitudes of the states. The off-diagonals are of the form $f_rf_s$ where $r\ne s$ and their sum over $j$ has the value,
\begin{equation}
    (1-p_d)^2+p_d^2-2(1-p_d)p_d = (1-2p_d)^2.
\end{equation}
The terms on the left are the probabilities that both $f_r$ and $f_s$ are positive, both negative, or one of each respectively.
Each of these terms is multiplied by the binomial sum of the probabilities of all the combinations of $\pm 1$ on all the other elements of $\ket{f}_j$ and not $r$ or $s$, which evaluates to 1. Note that this result holds for any dimension. In summary, the map that is performed by $D$ multiplies every off-diagonal element of $\rho$ by $(1-2p_{\mathrm{d}})^2$. This is a dephasing map.

If $p_{\mathrm{d}}=0$ none of the signs are flipped, and if $p_{\mathrm{d}}=1$ all of the signs are flipped. Since the QRW is invariant under a global phase flip, these two extremes reproduce an ideal QRW. When $0<p_{\mathrm{d}} <1$ dephasing is introduced into the system. A value of $p_{\mathrm{d}}=1/2$ corresponds to complete dephasing which causes the walker to behave classically. The classical results in this work were produced by using our QRW code with a value of $p_{\mathrm{d}}=1/2$. This was checked with purely classical simulations to verify that we are indeed obtaining a CRW. 

If we imagine an weak measurement of the QRW at every step where it is projectively measured with probability $p_{\mathrm{m}}$ or otherwise left alone, this map would describe dephasing by a dynamical matrix which multiplies all the off diagonal elements of $\rho$ by $1-p_{\mathrm{m}}$. So our dephasing process is equivalent to a measurement performed with a probability $p_{\mathrm{m}}=4(1-p_{\mathrm{d}})p_{\mathrm{d}}$.

\begin{figure}[h]
\begin{center}
\includegraphics[width=0.85\columnwidth]{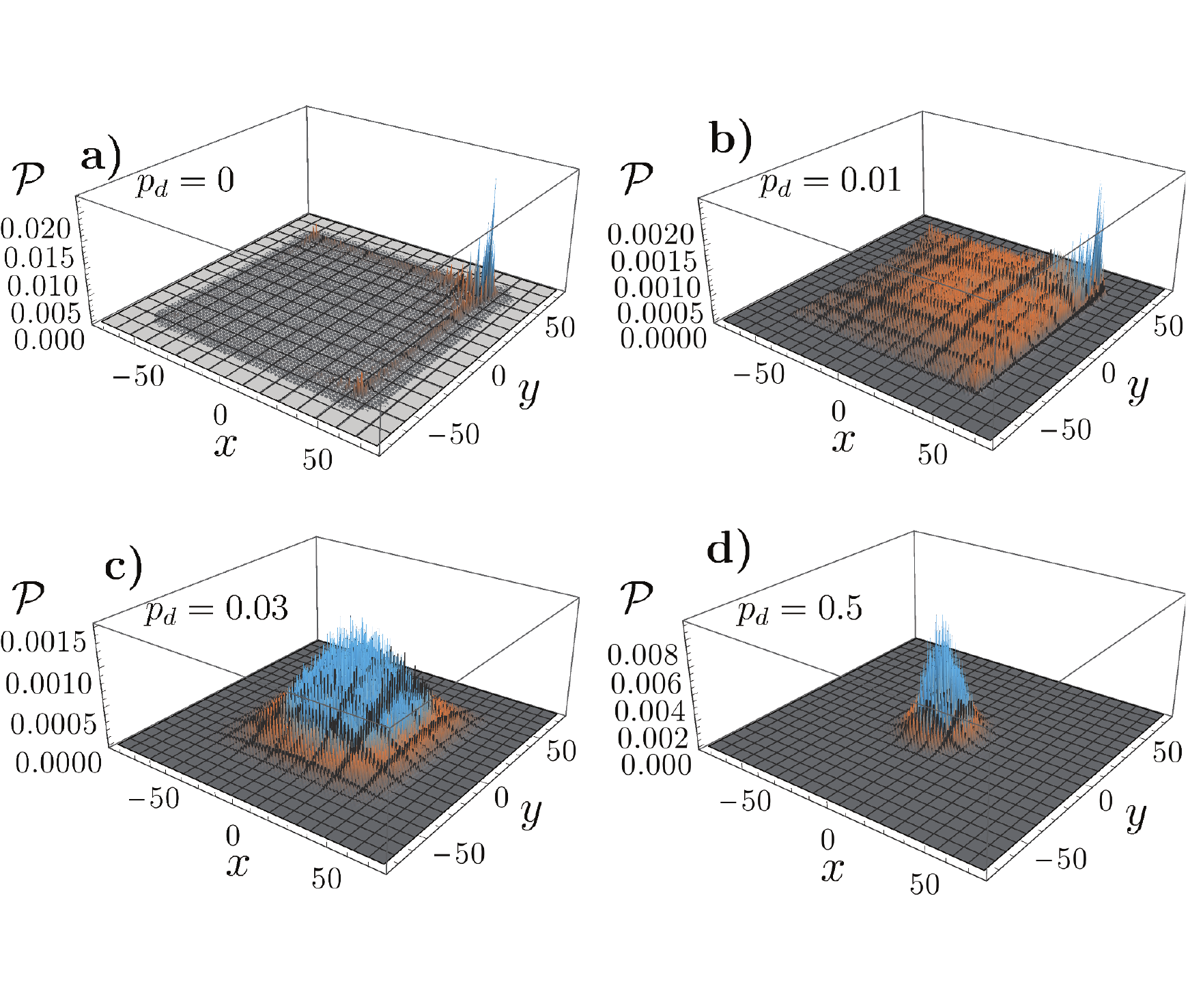}
\end{center}
\caption{\label{fig:AndersonNoDephasing} The QRW probability distribution shown at the final time step over a two-dimensional square lattice defined by $t_\mathrm{max}=75$ with no defects present. (a) The QRW with no dephasing $p_{\mathrm{d}}=0$ always yields a deterministic probability distribution with ballistic spreading. (b) The same QRW but with a dephasing probability of $p_{\mathrm{d}}=0.01$. It has a similar probability distribution but begins approaching classical statistics. (c) The same QRW again but with a dephasing probability of $p_{\mathrm{d}}=0.03$. Here the probability distribution becomes centred around the origin and begins to qualitatively look much like the statistics of a CRW. (d) The same QRW again but with a dephasing probability of $p_{\mathrm{d}}=0.5$. This is complete dephasing and the walk become identical to a CRW.}
\end{figure}
In this work, dephasing is a method for introducing quantum decoherence to the QRW. To illustrate the effect of dephasing in our model we plot the probability distribution at the final time $t_\mathrm{max}=75$ of various random walks in Fig.~\ref{fig:AndersonNoDephasing}. In Fig.~\ref{fig:AndersonNoDephasing}a the walk has no dephasing $p_{\mathrm{d}}=0$ and is thus completely deterministic. We see that this probability distribution has one main peak near the positive $x$ and positive $y$ direction, which is in the initialised direction of the coins, and is at the edge of the lattice. This strong peak is the result of constructive interference with walkers moving in this direction and destructive interference with walkers moving in other directions \cite{bib:kempe2003quantum}. This is in contrast to what occurs when dephasing is introduced. Fig.~\ref{fig:AndersonNoDephasing}b shows the same evolution again but with a dephasing probability of $p_{\mathrm{d}}=0.01$. With this value of dephasing the distribution retains most of its quantum behaviour. Fig.~\ref{fig:AndersonNoDephasing}c shows the same evolution again but with a dephasing probability of $p_{\mathrm{d}}=0.03$. With this value of dephasing the probability distribution loses much of its quantum behaviour and begins behaving like a CRW. Finally in Fig.~\ref{fig:AndersonNoDephasing}d we show the same evolution but with $p_{\mathrm{d}}=0.5$ and obtain the probability distribution of a CRW. 
 
We notice that with sufficiently strong dephasing the probability distribution becomes localized around the origin so that the QRW behaves like a CRW distribution. Note that the corresponding value of $p_\mathrm{d}$ that collapses the QRW to a CRW depends on $t_\mathrm{max}$. As $t_\mathrm{max}$ increases the underlying lattice has more sites where dephasing can occur and thus a smaller $p_\mathrm{d}$ will cause the corresponding collapse. By incrementing $p_\mathrm{d}$ we can smoothly interpolate between QRWs and CRWs, which is a key feature of this work.

\section{Congestion and Dephasing Combined} \label{ch:QRWcombined}

\begin{figure}[h]
\centering
\includegraphics[width=0.45\columnwidth]{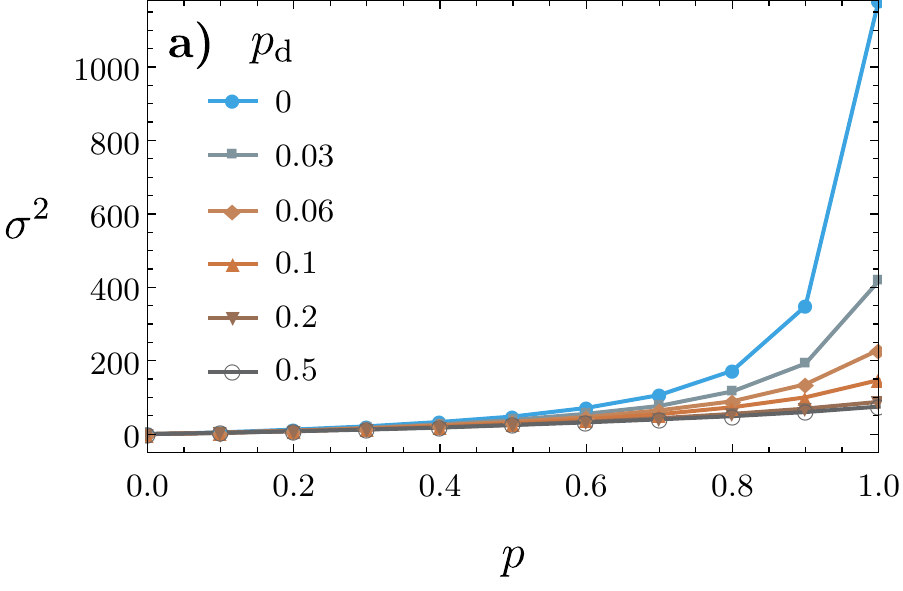}
\includegraphics[width=0.45\columnwidth]{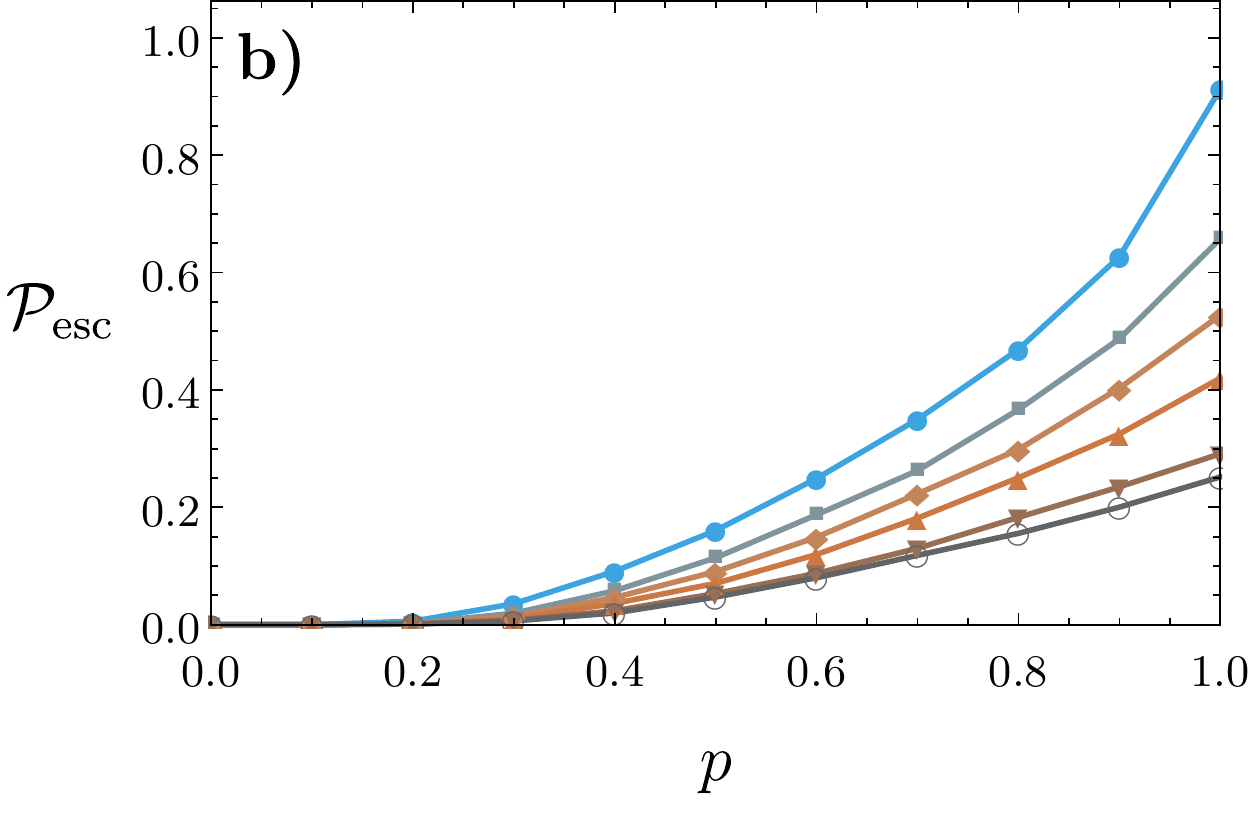}
\caption{\label{fig:DepPercVar} The variance $\sigma^2$ and escape probability $\mathcal{P}_{\mathrm{esc}}$ obtained at the final time step plotted against the congestion probability $1-p$ for varying values of the dephasing probability $p_{\mathrm{d}}$ on a square two-dimensional lattice of size given by $t_\mathrm{max}=75$. (a) The propagation of the walker decreases monotonically with the congestion rate for increasing values of dephasing $p_{\mathrm{d}}$. A quadratic behavior remains for small values of $p_{\mathrm{d}}$. (b) The escape boundary is at $t_\mathrm{b}=10$. With decreasing $p_{\mathrm{d}}$ the quantum walker has a larger chance to escape the boundary. As $p_{\mathrm{d}}$ increases the QRW enters the classical regime and quantum advantages are lost.}
\end{figure}

Next we combine congestion and dephasing and examine the joint effects. Fig.~\ref{fig:DepPercVar}a shows the variance obtained at the final time step of the QRW as a function of the congestion probability $1-p$ for varying values of the dephasing probability $p_{\mathrm{d}}$ on a two-dimensional square lattice of size given by $t_\mathrm{max}=75$. A monotonic decrease is observed in the variance for a given $p$ as $p_{\mathrm{d}}$ is increased and a quadratic rate of spreading is maintained for small values of $p_{\mathrm{d}}$.  Fig.~\ref{fig:DepPercVar}b shows $P_\mathrm{esc}$ with boundary $t_\mathrm{b}=10$ as a function of congestion probability $1-p$ for varying values of dephasing probabilities $p_{\mathrm{d}}$ on a two-dimensional square lattice defined by $t_\mathrm{max}=75$. When $p_{\mathrm{d}}=0$ the walk is fully quantum so more of the probability distribution escapes the boundary. When dephasing is increased process errors are introduced, reducing $\mathcal{P}_{\mathrm{esc}}$ for any given value of $p$. 

\section{Summary}

Quantum random walks are a promising route towards quantum information processing, exhibiting many unique features compared to the classical random walk as motivated in Sec. \ref{ch:QRWintro}. We review quantum random walks in Sec. \ref{ch:QRWQRW}. Then we introduced a model for adding congestion to the underlying lattice via the introduction of bit-flip coins in Sec. \ref{ch:QRWcongestion}. Congestions inhibits the spread of the classical and quantum walker, reducing the escape probability and variance metrics. We found that as a quantum random walk evolves it will suddenly and dramatically escape a finite boundary. It maintains this property even in the presence of congestion. 

We also introduce a dephasing error model in Sec. \ref{ch:QRWdephasing}. Dephasing errors are errors on the quantum walker caused by coupling to the environment as it evolves. In the limit of large dephasing the quantum random walk spatially localises and behaves like a classical random walk. The spread of the walker is sensitive to small amounts of dephasing in our dephasing model and becomes more sensitive as the size of the lattice increases. Dephasing also allows for a mapping between quantum and classical walks, via the coin operator, to allow for a direct comparison of the two. 

We studied the effects of spatial defects and dephasing together on the propagation of the walker in Sec. \ref{ch:QRWcombined} and found a monotonic decrease is observed in the variance and escape probability for a given congestion probability as the dephasing probability is increased. Our results indicate that a quantum walker on a lattice with defects and dephasing still exhibit a quadratic rate of spreading. Thus, as the quadratic spread of quantum walks is one of the key features that make them applicable to quantum information processing applications, such as the quantum search algorithm, we expect that quantum walkers on congested lattices may retain their advantage over CRWs despite the congestion.

In the coming chapters we will look at \BStwo, which can be regarded as a particular type of multi-walker quantum walk that simulates the output statistics of a linear optics network fed with multiple single-photon states.

\begin{savequote}[45mm]
If I have seen further it is by standing on the shoulders of giants. 
\qauthor{Sir Isaac Newton}
\end{savequote}

\chapter{An Introduction to \BS} \label{Ch:BSIntro}

\section{Synopsis}
\BS is a problem that studies the interference via linear optics of many bosonic particles. It can be thought of as a simplified model for quantum computing and it may hold the key to implementing the first ever post-classical quantum simulation. \BS is significantly more straightforward to implement than any universal quantum computer proposed so far. 

We begin this chapter in section \ref{sec:BSintroMotivation} by motivating \BS and discussing some of its history. We then, in section \ref{sec:BSintroFormalism}, summarize the \BS formalism giving a simplistic and detailed model, discuss what the permanent represents physically, and explain errors in \BStwo. In section \ref{sec:BSintroECTT} we discuss \BS and the implications it has on the extended Church-Turing thesis. In section \ref{sec:BSintroBuild} we discuss the feasibility of building a \BS device with a particular focus on photon sources, linear optical networks, and photodetection. 
In section \ref{sec:BSintroApplications} we go into some applications inspired by \BS one of which was our own work and, later, go into extensive detail in chapter \ref{Ch:MORDOR}.

\section{Motivation for \BS} \label{sec:BSintroMotivation}


Aaronson \& Arkhipov (AA) surprised the quantum optics community when they argued that a passive linear optical interferometer with Fock state inputs cannot likely be efficiently simulated by a classical computer \cite{bib:AaronsonArkhipov}. This has become known as the \BS problem. Since the first appearance of this work in 2010, research into \BS has exploded. There have been a number of experimental implementations that utilize three photons from spontaneous parametric down conversion (SPDC) sources \cite{bib:ralph2013quantum, bib:Broome2012, bib:Spring2, bib:anon, bib:Tillmann4, bib:crespi2013integrated, bib:he16} (although the validity of one of these experiments are under debate as not all three photons were heralded single photons \cite{bib:dowlingSchmampling}). There have also been many theoretical developments in \BS that consider the effects of loss, noise, decoherence, non-Fock inputs, scalability of SPDC sources, ion-trap implementations, and so forth \cite{bib:RohdeLowFid12, bib:RohdeRalphErrBS, bib:PhysRevA.88.044301, bib:motes13, bib:catSampling, bib:Kaushik14, bib:Pacs13, bib:Shc14a, bib:lou}. We will discuss and summarise some of these results in the sections and chapters below. 

Why is \BS attracting so much hype? What is it good for? \BS can be implemented on passive linear optical interferometers and thus it may be physically implemented with significantly reduced experimental overhead as compared to building a universal quantum computer. Yet, it still implements a computationally complex problem that no classical computer can efficiently simulate. It is the first interesting example of a realistically implementable post-classical computing problem although the potential of \BS is yet to be fully understood. 

To be more specific about what is computationally hard in \BS for a classical computer to simulate we must take a look at the output distribution. The output distribution has an exponentially large number of possible output configurations which are sampled using photon-number detectors. In addition to there being an exponentially large number of output configurations, the sample is computed by solving a matrix permanent, which has no known efficient algorithm unlike matrix determinants. In other words one cannot predict the outcome with a classical computer unless they were to wait an exponential amount of time or use an exponential amount of resources.

A logical question to ask is weather a passive linear optical interferometer can be used for anything interesting other than \BStwo? \BS itself has no known applications other than being able to efficiently sample the distribution of bosons that a classical computer cannot efficiently simulate. However, it was recently shown by MORDOR 
that passive linear optics, which captures the essence of \BStwo, can be applied to quantum metrology \cite{bib:MORDOR}. We go into this result in chapter \ref{Ch:MORDOR}. It was also shown almost simultaneously by Huh \emph{et al.} that a modification of \BS can be used for simulating molecular vibronic spectra \cite{bib:vibSpec}. 

\BS in this respect is similar to Feynman's work in the 1980s which had hypothesized that an ordinary quantum computer could be used to carry out certain physics simulations without the exponential overhead required on a classical computer. This hypothesis was not proven until much later in Lloyd's work in 1996 \cite{bib:feyn, bib:LloydSim}. The Deutsch-Jozsa algorithm of 1992 was the first exponential speedup advantage for a quantum computer but it solved a problem that had no practical applications \cite{bib:DeutschJozsa92}. \BS is similar to the Deutsch-Jozsa algorithm in that it solves no interesting problem but is a first example demonstration of passive linear optics doing something of interest from a computational complexity perspective. It shows that passive linear optical interferometers have some sort of hidden computational power. So the real question now is can \BS be shown to be good for more things besides the \BS problem and the two \BS inspired applications above? This potential is what has captured the imagination of many researchers in the field including my close collaborators and myself. 

Gard, \emph{et al.} independently reached the same conclusion as AA. They did this in the context of simulating multi-photon coincidence counts at the output of a linear optical implementation of a quantum random walk with multi-photon walkers \cite{bib:gard2013quantum}. In follow up work, Gard \emph{et al.} \cite{bib:gard2014inefficiency} argued using a physical instead of a computational complexity point of view that the difficulty to simulate such interferometers arose from two necessary requirements:  (1) The photons interact at the beamsplitters via a Hong-Ou-Mandel effect that leads to an exponentially large Hilbert space in the number-path degrees of freedom, which rules out a brute force simulation; and (2) That the simulation of the interferometer is linked to computing the permanent of a large matrix with complex entries, a problem known to be in the complexity class \textbf{\#P}\--hard. It is believed that this complexity class is intractable for classical computers as well as for universal quantum computers \cite{bib:Ryser63}. The first of these requirements is a necessary condition but is not sufficient to imply an intractable simulation. As a counterexample, the Gottesman-Knill theorem gives examples of classically simulatable quantum circuits where gates in the Clifford algebra generate exponentially large amounts of qubit entanglement. In some problems it is known that shortcuts through exponential Hilbert space exist. Intuitively, we expect that since \BS is tied to solving matrix permanents the computation is expected to not have a shortcut. In contrast, the equivalent sampling problem with fermions has an exponentially large Hilbert space but is known to be classically easy to simulate since the problem relates to matrix determinants rather than permanents, which are known to be in the complexity class \textbf{P} and classically simulatable \cite{bib:gard2014inefficiency} since a shortcut through Hilbert space exists using Gaussian elimination. 

An important point to note is that in almost all work on \BStwo, the interferometer is described as a passive \emph{linear} optical device with \emph{non-interacting} bosons. However, we know that the Hong-Ou-Mandel effect followed by a projective measurement imparts an effective nonlinearity and so there is effectively an interaction between the indistinguishable bosons at each beamsplitter \cite{bib:LOQC}. This exchange interaction between indistinguishable bosons arises simply from the multi-particle wavefunction needing to be properly symmetrized. This effect can give rise to quite noticeable effects. As an example, the bound state of the neutral hydrogen molecule, which is the most common molecule in our Universe, arises from a similar exchange interaction. It is therefore technically incorrect to describe these interferometers as linear devices with \emph{non-interacting} bosons. The exchange interaction is just as real as tagging on an additional term in a Hamiltonian. If one adds post-selection in \BStwo-like schemes an effective Kerr-like nonlinearity is imparted between the bosons \cite{bib:lapaire2003conditional}, but \BS itself remains linear as described by AA. 

True \BS certification may be done to distinguish it from uniform sampling or random-state sampling~\cite{bib:AA13response, bib:Spagnolo13, bib:Carolan13, bib:Molmer13, bib:QKUM14}. Various sources of error including mode-mismatch, spectra of the bosons, and spectral sensitivities of detectors of \BS have been studied \cite{bib:Shc14a, bib:Rohde15a, bib:RohdeLowFid12}. This has led to a theory of interference with partially indistinguishable particles, where any realistic imperfections in the source and detectors can be completely characterized \cite{bib:Shc14b, bib:Tich14}. Scalable implementations of \BS in optical systems, ion traps, and microwave cavities have been proposed \cite{bib:motes13, bib:Lund13, bib:motes2014scalable, bib:Shen14, bib:peropadre15}.

Another interesting observation is that there is evidence that \BS may not be efficiently distinguished from a classical device that efficiently produces samples from a suitable distribution. This interesting question leads to the possibility of a classical certification of \BS, which is still largely open \cite{bib:gogolin2013, bib:AaronsonArkhipov}. More specifically, \BS is not known to reside in the sampling-equivalent complexity class \textbf{NP}, i.e the class of problems that can be efficiently classically simulated. If that is indeed the case, this effectively rules out verification algorithms that distinguish \BS from any classical distribution with certainty.

In the next section we formally introduce \BStwo.

\section{The \BS Formalism} \label{sec:BSintroFormalism}

We begin by reviewing the \BS model using single photons. For an elementary introduction to \BStwo, see \cite{bib:introChapter}. Unlike universal LOQC, which requires active elements (specifically fast-feedforward), the \BS model is strictly passive, requiring only single-photon sources, passive linear optics (i.e beamsplitters and phase-shifters), and photodetection. No quantum memory or feedforward is required.

In section \ref{sec:BSIntroSimplisticModel} we provide a very simplistic view of \BS for the casual reader while in section \ref{sec:BSIntroDetailedModel} we provide a detailed overview of the \BS formalism looking at the input state, the evolution, the output state, measurement, and discussing why it is inefficient to classically simulate. In section \ref{sec:BSintroPhysically} we look into what the permanent is representing physically, we compare the permanent versus the determinant, we visualize what the unitary does physically, and looking at how the permanent arises in the first place. Finally, in section \ref{sec:BSintroErrors} we discuss an error threshold such that \BS  remains classically intractable to simulate. 

\subsection{Simplistic Model} \label{sec:BSIntroSimplisticModel}

\begin{figure}[!htb]
\centering
\includegraphics[width=0.3\columnwidth]{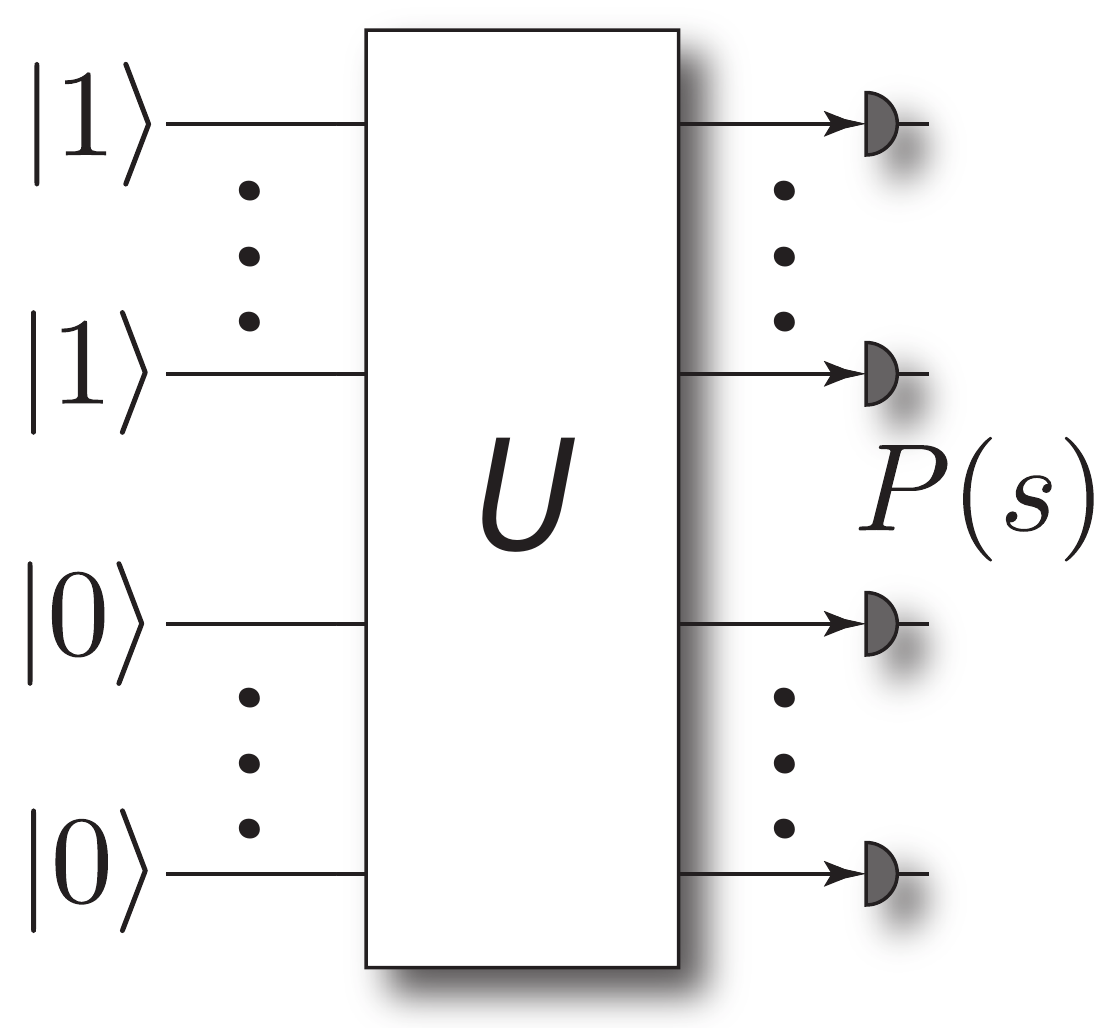}
\caption{A simplistic view of the \BS model. An input state is prepared, as shown on the left, comprising a number of single-photon Fock states and vacuum states in $m$ total modes. The input state passes through a passive linear optics network $U$ comprised of beamsplitters and phase-shifters that transforms the input state. Finally the output photon number statistics $P(s)$ are sampled via coincidence number-resolving photodetection as shown on the right, where $s$ is a particular configuration at the output.} \label{fig:modelSimple}
\end{figure}

\BS can be thought of in a very simplistic way as illustrated in Fig.~\ref{fig:modelSimple}.
\BS is comprised of three particular elements. First, there are $m$ input modes where the input state is inserted as shown on the left. The first $n$ modes have single photon Fock states $\ket{1}$ while the remaining $m-n$ have vacuum states $\ket{0}$. Second there is the evolution of the input state as shown as a box with the $U$, which is comprised of linear optical elements i.e. beam-splitters and phase-shifters. Finally, there are the $m$ output modes where the output state is detected using photo-detectors. In the next section we go into this in more detail.

\subsection{Detailed Model} \label{sec:BSIntroDetailedModel}
\begin{figure}[!htb]
\centering
\includegraphics[width=.4\columnwidth]{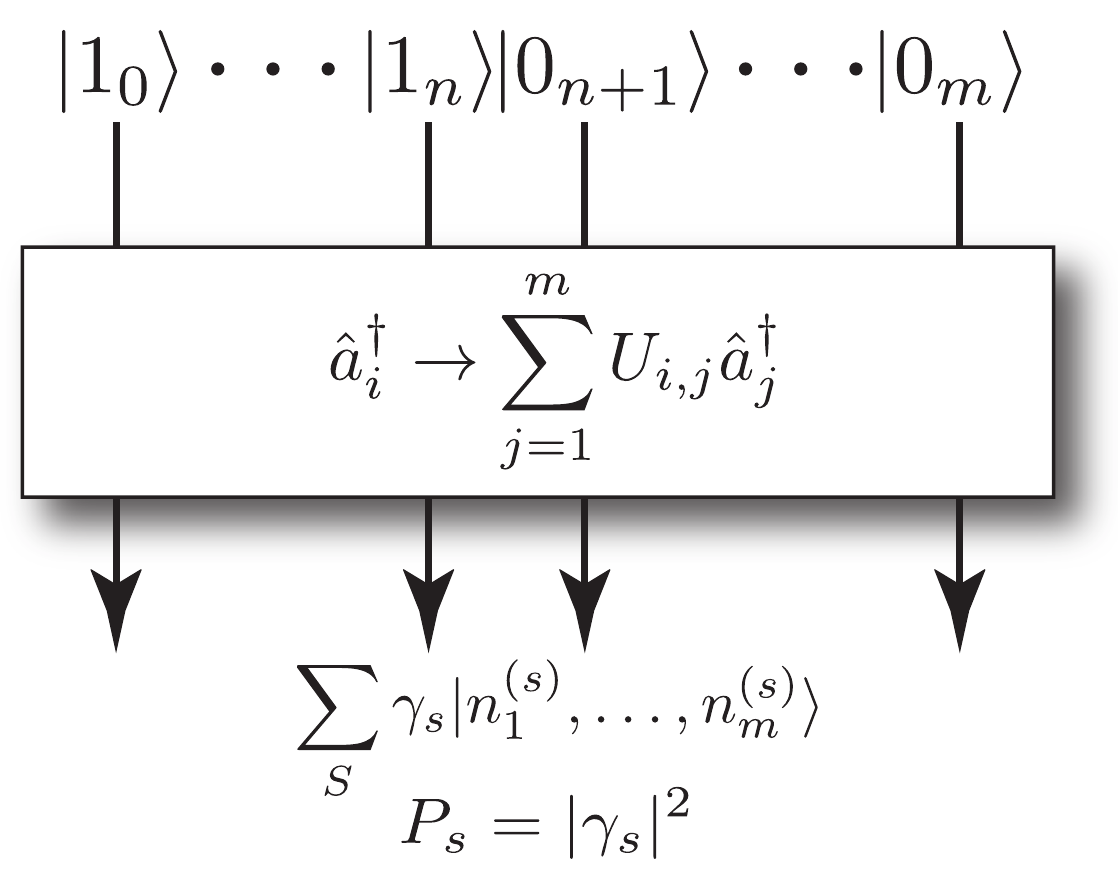}
\caption{A detailed \BS model. $n$ single photons are prepared in $m$ optical modes. These are evolved via a passive linear optics network $\hat{U}$ which transforms the photon creation operators. Finally the output statistics are sampled via coincidence photodetection. The experiment is repeated many times, partially reconstructing the output distribution.} \label{fig:model}
\end{figure}

A full detailed model of \BS is illustrated in Fig.~\ref{fig:model}. In this diagram the input state is inserted into $m$ modes at the top, evolves via the photon creation operaters of the unitary evolution in the box from input modes $i$ to output modes $j$, and is outputted at the bottom where the photo statistics are taken. In this section we review this process in detail discussing the input state, the evolution, the output state, measurement, and discussing the inefficiency of simulating this classically. 

\subsubsection{Input State}
First we prepare an $m$-mode input state $\ket{\psi_\mathrm{in}}$, where the first $n$ modes are prepared with the single photon Fock state, and the remaining \mbox{$m-n$} modes are injected with the vacuum state $\ket{0}$,
\begin{eqnarray} \label{eq:BSintroInputState}
\ket{\psi_\mathrm{in}} &=& \ket{1_1,\dots,1_n,0_{n+1},\dots,0_m} \nonumber \\
&=& a_1^\dag\dots a_n^\dag \ket{0_1,\dots,0_m},
\end{eqnarray}
where $a_i^\dag$ is the photon creation operator on the $i$th mode. It is assumed that the number of modes scales quadratically with the number of photons, \mbox{$m=\Omega(n^2)$}. For a large $n$, AA conjectured that the number of modes $m=\Omega(n^2)$ sufficiently ensures that, to a high probability, no more than a single photon arrives per output mode. This is sometimes referred to as the ``bosonic birthday paradox'' \cite{bib:birthday}. This implies that in this regime on/off (or `bucket') detectors will suffice, and photon-number resolution is not necessary, a further experimental simplification compared to full-fledged LOQC. This does not imply that no two-photon or higher-photon interference takes place since the output statistics and complexity arguments rely on the fact that bosonic interference occurs. In fact, many-photon interference is taking place, particularly during the beginning stages of the beamsplitter array.  

A variation to this input state is when $n$ photons are randomly inputed into the $m$ modes. This has become colloquially known as ``scattershot \BStwo.'' This variation is advantageous in that the input state is easier to prepare. Lund \emph{et al.} showed that the sampling problem remains classically difficult in the case of `scattershot \BStwo' \cite{bib:Lund13}.

\subsubsection{Evolution}

Next we propagate this input state through a passive linear optics network, which implements a unitary map on the photon creation operators,
\begin{equation} \label{eq:Utransform}
\hat{U}\hat{a}_i^\dag\hat{U}^\dag = \sum_{j=1}^m U_{i,j} \hat{a}_j^\dag,
\end{equation} 
where $\hat{U}$ is an $m\times m$ unitary matrix characterizing the linear optics network given by the Haar class. It was shown by Reck \emph{et al.} \cite{bib:Reck94} that any $\hat{U}$ may be efficiently constructed using $O(m^2)$ linear optics elements.

\subsubsection{Output State}
The output state is a superposition of the different configurations of how the $n$ photons could have arrived at the $m$ output modes. In a photon-number representation, the output state is a superposition of every possible $n$-photon-number configuration of the form,
\begin{equation} \label{eq:outputState}
\ket{\psi_\mathrm{out}} = \hat{U}\ket{\psi_\mathrm{in}} = \sum_S \gamma_s \ket{n_1^{(s)},\dots,n_m^{(s)}},
\end{equation}
where $S$ is the set of all possible photon number configurations, $s$ is a particular photon number configuration, $n_i^{(s)}$ is the number of photons in the $i$th mode associated with configuration $s$, and \mbox{$\gamma_s\in \mathbb{C}$} is the amplitude associated with configuration $s$. The total photon-number is conserved, thus \mbox{$\sum_i {n_i^{(s)}} = n$} for all $s$, which is a way to account for any losses due to inefficiencies in the experiment by post-selecting on all $n$ photons. Also, the probability of measuring configuration $S$ is given by \mbox{$P_s = |\gamma_s|^2$}. Note that a large class of distributions of photons at the output of a multi-mode interferometer with Gaussian inputs was investigated by Kok and Braunstein \cite{bib:kok2001multi}, who gave an analytic form of the output state even with post-selection. 

\subsubsection{Measurement}
Finally, we perform number-resolved photodetection \cite{bib:RosenbergTES}, which are described by projection operators $\hat{\Pi}(n) = \ket{n}\bra{n}$, on the output distribution, obtaining a sample from the distribution $P_s=|\gamma_s|^2$. Each time we obtain an $m$-fold coincidence measurement outcome with a total of $n$ photons. The experiment is repeated many times, building up statistics of the output distribution yielding the so-called \emph{sampling problem}, whereby the goal is to sample a statistical distribution using a finite number of measurements. This is in contrast to well-known \emph{decision problems}, such as Shor's algorithm \cite{bib:Shor97}, which provide a well defined answer to a well posed question. Because \BS is a sampling problem, finding a computational application is further complicated \textemdash if every time we run the device we obtain a different outcome, how does the outcome answer a well-defined question, and how do we map it to a problem of interest? This is one of the central challenges of \BS \textemdash what can we do with it?

\subsubsection{Discussion of Inefficient Classical Simulation}
The number of configurations $|S|$ grows exponentially with the number of photons and modes, 
\begin{eqnarray}
|S|&=&\binom{n+m-1}{n} 
\end{eqnarray}
Thus, with an `efficient' (i.e. polynomial) number of trials, we are unlikely to sample from a given configuration more than once. This implies that we are unable to determine any given $P_s$ with more than binary accuracy. Thus, \BS does \emph{not} let us calculate matrix permanents, as doing so would require determining amplitudes with a high level of precision, which would require a super-exponentially large number of measurements. This is evidence towards the hardness of simulating \BS as the Hilbert space is super-exponentially large. Much stronger evidence arose when Aaronson \& Arkhipov showed that this sampling problem likely cannot be efficiently simulated classically \cite{bib:AaronsonArkhipov}. The intuitive explanation for this supposed classical hardness is that each of the amplitudes $\gamma_S$ is proportional to an $n\times n$ matrix permanentas shown by Scheel \cite{bib:Scheel04perm},
\begin{equation}
\gamma_S = \frac{\mathrm{Per}(U_s)}{\sqrt{n_1^{(s)}!\dots n_m^{(s)}!}},
\end{equation}
where $U_s$ is an \mbox{$n\times n$} sub-matrix of $U$ that depends on the specific configuration $s$, and \mbox{$\mathrm{Per}(U_s)$} is the permanent of $U_s$. Permanents are believed to be classically hard to calculate, residing in the complexity class \mbox{\textbf{$\#$P}-hard}, the class of counting problems on polynomial-time algorithms. The best known classical algorithm for calculating matrix permanents is by Ryser \cite{bib:Ryser63}, requiring $O(2^n n^2)$ time steps. Because this requires exponential time to evaluate and because $\gamma_s$ is proportional to a permanent, sampling from the output distribution $P_s$ is a classically hard problem and thus so is \BStwo. 

Exact \BS by a polynomial-time classical probabilistic algorithm would imply a collapse of the polynomial hierarchy, while non-collapse of the polynomial hierarchy is generally believed to be a reasonable conjecture~\cite{bib:AaronsonArkhipov}. Gard \emph{et al.} gave an argument that classical computers likely cannot efficiently simulate multimode linear-optical interferometers with arbitrary Fock-state inputs \cite{bib:GardCrossDowling}. AA~\cite{bib:AaronsonArkhipov} gave a full complexity proof for the exact case where the device is noiseless. For the noisy case, a partial proof was provided which requires two conjectures that are believed to be true.

Importantly, \BS is not believed to be capable of efficiently simulating full quantum computation. Nonetheless, it is a relatively simple scheme that is experimentally viable with currently available technology that in the near future can sample bosonic statistics that the worlds best classical super computer today cannot. It is as though nature can easily do the simulation for us. Thus \BS is an attractive post-classical quantum computation scheme. It was shown by Rohde \& Ralph that \BS may implement a computationally hard algorithm even in the presence of high levels of loss \cite{bib:RohdeRalphErrBS} and mode-mismatch \cite{bib:RohdeLowFid12}, although formal hardness proofs are still lacking.

\subsection{What the Permanent Represents Physically} \label{sec:BSintroPhysically}

\subsubsection{Permanent versus Determinant}

Let us examine this relationship with the permanent more closely. The permanent of a matrix $U$ is given by
\begin{equation}
\mathrm{Per}(\mathrm{U}) = \sum_{\sigma \in \mathrm{S}_{n}} \prod_{i=1}^{n}\mathrm{U}_{i,\sigma(i)},
\end{equation}
where the sum is over all elements $\sigma$ of the symmetric group $S_n$, or in other words it is the sum over all permutations of the numbers ranging from one to $n$. It can be seen that this form for the permanent is closely related to the more familiar determinant
\begin{equation}
\mathrm{Det}(\mathrm{U}) = \sum_{\sigma \in \mathrm{S}_{n}} \mathrm{sgn}(\sigma) \prod_{i=1}^{n}\mathrm{U}_{i,\sigma(i)},
\end{equation}
where $\mathrm{sgn}(\sigma)$ is the signature of the permutation $\sigma$ and is $1$ if the signature has an even number of permutations and $-1$ if the signature has an odd number of permutations. The main difference between the permanent and the determinant is that the determinant has alternating plus and minus signs along the entries while the permanent has all plus signs. With the determinant a familiar trick called Gaussian elimination may be used to simplify the problem into one that can be solved in polynomial time making it efficiently solvable using a classical computer. There is no known trick for simplifying the permanent. 

\subsubsection{Convenient Visualization of the Unitary}
Now let us look more closely at the unitary $U$. It has the following form in its matrix representation
\begin{eqnarray}
U=\left( {\begin{array}{cccc}
   U_{1,1} & U_{1,2} & \cdots & U_{1,m} \\
   U_{2,1} & U_{2,2} & \cdots & U_{2,m}  \\
   \vdots & \vdots & \ddots & \vdots \\
   U_{m,1} & U_{m,2} & \cdots & U_{m,m} \\
  \end{array} } \right).
\end{eqnarray}
A conveinent way to visualize this matrix is to think of the rows of this matrix as the input modes and the columns of this matrix as the output modes to the \BS device. Then a particular entry $U_{i,j}$ is the probability amplitude a photon went into the $i$th mode and exited the $j$th mode. Then the classical probability of this happening is $|U_{i,j}|^2$. 

Next we will take a look at the case of multiple photons and see how the permanent arises for two and three photons.

\subsubsection{How the Permanent Arises in \BS}

\begin{figure}[!htb]
\centering
\includegraphics[width=\imageWidthThree\columnwidth]{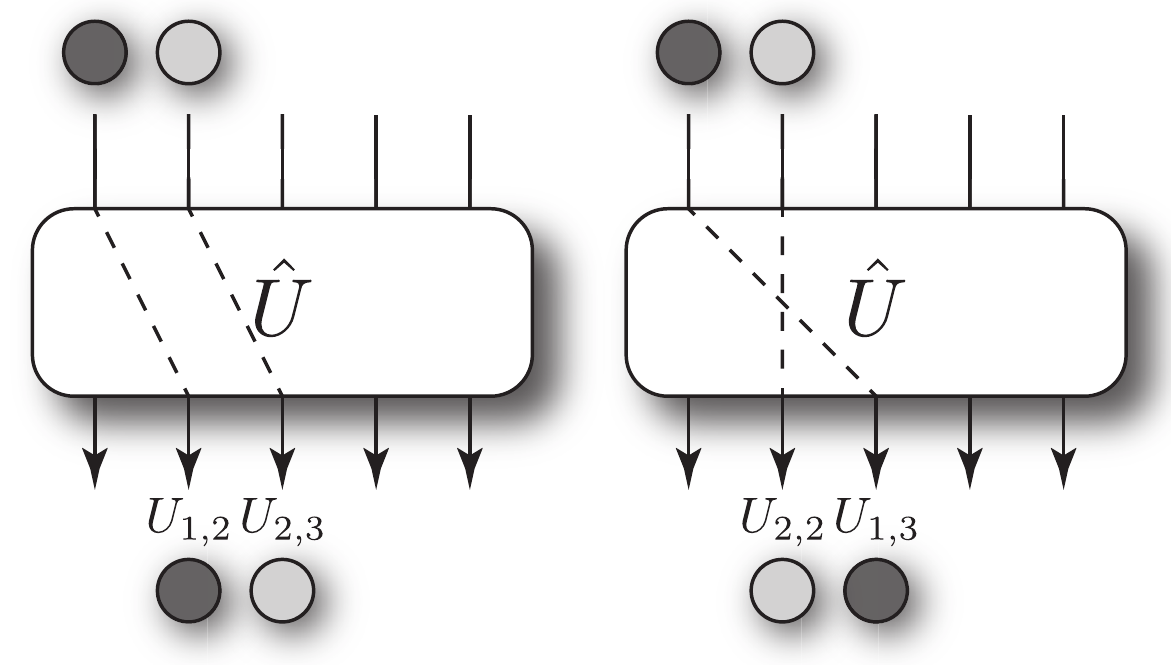}
\caption{Two-photon \BStwo, where we wish to calculate the amplitude of measuring a photon at each of the output modes 2 and 3. There are two ways in which this may occur which yields a sum of two paths leading us to the matrix permanent.} \label{fig:two_photon_perm}
\end{figure}

Let us first consider Fig.~\ref{fig:two_photon_perm}. Here the first two modes have single photons, with the remaining modes in the vacuum state. Let us consider the case where one of the two input photons arrive at output mode 2 while the other arrives at output mode 3. There  are two ways in which this could occur. Either the first photon arrives at mode 2 and the second at mode 3, or vice versa. This implies that there are \mbox{$2!=2$} ways in which the photons could reach the outputs. We may write the amplitude as
\begin{eqnarray} \label{eq:coinProbEx}
\gamma_{\{2,3\}} &=& \underbrace{U_{1,2}U_{2,3}}_{\mathrm{photons\ do\ not\ swap}} + \underbrace{U_{1,3}U_{2,2}}_{\mathrm{photons\ swap}} \nonumber \\
&=& \mathrm{Per} \left[ {\begin{array}{cc}
   U_{1,2} & U_{2,2} \\
   U_{1,3} & U_{2,3} \\
  \end{array} } \right],
\end{eqnarray}
which is a \mbox{$2\times 2$} matrix permanent.

\begin{figure}[!htb]
\centering
\includegraphics[width=\imageWidthThree\columnwidth]{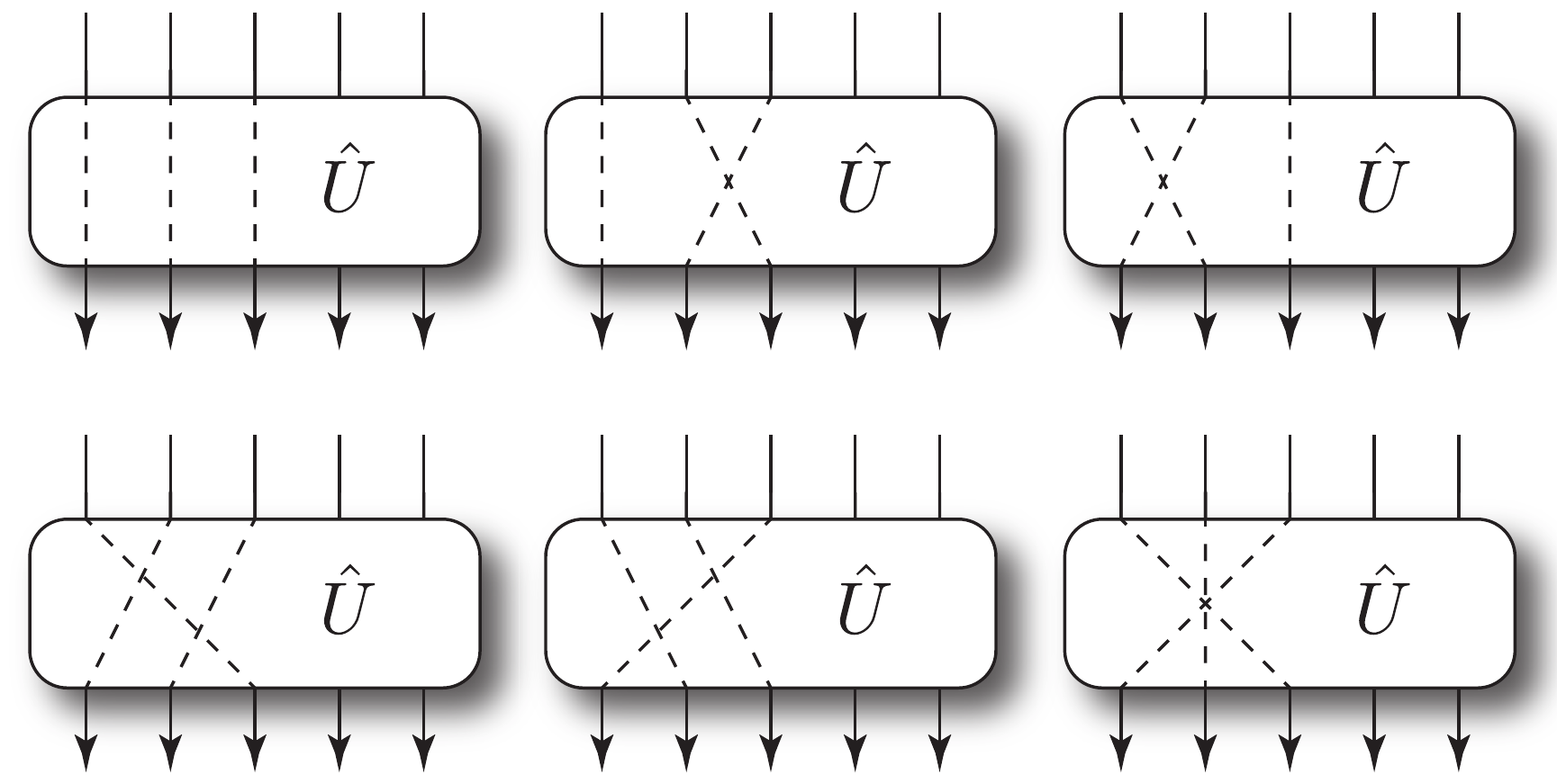}
\caption{Three photon \BStwo, where we wish to calculate the amplitude of measuring a single photon at each of the output modes 1, 2, and 3 where we begin with a single photon in input modes 1, 2, and 3. There are now \mbox{$3!=6$} possible ways for this to occur.} \label{fig:three_photon_perm}
\end{figure}

As a slightly more complex example, consider the three photon case shown in Fig.~\ref{fig:three_photon_perm}. We are considering the case where we begin with a single photon in each of the first three modes and consider the outcome where a single photon arrives in each of the first three output modes. Now we see that there are \mbox{$3!=6$} ways in which the three photons could reach the outputs. The associated amplitude is given by a \mbox{$3\times 3$} matrix permanent,
\begin{eqnarray} \label{eq:coinProbEx3}
\gamma_{\{1,2,3\}} &=& U_{1,1}U_{2,2}U_{3,3} + U_{1,1}U_{3,2}U_{2,3} \nonumber \\
&+& U_{2,1}U_{1,2}U_{3,3} + U_{2,1}U_{3,2}U_{1,3} \nonumber \\
&+& U_{3,1}U_{1,2}U_{2,3} + U_{3,1}U_{2,2}U_{1,3}
\nonumber \\
&=& \mathrm{Per} \left[ {\begin{array}{ccc}
   U_{1,1} & U_{2,1} & U_{3,1} \\
   U_{1,2} & U_{2,2} & U_{3,2} \\
   U_{1,3} & U_{2,3} & U_{3,3} \\
  \end{array} } \right].
\end{eqnarray}

In general, with $n$ photons, there will be $n!$ ways in which the photons could reach the outputs, assuming they all arrive at distinct outputs, and the associated amplitude will relate to an \mbox{$n\times n$} matrix permanent. If multiple photons arrive in the same output mode then multiple copies of the corresponding column of $U$ would be used to solve the matrix permanent. In the case where multiple photons enter into the same mode then multiple rows corresponding to the input row of $U$ would be used to solve the matrix permanent. Calculating matrix permanents, in general, is known to be \textbf{\#P}-hard, even harder than \textbf{NP}-complete, and the best known algorithm is by Ryser \cite{bib:Ryser63}, requiring \mbox{$O(2^n n^2)$} runtime. Thus, we can immediately see that if \BS were to be classically simulated by exactly calculating the matrix permanents, it would require exponential classical resources.

\subsection{Errors in \BS} \label{sec:BSintroErrors}

In the original \BS work AA provided a detailed analysis of the robustness of their result in the presence of error. This is important because a physical system such as \BS is bound to have some kind of error with all of the subtleties involving creating single photons that all match perfectly, evolving them accurately with optical elements, and detecting the distribution with imperfect photodetectors. With all of this to account for one could never experimentally achieve the true distribution.

In \BS one may like to consider estimation of a distribution. The exact permanent of a binary matrix is known to be \textbf{\#P}-hard but one can efficiently \textit{estimate} the permanent of a matrix that has real, non-negative entries \cite{bib:Valiant79, bib:Jerrum04}. Since \BS has complex-valued entries in the matrix it cannot be efficiently estimated \cite{bib:Jerrum04}. In fact if this complex case could be estimated then \BS would be an anomaly occuring only when trying to calculate the exact value of the permanent. This does not seem to be the case however so estimating the output distribution of a \BS machine is most likely computationally hard. 

There is a limit to how much error can be tolerated so that we do not deviate too far from the desired distribution or require so many samples that the algorithm becomes inefficient. One method to remove errant samples would be to post-select to remove statistics where photon-loss occurred.  However, if the post-selection probability scales as $O(1/\mathrm{exp}(n))$, then we would never be able to scale the \BS problem to large $n$ regimes.

What is an acceptable level of error? Let us apply an error threshold $\epsilon$, where $\epsilon$ is the maximum allowable variation distance from the exact solution assuming some clever statistical distance metric of which many are appropriate. Then we would need a success probably $P$ where 
\begin{equation}
P > 1/\mathrm{poly}(n),
\end{equation}
to correctly and efficiently sample a \BS device with arbitrarily large photon number $n$. If we would like to scale $\epsilon$ smaller, then
\begin{equation}
P > 1/\mathrm{poly}(n,1/\epsilon),
\end{equation}
\cite{bib:AaronsonArkhipov}. This is an interesting point because it relates to the Extended Church-Turing Thesis and cannot be experimentally verified for asymptotic $n$ as discussed in the next section.

\section{\BS and the Extended Church-Turing Thesis} \label{sec:BSintroECTT}

Any model for quantum computation is subject to errors of some type including dephasing, photon loss, and mode-mismatch among others. Here we consider a realistic generalised error model for \BS such that the desired single photon states are correct with probability $p$ and incorrect in some erroneous state with probability $1-p$ \cite{bib:BSECT}. This error model may generally include incorrect photon number, such as loss or higher order excitations, or mode-mismatch. We can then write our input state as
\begin{equation} \label{eq:error_model}
\hat\rho_\mathrm{in} =\left(\bigotimes_{i=1}^n[p\ket{1}\bra{1} + (1-p)\hat\rho_\mathrm{error}^{(i)}]\right) \otimes [\ket{0}\bra{0}]^{\otimes^{m-n}},
\end{equation}
where \mbox{$\hat\rho_\mathrm{error}^{(i)}$} may be unique for each input mode $i$. This error model is independent such that each state is independently subjected to an error channel. The fidelity of the single photon states are given by our measure of $p$. It is a dial that may be tuned such that when \mbox{$p=1$}, the input state is perfect, and when \mbox{$p<1$}, the state contains terms with error. We desire to sample from the distribution of Eq.~\ref{eq:BSintroInputState}, whereby the input state is perfect. This occurs with probability $p^n$.

Now we let $P$ be the probability that we have sampled from the correct distribution. By following the complexity proof for errors in \BS provided by AA where \mbox{$P>1/\mathrm{poly}(n)$}, we thus find that for computational hardness our generalized error model requires that \mbox{$p^n > 1/\mathrm{poly}(n)$}. Clearly this bound can never be satisfied for any $p<1$ in the asymptotic limit of large $n$ since we have an exponential dependance on the left side and a polynomial dependance on the right side. Therefore \BS will always fail in the asymptotic limit with this error model.

Various authors  \cite{bib:Broome2012, bib:Shen14, bib:AA13response, bib:Shchesnovich13, bib:Molmer13} have claimed that large-scale demonstrations of \BS could provide elucidation on the validity of the Extended Church-Turing (ECT) thesis \textemdash\ the statement that any physical system may be efficiently simulated on a Turing machine. The ECT thesis though is an asymptotic statement about arbitrarily large systems. We have shown that the required error bound for \BS is never satisfied for arbitrarily large systems, therefore \BS cannot elucidate the validity of the ECT thesis since asymptotically large \BS devices fail under any realistic error model.

There seem to be two obvious ways that our claim about the ECT thesis may be overcome: (1) It may be shown that the error bound can be loosened to $1/\mathrm{exp}(n)$, or (2) fault-tolerance techniques for \BS may be developed that allow arbitrarily large scaling of \BStwo. No such developments have been made; therefore, based on current understanding, \BS will not illuminate whether the ECT thesis is correct or not. However, \BS may still yield an interesting \emph{post-classical} computation since this only requires a finite sized device that can out perform the best classical computers.

\section{How to Build a \BS Device} \label{sec:BSintroBuild}

In this section we explain the basic components required to build a \BS device. This device consists of three basic components: (1) single-photon sources; (2) linear optical networks; and, (3) photodetectors. Each of these present their own engineering challenges and there are a range of technologies that could be employed for each of these components. However, although \BS is much easier to implement than full-scale LOQC, it remains challenging to build a post-classical \BS device. While challenging, a realizable post-classical \BS device is foreseeable in the near future. 

\subsection{Photon sources}

The first engineering challenge is to prepare an input state of the form of Eq.~\ref{eq:BSintroInputState}. This state may be generated using various photon source technologies. For a review of many of the photon sources see Ref.~\cite{bib:SourceAndDetectorReview}. Presently, the most commonly employed photon source technology is spontaneous parametric down-conversion (SPDC). The topic of Ch. \ref{Ch:SPDC} focuses on \BS with SPDC sources. Another viable source is quantum dots \cite{bib:Moreau01}, which has been used to successfully implement a four photon \BS experiment in a temporal architecture \cite{bib:he16} proposed by Motes \emph{et al.} \cite{bib:PRLFiberLoop}. 

\subsection{Linear optics networks}

After the input state has been prepared it is evolved via a linear optics network, $\hat{U}$. $\hat{U}$ transforms the input state as per Eq.~\ref{eq:Utransform} and may be completely characterized before the experiment using coherent state inputs \cite{bib:Reck94}. $\hat{U}$ is composed of an array of discrete elements, namely, beamsplitters and phase-shifters. A beamsplitter with phase-shifters may be represented as a two-mode unitary of the form \cite{bib:GerryKnight05},
\begin{equation} \label{eq:BS}
\hat{U}_{\mathrm{BS}}(t) = \left( \begin{array}{cc}
e^{i(\alpha-\frac{\beta}{2}-\frac{\gamma}{2})}\mathrm{cos}\left(\frac{\delta}{2}\right) & -e^{i(\alpha-\frac{\beta}{2}+\frac{\gamma}{2})}\mathrm{sin}\left(\frac{\delta}{2}\right)  \\
e^{i(\alpha+\frac{\beta}{2}-\frac{\gamma}{2})}\mathrm{sin}\left(\frac{\delta}{2}\right) & e^{i(\alpha+\frac{\beta}{2}+\frac{\gamma}{2})}\mathrm{cos}\left(\frac{\delta}{2}\right)
\end{array} \right), 
\end{equation}
where \mbox{$0\leq\alpha\leq2\pi$} and \mbox{$0\leq\{\beta,\gamma,\delta\}\leq\pi$} are arbitrary phases.

For a $\hat{U}$ that implements a classically hard problem, one would need hundreds of discrete optical elements. Constructing an arbitrary $\hat{U}$ using the traditional linear optics approach of setting and aligning each optical element would be extremely cumbersome. Thus, using discrete optical elements is not a very promising route towards scalable \BStwo.

One method to simplify the construction of the linear optics network is to use integrated waveguides. Quantum interference was first demonstrated with this technology by Peruzzo \emph{et al.} \cite{bib:Peruzzo11b}. This technology requires more frugal space requirements, is more optically stable, and far easier to manufacture, allowing the entire linear optics network to be integrated onto a small chip \cite{bib:Politi02052008, bib:matthews2009, bib:Politi04092009}. The main issue with integrated waveguides is achieving sufficiently low loss rates inside of the waveguide and in the coupling of the waveguide to the photon sources and photodetectors. Presently, the loss rates in these devices are extremely high and thus post-selection upon $n$ photons at the output occurs with very low probability. It is foreseeable that photon sources and photodetectors will eventually be integrated into the waveguide which would eliminate coupling loss rates, substantially improving scalability.   

Another potential route to simplifying the linear optics network is to use time-bin encoding in a loop architecture based on the work by Motes \emph{et al.} \cite{bib:motes2014scalable}. The major advantage of this architecture is that it only requires two delay loops, two on/off switches, and one controllable beam splitter. This possibility eliminates the problem of aligning hundreds of optical elements and has fixed experimental complexity, irrespective of the size of the \BS device. A major problem with this architecture however is that it remains difficult to control a dynamic beamsplitter with high fidelity at a rate that is on the order of the time-bin width $\tau$. Nonetheless this architecture has since been successfully experimentally implemented by He \emph{et al.} \cite{bib:he16} performing four photon \BStwo, which is the largest instance of \BS to be performed to date. 
Furthermore they claim that they can do more than 20 photon \BS with further refinement of system efficiency. This temporal architecture is described in detail in Chapter \ref{Ch:FiberLoop}. 

\subsection{Photodetection}

The final requirement in the \BS device is sampling the output distribution as shown in Eq. \ref{eq:outputState}. With linear optics this is done using photodetectors. For a review on various types of photodetection see Ref.~\cite{bib:SourceAndDetectorReview, bib:kok2001detection}.  

There are two general types of photodetectors \textemdash photon-number resolving detectors and bucket detectors. The former counts the number of incident photons. These are much more difficult to make and more expensive in general than bucket detectors. Bucket detectors, on the other hand, simply trigger if any non-zero number of photons are incident on the detector. As discussed earlier, in the limit of large \BS devices, we are statistically guaranteed that we never measure more than one photon per mode, since the number of modes scales as \mbox{$m=O(n^2)$}. Thus, bucket detectors are sufficient for large \BS devices, a significant experimental simplification compared to universal LOQC protocols. 

The predominant mechanism for experimentally realising single-photon counting is making use of beam-splitter cascades \cite{bib:kok2001detection}, which uses much the same technology as \BStwo. More recently photodetector designs use superconductivity to measure photons. Superconductivity is an extreme state where electrical current flows with zero resistance. It occurs in conductive materials when a certain critical temperature is reached. This critical temperature is far from occurring naturally on Earth and thus high-tech and expensive lab equipment is required. For many materials this temperature is close to absolute zero. 

\begin{figure}[!htb]
\centering
\includegraphics[width=0.3\columnwidth]{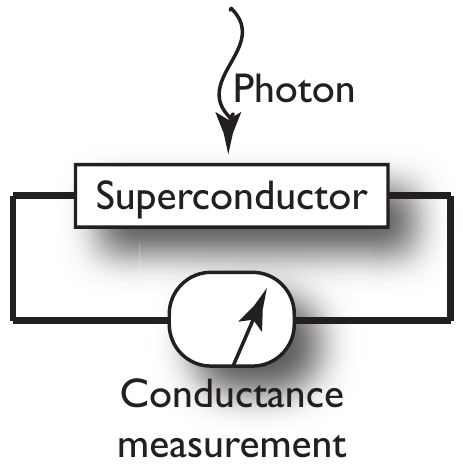}
\caption{Basic design for a superconducting photon detector. A current is passed through a superconductor and the conductance is monitored. Photons impart energy on the superconductor that is just cooled to its critical temperature. This added photon energy causes a measurable change in the conductance allowing for the detection of photons.}
\label{fig:superconducting_detector}
\end{figure}

While there are several variations of superconductive photodetectors, many follow the same principles as the one shown in Fig. \ref{fig:superconducting_detector}. The idea is that a superconductor is cooled to a point just below its critical temperature. Current is then applied through the superconductor which experiences zero resistance. If there is no resistance, then there is no voltage drop across the superconductor and a conductance measurement reads infinity. Then the photon or photons that are to be measured will hit the superconductor and be absorbed. Each photon that is absorbed by the superconductor imparts energy $h\nu$ onto it, where $h$ is Planck's constant and $\nu$ is the frequency of the photon. This heats the superconductor above its critical temperature. The conductance measurement will then change according to the absorbed photon, thus informing the measurer that a photon was detected. This scheme may be able to count several photons since the conductance will change proportionally to the number of absorbed photons. However, if too many photons are absorbed all properties of superconductivity are lost and thus number-resolution is lost. For a more extensive introduction to this topic, see chapter four of Ref. \cite{bib:KokLovett11}.

Photodetectors may be used to help overcome the problem of temporal mismatch. Such detectors must have the ability to record the time at which the photon arrived. If we post-select upon detecting all $n$ output photons in the same time-window $\Delta t$ then we can assume that their temporal distribution overlaps sufficiently well to yield a classically hard sampling problem. This method however is not reliable for scalable \BStwo. If the temporal distributions are not sufficiently overlapping, then the probability of post-selecting all $n$ photons in the same time-window decreases exponentially with $n$. However, if the sources are producing nearly identical photons in the time domain then this method would be a practical cross check.

As the distinguishability of photons varies the complexity of sampling the output distribution also varies. A theoretical framework was developed by Tillmann \emph{et al.} \cite{bib:tillmann2014BS} that describes the transition probabilities of photons with arbitrary distinguishability through the linear optical network. The output distribution of \BS with distinguishable photons is then given by matrix immanants, thus affecting the computational complexity of the output distribution. They also test this experimentally by tuning the temporal mismatch of their input photons. This \BS experiment is unique in that it is the first to use distinguishable photons at the input. A similar situation, of \BS with photons of arbitrary spectral structure, was considered by Rohde \cite{bib:Rohde15a}.

\section{Applications Inspired by \BS} \label{sec:BSintroApplications}

For the first four years of \BS there was no application for it. It was interesting simply because the statistics at the output of the device could not be simulated efficiently with a classical computer but the device itself could relatively easily be built where nature can do the simulating for us. With a slight alteration to the protocol where we harness the physics of the device as opposed to the computational complexity we found an application inspired by \BStwo, which is in quantum metrology \cite{bib:MORDOR}. This work was released almost simultaneously with another research group where they presented a \BS inspired application in generating molecular vibronic spectra \cite{bib:vibSpec}. In Ch. \ref{Ch:MORDOR} I present in detail how we developed one of two world first applications inspired by \BStwo. 

\section{Summary}

In this chapter we have given an introduction to the rapidly evolving field of \BS \textemdash\ a scheme that has inspired many people in the quantum information processing community in the last couple of years. It is intriguing because it can easily be realised in the lab as compared to building a universal quantum computer and at the same time actually sample from a distribution that no computer on Earth can efficiently simulate. It is the first known device that can do such a thing.

We began this chapter by motivating \BS and discussing some of its history in section \ref{sec:BSintroMotivation}. We then went into detail discussing \BS by first giving a very simplistic model followed by a more detailed model and discussing why it is inefficient to classically simulate, what the permanent represents physically, and errors in \BStwo, all in section \ref{sec:BSintroFormalism}. Next, in section \ref{sec:BSintroECTT}, we went into the implications of \BS to the extended Church-Turing thesis and argued that \BS can not be used to prove or disprove this. In section \ref{sec:BSintroBuild} we discussed the technology required to build a \BS device including the photon sources, linear optical networks, and photodetection. 
Finally, in section \ref{sec:BSintroApplications} we discussed two \BS inspired applications, but focus on our own metrology result.

\begin{savequote}[45mm]
Physics is like sex: sure, it may give some practical results, but that's not why we do it.
\qauthor{Richard Feynman}
\end{savequote}

\chapter{Spontaneous Parametric Down-Conversion Photon Sources for \BS} \label{Ch:SPDC}

\section{Synopsis}
\BS has emerged as a promising avenue towards post-classical optical quantum computation, and numerous elementary demonstrations have recently been performed. One of the challenges for realising \BS is the creation of single photons. Spontaneous parametric down-conversion (SPDC) is the most common method for single-photon state preparation and is employed in most optical quantum information processing experiments. 

We motivate this work in detail in section \ref{sec:SPDCmotivation}. In section \ref{sec:SPDCSPDC} we describe what spontaneous parametric down conversion is. We present a simple architecture for \BS based on multiplexed SPDC sources, as shown in section \ref{sec:SPDCarcitecture}, and demonstrate in section \ref{sec:SPDCscalability} that the architecture is limited only by the post-selection detection efficiency assuming that other errors, such as spectral impurity, dark counts, and interferometric instability are negligible. For any given number of input photons, there exists a minimum detector efficiency that allows post selection. If this efficiency is achieved, photon-number errors in the SPDC sources are sufficiently low as to guarantee correct \BS much of the time. In this scheme the required detector efficiency must increase exponentially in the photon number. Thus, we show that idealised SPDC sources will not present a bottleneck for future \BS implementations. Rather, photodetection efficiency is the limiting factor and thus future implementations may continue to employ SPDC sources.

\section{Motivation} \label{sec:SPDCmotivation}

In this chapter we show that large scale \BS can be implemented provided that detection efficiencies, which must increase exponentially with photon number, are sufficient to guarantee post-selection with high probability. Increasing input photon number will thus yield a larger required detection efficiency. 

Spontaneous parametric down-conversion (SPDC) has become the main method for single-photon state preparation, is widely used in optical quantum information processing, and was employed in all of the recent experimental \BS implementations. A pressing question for future larger-scale implementations is scalability. Scalability in this context refers to increasing the input photon number into the \BS device provided that the error in the single photon photo-detectors, which scales exponentially with input photon number, is sufficiently low to ensure successful implementation of \BS most of the time. That is, what are the limitations and requirements on physical resources to implement a scalable device? In particular, will SPDC sources suffice, or will we have to transition to other photon source technologies? 

We consider a general architecture for the experimental implementation of \BStwo, where multiplexed SPDC sources are employed for state preparation. We show that in such an architecture the device is limited only by the post-selection probability. In other words, the architecture is scalable provided that detector efficiencies are sufficiently high to enable post-selected computation. In this regime, the quality of current SPDC states is sufficient to enable large-scale \BStwo. Thus, it is photodetection, not SPDC sources, that provide the bottleneck to larger-scale demonstrations.

\section{Spontaneous Parametric Down-Conversion} \label{sec:SPDCSPDC}

\begin{figure}[!htb]
\centering
\includegraphics[width=0.4\columnwidth]{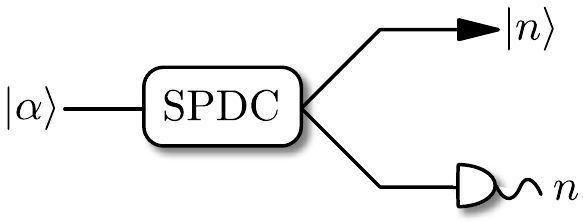}
\caption{Spontaneous parametric down-conversion (SPDC) source. A crystal with a second order non-linearity, $\chi^{(2)}$, is pumped with a classical coherent light source $\ket\alpha$. The source then probabilistically emits photon pairs into the signal and idler modes, including vacuum $\ket{0}\ket{0}$ and higher order terms where multiple pairs are emitted. The idler mode is detected revealing how many photons are present in the signal mode.}
\label{fig:SPDC}
\end{figure}

The SPDC source works by first pumping a non-linear crystal with a coherent state $\ket{\alpha}$ as shown in Fig. \ref{fig:SPDC}. A coherent state is well approximated by a laser source. This evolution is given by a Hamiltonian of the form
\begin{align}
\hat{H}_\mathrm{SPDC} = \chi\hat{a}_\mathrm{pump} \hat{a}_\mathrm{signal}^\dag \hat{a}_\mathrm{idler}^\dag + \mathrm{h.c.},
\end{align}
where $\chi$ is the interaction strength, which depends on the non-linear material and h.c. stands for the hermitian conjugate of the first term (e.g. the hermitian conjugate of an operator $\hat{A}$ is $\hat{A}^{\dag}$). With some probability one of the laser photons interacts with the crystal and emits an entangled superposition of photons across two output modes, the \emph{signal} and \emph{idler}. The output of an SPDC source is of the form \cite{bib:GerryKnight05},
\begin{equation} \label{SPDC}
\ket{\Psi_\mathrm{SPDC}} = \sum_{n=0}^{\infty}\lambda_n\ket{n}_s\ket{n}_i,
\end{equation}
where $n$ is the number of photons, $s$ represents the signal mode, and $i$ represents the idler mode. The photon-number distribution is given by
\begin{equation} \label{eq:SPDCprobabilitydistribution}
 \lambda_n = \frac{1}{\mathrm{cosh}(r)}(-1)^ne^{i n \theta}\left(\mathrm{tanh}(r)\right)^n,
\end{equation}
where $r$ is the squeezing parameter and $\theta$ is the phase shift between the two modes.

For \BStwo, we are interested in the $\ket{1}_s\ket{1}_i$ term of this superposition since we require single photons at the input of the first $n$ modes. The signal photons are measured by a photodetector and because of the correlation in photon-number, we know that a photon is also present in the idler mode. The idler photons are then routed into one of the input ports of the \BS device using a multiplexer \cite{bib:migdall2002tailoring, bib:LPOR201400027, bib:ma2011experimental, bib:kok2001detection}.

There are several problems associated with SPDC sources, which limit the scalability of \BStwo. The major problem is higher order photon-number terms. In the \BS model we only want the $\ket{1}_s\ket{1}_i$ term, which is far from deterministic. The SPDC source is going to emit the zero-photon term with highest probability and emit higher order terms with exponentially decreasing probability. If the heralding photodetector does not have unit efficiency, then the heralded mode may contain higher order photon-number terms.


Another problem is that photons from SPDC sources have uncertainty in their temporal distribution. If a \BS device is built using multiple SPDC sources it is difficult to temporally align each of the $n$ photons entering the device. This is called temporal mismatch. The error term associated with this scales exponentially with $n$, yielding an error model consistent with Eq.~(\ref{eq:error_model}), which undermines operation in the asymptotic limit.

\section{Spontaneous Parametric Down-Conversion Multiplexing Architecture for \BS} \label{sec:SPDCarcitecture}

Given that SPDC is the most widely used and readily accessible source for single-photon state preparation, we will present a simple architecture for \BS based on SPDC sources. In an ideal \BS implementation one would employ deterministic photon sources that produce exactly one photon on demand. SPDC sources, on the other hand, coherently prepare photon pairs in two modes with a correlated Poisson probability distribution. By measuring one of the modes and post-selecting upon detecting one photon in that mode, a single photon is guaranteed to appear in the other mode. This method provides us with a probabilistic, heralded single-photon source. It is critical that each photon is heralded to ensure a pure set of Fock-state inputs. 
The photon number probability distribution is given by \cite{bib:GerryKnight05},
\begin{equation} \label{eq:PDC_dist}
P^\mathrm{SPDC}(s) = |\lambda_{s}|^2 = \frac{\mathrm{tanh}^{2s}r}{\mathrm{cosh}^{2}r},
\end{equation}
where $s$ is the photon number (per mode). Thus, 
The SPDC source most often emits the vacuum state, and sometimes higher-order pairs with exponentially decreasing probability. For small squeezing parameters the higher-order terms can be made small, yielding a heralded source that produces single-photon pairs with high confidence that the heralded state has only a single photon. To herald a single photon, we detect one arm of a single SPDC source using an inefficient number-resolving photodetector. Such a detector can be characterised by the conditional probability of detecting $t$ photons given that $s$ photons were present. For a simple inefficient detector this is given by,
\begin{equation} \label{eq:det_mn}
P_\mathrm{D}(t|s) = \binom{s}{t} \eta^t (1-\eta)^{s-t},
\end{equation}
where $\eta$ is the detection efficiency. Thus, in the presence of loss, the detector exhibits ambiguity in the measured photon number, sometimes detecting fewer photons than were present. Dark counts, the other dominant source of imperfection in photodetection, are measurement events from extraneous photons from the environment and could also be incorporated into the model, but this effect can be made very small with time-gating.

We specifically consider heralded SPDC states, using just one mode of the SPDC for the computation rather than both, to ensure that the state entering $U$ closely approximates Eq.~(\ref{eq:BSintroInputState}). Without the heralding, the SPDC state is Gaussian, which is inconsistent with the \BS model and not known to implement a classically hard algorithm \cite{bib:bartlett2003requirement, bib:gogolin2013}.

Combining Eqs.~(\ref{eq:PDC_dist}) \& (\ref{eq:det_mn}) we obtain the probability of detecting $t$ photons in the heralding arm of a single SPDC source,
\begin{eqnarray} \label{eq:pdc_det}
P^\mathrm{SPDC}_\mathrm{D}(t) &=& \sum_{i\geq t} P_\mathrm{D}(t|i) P^\mathrm{SPDC}(i) \nonumber \\
&=& \sum_{i\geq t} \binom{i}{t} \eta^t (1-\eta)^{i-t} P^\mathrm{SPDC}(i).
\end{eqnarray}
Thus the probability of detecting a single photon in the heralding arm is simply,
\begin{equation}
P^\mathrm{SPDC}_\mathrm{D}(1) = \sum_{i\geq 1} i \, \eta (1-\eta)^{i-1} P^\mathrm{SPDC}(i).
\end{equation}

\begin{figure}[!htb]
\centering
\includegraphics[width=0.5\columnwidth]{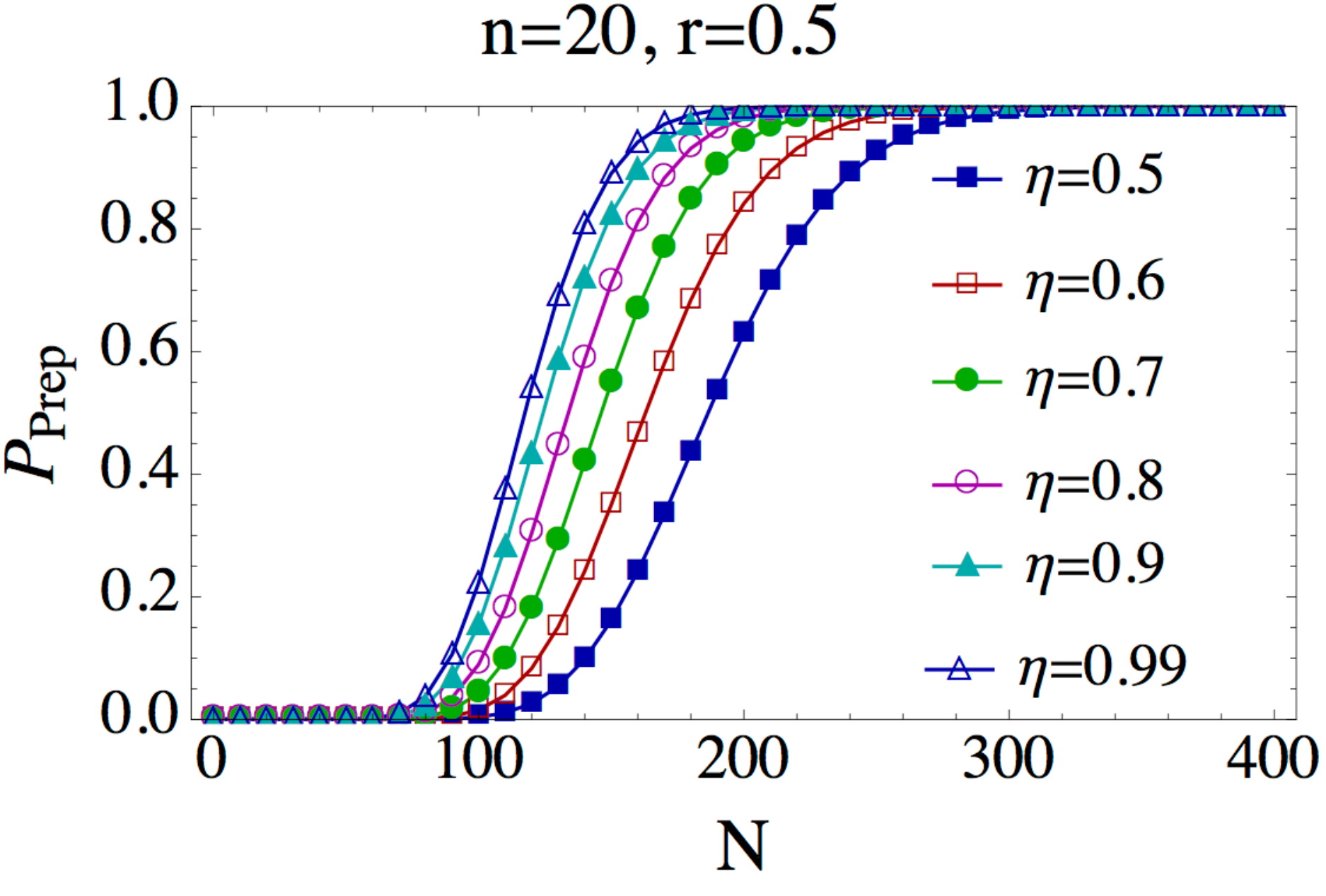}
\caption{Asymptotic behaviour of the state preparation success probability as a function of the number of SPDC sources, $N$, and detector efficiency, $\eta$, in the case where we are required to successfully herald \mbox{$n=20$} photons with a squeezing of $r=0.5$. In the limit of large $N$, $P_\mathrm{prep}$ approaches unity.} \label{fig:Pprep_asymptote}
\end{figure}

We will operate $N$ such heralded sources in parallel, where \mbox{$N\gg n$} and $n$ is the number of single photon Fock states required for \BStwo. The probability that at least $n$ of the SPDC sources successfully herald is given by,
\begin{eqnarray} \label{eq:P_prep}
P_\mathrm{prep}(n) &=& \sum_{i \geq n} \binom{N}{i} \left[{P^\mathrm{SPDC}_\mathrm{D}(1)}\right]^{i} \left[1 - P^\mathrm{SPDC}_\mathrm{D}(1)\right]^{N-i}. \nonumber \\
\end{eqnarray}
In the limit of large $N$ this asymptotes to unity,
\begin{equation}
\lim_{N\to \infty}P_{\mathrm{prep}}(n)=1.
\end{equation}
The asymptotic behaviour of $P_{\mathrm{prep}}$ is illustrated in Fig.~\ref{fig:Pprep_asymptote}. Clearly, with a sufficiently large number of SPDC sources operating in parallel, we are guaranteed to successfully herald the required $n$ single photons.

Having successfully heralded at least $n$ SPDC sources, we employ a dynamic multiplexer \cite{bib:migdall2002tailoring} to route $n$ of the heralded states to the first $n$ modes of the \BS interferometer $U$. We will assume the multiplexer is ideal in our analysis, although losses could be absorbed into the detector efficiency. Experimental progress has been made recently in developing active multiplexers \cite{bib:ma2011experimental, bib:LPOR201400027, bib:kok2001detection}. These multiplexers rely on optical switches, which are the topic of much experimental investigation \cite{bib:beaujean2016oxazines, bib:kawashima2016multi}.  

Following the unitary network, number-resolving photodetection is applied. Because the photodetectors do not have unit efficiency we must post-select on events where all $n$ photons are detected. The post-selection probability scales as,
\begin{equation} \label{eq:SPDCexp_scaling}
P_\mathrm{post}(n) = \eta^n.
\end{equation}
Thus, the required detection efficiency exponentially asymptotes to unity for large $n$. This necessitates that future large-scale \BS implementations will require extremely high efficiency photodetectors.

\begin{figure}[!htb]
\centering
\includegraphics[width=0.5\columnwidth]{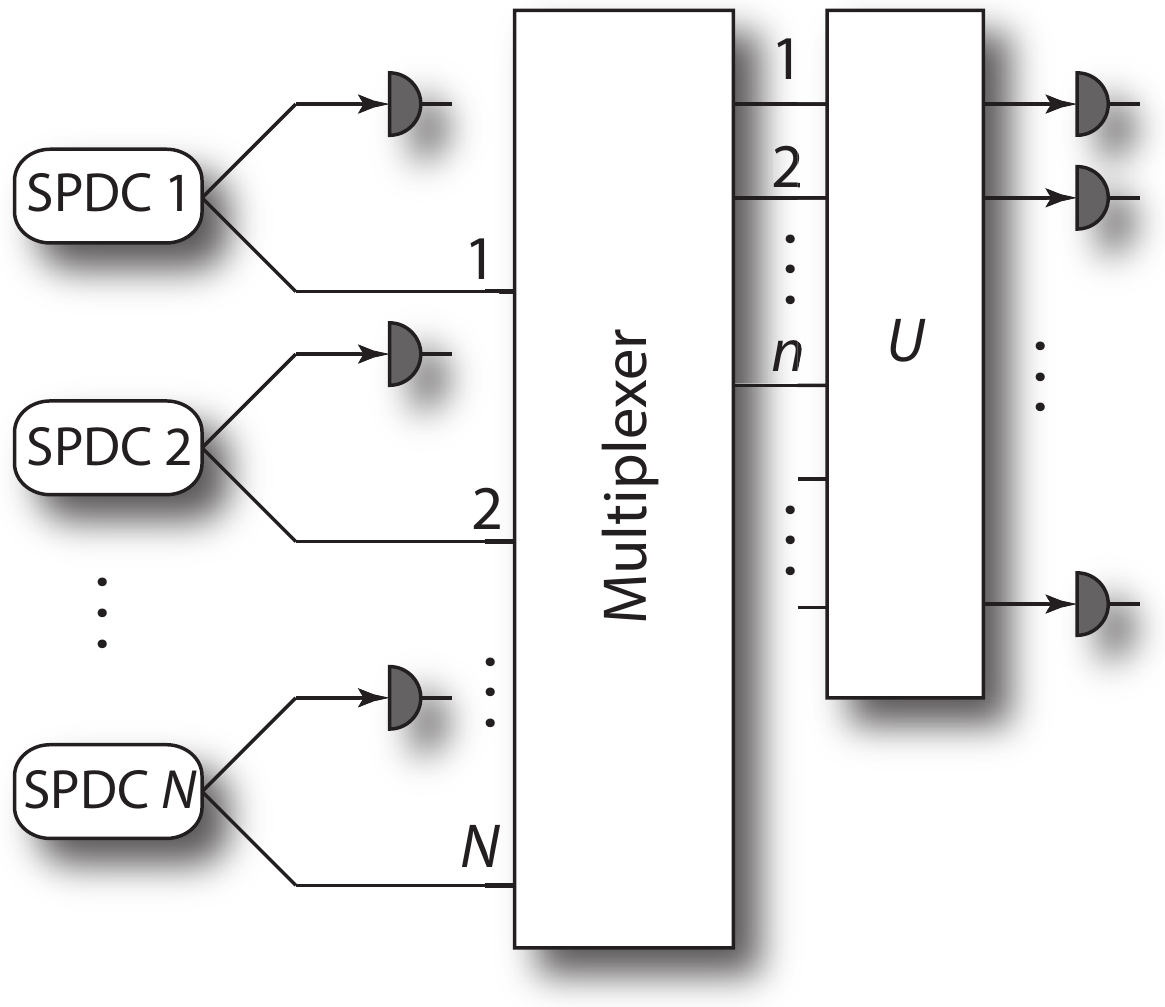}
\caption{Architecture for \BS with SPDC sources. $N$ sources operate in parallel, each heralded by an inefficient single-photon number-resolving detection. It is assumed that \mbox{$N\gg n$}, which guarantees that at least $n$ photons will be heralded. The multiplexer dynamically routes the successfully heralded modes to the first $n$ modes of the unitary network $U$. Finally, photodetection is performed and the output is post-selected on the detection on all $n$ photons.} \label{fig:SPDCarchitecture}
\end{figure}

The full architecture is illustrated in Fig.~\ref{fig:SPDCarchitecture}. Note that the multiplexer is \emph{critical} to the operation of the device for the original implementation of \BS as presented by AA. Without the multiplexer we still have high likelihood of sampling from at least an $n$-photon input distribution; however, every time the device is run we are likely to sample from a different permutation of the vacuum and single photon states at the input, making it impossible to perform sampling on a consistent input. Thus, the multiplexing ensures that the input state is consistently of the form of Eq.~(\ref{eq:BSintroInputState}) if the photodetectors have perfect efficiency. Results published after this work by Lund \emph{et al.} \cite{bib:Lund13} showed that the sampling problem remains classically difficult in the case of `scattershot \BStwo', whereby $n$ photons can be input into any random configuration to the \BS device as long as where they went into the device is known, which is the case if you are using SPDC sources and heralding. This means we can now further simplify our architecture by getting rid of such a complicated multiplexer. The realistic case of inefficient photodetectors is presented next.

%
%

\section{Scalability of the Architecture} \label{sec:SPDCscalability}

Having described a general architecture for \BS based on SPDC sources, the pressing question is its scalability. The obvious scaling issue arises from Eq.~(\ref{eq:SPDCexp_scaling}), whereby the photodetection efficiency must be exponentially close to unity. Unless error correction mechanisms are introduced, this scaling is inevitable and post-selection is the only avenue to guarantee successful operation of the device. However, no error correction has been described in the context of \BStwo. Thus, we will focus on post-selected operation of the device, and address the question as to whether the device acts correctly in that context.

In the described architecture, the dominant error source is incorrect heralding of the SPDC states. In the limit of perfect detectors we are guaranteed to have prepared single-photon states. However, inefficient detectors introduce ambiguity in the heralding, creating a situation where higher-order photon number terms are perceived as single photon terms. For example, if a single photon is lost via detection inefficiency, the two photon state will be interpreted as a single photon state. This will corrupt the input state to the interferometer, yielding an input state different than Eq.~(\ref{eq:BSintroInputState}).

For a single detector, the probability that we have prepared the $s$-photon Fock state, given that the detector has outcome $t$, is given by Bayes' rule,
\begin{eqnarray}
P_\mathrm{corr}(s|t) &=& \frac{P_\mathrm{D}(t|s) P^\mathrm{SPDC}(s)}{P^\mathrm{SPDC}_\mathrm{D}(t)} \nonumber \\
&=& \frac{\binom{s}{t} (1-\eta)^{s-t} \mathrm{tanh}^{2s} r}{\sum_{i\geq t} \binom{i}{t} (1-\eta)^{i-t} \mathrm{tanh}^{2i}r}.
\end{eqnarray}
We are interested in the case where we herald a single photon. Thus,  
\begin{equation}
P_\mathrm{corr}(1|1) = [1 - (1 - \eta)\,\mathrm{tanh}^{2}r]^2.
\end{equation}
$P_\mathrm{corr}(1|1)$ can be interpreted as the conditional probability that we have prepared the correct single photon state given that heralding was successful. For small pump powers (\mbox{$r\approx 0$}) the \emph{unconditional} probability of detecting a single photon approaches zero, although the \emph{conditional} probability approaches unity since there are negligible higher photon-number contributions.

The probability that a single photon is correctly heralded $n$ times in parallel, thereby preparing the $n$ copies of a single photon Fock state as per Eq. (\ref{eq:SPDCexp_scaling}) is
\begin{eqnarray} \label{eq:P_par}
P_\mathrm{par}(n) &=& [P_\mathrm{corr}(1|1)]^n \nonumber \\
&=& [1 - (1 - \eta)\,\mathrm{tanh}^{2}r]^{2n}.
\end{eqnarray}

We will require that, given $n$ heralded SPDC states, upon post-selection we correctly detect exactly $n$ photons the majority of the time. We will arbitrarily require that $P_\mathrm{post}(n) > \epsilon$, where $\epsilon$ is the lower bound on the probability that $n$ single photons are successfully detected at the output of the \BS device. This puts a lower bound on the required photodetection efficiency of
\begin{equation}
\eta = \sqrt[n]{\epsilon}.
\end{equation}

\begin{figure}[!htb]
\centering
\includegraphics[width=0.5\columnwidth]{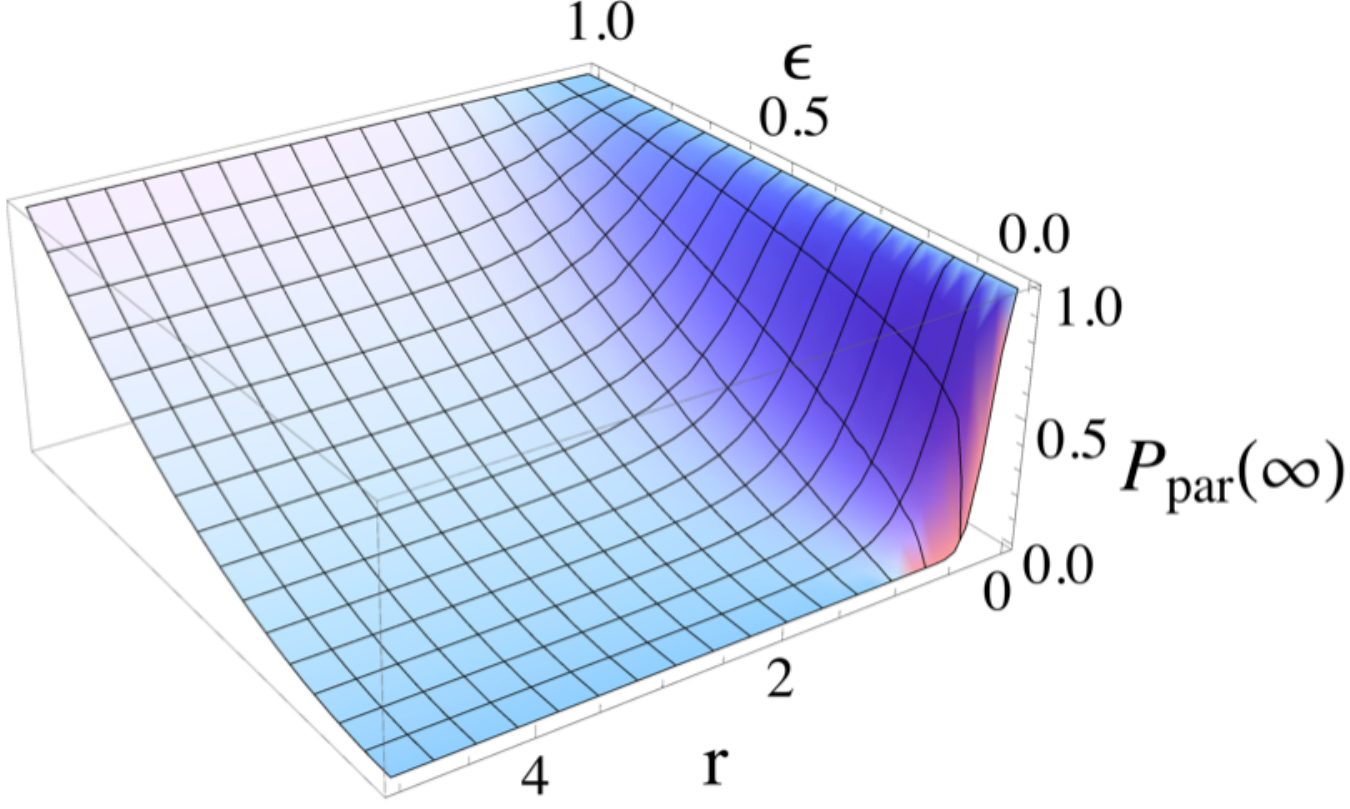}
\caption{(Color online) Probability that we are sampling from the correct input distribution in the limit of large $n$, obtained from Eq.~(\ref{eq:Ppar}), plotted against the SPDC squeezing parameter $r$ and the lower bound on the probability that $n$ single photons are successfully detected at the output of the \BS device $\epsilon$.} \label{fig:Ppar}
\end{figure}

Next we will assume that all photodetectors in the architecture have the same efficiency. Thus, we obtain that the probability of correctly preparing all $n$ photons via post-selected SPDC is,
\begin{equation}
P_\mathrm{par}(n) = [1 + (\sqrt[n]{\epsilon}-1)\,\mathrm{tanh}^{2}r]^{2n}.
\end{equation}
In the limit of large $n$ (i.e. large instances of \BS), this asymptotes to,
\begin{equation} \label{eq:Ppar}
\lim_{n\to \infty} P_\mathrm{par}(n) = \epsilon^{2 \,\mathrm{tanh}^2 r}.
\end{equation}
For small $r$ this approaches unity, and in the limit of large $r$ to $\epsilon^2$. Thus, for $\epsilon = 1/2$, in the worst case scenario, we are sampling from the correct distribution in $1/4$ of the trials. This is shown in Fig.~\ref{fig:Ppar}.

Equation~(\ref{eq:Ppar}) specifies the asymptotic probability of sampling from the correct input distribution, given that post-selection was successful. For small squeezing we sample from the correct input distribution most of the time, due to the lower probability of higher-order terms occurring. Thus, for experimentally realistic SPDC sources, provided that detector efficiencies are sufficiently high to enable post-selection, we have a high likelihood of correct \BS and SPDC photon-number errors are negligible.

Conversely, we could require that $P_\mathrm{par} > \epsilon'$ from Eq. (\ref{eq:P_par}), where $\epsilon'$ is the lower bound on the probability that a single photon is correctly heralded $n$ times in parallel before entering the multiplexer. Solving this for $\eta$ yields,
\begin{equation} \label{eq:eta2}
\eta = 1 + (\sqrt[2n]{\epsilon'} - 1)\,\mathrm{coth}^2 r.
\end{equation}
From Eq. (\ref{eq:SPDCexp_scaling}) we obtain an expression for the post-selection probability under the condition that we require a certain fidelity on the SPDC heralding,
\begin{equation} \label{eq:Ppost}
P_\mathrm{post}(n) = [1 + (\sqrt[2n]{\epsilon'} - 1)\,\mathrm{coth}^2 r]^n.
\end{equation}

\begin{figure}[!htb]
\centering
\includegraphics[width=0.4\columnwidth]{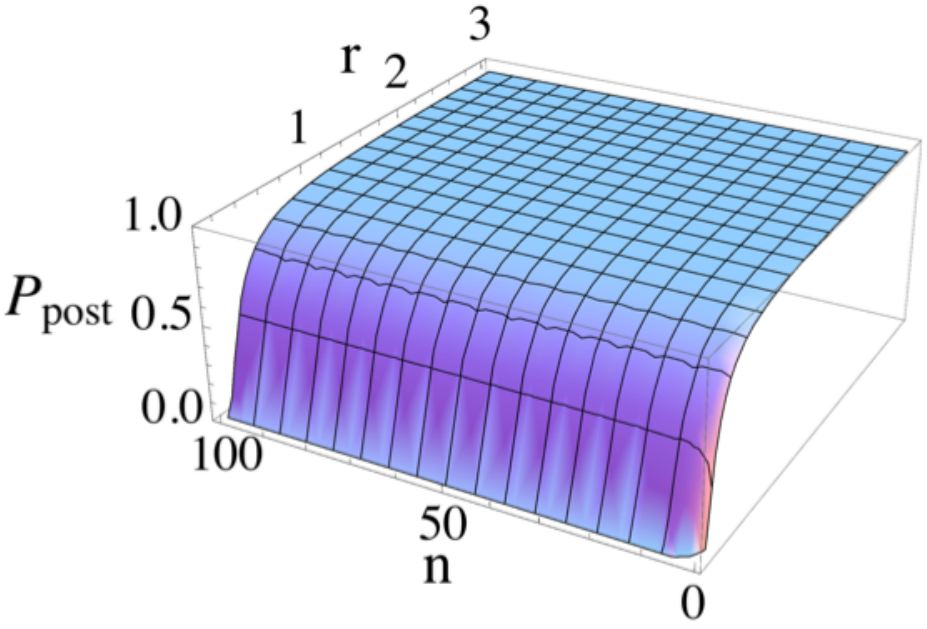}
\caption{(Color online) The post-selection probability $P_\mathrm{post}$ from Eq.~(\ref{eq:Ppost}) presented as a function of the squeezing parameter $r$ and $n$ single photons being correctly heralded in parallel before entering the multiplexer. Here we assume a fidelity of $\epsilon'=0.9$.} \label{fig:Ppost}
\end{figure}

Fig.~\ref{fig:Ppost} illustrates $P_\mathrm{post}(n)$ as a function of the squeezing parameter $r$ and the number of successfully routed photons $n$. We observe that for large $n$, post-selection is highly likely to succeed if the SPDC state preparation was successful to within error $\epsilon'=0.9$. We observe that in the limit of large $n$ and experimentally realistic values of \mbox{$r\approx1/2$}, \BS using \mbox{$N\gg n$} SPDC sources is scalable. 

%
%

\section{Summary}

We first explained why spontaneous parametric down conversion (SPDC) is of interests for linear optical applications such as \BS in section \ref{sec:SPDCmotivation}. Next, in section \ref{sec:SPDCSPDC}, we explain spontaneous parametric down conversion. We presented a simple architecture for \BS via multiplexed SPDC sources in section \ref{sec:SPDCarcitecture}. We demonstrated that the SPDCs do not limit the scalability of the architecture. Rather, the single-photon detectors, whose efficiencies must increase exponentially with input photon number, limit the scalability. That is provided that detection efficiencies are sufficiently high to enable post-selected operation (section \ref{sec:SPDCscalability}), the SPDCs will produce Fock states of sufficient fidelity to implement correct \BS with high probability. Conversely, if detection efficiencies are sufficiently high to guarantee SPDC heralding with high fidelity, post-selection will succeed with high probability.

Thus, SPDC sources are a viable photon source technology for future large-scale demonstrations of \BStwo. Additionally, existing SPDC sources will likely need significant improvement to increase squeezing purity and mode-matching. While post-selection guarantees correct operation of a \BS device, the required detection efficiencies scale unfavourably. Thus, future work should further address the question as to whether lossy \BS is computationally hard \cite{bib:RohdeRalphErrBS}, as this could significantly reduce physical resource requirements. Other error models, such as mode-mismatch \cite{bib:RohdeLowFid12}, should also be investigated further.

The analysis presented could be applied to other post-selected linear optics protocols employing SPDCs as heralded Fock state sources.

\begin{savequote}[45mm]
Life is extraordinary don't let it be ordinary!
\qauthor{Keenan Crisp}
\end{savequote}

\chapter{Scalable \BS with Time-Bin Encoding Using a Loop-Based Architecture} \label{Ch:FiberLoop}

\section{Synopsis}
It was recently shown by Motes, Gilchrist, Dowling \& Rohde \cite{bib:motes2014scalable} that a time-bin encoded fiber-loop architecture can implement an arbitrary passive linear optics transformation and can perform arbitrarily scalable \BStwo. We being by motivating this work in section \ref{sec:FLmotivation}. The architecture, as presented in section \ref{sec:FLarchitecture}, has fixed experimental complexity, irrespective of the size of the desired interferometer, whose scale is limited only by fiber and switch loss rates. The architecture employs time-bin encoding, whereby the incident photons form a pulse train, which enters the loops. Dynamically controlled loop coupling ratios allow the construction of the arbitrary linear optics interferometers required for \BStwo. The architecture employs only a single point of interference and may thus be easier to stabilize than other approaches. The scheme has polynomial complexity and could be realized using demonstrated present-day technologies. In section \ref{sec:FLsimplicity} we discuss a simplification to the architecture whereby we allow for each instance of the inner loop to have a fixed beam-splitter ratio. In section \ref{sec:FLdiscussionofarchitecture} we discuss the advantages of our architecture over other implementations. 

The original work showed the case of an ideal scheme whereby the architecture has no sources of error \cite{bib:motes2014scalable}. In any realistic implementation, however, physical errors are present, which corrupt the output of the transformation. We later   investigated the dominant sources of error in this architecture \cite{bib:motes15} --- loss and mode-mismatch --- which are presented in section \ref{sec:LossErrors} and \ref{sec:ModeMismatch} respectively and consider how it affects the \BS protocol, a key application for passive linear optics. For our loss analysis we consider two major components that contribute to loss --- fiber and switches --- and calculate how this affects the success probability and fidelity of the device. Interestingly, we find that errors due to loss are not uniform (unique to time-bin encoding), which asymmetrically biases the implemented unitary. Thus, loss necessarily limits the class of unitaries that may be implemented, and therefore future implementations must prioritise minimising loss rates if arbitrary unitaries are to be implemented. Our formalism for mode-mismatch is generlized to account for various phenomenon that may cause mode-mismatch, but we focus on two --- errors in fiber-loop lengths, and time-jitter of the photon source. These results provide a guideline for how well future experimental implementations might perform in light of these error mechanisms. If any experimentalist would like to implement this architecture I have code that can be reconfigured to analyse more specific error models so feel free to contact me. In section \ref{sec:FLerrorModels} we discuss some realistic parameters for losses and mode-mismatch that are currently the best achieved experimentally.

\section{Motivation} \label{sec:FLmotivation}


The remaining central challenge in \BS is constructing linear optics networks $\hat{U}$. It was shown by Reck \emph{et al.} \cite{bib:Reck94} that arbitrary networks of this form can be decomposed into a sequence of \mbox{$O(n^2)$} beamsplitters. In present-day experiments this type of decomposition is implemented using waveguides or discrete optical elements, but using these spatial techniques might require thousands of optical etchings in the waveguide or thousands of discrete optical elements, which must all be simultaneously aligned, so constructing the required linear optical interferometer is challenging. Two demonstrated ways to overcome the alignment problem are to use the time-bin encoded scheme by Motes, Gilchrist, Dowling \& Rohde (MGDR) \cite{bib:PRLFiberLoop} or time-dependent dispersion techniques as presented by Pant \& Englund \cite{bib:pant2015high}. Both methods do away with the hundreds or perhaps thousands of optical elements, requiring only a single pulsed photon-source and a single time-resolved photo-detector. An attractive feature of the former architecture is that there is only a single point of interference, and may therefore be much easier to align than conventional approaches. Additionally, the experimental complexity of these schemes are fixed, irrespective of the size of the desired interferometer. Although the fiber-loop scheme was initially presented for the purposes of \BStwo, Rohde recently demonstrated that with minor modifications the scheme can be made universal for quantum computing \cite{bib:RohdeUniversal}. Here, however, we will focus on the application of this scheme to \BStwo, or purely passive linear optics applications more generally.

MGDR originally showed that, using this fiber loop architecture, arbitrary linear optics transformations can be implemented on a pulse-train of photons \cite{bib:motes2014scalable}. However, the original work assumes that the architecture has no sources of error. When errors are present the scheme no longer implements an arbitrary unitary transformation, but is constrained by the error model. In followup work we analyse in detail various sources of error in the MGDR protocol \cite{bib:motes15}. Specifically, we analyse the effects of lossy elements in the architecture and the effects of mode-mismatch caused by imperfect fiber-loop lengths and time-jitter in the source. These effects accommodate the main challenges facing future experimental implementation \BStwo. 

\section{Fiber-loop Architecture} \label{sec:FLarchitecture}

\begin{figure}[!htb]
\centering
\includegraphics[width=0.45\columnwidth]{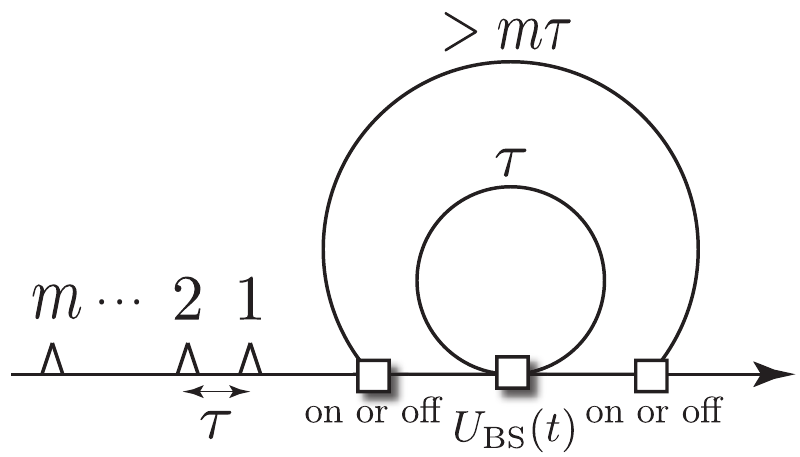}
\caption{The complete fiber-loop architecture fed by a pulse-train of photonic modes, each separated in time by $\tau$. The squares represent optical switches. A single length $\tau$ inner fiber-loop is embedded inside a length \mbox{$>m\tau$} outer fiber-loop. The outer loop allows an arbitrary number of inner loops to be applied consecutively. When \mbox{$m-1$} inner loops are implemented this architecture realises an arbitrary unitary transformation on $m$ modes given that no loss is present.} \label{fig:full_architecture}
\end{figure}

In this architecture, shown in Fig. \ref{fig:full_architecture}, a pulse-train of photonic modes consisting of, in general, Fock states and vacuum, are each separated by time $\tau$ and sent into an embedded fiber-loop. The $i$th time-bin corresponds to the $i$th mode in a conventional spatially-encoded scheme. 

This architecture assumes lossless components and perfect mode-matching at the central beamsplitter. In any realistic implementation this will not be the case, which we consider in sections \ref{sec:LossErrors} and \ref{sec:ModeMismatch}. Next, to help better understand the architecture we break down the architecture into its components by discussing what happens in a single inner loop and how multiple iterations of the single loop can implement arbitrary unitary transformation and thus \BStwo.

\subsection{Single Inner Loop} \label{sec:FLsingleinnerloop}

\begin{figure}[!htb]
\centering
\includegraphics[width=0.45\columnwidth]{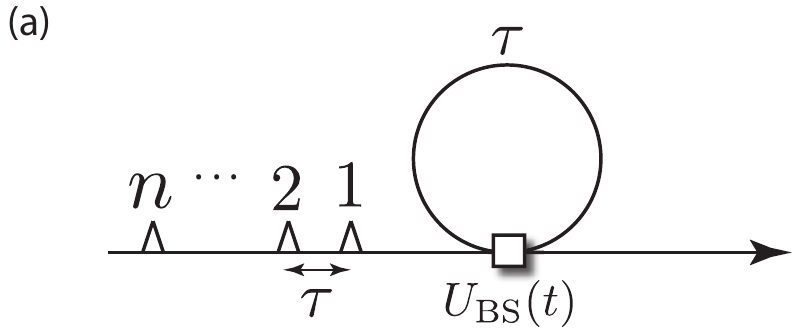}
\includegraphics[width=0.45\columnwidth]{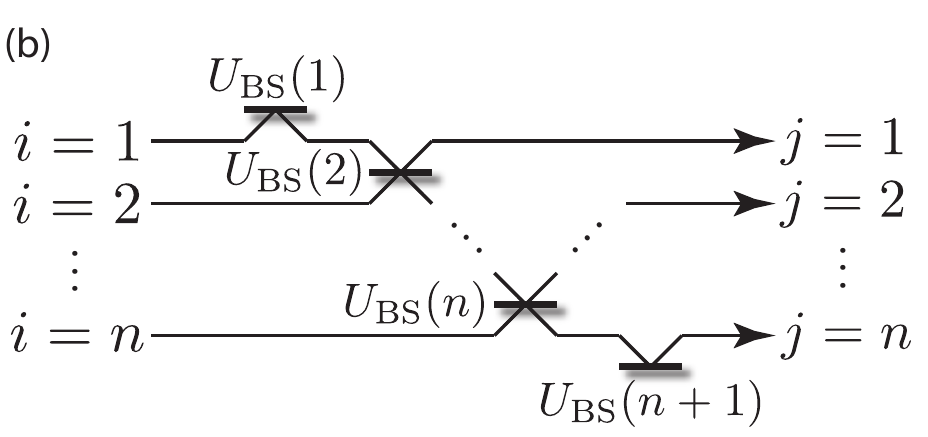} 
\caption{(a) A fiber loop fed by a pulse train of single photons, each separated in time by $\tau$. The box represents a dynamically controlled, variable reflectivity beamsplitter. The switching time of the beamsplitter must be less than $\tau$ to allow each time-bin to be individually addressed. (b) Expansion of the fiber loop architecture into its equivalent beamsplitter network.} \label{fig:singleloop}
\end{figure}

We begin by triggering a single photon source at time intervals $\tau$ (the source's repetition rate), which prepares a pulse train of $n$ single photons across a length of fiber. The first step in our architecture is to propagate the pulse train through a fiber loop with dynamically controlled coupling ratios, as shown in Fig.~\ref{fig:singleloop}(a). 
Each pulse takes time $\tau$ to traverse the inner loop so that it will interfere with the next time-bin at the central beamsplitter. Between each pulse a dynamically controlled beamsplitter, $\hat{U}_{\mathrm{BS}}(t)$ of the form, 
\begin{eqnarray} \label{eq:VarBS}
\hat{U}_{\mathrm{BS}}(t) &=& \left( \begin{array}{cc}
u_{1,1}(t) & u_{1,2}(t)  \\
u_{2,1}(t) & u_{2,2}(t) 
\end{array} \right),
\end{eqnarray}
where $\hat{U}_\mathrm{BS}$ is an arbitrary, time-dependent SU(2) operation, is applied at time $t$, which splits the incident field into a component entering the loop and a component exiting the loop. Here $u_{i,j}$ is the amplitude of input mode $i$ reaching output mode $j$. \mbox{$i=1$} (\mbox{$i=2$}) represents the mode entering from the source (inner loop), and \mbox{$j=1$} (\mbox{$j=2$}) represents the mode exiting the loop (entering the loop). When a mode enters the loop it progresses to the next time-bin. The component entering the loop takes time $\tau$ to transverse the loop such that it coincides with the subsequent pulse. In order for the first photon to interfere with every photon pulse it will traverse this loop \mbox{$m$} times. The second photon will traverse the loop \mbox{$m-1$} times and so on. The dynamics of photons propagating through the loop architecture may be `unravelled' into an equivalent series of beamsplitters acting on spatial modes, as shown in Fig.~\ref{fig:singleloop}(b). This elementary network is the basic building block employed by our architecture.

The boundary conditions of the protocol are that the first time-bin is coupled completely into the inner loop and the last time-bin coupled completely out of the inner loop (after it traverses the inner loop once), such that the implemented unitary is bounded as an \mbox{$m\times m$} matrix, where $m$ is the length of the pulse-train. This can be obtained with, 
\begin{equation} \label{eq:VarBSboundary}
\hat{U}_{\mathrm{BS}}\left(1\right) = \hat{U}_{\mathrm{BS}}\left(m+1\right)=\left( \begin{array}{cc}
0 & 1  \\
1 &0 
\end{array} \right),
\end{equation}
where $U_\mathrm{BS}(i)$ is the unitary associated with the central beamsplitter at time $i$. 

After the entire pulse-train exits the inner loop the unitary map $\hat{V}$ is implemented, 
\begin{equation} \label{eq:UnitaryMap}
V_{i,j} = \left\{ \begin{array}{ll}
 0 & i>j+1 \\
 u_{1,1}(i) & i=j+1 \\
 u_{1,2}(i)u_{2,1}(j+1) \prod_{k=i+1}^{j}{u_{2,2}}(k) & i<j+1
 \end{array} \right. ,
\end{equation}
where \mbox{$i\in\{1,m\}$} and \mbox{$j\in\{1,m\}$} represent input and output modes respectively. We see that the probability of finding a photon in the $j$th mode decays exponentially with $j$. In quite special cases, such as the reflectivity going to zero in the beam-splitter, this probability would not decay exponentially. When $i>j+1$ the $i$th input mode does not have access to the $j$th output mode so this matrix element is zero. When $i=j+1$ the modes do not enter the inner loop and travel strait through to the detector picking up a factor of $\gamma_{2,1}(i)$. When $i<j+1$ the modes traverse the loop $j-i+1$ times. Note that we have employed a slightly different, but equivalent, indexing convention to the original MGDR proposal. 

Evidently, the network shown in Fig.~\ref{fig:singleloop}(b) is not sufficient for universal linear optics networks as it contains many zero elements. To make the scheme universal we must show that the ingredients necessary to perform a full Reck \emph{et al.} type decomposition are available as we show in the next section. 

\subsection{Multiple Inner Loops} \label{sec:FLmultipleloops}

The inner loop alone cannot implement an arbitrary unitary transformation so additional loops are required. The outer loop allows for an arbitrary number of applications of the  inner loop to be implemented. The net unitary $\hat{U}$ after $L$ consecutive inner loops becomes, 
\begin{equation}
\hat{U}=\prod_{l=1}^{L}\hat{V}(l),
\end{equation}
where $l$ denotes the $l$th iteration of the inner loop and $\hat{V}$ is given by Eq. (\ref{eq:UnitaryMap}). The pulse-train will traverse the inner loop \mbox{$L=m-1$} times and the outer loop \mbox{$m-2$} times for an arbitrary unitary transformation to be implemented. The outer loop must have round trip time \mbox{$>m\tau$} so that the pulse-train does not interfere with itself for a particular instance of the inner loop $\hat{V}(l)$. The pulse-train is coupled in and out of the outer loop via on/off switches. Once the desired transformation is performed, the pulse-train exits both loops and is measured via time-resolved photo-detection, where the time-resolution of the detector must be greater than $\tau$. The $j$th time-bin at the output corresponds to the $j$th spatial mode in the standard \BS model. 

\begin{figure}[!htb]
\centering
\includegraphics[width=0.45\columnwidth]{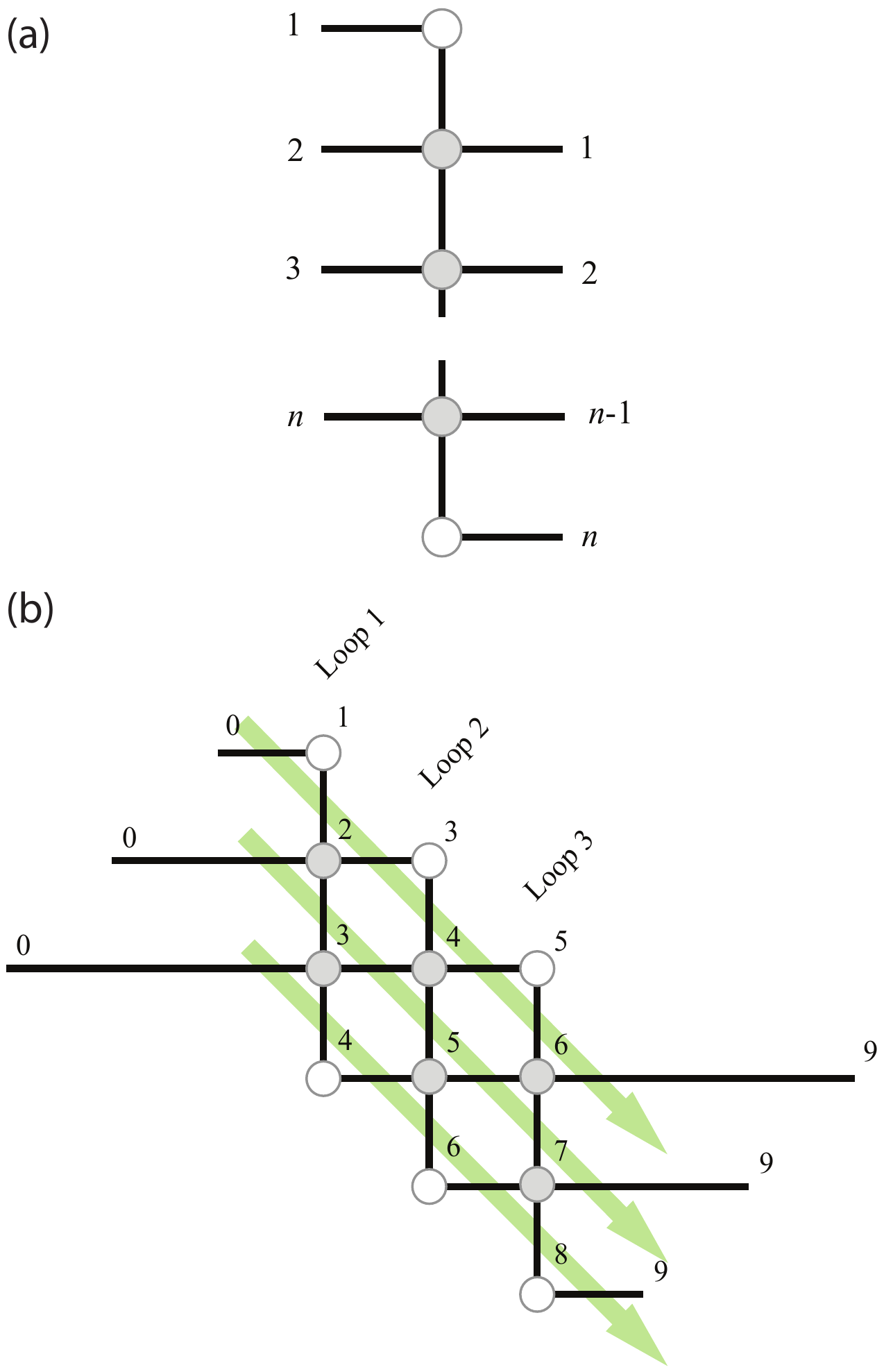}
\caption{The equivalent beam-splitter representation for the fiber loop architecture. (a) A single loop is represented for $n$ photons in the pulse train. Here the numbers represent modes. (b) The equivalent beamsplitter network of three consecutive loops with three input modes. Here the number represent time-bins.} \label{fig:AlexeiProof}
\end{figure}

To understand the equivalent beamsplitter representation of a single loop, consider Fig. \ref{fig:AlexeiProof}(a). The pulse train enters the loop, where the numbers on the left represent the corresponding time-bin. The first photon is deterministically coupled into the loop as depicted by an open circle. After the first and second photons interact some of the amplitude may escape the loop, which corresponds to the first output time-bin. The pulse train continues to interact through the loop via beamsplitter operations, which are represented as closed circles. After the $n$th photon transverses the loop any remaining amplitude deterministically leaves the loop, which corresponds to the $n$th output time-bin.

\begin{figure}[!htb]
\centering
\includegraphics[width=\imageWidth\columnwidth]{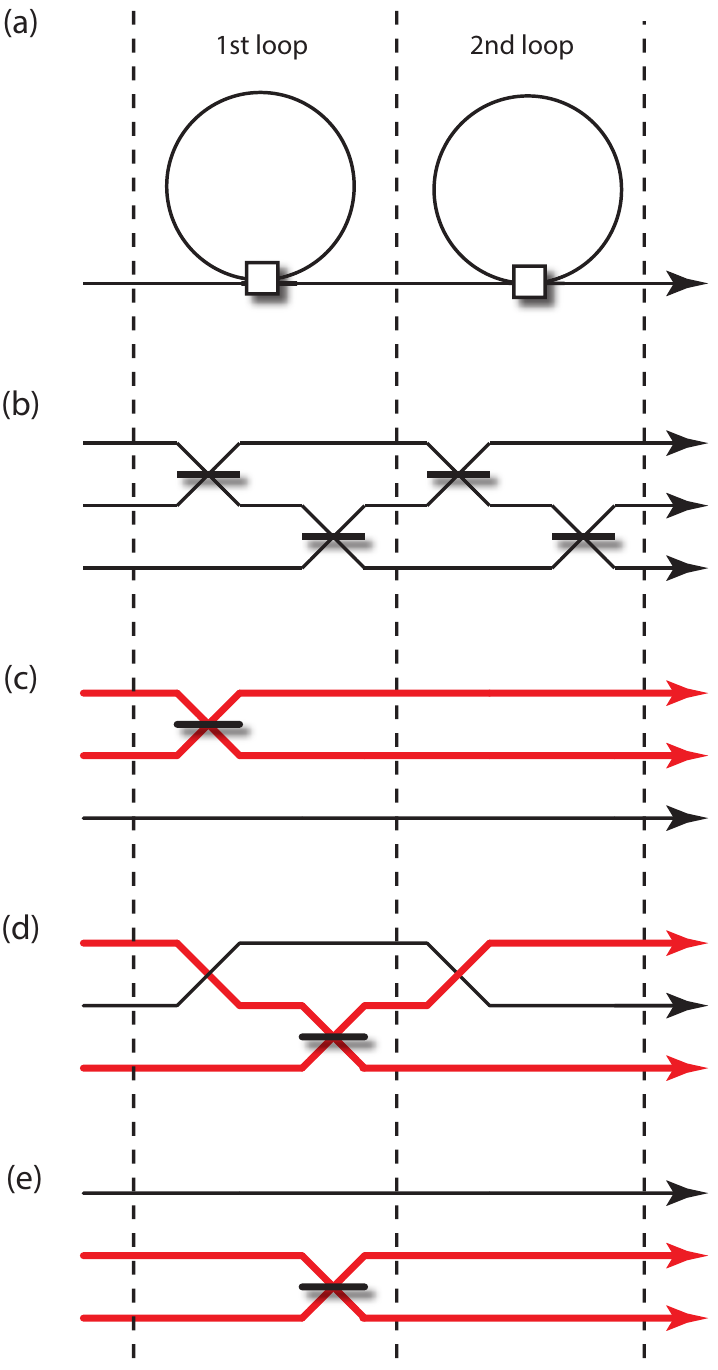} 
\caption{(a) Two consecutive fiber loops in series. (b) The equivalent beamsplitter expansion for number of modes \mbox{$m=3$}. By setting the beamsplitter ratios appropriately the two loops can implement an arbitrary beamsplitter between any pair of modes. These show the respective paths for implementing an arbitrary beamsplitter between modes one, two, and three: (c) modes 1 and 2, (d) modes 1 and 3, (e) modes 2 and 3.} \label{fig:universality}
\end{figure}

\begin{figure}[!htb]
\centering
\includegraphics[width=0.45\columnwidth]{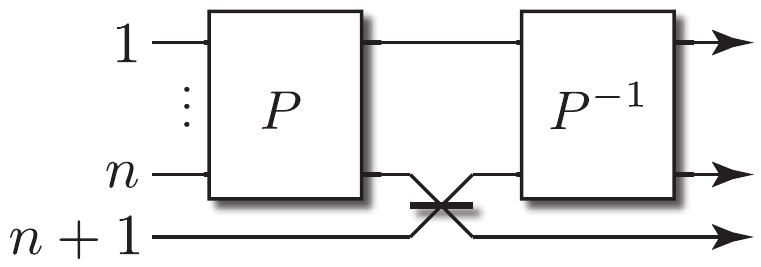} 
\caption{Generalizing the universality argument presented in Fig.~\ref{fig:universality} to arbitrary $m$. We choose a permutation that the first of the desired modes be routed to the beamsplitter, which interacts with mode \mbox{$n+1$}. Then the inverse permutation is applied, leaving us with a network that implements an arbitrary beamsplitter operation between one of the first $n$ modes and the \mbox{$(n+1)$th} mode. It follows inductively that an arbitrary beamplsitter operation can be applied between any pair of modes for any $m$.} \label{fig:induction}
\end{figure}

Now consider Fig.~\ref{fig:AlexeiProof}(b), which depicts how three consecutive loops in series with three input photons produce an equivalent beamsplitter network. The lengths of the black lines represent time in units of $\tau$. The three modes on the left represent the pulse train of photons at the input of the device at the first round-trip. The first photon reaches the first beamsplitter at \mbox{$\tau=1$}, the second photon reaches it at \mbox{$\tau=2$}, and so on. The photons travel through the fiber loop network interacting arbitrarily, which yields an arbitrary Reck \emph{et al.} style decomposition. Evidently, an $n$-mode unitary  can be built using $n$ modes and $n-1$ loops. In Figs.~\ref{fig:universality} \& \ref{fig:induction} we show an alternate proof based on an inductive argument.

We have shown that a series of consecutive fiber loops can implement an arbitrary sequence of pairwise beamsplitter operations. Next, we recognize that each of these fiber loops requires exactly the same physical resources, only differing by the switch's control sequence. We need not physically build each of these identical loops. Rather, we will embed the loop into a larger fiber loop of length \mbox{$>n\tau$}, as shown in Fig.~\ref{fig:full_architecture}. The larger loop is controlled by another two switches, which control the number of round trips in the larger loop. From the result of Reck \emph{et al.} we know that \mbox{$O(n^2)$} optical elements are required to construct an arbitrary \mbox{$n\times n$} interferometer. Thus, the number of round trips of the outer loop is $O(n^2)$. 

\section{Fixing Beam-Splitter Ratios for Simplicity} \label{sec:FLsimplicity}
An experimental simplification is when we do not require full dynamic control over the beamsplitter ratio. Although this scenario is not universal, it may be possible to construct useful classes of unitaries. We will consider the situation where the beamsplitter in each iteration of the single loop can be toggled between two settings --- completely reflective, or some other arbitrary fixed ratio. The former is required to allow that the time-bins be restricted to a finite time-window, whilst the latter implements the `useful' beamsplitter operations. We may have an arbitrary number of such loops in series, each with a potentially different fixed beamsplitter ratio.

Intuitively, we expect that a `maximally mixing' unitary (i.e. one with equal amplitudes between every input to output pair) would implement a classically hard \BS instance, as it maximizes the combinatorics associated with calculating output amplitudes. If, for example, a unitary is heavily biased towards certain output modes, or is sparse, the combinatorics are reduced. Specifically, we define a balanced unitary as, \mbox{$|U'_{i,j}|^2=1/n \,\,\forall \,\,i,j$},
such that, up to phase, all amplitudes are equal.

\begin{figure}[!htb]
\centering
\includegraphics[width=0.5\columnwidth]{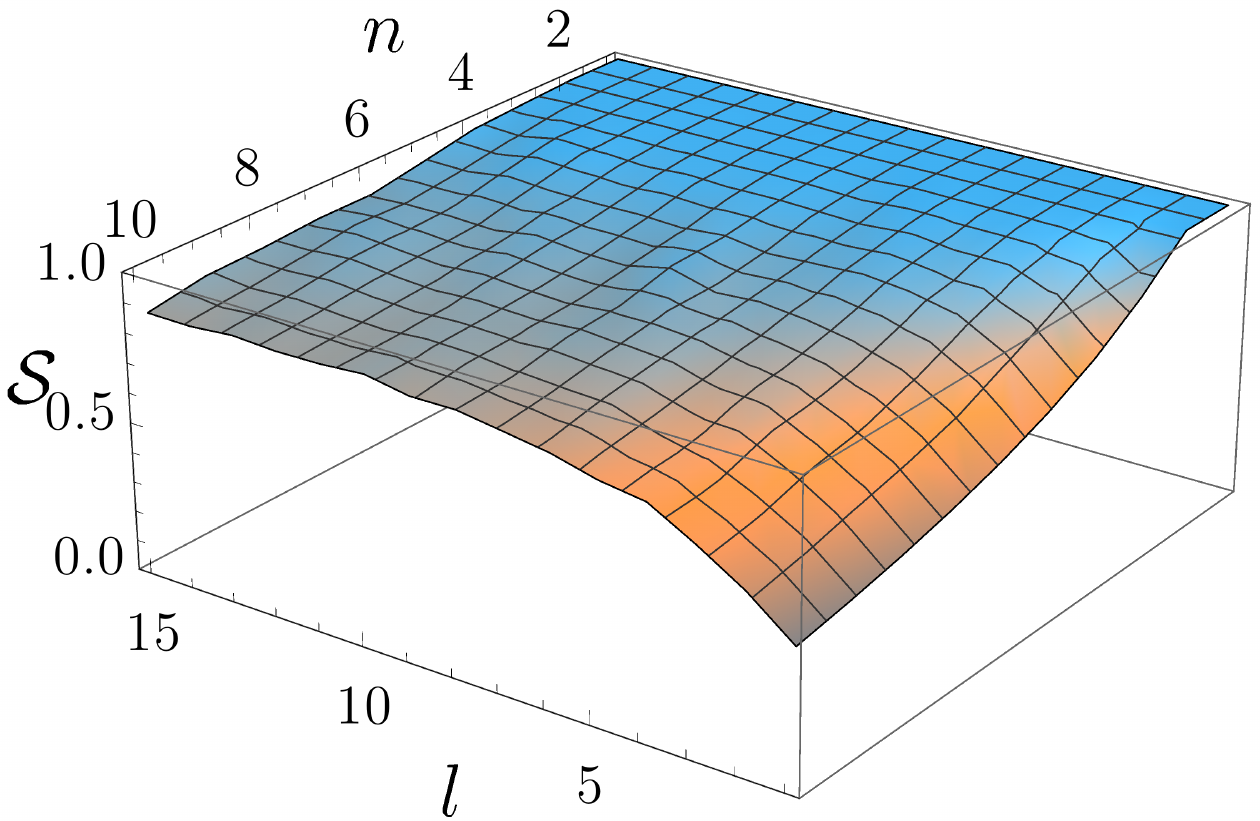} 
\caption{The maximum similarity $\mathcal{S}$ between $\hat{U}$ and the uniform unitary $\hat{U}'$ after $l$ loops with $n$ input photons. The beam-splitter ratio is fixed for each loop but independently, randomly chosen for each loop. This demonstrates that near-uniform unitaries may be constructed with sufficient loops.} \label{fig:similarity}
\end{figure}

In Fig.~\ref{fig:similarity} we take the unitary implemented by a series of $l$ fixed-ratio fiber loops, and compare it with the balanced unitary $\hat{U}'$. We characterize the uniformity of the obtained unitary using the similarity metric,
\begin{eqnarray}
\mathcal{S} &=& \max_{\hat{U}_\mathrm{BS}(t) \,\forall\, t}\left[\frac{\left(\sum_{i,j}\sqrt{|U_{i,j}|^2 \cdot |U'_{i,j}|^2}\right)^2}{\left(\sum_{i,j}|U_{i,j}|^2\right) \cdot \left(\sum_{i,j}|U'_{i,j}|^2\right)}\right] \\ \nonumber
&=& \max_{\hat{U}_\mathrm{BS}(t)\,\forall\, t}\left[\frac{1}{n^3}\left(\sum_{i,j}|U_{i,j}|\right)^2\right],
\end{eqnarray}
where we maximize $S$ by performing a Monte-Carlo simulation over different beamsplitter ratios, $\hat{U}_\mathrm{BS}$. That is, $\mathcal{S}$ tells us how close $\hat{U}$ is to uniform, with \mbox{$\mathcal{S}=1$} being completely uniform up to phase. With a sufficient number of loops in series, we obtain very high similarities, suggesting that the simplified architecture may implement hard instances of \BStwo.

\section{Discussion of Architecture} \label{sec:FLdiscussionofarchitecture}
Because there is only a single point of interference, this architecture may be significantly easier to stabilize and mode-match than conventional approaches, where \mbox{$O(n^2)$} independent beamsplitters must be simultaneously aligned and stabilized. At this point of interference, the dominant source of error will be temporal mode-mismatch \cite{bib:RohdeRalph05}, which is caused by errors in the lengths of the fiber loops, or time-jitter in the photon sources. Temporal mismatch may be regarded as a displacement in the temporal wavepacket of the photons \cite{bib:spectralStructure}. Let us assume that at each round trip the photon exiting the inner loop is mismatched by time $\Delta$. Over short time scales this yields dephasing \cite{bib:RohdeRalph06}, and over longer time scales, ambiguity as to which time-bin the photon resides in. The worst case is that a given photon undergoes temporal mismatch of magnitude $n\Delta$. Time-bin ambiguity occurs when \mbox{$n\Delta \geq \tau$}, which yields the requirement that \mbox{$n<\tau/\Delta$}. Over shorter timescales, temporal mode-mismatch is equivalent to dephasing as mismatched photons yield which-path information. This leads to the constraint that \mbox{$n\Delta \ll \sigma$}, where $\sigma$ is the width of the photons' wave packets. Thus, time-jitter or temporal mode-mismatch must be kept small relative to the scale of the photons' wavepackets. The current switching rates of state-of-the-art dynamically controlled switches is on the order of GHz \cite{bib:winzer2010spectrally, bib:schindler2014monolithic, bib:prosyk2012high, bib:PCMichaelSteel} and the temporal spacing of photons is on the order of nanoseconds. Whilst these switches are fast enough, they require additional coupling that involves high loss. This will encourage further development of these type of technologies which is also required for LOQC architectures.

Integrated waveguides are gaining popularity in photonics as they are inherently very stable. However, although interferometrically stable after the fabrication process, there are nonetheless $O(n^2)$ points of interference, which must be carefully aligned. On the other hand, the fiber-loop architecture has only a single point of interference that needs to be aligned. Another advantage of our architecture is that only one photon source (such as a quantum dot or SPDC source with high repetition rate) could be employed, whereas bulk-optics or waveguide implementations would require an array of sources operating in parallel, further reducing experimental overhead.

The experimental viability of loop-based photonic architectures was validated by recent quantum walk \cite{bib:ADZ} experiments by Schreiber \emph{et al.} \cite{bib:Schreiber12, bib:Schreiber10}, where quantum memories were implemented via delay lines in free-space. It was also shown by Donohue \emph{et al.} \cite{bib:Resch} that transmitting time-bin encoded photons in optical fibers is a robust form of optical quantum information given that the separation of time-bins is larger than the time resolution of the detector.

In principle, the fiber loops could be replaced by any quantum memory or delay line such as propagation in free-space, which would be significantly less lossy. In this case, the dominant source of loss would be in the dynamic switches, which, using present-day technology, have high loss rates.

The presented universal architecture is in principle arbitrarily scalable, provided the length of the larger loop is sufficiently high (\mbox{$>n\tau$}). However, in practice, fiber is lossy with present-day technology. If we let $\eta_\mathrm{inner}$ be the net efficiency of the inner loop (i.e. the probability that an incident photon will reach the output), and $\eta_\mathrm{outer}$ be the net efficiency of the outer loop, then the worst case net efficiency of the device is,
\mbox{$\eta_\mathrm{net} = ({\eta_\mathrm{inner}}^n \eta_\mathrm{outer})^{O(n^2)}$},
which scales exponentially with $n$. Thus, to construct large interferometers using this architecture will require exponentially low loss rates. This is also the case for conventional spatially encoded implementations. However, it was shown by Rohde \& Ralph \cite{bib:RohdeRalphAA12} that \BS might remain a computationally hard problem even in the presence of high loss rates. Other error models, such as dephasing or mode-mismatch \cite{bib:RohdeLowFid12}, exhibit similar scaling characteristics. Our architecture has the same efficiency scaling as conventional bulk-optics or waveguide implementations. If the worst-case photon efficiency (combining state preparation, evolution and photodetection) is $\zeta$, then with $n$ photons the net efficiency is lower bounded by $\zeta^n$.

Next, we discuss the major sources of error that would challenge an experimental implementation of our architecture including an anaysis of loss and mode-mismatch. 

 
\section{Loss Errors} \label{sec:LossErrors}

In an implementation of a passive linear optics network, whereby the loss between each input/output pair of modes is uniform, loss simply amounts to a reduced success probability upon post-selecting on detecting all photons. Loss in the unitary transformation is a problem but there is evidence that even lossy systems or systems with mode-mismatch are still likely hard to simulate given that the errors are sufficiently small \cite{bib:RohdeRalphErrBS, bib:RohdeLowFid12}. Recently, Aaronson and Brod investigated the complexity of \BS under photon losses \cite{bib:aaronson2016bosonsampling}. In the fiber-loop architecture, the different paths traverse the inner loop a different number of times leading to non-uniform loss. This biases the unitary transformation resulting in a unitary that is not the desired one, even after post-selecting upon measuring all photons. That is, the effects of loss cannot be simply factored out of the unitary. In some architectures, asymmetric losses may be compensated for by artificially adding losses that rebalance the circuit, at the expense of overall success probability. In the fiber-loop architecture this turns out not to be the case. 

In Sec. \ref{sec:LossMetrics} we introduce the metrics that we will use to analyse loss. In Sec. \ref{sec:innerLoopLoss} we  determine the effect of loss due to the lossy switch and lossy fiber in the inner loop. Then in Sec. \ref{sec:OuterLoopLoss} we analyse the net loss combining the inner loop losses with the outer loop losses. We denote quantities here that have loss with a prime.

\subsection{Loss Metrics} \label{sec:LossMetrics}

We consider two metrics: similarity and post-selection probability. 

\subsubsection{Similarity}

An interesting question is how small does loss need to be such that a particular unitary transformation is implemented with a particular error bar. The answer to this question is highly dependent on which unitary we wish to implement --- some unitaries will suffer more asymmetric bias than others, depending on the switching sequence that is required to implement them. Thus, the first question to ask is which unitary to consider. In the work of MGDR, a so-called `uniform' unitary was considered. This is a unitary where the amplitude (but not necessarily phases) of each element of the unitary are equal. That is, the magnitude of the amplitude between each input/output pair of modes is the same. This class of unitaries was considered as an example of `non-trivial' matrices, which uniformly mix every input mode with every output mode. However, it is still an open question as to exactly what classes of unitaries yield hard sampling problems in the context of \BStwo. We will here consider the same setting. We will explore this by using the similarity metric, $\mathcal{S'}$, which compares the implemented map with the uniform map,
\begin{eqnarray}
\mathcal{S'} &=& \max_{\hat{U}_\mathrm{BS}(t) \,\forall\, t}\left[\frac{\left(\sum_{i,j=1}^{m}\sqrt{|U_{i,j}|^2 \cdot |\mathcal{W}_{i,j}|^2}\right)^2}{\left(\sum_{i,j=1}^{m}|U_{i,j}|^2\right) \cdot \left(\sum_{i,j=1}^{m}|\mathcal{W}_{i,j}|^2\right)}\right] \nonumber \\
&=&\max_{\hat{U}_\mathrm{BS}(t)\,\forall\, t}\left[\frac{1}{m^2}\frac{\left(\sum_{i,j=1}^{m}|U_{i,j}|\right)^2}{\sum_{i,j=1}^{m}|U_{i,j}|^2}\right],
\end{eqnarray}
where $\mathcal{W}_{i,j}$ is an \mbox{$m\times m$} uniform unitary given by \mbox{$|\mathcal{W}_{i,j}|^2=1/m$}. $\mathcal{S'}$ is maximised by performing a Monte-Carlo simulation over different beamsplitter ratios so as to find the optimal switching sequence to make the map as uniform as possible. In this analysis we use $\mathcal{S'}$ instead of $\mathcal{S}$ to denote the similarity metric with loss present. 

\subsubsection{Post-selection Probability}

Another interesting question is how the probability of post-selecting upon all $n$ photons is affected by loss, i.e the total success probability of the device. This is of especial importance experimentally, as it directly translates to count rates. The post-selection probability of detecting all $n$ photons at the output is,
\begin{equation} 
\mathcal{P}_{\mathrm{S}} = \prod_{i=1}^{m}\left(\sum_{j=1}^{m}|U_{i, j}|^{2}\right)^{k_i},
\label{eq:Psel}
\end{equation}
where $\{k\}$ is an integer string of length $m$ that represents a known input configuration of photons and $k_i$ is the number of photons in mode $i$. This equation is intuitively derived as follows. For a single photon the probability of entering mode $i$ and exiting mode $j$ is $|U_{i,j}|^2$. Then the total probability that the $i$th photon exits the architecture is the sum of this over all $j$ possible output ports, i.e. \mbox{$\sum_{j=1}^{m}|U_{i,j}|^2$}. Thus the probability of detecting all $n$ photons at the output beginning in a particular configuration $\{k\}$ is the product of this probability over all modes $i$ where \mbox{$k_i\neq0$}, as per Eq. (\ref{eq:Psel}). This generalisation, by allowing arbitrary strings $\{k\}$, allows for implementations such as randomised \BS as described by Lund \emph{et al.} \cite{bib:Lund13}. 

With losses present, $\hat{U}$ is in general no longer unitary. Rather, it is a mapping of input-to-output amplitudes, and will not be normalised. When there is no loss in the architecture \mbox{$\mathcal{P}_\mathrm{S}=1$}, and with loss strictly \mbox{$\mathcal{P}_\mathrm{S}<1$}, dropping exponentially with the number of photons. Implementing the required \mbox{$m-1$} loops will have exponentially worse loss than a single loop. 



\subsection{Inner Loop Loss} \label{sec:innerLoopLoss}

\begin{figure}[!htb]
\centering
\includegraphics[width=0.45\columnwidth]{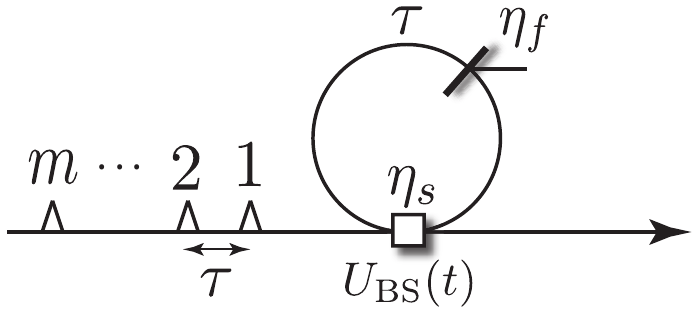}
\caption{A lossy inner fiber-loop fed by a pulse-train of photonic modes, each separated in time by $\tau$. We model the loss of the loop with a beamsplitter of reflectivity $\eta_f$ and the loss of the switch with an efficiency $\eta_s$. Each mode experiences different amounts of loss, i.e. the first mode traverses the loop up to $m$ times, the second up to \mbox{$m-1$} times, \dots, and the $m$th mode at most once.} \label{fig:singlelooperror}
\end{figure}

We will model loss inside of the inner fiber-loop with a beamsplitter of reflectivity $\eta_f$ and loss in the switch as $\eta_s$ as shown in Fig. \ref{fig:singlelooperror}. When \mbox{$\eta_f=\eta_s=1$} the device has perfect efficiency. Before and after the inner loop the loss experienced by each mode in the fiber is negligible, since it may be arbitrarily short. Taking these losses into account, the implemented map of Eq. (\ref{eq:UnitaryMap}) becomes,
\begin{equation} \label{eq:lossyUnitaryMap}
V'_{i,j} = \eta_s\left\{ \begin{array}{ll}
 0 & i>j+1 \\
 u_{1,1}(i) & i=j+1 \\
 \eta^{j-i+1}u_{1,2}(i)u_{2,1}(j+1)\cdot \\ \prod_{k=i+1}^{j}{u_{2,2}}(k) & i<j+1
 \end{array} \right. ,
\end{equation}
for a given loop, where \mbox{$\eta=\eta_f\eta_s$}. Note that this mapping is no longer a unitary matrix when \mbox{$\eta_f<1$} or \mbox{$\eta_s<1$}. This uneven distribution of losses in the input-to-output mapping causes a skew in the matrix which prevents it from implementing the desired unitary transformation, even after post-selection. 

\subsection{Outer Loop Loss} \label{sec:OuterLoopLoss}

\begin{figure}[!htb]
\centering
\includegraphics[width=0.5\columnwidth]{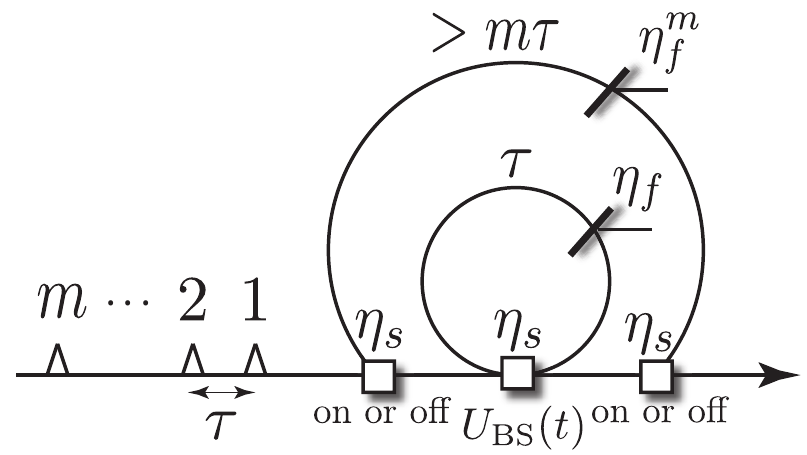}
\caption{The full architecture which implements the lossy transformation $\hat{U}'$. Each mode experiences ${\eta_f}^m$ loss per outer loop since they each take time $m\tau$ to traverse the outer loop. For an arbitrary unitary to be implemented in the ideal case the photons will traverse the outer loop $L-1$ times. This yields a net fiber loss from the outer loop of ${\eta_f}^{m(L-1)}$ that can be factored out of $\hat{U}'$, since it affects all paths equally. The net switch loss from the outer two switches is ${\eta_s}^{2(L-1)}$ and can also be factored out of $\hat{U}'$. The losses within the inner loop, on the other hand, affect different paths differently, and in general cannot be factored out.} \label{fig:FullLossError}
\end{figure}

In the full fiber-loop architecture $L$ inner loops are implemented via \mbox{$L-1$} round-trips of the outer loop, before being coupled out to the detector. This architecture can implement an arbitrary unitary transformation when \mbox{$L=m-1$} if there are no errors present. The outer loop and outer switches cause a uniform loss on the entire pulse-train, since every path through the interferometer passes through these elements the same number of times. Hence, these factor out of $\hat{U}$. The full lossy transformation that occurs is then, 
\begin{equation}
\hat{U}'={\eta_{f}}^{m(L-1)}{\eta_{s}}^{2(L-1)}\prod_{l=1}^{L}\hat{V}'(l),
\end{equation}
where $L=m-1$ if an arbitrary transformation is desired, and $\hat{V}'$ is given by Eq. (\ref{eq:lossyUnitaryMap}). 
The ${\eta_{f}}^{m(L-1)}$ occurs because the pulse-train traverses an $m\tau$ length of fiber in the outer loop $L-1$ times (i.e $\eta_f$ can be regarded as the efficiency per unit of fiber of length $\tau$), and the ${\eta_{s}}^{2(L-1)}$ occurs because the pulse-train passes through the two outer switches \mbox{$L-1$} times. Fig. \ref{fig:FullLossError} shows the entire architecture with these loss errors. For an example of loop bias due to loss see App. \ref{app:MultipleLoopBiasExample}. Extending from this loop bias example we generalize the loss matrix denoted as $\hat{\mathcal{L}}$, which represents the accumulation of losses in the fiber-loop architecture, as a function of the number of loops $L$ for an arbitrarily sized \mbox{$m\times m$} transformation,
\begin{equation} \label{eq:lossyMatrix}
\mathcal{L}_{i,j}(L) = \eta_s^L \eta^{L+j-i},
\end{equation}
again where $\eta=\eta_f\eta_s$. Now the lossy map $\hat{U}'$ may be written as an element wise product of the ideal unitary $\hat{U}$ and the loss matrix $\hat{\mathcal{L}}$,
\begin{eqnarray}
\hat{U}' &=& \hat{U}\circ\hat{\mathcal{L}}.
\end{eqnarray}
Elements of $\hat{\mathcal{L}}$ that have no losses in them due to input modes not reaching output modes when $L<m-1$ will be accounted for appropriately when $\hat{\mathcal{L}}$ is multiplied by $\hat{U}$ by making the cooresponding matrix element in $\hat{U}'$ go to zero.

\begin{figure}[!htb]
\centering
\subfloat[Part 1][]
{\includegraphics[width=0.5\columnwidth]{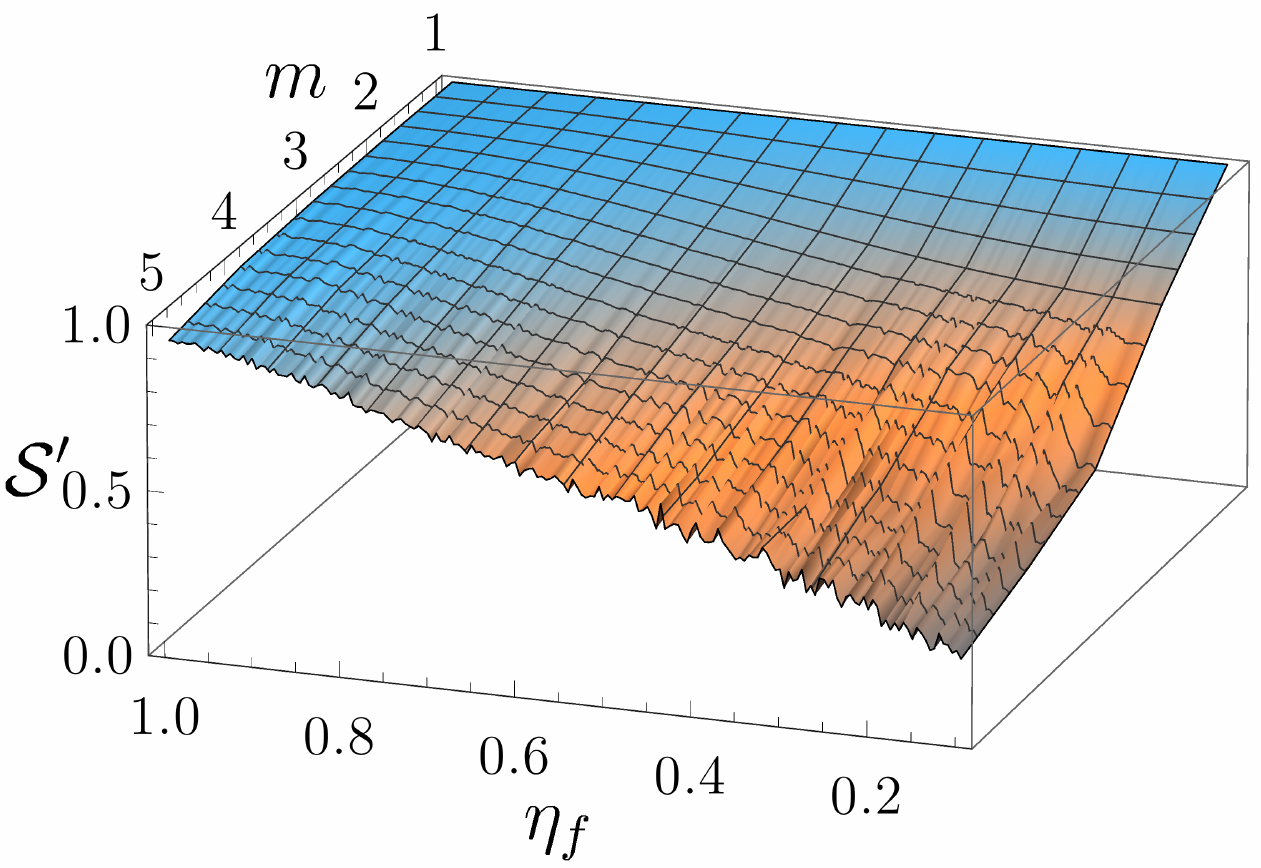} \label{fig:SL}}
\subfloat[Part 2][]
{\includegraphics[width=0.5\columnwidth]{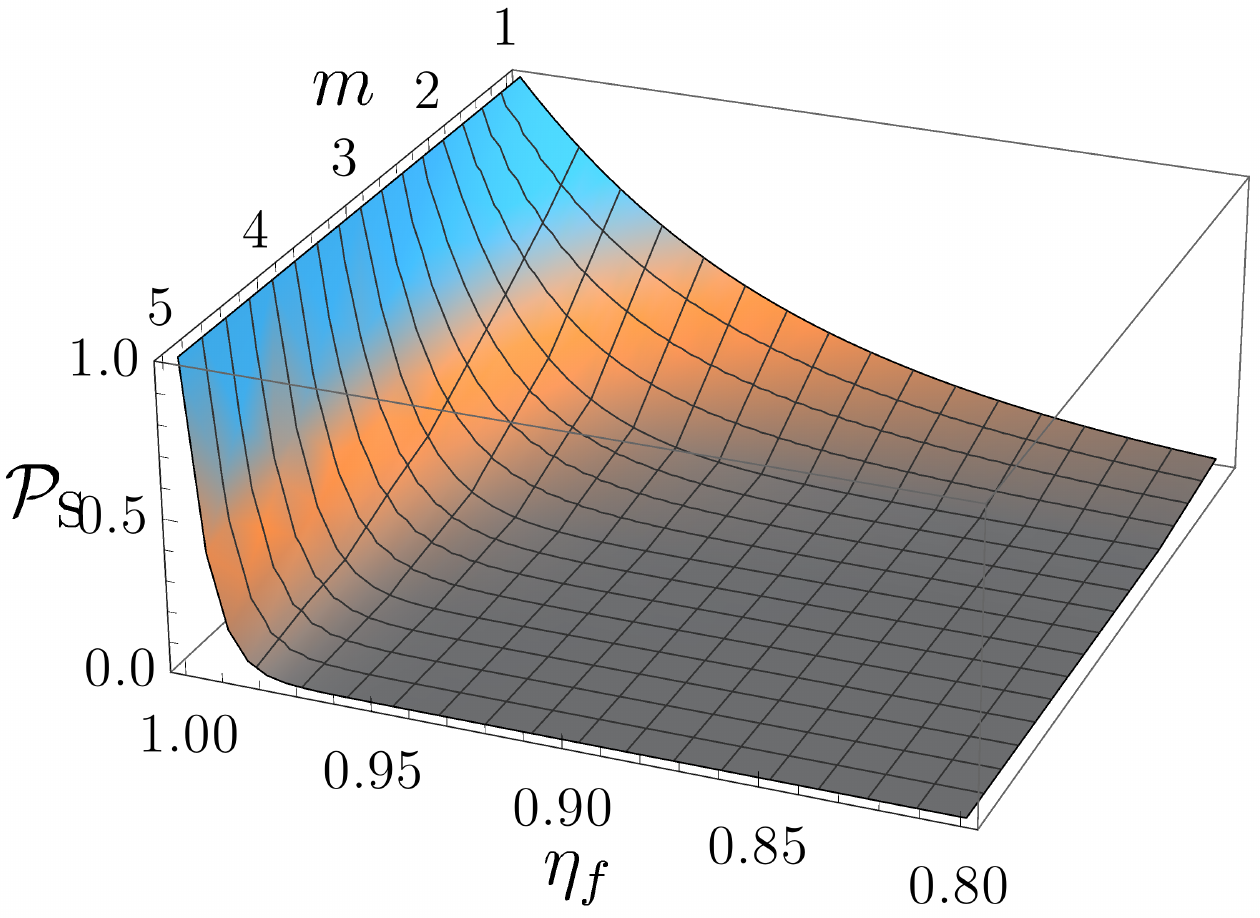} \label{fig:PSL}}
\caption{(a) Similarity with noise $\mathcal{S'}$ versus mode/photon number \mbox{$m=n$} and loop efficiency $\eta_f$ for \mbox{$m-1$} loops. The map remains similar to the uniform unitary for low loss rates, implying that non-trivial unitary transformations may be implemented. \mbox{$m-1$} loops are considered because this is the number of loops required to implement an arbitrary unitary transformation in the lossless case. The wiggles are due to sampling noise. (b) Post-selection probability $\mathcal{P}_\mathrm{S}$ versus mode/photon number \mbox{$m=n$} and loop efficiency $\eta_f$ for \mbox{$m-1$} loops. These two plots are related in that each point in $\mathcal{P}_\mathrm{S}$ was calculated from the switching sequence $\hat{U}'$ corresponding to that which maximises $\mathcal{S'}$. In both (a) and (b) the data was averaged over 1750 Monte-Carlo iterations and we let \mbox{$\eta_s=1$}, i.e the switches are ideal but the fibers are not.}
\label{fig:SPSL}
\end{figure}

In Fig. \ref{fig:SPSL}(a) we show how the optimised similarity with the uniform distribution varies with $\eta_f$ and $m$ for \mbox{$L=m-1$} inner loops, one photon in all $m$ modes, and \mbox{$\eta_s=1$}. With low loss rates (\mbox{$\eta_f \approx 1$}) the implemented unitary remains highly uniform. However, with several loops the success probability of detecting all $n$ photons at the output decays exponentially as shown is Fig. \ref{fig:SPSL}(b). For these plots the randomly generated $\hat{U}'$ that maximises $\mathcal{S}$ for each $\eta_f$ and $m$ is used to calculate the corresponding $\mathcal{P}_\mathrm{S}$. 

\begin{figure}[!htb]
\centering
\subfloat[Part 1][]
{\includegraphics[width=0.5\columnwidth]{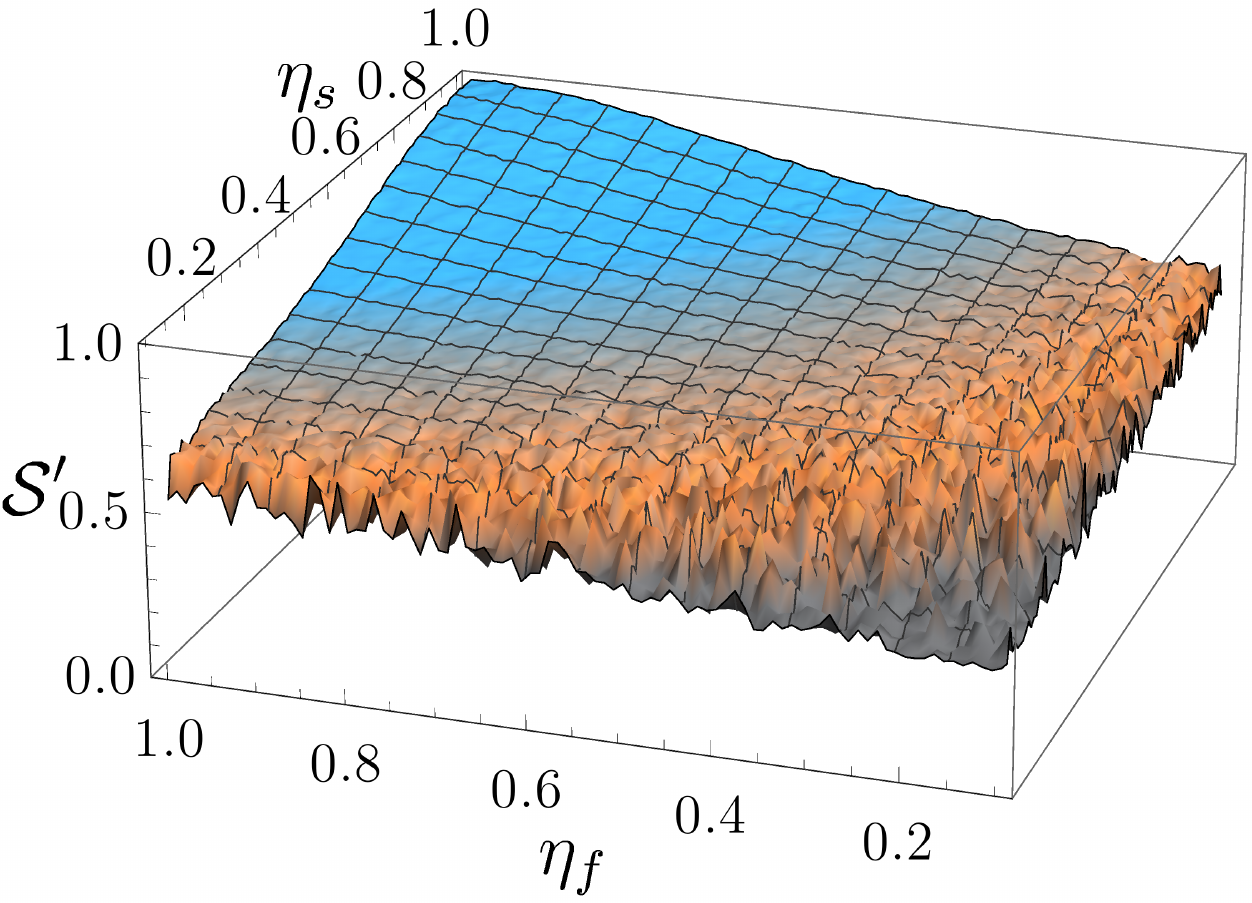} \label{fig:SLL}}
\subfloat[Part 2][]
{\includegraphics[width=0.5\columnwidth]{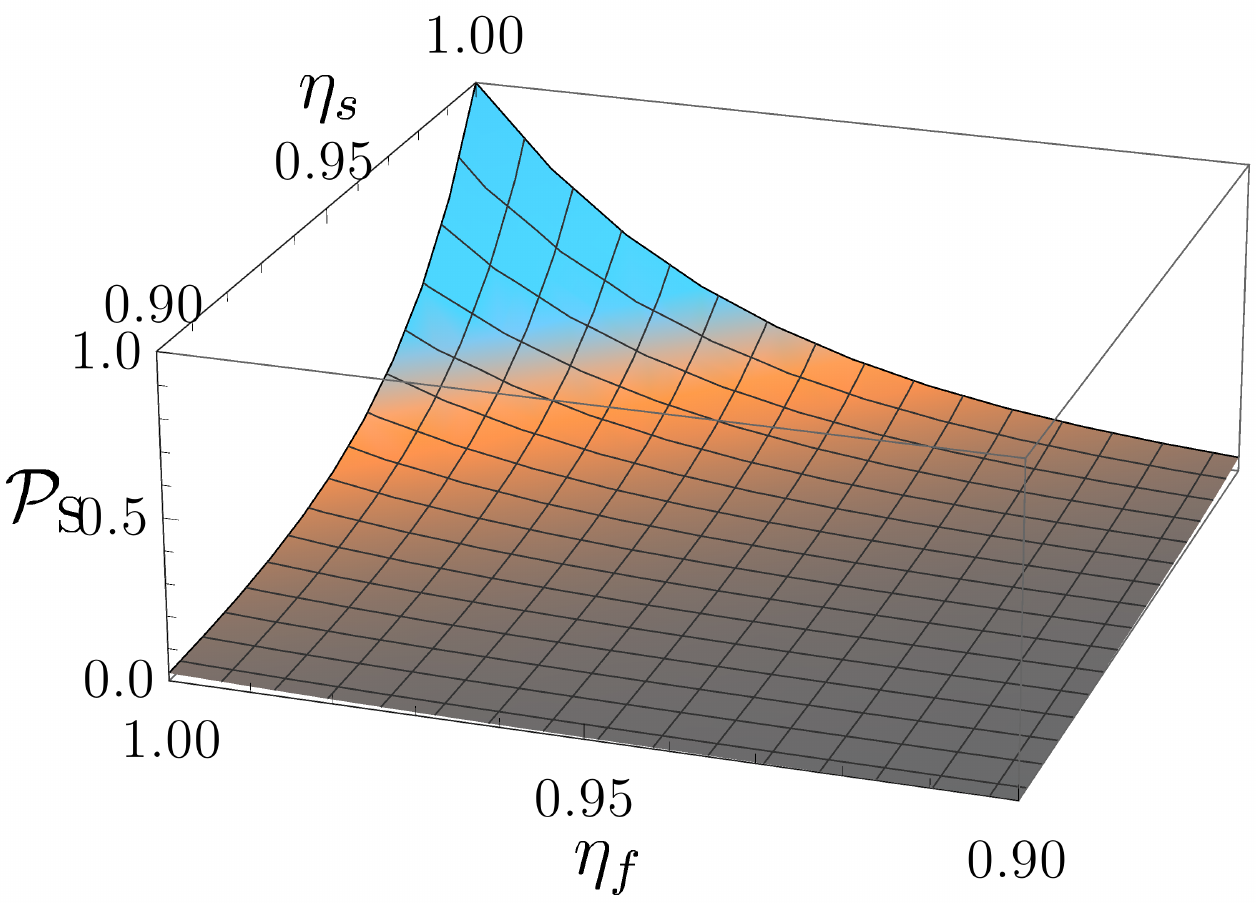} \label{fig:PSLL}}
\caption{(a) Similarity with noise $\mathcal{S'}$, and (b) post-selection probability $\mathcal{P}_\mathrm{S}$ versus loop efficiency $\eta_f$ and switch efficiency $\eta_s$ with \mbox{$m=3$} modes, one photon per input mode, and \mbox{$m-1$} loops. These two plots are again related in that $\mathcal{P}_\mathrm{S}$ is calculated from the switching sequence that maximises $\mathcal{S'}$. This data was averaged over 1750 iterations.}
\label{fig:SPSLL}
\end{figure}

Now we consider how $\mathcal{S}$ and $\mathcal{P}_\mathrm{S}$ are affected in Fig. \ref{fig:SPSLL} with both the fiber loss and switch loss. We show this for the case of \mbox{$m=3$} and one photon per input mode, which is in the regime of present-day demonstrations. 

\section{Mode-mismatching Errors} \label{sec:ModeMismatch}

In any interferometric experiment it is inevitable that mode-mismatch will occur and is thus an essential source of error that we will consider in this section. There are many factors that may contribute to mode-mismatch in this architecture, such as incorrect fiber lengths, time-jitter in the sources, beamsplitter misalignment, and dispersion of the wave-packets. In this section we will focus on two major sources of mode-mismatch: incorrect fiber lengths and source time-jitter. The former results in reduced Hong-Ou-Mandel visibility at the central beamsplitter, owing to mismatched arrival times of photons. The latter effectively results in randomisation of the preparation times of the photons.

We consider how mode-mismatch affects our protocol by calculating the fidelity, $\mathcal{F}$, between the ideal output state $\ket{\psi_i}$ that one expects theoretically with no errors present, and the actual experimentally obtained output state $\ket{\psi_a}$. Imperfect fiber lengths and time-jitter both cause temporal shifts in the centre of the wave-packet, which will affect the output by both introducing uncertainty into the timing of the bins reaching the detector, and undermining the Hong-Ou-Mandel visibility at the central beamsplitter. To calculate $\mathcal{F}$ then we need to calculate the temporal overlap between $\ket{\psi_i}$ and $\ket{\psi_a}$. Therefore, we need to consider the temporal structure of the photons.

We will model the temporal structure of photons using the formalism of Rohde \emph{et al.} \cite{bib:spectralStructure}. We only consider the inner loop in this analysis because there is no interference at any point in the outer loop. We obtain lower and upper bounds on $\mathcal{F}$ by performing a Monte-Carlo search over different randomly generated unitaries $V$. We could also instead consider $V'$ in this formalism to also jointly include losses. But we will treat losses separately from mode-mismatch for simplicity.

\subsection{Temporal Structure of Photons}

The temporal structure of a photon can be represented using a mode operator,
\begin{equation} \label{eq:TPCO}
\hat{\mathcal{A}}^{\dag}(t',\Delta)=\int_{-\infty}^{\infty}\psi(t'-\Delta)\hat{a}^{\dag}(t')dt',
\end{equation}
where \mbox{$\psi(t-\Delta)$} is the temporal density function centered at time $t$, $\Delta$ is a shift of the temporal centre of the photon, and $\hat{a}^{\dag}(t)$ is the time-dependent photon creation operator. This operator \mbox{$\hat{\mathcal{A}}^{\dag}(t,\Delta)$} acts on the vacuum $\ket{0}$ to create a photon with normalised Gaussian spectral density function,
\begin{equation} \label{eq:psi}
\psi(t-\Delta)= \frac{1}{\sqrt{\omega\sqrt{\pi}}} e^{-\frac{(t-\Delta)^2}{2\omega^2}},
\end{equation}
where $\omega/\sqrt{2}$ is the standard deviation. We assume that \mbox{$\tau\gg\Delta$}, in which case $t$ denotes a time-bin, and $\Delta$ denotes a small mismatch within the respective time-bin, not large enough to cause a photon to `jump' from one time-bin to the next. Thus, both $t$ and $\Delta$ represent shifts in the centre of the photon's wavepacket, but the former is of the order of the time-bin separation, while the latter is of much smaller order than the time-bin separation. The units of $\tau$, $t$, $\Delta$, and $\omega$ are all units of time of which the magnitude depends on the properties of the photon source used.

\subsection{Formalism for Analysis of Mode-Mismatch}
\begin{figure}[!htb]
\centering
\includegraphics[width=0.5\columnwidth]{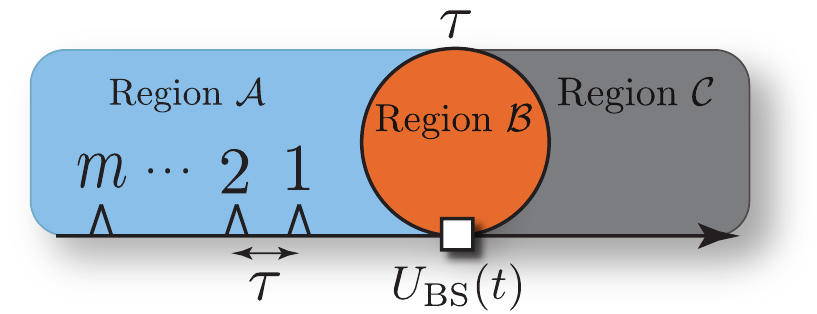}
\caption{The three regions we consider in the mode-mismatch formalism. Region $\mathcal{A}$ corresponds to the modes coming from the source, region $\mathcal{B}$ to the modes inside the inner loop, and region $\mathcal{C}$ to the modes exiting the loop.} \label{fig:Region}
\end{figure}

To analyse mode-mismatch we will consider three regions of the architecture we label as $\mathcal{A}$, $\mathcal{B}$, and $\mathcal{C}$ as shown in Fig. \ref{fig:Region}. Region $\mathcal{A}$ corresponds to the modes that are input into the architecture from the source, region $\mathcal{B}$ corresponds to pulses inside the inner loop, and region $\mathcal{C}$ corresponds to pulses that exit the dynamic beamsplitter towards the detector. We introduce mode operators associated with each of these distinct regions --- \mbox{$\hat{\mathcal{A}}^{\dag}(t,\Delta)$}, \mbox{$\hat{\mathcal{B}}^{\dag}(t,\Delta)$}, and \mbox{$\hat{\mathcal{C}}^{\dag}(t,\Delta)$} --- each of the form of Eq. (\ref{eq:TPCO}).  

Since every pulse begins in region $\mathcal{A}$ the input state is a tensor product of pure states of the form,
\begin{equation} \label{eq:inputState}
\ket{\Psi_\mathrm{in}}= \bigotimes_{i=1}^{m} \frac{1}{\sqrt{k_i!}}\hat{\mathcal{A}}^{\dag}(t_i,\Delta_i)^{k_i}\ket{0}_i,
\end{equation} 
where the tensor product is taken over all $m$ modes, $\{k\}$ is a known string representing the input photon-number configuration, and $k_i$ is number of photons in the $i$th input mode.

Next, the input state is transformed by the dynamic beamsplitter, which takes the mode-operators from region $\mathcal{A}$ into superpositions of regions $\mathcal{B}$ and $\mathcal{C}$, 
\begin{eqnarray} \label{eq:BSA}
\hat{U}_{\mathrm{BS}}(t)\hat{\mathcal{A}}^{\dag}(t,\Delta)\hat{U}_{\mathrm{BS}}^{\dag}(t)&\to& u_{1,2}(t)\hat{\mathcal{B}}^{\dag}(t+1,\Delta) \nonumber \\
&+& u_{1,1}(t)\hat{\mathcal{C}}^{\dag}(t,\Delta),
\end{eqnarray}
and pulses from region $\mathcal{B}$ to superpositions of regions $\mathcal{B}$ and $\mathcal{C}$, 
\begin{eqnarray} \label{eq:BSB}
\hat{U}_{\mathrm{BS}}(t)\hat{\mathcal{B}}^{\dag}(t,\Delta)\hat{U}_{\mathrm{BS}}^{\dag}(t)&\to& u_{2,2}(t)\hat{\mathcal{B}}^{\dag}(t+1,\Delta) \nonumber \\
&+& u_{2,1}(t)\hat{\mathcal{C}}^{\dag}(t,\Delta),
\end{eqnarray}
where we have used Eq. (\ref{eq:VarBS}) for the elements of the dynamic beamsplitter at time $t$. $\hat{U}_{\mathrm{BS}}(t)$ only acts on photons arriving at the beamsplitter at time \mbox{$t\pm\Delta$} since \mbox{$\tau \gg \Delta$}. When a photon enters the loop \mbox{$t\to t+1$} as it advances to the next time-bin and will interfere with the next temporal mode. After this evolution, the entire pulse-train is coupled out of the loop such that the entire output state is a superposition of all possible output configurations.

Now we model the state of the pulse train after $t$ beam-splitters have been implemented,
\begin{eqnarray} \label{eq:outevo}
\ket{\Psi(t)} &=& \left[\prod_{i'=1}^{t}\hat{U}_{\mathrm{BS}}(i')\right]\cdot \ket{\Psi_\mathrm{in}}  \nonumber \\
&=& \left[\prod_{i'=1}^{t}\hat{U}_{\mathrm{BS}}(i')\right] \cdot \left[\prod_{i=1}^{m} \frac{1}{\sqrt{k_i!}}\hat{\mathcal{A}}^{\dag}(t_{i},\Delta_i)^{k_i}\right]\cdot \left[\prod_{i'=1}^{t}\hat{U}_{\mathrm{BS}}(i')\right]^\dag \ket{0}^{\otimes m}
\end{eqnarray}
where the integer values of $t$ denote the distinct time-bins. We note that there are $m+1$ total beam-splitters in a single implementation of the inner loop since there are \mbox{$m-1$} beamsplitters to interfere the modes and another two beamsplitters to account for the initial and final boundary conditions of the MGDR protocol. Given how we modelled how the mode operators are transformed by $\hat{U}_{\mathrm{BS}}(t)$ in Eqs. (\ref{eq:BSA}) and (\ref{eq:BSB}) the $t$th beam-splitter acts on the mode operators only in modal position $t$. Since a pulse coming out of the inner loop exits at beam-splitter $t$ its modal position is $m=t-1$ which accounts for there being $m+1$ beam-splitters and $m$ modes.

In general the final evaluated form of $\ket{\Psi_\mathrm{out}}$ may be expressed as a superposition of all possible output photon-number configurations $S$, and their associated temporal configurations $T(S)$,
\begin{equation}
\ket{\Psi_\mathrm{out}}= \sum_{S}\sum_{T(S)}\bigg[\gamma_{S,T}\prod_{i=1}^{n} \hat{\mathcal{C}}^{\dag}\left(t_{S_i},\Delta_{T(S_i)}\right)\bigg] \ket{0}^{\otimes m},
\end{equation}
where $\gamma_{S,T}$ is the probability amplitude associated with photon time-bin configuration $S$ and temporal shift configuration $T(S)$, $t_{S_i}$ denotes the time-bin of the $i$th photon, $T(S)$ denotes a configuration of temporal shifts associated with the configuration $S$, and $\Delta_{T(S_i)}$ is the temporal shift of the $i$th photon associated with configurations $S$ and $T$. This is the most general representation of a configuration of photons across time-bins with associated shifts. The probability of measuring a particular configuration is $|\gamma_{S,T}|^2$, and to evaluate these probabilities we must fully characterise spectrum of time-bin and temporal shift configurations, $S$ and $T$. Finding analytic forms for these expressions is largely prohibitive, and we calculate the $\gamma_{S,T}$ via brute-force simulation of the evolution of the mode-operators through the network as described earlier.



\subsection{Fidelity Metric} \label{sec:ErrorFidelityMetric}
We analyse the results of this section by calculating the fidelity $\mathcal{F}$ between the ideal output state and the actual output state, given by,
\begin{equation} 
\mathcal{F}= |\overlap{\Psi_\mathrm{i}}{\Psi_\mathrm{a}}|^2,
\end{equation}
where $\ket{\Psi_\mathrm{i}}$ is the ideal output state with no mode-mismatch (\mbox{$\Delta \to 0$}) and $\ket{\Psi_\mathrm{a}}$ is the actual output state obtained with mode-mismatch. $\ket{\Psi_\mathrm{a}}$ reduces to $\ket{\Psi_\mathrm{i}}$ in the limit of no errors yielding \mbox{$\mathcal{F}=1$}. Calculating this overlap but letting $\ket{\Psi_\mathrm{i}}$ have general temporal mode mismatch until the end of the calculation we obtain,
\begin{eqnarray} \label{eq:Fidai}
\mathcal{F}&=& \bigg|\underbrace{\bra{0}^{\otimes m}\sum_{S'}\sum_{T'(S')}\bigg[\gamma_{S',T'}\prod_{i'=1}^{m} \hat{\mathcal{C}}\big(t_{S'_{i'}},\Delta_{T'(S_{i'}')}\big)\bigg]}_{\bra{\Psi_\mathrm{i}}} \underbrace{\sum_{S}\sum_{T(S)}\bigg[\gamma_{S,T}\prod_{i=1}^{m} \hat{\mathcal{C}}^{\dag}\big(t_{S_i},\Delta_{T(S_i)}\big) \bigg]\ket{0}^{\otimes m}}_{\ket{\Psi_\mathrm{a}}} \bigg|^2  \nonumber \\
&=& \bigg| \sum_{S',S,}\,\,\sum_{T'(S'),T(S)}\bigg[\gamma_{S',T'}\gamma_{S,T} \bra{0}^{\otimes m} \prod_{i'=1}^{m} \hat{\mathcal{C}}\big(t_{S'_{i'}},\Delta_{T'(S_{i'}')}\big) \prod_{i=1}^{m} \hat{\mathcal{C}}^{\dag}\big(t_{S_i},\Delta_{T(S_i)}\big) \ket{0}^{\otimes m} \bigg] \bigg|^2. \nonumber \\
\end{eqnarray}
To simplify this expression further we use the formalism of second quantisation \cite{bib:berazin2012}, which describes how the indistinguishability of particles in quantum mechanics undergo symmetrisation. Here we use the exchange symmetry of the bosonic Fock states, which accounts for how each temporal photon annihilation operator $\hat{\mathcal{C}}\big(t_{S'_{i'}},\Delta_{T'(S_{i'}')}\big)$ overlaps with each temporal photon creation operator $\hat{\mathcal{C}}^{\dag}\big(t_{S_{i}}, \Delta_{T(S_i)}\big)$. Using bosonic exchange symmetry we sum over all $m!$ permutations of $\bigotimes_{i'=1}^{m} \hat{\mathcal{C}}\big(t_{S'_{i'}},\Delta_{S_{i'}'}\big) \bigotimes_{i=1}^{m} \hat{\mathcal{C}}^{\dag}\big(t_{S_i},\Delta_{S_i}\big)$. Then Eq. (\ref{eq:Fidai}) becomes,
\begin{eqnarray} \label{eq:Fidai2}
\mathcal{F}&=& \Bigg| \sum_{S',S}\,\,\sum_{T'(S'),T(S)}\Bigg[\gamma_{S',T'}\gamma_{S,T}\sum_{\sigma} \bigg[\prod_{i=1}^{m} \bra{0}\hat{\mathcal{C}}\big(t_{S'_{\sigma_{i'}}},\Delta_{T'(S_{\sigma_{i'}}')}\big) \hat{\mathcal{C}}^{\dag}\big(t_{S_i},\Delta_{T(S_i)}  \big) \ket{0} \bigg] \Bigg] \Bigg|^2, \nonumber \\
\end{eqnarray}
where $\sigma$ are the permutations over $m$ elements.

Finally, to calculate $\mathcal{F}$ we must find the wave packet simplification for 
\begin{equation}
\bra{0}\hat{\mathcal{C}}\big(t_{S'_{\sigma_{i'}}},\Delta_{T'(S_{\sigma_{i'}}')}\big) \hat{\mathcal{C}}^{\dag}\big(t_{S_i},\Delta_{T(S_i)} \ket{0},
\end{equation}
which we perform in App. \ref{app:WavePacketSimp}. Using this result we obtain,
\begin{eqnarray} \label{eq:Fidai3}
\mathcal{F} = \Bigg| \sum_{S',S}\,\,\sum_{T'(S'),T(S)}\Bigg[\gamma_{S',T'}\gamma_{S,T}\sum_{\sigma}\bigg[\prod_{i=1}^{m} \mathrm{exp}\bigg(-\frac{\Big(\Delta_{T'(S'_{\sigma_{i'}})} -\Delta_{T(S_{i})}\Big)^2}{4 \omega^2}\bigg) \bigg] \Bigg] \Bigg|^2.
\end{eqnarray}
Letting the ideal state $\ket{\Psi_\mathrm{i}}$ have no temporal shifts, \mbox{$\Delta \to 0$}, this reduces to,
\begin{eqnarray} \label{eq:Fidai4}
\mathcal{F} &=& \Bigg| \sum_{S',S} \,\, \sum_{T'(S'),T(S)}\gamma_{S',T'}\cdot \gamma_{S,T} \cdot m! \cdot \mathrm{exp}\bigg(-\frac{m\Delta_{T(S_{i})}^2}{4 \omega^2}\bigg) \Bigg|^2.
\end{eqnarray}
This derivation assumes the width of all wave-packets remain the same, i.e the photons are identical up to a temporal displacement. The width of the wave-packets may broaden due to dispersion but under the relatively short lengths of fiber-loop required for small $m$ the effect of dispersion may be neglected; however, as a very crude approximation, this formalism may be easily modified to include dispersion by creating an operator that broadens the wave-packet width $\omega$ as a function of the length of the fiber the wave-packet has traversed.

Next we consider two types of mode-mismatch: non-ideal lengths of the inner loop, and time-jitter at the input source.

\subsection{Imperfect Inner Loop Length}


Here we analyse errors in the MGDR fiber-loop architecture caused by a non-ideal length of inner fiber-loop 
We let the length of the inner loop have some length \mbox{$\tau + \delta$}, where $\delta$ is the error in the intended length $\tau$ and may be positive or negative. Thus every photon that traverses the inner loop acquires a temporal shift of $\delta$ from its expected centre. We ignore imperfect lengths of the outer loop because every mode will traverse the outer loop an equal number of times creating a global temporal shift with no impact on interference at the central beamsplitter.

\begin{figure}[!htb]
\centering
\subfloat[Part 1][]
{\includegraphics[width=0.5\columnwidth]{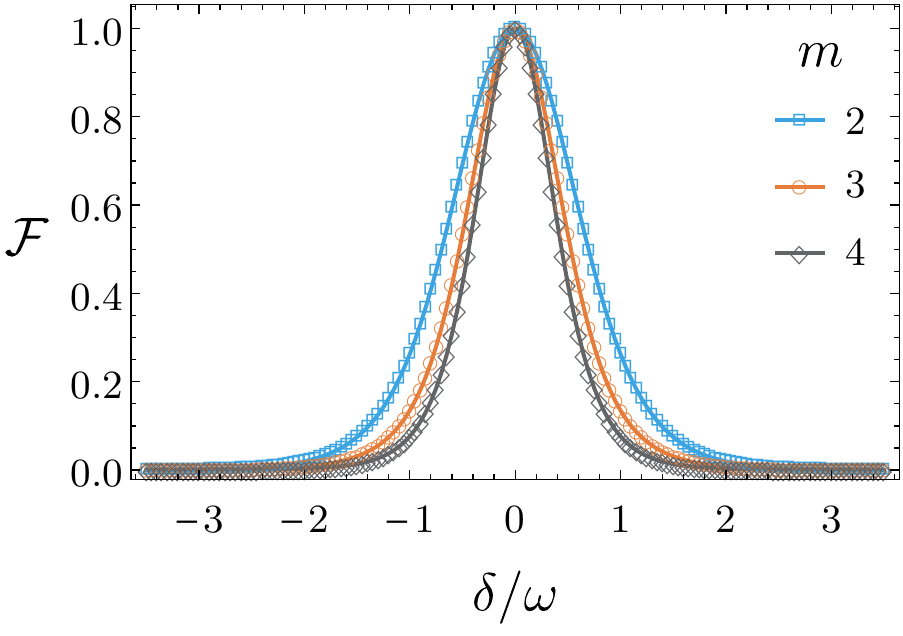} \label{fig:FidInnerLoopDelay}}
\subfloat[Part 2][]
{\includegraphics[width=0.5\columnwidth]{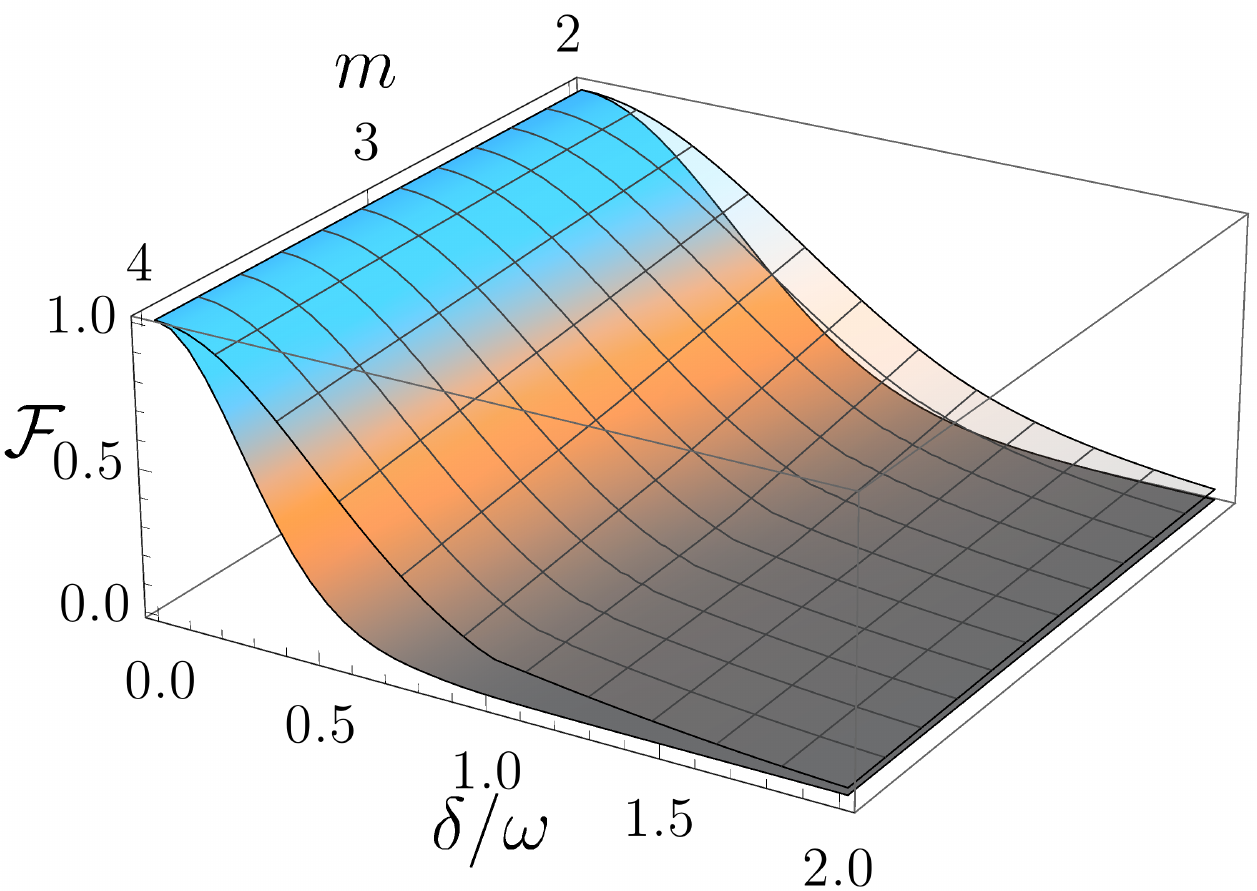} \label{fig:BestWorstFidJitter}}
\caption{(a) The average fidelity $\mathcal{F}$ between the ideal state $\ket{\Psi_\mathrm{i}}$ and the actual experimental state $\ket{\Psi_\mathrm{a}}$ versus the error in the intended length of the inner loop $\delta$. (b) The worst (bottom) and best (top) case fidelity $\mathcal{F}$ between the ideal state $\ket{\Psi_\mathrm{i}}$ and the actual experimental state $\ket{\Psi_\mathrm{a}}$ versus the error in the intended length of the inner loop $\delta$ and number of modes $m$. In (a) and (b) there are $m$ modes with one photon per mode and the data was obtained over 250 implementations each with a unique randomly generated unitary.}
\label{fig:Fidcombined1}
\end{figure}

The input state is given by Eq. (\ref{eq:inputState}) where \mbox{$\Delta_i=0\ \forall\ i$}. This models an ideal input state with no time-jitter or other errors in the source. To account for the unwanted time-delay $\delta$ we introduce the time-delay operator \mbox{$\hat{\mathrm{T}}(\delta)$},
\begin{equation}
\hat{\mathrm{T}}(\delta)\hat{\mathcal{B}}^{\dag}(t,\Delta)\hat{\mathrm{T}}^{\dag}(\delta)= \hat{\mathcal{B}}^{\dag}(t,\Delta+\delta),
\end{equation}
which acts only in region $\mathcal{B}$ -- the region inside the inner loop. This adds a small temporal displacement, not enough to confuse time-bins. Thus it affects $\Delta$ but not $t$. It has no effect on the mode-operators $\hat{\mathcal{A}}$ and $\hat{\mathcal{C}}$. Using the boundary conditions shown in the MGDR protocol, the first photon is coupled completely into the loop so it picks up a time-delay of $\delta$. Afterwards the pulse-train interacts at the beamsplitter described in Eqs. (\ref{eq:BSA}) and (\ref{eq:BSB}), where it is sent into a superposition of regions $\mathcal{B}$ and $\mathcal{C}$. As the state evolves all amplitudes entering the inner loop (region $\mathcal{B}$) will acquire a time-shift of $\delta$. After the last mode traverses the inner loop the state is coupled completely out as per the MGDR protocol. 
The output state is given by,
\begin{eqnarray} \label{}
\ket{\Psi}_{\mathrm{out}} &=& \left[\prod_{i'=1}^{t}\hat{\mathrm{T}}(\delta)\hat{U}_{\mathrm{BS}}(i')\right] \cdot \left[\prod_{i=1}^{m} \frac{1}{\sqrt{k_i!}}\hat{\mathcal{A}}^{\dag}(t_{i},\Delta_i)^{k_i}\right]\cdot \left[\prod_{i'=1}^{t}\hat{\mathrm{T}}(\delta)\hat{U}_{\mathrm{BS}}(i')\right]^\dag \ket{0}^{\otimes m},
\end{eqnarray}
where we have inserted the time-delay operator appropriately in Eq. (\ref{eq:outevo}). Fig. \ref{fig:FidInnerLoopDelay} shows how the fidelity $\mathcal{F}$ scales with $m$, $n$, and $\delta$ and Fig. \ref{fig:BestWorstFidJitter} shows the worst- and best-case fidelities, where we have searched over switching sequences.

\subsection{Time-Jitter from Input Source}
\begin{figure}[!htb]
\centering
\subfloat[Part 1][]
{\includegraphics[width=0.5\columnwidth]{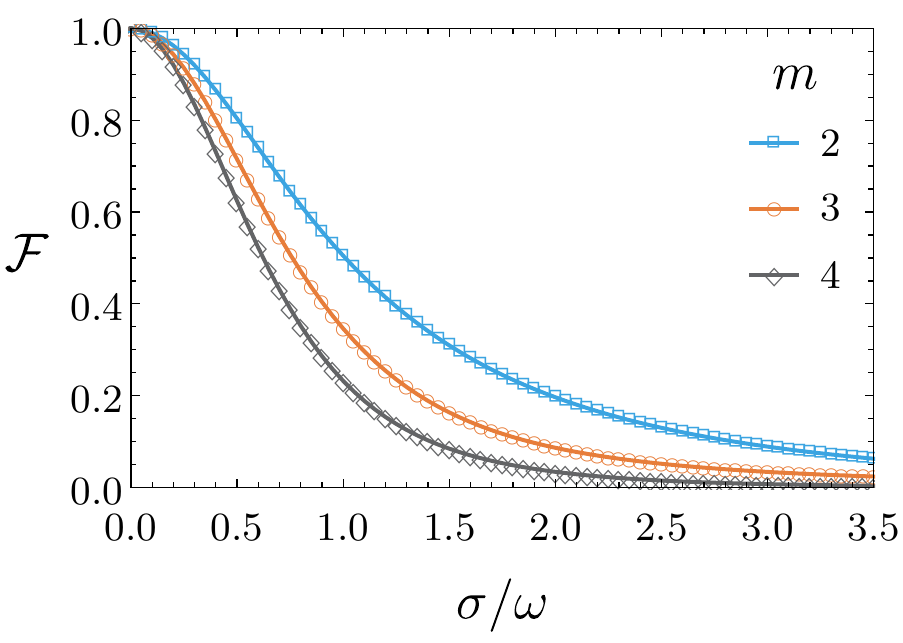} \label{fig:FidJitter}}
\subfloat[Part 2][]
{\includegraphics[width=0.5\columnwidth]{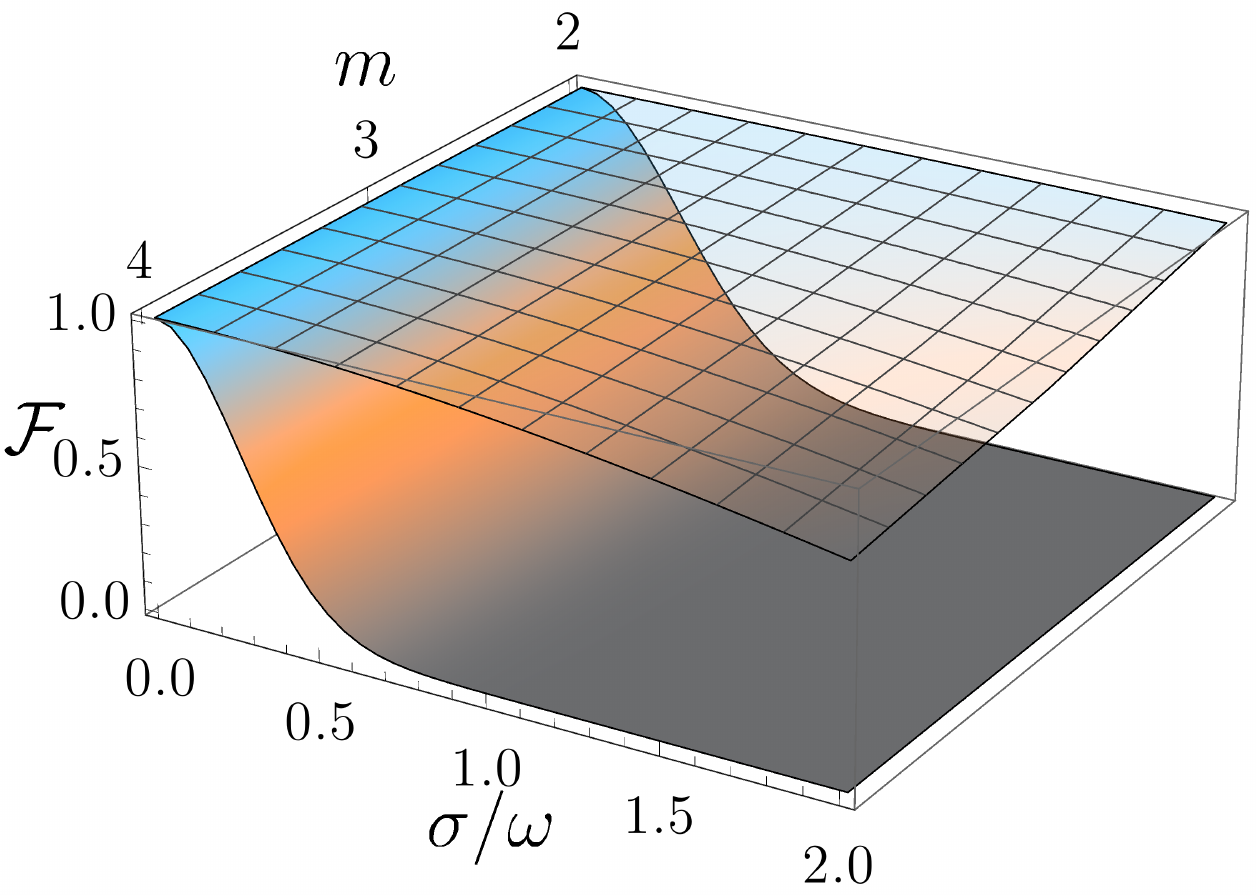} \label{fig:BestWorstFidJitter}}
\caption{(a) The fidelity $\mathcal{F}$ between the ideal state $\ket{\Psi_i}$ and the actual experimental state $\ket{\Psi_a}$ with random time-jitter in the input source versus modes $m$ and standard deviation $\sigma$ with no fiber length error $\delta=0$. (b) The worst (bottom) and best (top) case fidelity $\mathcal{F}$ between the ideal state $\ket{\Psi_\mathrm{i}}$ and the actual experimental state $\ket{\Psi_\mathrm{a}}$ with time-jitter. In (a) and (b) there is one photon per mode, the data was averaged over 250 implementations each with a unique randomly generated unitary, and the time-jitter was drawn from the normal distribution.}
\label{fig:Fidcombined2}
\end{figure}

A major source of error in the time-bin architecture is time-jitter of the input source. Ideally each mode will be separated by time $\tau$ but in reality non-ideal sources will randomly shift modes from their desired centre of time $t_i$ in mode $i$. To model time-jitter we let the temporal shift of input mode $i$ be a Gaussian random variable $\epsilon_i$ drawn from the normal distribution,
\begin{eqnarray}
\mathcal{N}_i(\epsilon_i)=\frac{1}{\sigma\sqrt{2\pi}}\mathrm{exp}\left(-\frac{(t_i-\epsilon_i)^2}{2 \sigma^2}\right),
\label{eq:NormalJitter}
\end{eqnarray}
centered in mode $i$ at time $t_i$ and with a standard deviation of $\sigma$. The input state of Eq. (\ref{eq:inputState}) becomes,
\begin{equation}
\ket{\psi_\mathrm{in}}= \bigotimes_{i=1}^{m}\frac{1}{\sqrt{k_i!}} \hat{\mathcal{A}}^{\dag}\big(t_i, \epsilon_i\big)^{{k_i}}\ket{0}_i.
\end{equation}
We assume that the shifts caused by time-jitter are much less than the time-bin separation $\tau$, such that the probability of time-bin confusion remains negligible, i.e. \mbox{$\mathcal{N}(\epsilon_i) \ll \tau$}. Fig. \ref{fig:FidJitter} shows how the fidelity $\mathcal{F}$ scales with $m$, $n$, and $\sigma$. Fig. \ref{fig:BestWorstFidJitter} shows the worst- and best-case $\mathcal{F}$, searching over many switching sequences.

\section{Discussion of Using Realistic Devices} \label{sec:FLerrorModels}
With the two types of errors we have considered, loss and mode-mismatch, the final wavefunction is affected in predictable ways. In the case of loss the unitary transformation implemented on the initial input state becomes biased. This causes the corresponding probabilities of obtaining particular output wavefunctions to also become biased. There are two quantities to consider in our work when considering loss, the loss in the fiber $\eta_f$ and the loss from the switches $\eta_s$. Today the best known efficiencies for these are $\eta_f=0.99$ \cite{bib:PCAndrewWhite} and $\eta_s\approx0.8$ \cite{bib:Han15}. With these efficiencies the post-selection probability $\mathcal{P}_\mathrm{S}$ is too small to implement a \BS experiment of interesting size as can be seen in Fig. \ref{fig:SPSLL}. Current fiber technology is efficient enough to implement our architecture but the switches are too lossy.  

The second type of loss we have considered is mode-mismatch. Our model can account for any error that can shift the temporal location of the wave packets of the photons. We focused on two such errors: 1) an imperfect length of the inner loop and 2) time-jitter from the input source. In all cases this causes the experimentally obtained output wavefunction to differ from the theoretically desired output wavefunction. We used the Fidelity metric $\mathcal{F}$ to characterize this difference. We suspect that the error in the length of the inner loop $\delta$ can be made extremely small by carefully characterizing the length of the inner loop using coherent states of light so this should not be a considerable issue. The second error source we considered, time-jitter, however is a problem in any experiment that requires single photons. Ideally the standard devitaion $\sigma$ in Eq. (\ref{eq:NormalJitter}) would approach zero such that there is no time-jitter. Today some of the best experimentally obtained values for this $\sigma/\omega$ are found in references \cite{bib:PhysRevB.89.161303, bib:scheidl2014crossed, bib:kaltenbaek2006experimental}. In these works $\sigma/\omega\geq 5$. Comparing this value to Fig. \ref{fig:Fidcombined2} we see that source jitter needs to improve by about an order of magnitude for a time-bin architecture to be feasable.  

It seems that in the near future a \BS experiment with more modes and more photons than has ever been implemented could be carried out using this architecture due to the fast rate at which quantum technologies are being researched and developed; however, it may be much longer before this architecture has sufficiently low errors that it could be used to implement a \BS experiment that is in the classically hard regime to simulate. 

\section{Summary}
We have presented the original work of Motes, Gilchrist, Dowling \& Rohde \cite{bib:motes2014scalable} which is an arbitrarily scalable architecture for universal \BS based on two nested fiber loops as motivated in section \ref{sec:FLmotivation}. The complexity of the architecture (which is described in detail in section \ref{sec:FLarchitecture}) is constant, independent of the size of the interferometer being implemented. Scalability is limited only by fiber and switch transmission efficiencies as well as source efficiencies. There is only one point of interference in the architecture, which suggests that it may be significantly easier to stabilize than traditional approaches based on waveguides or discrete elements. We also considered an experimental simplification in section \ref{sec:FLsimplicity} where full dynamic control is not required and showed that, while not universal, with sufficient loops the unitary approximates a maximally mixing unitary. While we have specifically considered this architecture in the context of \BStwo, the same scheme, or variations on it, may lend themselves to other linear optics applications, such as interferometry, metrology, or full-fledged LOQC. In section \ref{sec:FLdiscussionofarchitecture} we discuss many of the advantages of our architecture. 

Following the work presented in \cite{bib:motes2014scalable} we analysed sources of error in the fiber loop architecture. Specifically we have analysed loss and mode-mismatch in section \ref{sec:LossErrors} and \ref{sec:ModeMismatch} respectively. In the loss analysis we examined how lossy fibers and switches affect the operation of the architecture in both the inner and outer loops. We found that loss causes an asymmetric bias in the desired unitary, unique to a temporally implemented unitary transformation. That is, even upon post-selection the operation of the device is erroneous. Additionally, like all linear optical architectures, our scheme has exponential dependence on loss, thereby reducing the post-selection success probability of detecting all $n$ photons. In the mode-mismatch analysis we analysed only the inner loop since no interference occurs in the outer loop. We examined two types of mode-mismatch including an imperfect length of fiber in the inner loop, and time-jitter of the photon source. This analysis provides a guideline for future experimental implementations, to provide insight into how such a device might realistically behave in the presence of loss and mode-mismatch, the two dominant error mechanisms affecting this protocol. In section \ref{sec:FLerrorModels} we discuss our architecture using todays lowest reported loss rates and amount of mode-mismatch.

\begin{savequote}[45mm]
What we think, we become.
\qauthor{Buddha}
\end{savequote}

\chapter{\BS with Other Quantum States of Light} \label{Ch:sampOther}

\section{Synopsis}

It is known that the \BS sampling problem likely cannot be efficiently classically simulated. This raises the question as to whether or not other similar systems implement a sampling problem that is also computationally hard. One such variation to consider is what happens when other quantum states of light are used at the input to the \BS device other than single-photon Fock states. Are there other quantum states of light that are also computational hard to simulate? We answer this in the affirmative. 

In section \ref{sec:SampOtherMotivation} we provide further motivation for this question. We have investigated the \BS problem with three different quantum states of light other than single-photon Fock states which are published \cite{bib:Kaushik14, bib:Olson14, bib:catSampling} and presented in this chapter:
\begin{enumerate}
\item In section \ref{sec:SampOtherSPACS} we consider single-photon-added coherent states (SPACS). We show that the associated \BS problem with displaced single-photon Fock states at the input and using a displaced photon-number detection scheme is in the same complexity class as \BS for all values of displacement. We then show that the associated \BS problem with SPACS and using a displaced photon-number detection scheme has an interesting computational complexity transition. The transition is from computationally hard when the coherent amplitudes are sufficiently small to computionally easy when the coherent amplitudes become large. The intuitive explanation is that with small coherent amplitudes we are approximating single photons while with large coherent amplitudes we are approximating a more classical state like a laser beam. 

\item In section \ref{sec:SampOtherPASSV} we show that the \BS problem with photon-added or -subtracted squeezed vacuum (PASSV) states are in the same complexity class as \BS when sampling at the output is performed via parity measurements. Here we found an exact proof that works for an arbitrary amount of squeezing. 

\item In section \ref{sec:SampOtherCATS} we present the \BS problem with a very broad class of quantum states of light --- arbitrary superpositions of two or more coherent states --- at the input. We were not able to find a full complexity proof for the hardness of this problem but have strong evidence that when this state is evolved via passive linear optics and sampled with number-resolved photodetection it is likely classically hard to simulate like \BStwo.
 
\end{enumerate}

\section{Motivation} \label{sec:SampOtherMotivation}

The linear optical community was surprised when they found out about the computational complexity of \BStwo. Simply having a linear optical system with Fock states as input and sampling using a suitable detection strategy at the output is computationally complex. This fact opened inquiry into the complexity of other linear optical systems. Understanding these systems could shed light in the field of computational complexity increasing our understanding of complexity classes and computation in general which would help lead us to a quantum computer. It may also help us understand more systems that can be simulated through experiment but are not efficient to simulate on a classical computer.   

A modification to consider is whether or not other states of light input to a linear optical network using similar output detection strategies are also of similar computational complexity as \BStwo. It is known that passive linear optics may be efficiently simulated with Gaussian inputs and non-adaptive Guassian measurements \cite{bib:Bartlett02, bib:Bartlett02b}. However, the more general question as to which quantum states of light may be efficiently simulated with number-resolved measurements is an open question. Recent results include that the case of sampling with Gaussian states in the photon number basis can be just as hard as \BS \cite{bib:Lund13}. Sampling with thermal states can be simulated efficiently on a classical computer~\cite{bib:rahimi2014can}. It has also been shown in some special cases that sampling two-mode squeezed vacuum states is likely hard to efficiently simulate classically~\cite{bib:PhysRevA.88.044301, bib:Lund13}. Other quantum states of light considered were single-photon-added coherent states (SPACS) by Seshadreesan \emph{et al.} \cite{bib:Kaushik14}, photon-added or -subtracted squeezed vacuum (PASSV) states by Olson \emph{et al.} \cite{bib:Olson14}, and generalized cat states (i.e. arbitrary superpositions of coherent states) by Rohde \emph{et al.} \cite{bib:catSampling}. These three states evolved using linear optics and using photon number detection, have been analysed and are presented in this chapter in sections \ref{sec:SampOtherSPACS}, \ref{sec:SampOtherPASSV}, and \ref{sec:SampOtherCATS} respectively. Although these input states are more difficult to prepare than the single-photon Fock state, the analysis of their computational complexity allows us to demonstrate interesting phenomenon. We provide clarity on the theory of classifying the sampling complexity of various quantum states and we demonstrate that Fock states are not unique in their sampling complexity as there are a plethora of other quantum states of light which yield sampling problems with similar complexity to \BStwo.

\section{Single-Photon-Added Coherent State Sampling} \label{sec:SampOtherSPACS}

In this section we look into the computational complexity of \BS with, instead of single-photon Fock states at the input, displaced single-photon Fock states (DSPFS) and single-photon-added coherent states (SPACS). We do this with a displaced photon number detection. The displacement operator can be written as
\begin{equation}
\label{disop}
\hat{D}(\alpha)=\exp\left(\alpha \hat{a}^{\dagger}-\alpha^*\hat{a}\right),
\end{equation}
where $\alpha$ is a complex amplitude that indicates displacement in phase space, and $\hat{a}^\dagger$ ($\hat{a}$) is the mode photon creation (annihilation) operator. The DSPFS is the state $\hat{D}(\alpha)\hat{a}^\dagger|0\rangle$, while the SPACS is $\propto\hat{a}^\dagger\hat{D}(\alpha)|0\rangle$.
In this section we show three things. Firstly we show in section \ref{sec:SPACS_DSPFS} that sampling with DSPFS is in the same complexity class as AA \BS for any displacement $\alpha$. In section \ref{sec:SPACS_SPACS} we show the modified \BS protocol with SPACS, which differ from DSPFS by the ordering of the operators. The results here are interesting becuase the sampling problem with SPACS is just as hard as AA's \BS when the input coherent amplitudes are sufficiently small but when the input coherent amplitudes become larger the problem transitions from hard to simulate classically to easy to simulate classically. This interesting transition is discussed in section \ref{sec:SPACStransition}. 

\subsection{Sampling Displaced Single-Photon Fock states (DSPFS)}
\label{dfssampling} \label{sec:SPACS_DSPFS}

In this section we use DSPFS instead of single-photon Fock states as per Eq. (\ref{eq:BSintroInputState}) at the input to the linear-optical interferometer. This input state has the form
\begin{equation} \label{eq:input_state_DFS}
\ket{\psi_\mathrm{in}}^{\mathrm{DSPFS}} =\left( \prod_{i=1}^{n}\hat{D}_i\left(\alpha^{(i)}\right)\hat{a}_i^\dagger\right) \ket{0_1,\dots,0_m},
\end{equation}
where $\hat{D}_i\left(\alpha^{(i)}\right)$ is the displacement operator of the $i$th mode, and $\alpha^{(i)}$ is the complex coherent amplitude for the displacement. The input states reaches the linear optical interferometer $\hat{U}$ where a unitary operation transforms the state into
\begin{align}
\label{UxformDFS}
|\psi_{\rm out}\rangle^{\mathrm{DSPFS}} &=\hat{U} \left(\prod_{i=1}^{n}\hat{D}_i\left(\alpha^{(i)}\right)\hat{a}_i^\dagger\right)\hat{U}^{\dagger}\hat{U}\ket{0_1,\dots,0_m},\nonumber\\
&=\hat{U} \left(\prod_{i=1}^{n}\hat{D}_i\left(\alpha^{(i)}\right)\right)\hat{U}^{\dagger}\hat{U}\left(\prod_{k=1}^{n}\hat{a}_k^\dagger\right)\hat{U}^{\dagger}\ket{0_1,\dots,0_m}\nonumber\\
&= \prod_{i=1}^{n}\left(\hat{U}\hat{D}_i\left(\alpha^{(i)}\right)\hat{U}^{\dagger}\right)\prod_{k=1}^{n}\left(\hat{U}\hat{a}_k^\dagger\hat{U}^{\dagger}\right)\ket{0_1,\dots,0_m}\nonumber\\
&=\left(\prod_{j=1}^{m}\hat{D}_j\left(\beta^{(j)}\right)\right)\left(\sum_{S}\gamma_S (\hat{b}_1^{\dagger})^{s_1}(\hat{b}_2^{\dagger})^{s_2}\dots(\hat{b}_m^{\dagger})^{s_m}\right)\nonumber\\
&\quad\times\ket{0_1,\dots,0_m},
\end{align}
where $\beta^{(j)} = \sum_i U_{i,j} \alpha^{(i)}$ is the new displacement amplitude in the $j$th mode, $\hat{b}_k^{\dagger}$ is the photon-creation operator of the  $k$th mode, and $s_k$ is the number of photons in the $k$th mode, associated with configuration $S$ at the output such that $\sum_{k=1}^m s_k=n$ for each $S$. To derive this expression we have used: $\hat{U}^{\dagger}\hat{U}=I$, $\hat{U}\ket{0_1,\dots,0_m}=\ket{0_1,\dots,0_m}$, Eq. (\ref{eq:Utransform}) and Eq. (\ref{eq:outputState}). We have also invoked that a unitary on a tensor product of coherent states is another tensor product of coherent states as shown in App. \ref{app:coherent_map}. The result is a displaced version of AA's original \BS output state.

Now the question is does this output state have computational complexity similar to that of \BStwo? The answer is yes. This is because the the new complex displacement amplitudes $\beta^{(j)}$ can be efficiently computed for any unitary operator $U$. Also, a counter-displacement with amplitudes $-\beta^{(j)}$ could simply be applied to the $m$ output modes since $D(-\alpha)D(\alpha)=I$, which is classically efficient and can be performed using unbalanced homodyning~\cite{bib:BW99, bib:WLD07}. With this counter displacement we are left with exact \BS and thus sampling DSPFS using a measurement scheme at the output that is comprised with an inverse displacement followed by coincidence photon-number detection is in the same complexity class as \BStwo. This demonstrates that an entire class of quantum states of light may be used to achieve a sampling problem of equal complexity to \BStwo.

\subsection{Sampling Single-Photon-Added Coherent States (SPACS)}
\label{sec:SPACS_SPACS}

SPACS differ from DSPFS in the ordering of the operators. Since the displacement operator of Eq. (\ref{disop}) does not commute with the photon creation operator $\hat{a}^\dagger$, these states are sufficiently different. A $k$-photon-added coherent state may be written as
\begin{equation}
\label{pacsdefinition}
|\alpha,k\rangle=\mathcal{N}_k \hat{a}^{\dagger^{k}}|\alpha\rangle,
\end{equation}
with normalization
\begin{equation}
\mathcal{N}_k=\frac{1}{\sqrt{k!L_k (-|\alpha|^2)}},
\end{equation}
where $L_k$ is the Laguerre polynomial of order $k$. These states were first described by Agarwal \& Tara \cite{bib:Agarwal91}. The state we are interested in this work is the SPACS whereby we consider $\ket{\alpha,1}$ in Eq. (\ref{pacsdefinition}). 

\begin{figure}
\centering
\includegraphics[width=0.3\columnwidth]{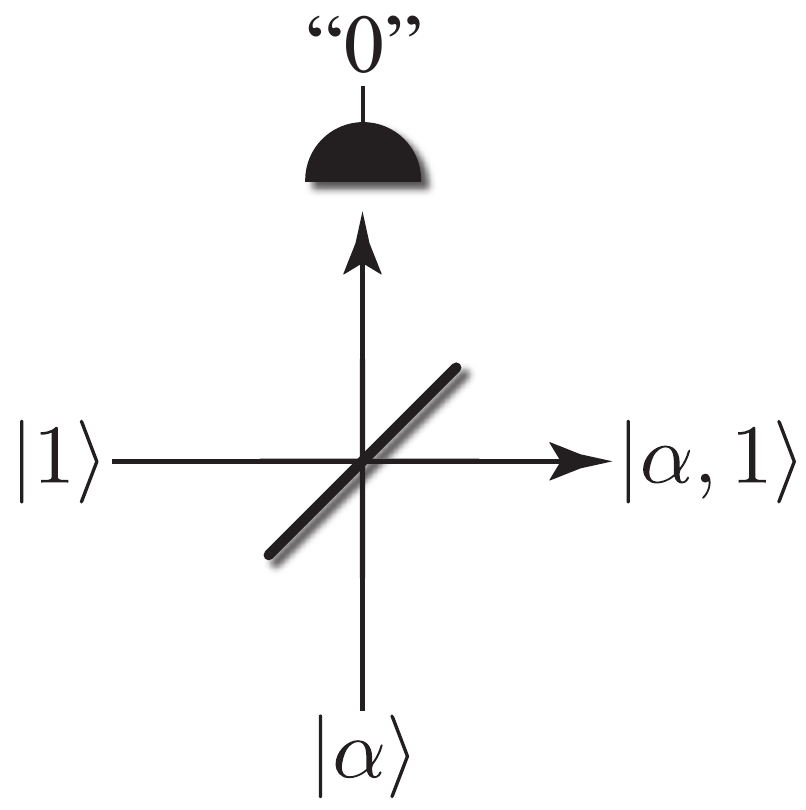}
\caption{A SPACS may be made by mixing a coherent state and a single photon state on a highly reflective beamsplitter. When no photon is detected in the transmitted mode, a SPACS is heralded in the transmitted mode.} \label{fig:PACS_prep}
\end{figure}

A SPACS may be created by mixing a single photon (perhaps prepared with spontaneous parametric down-conversion) with a coherent state on a highly reflective beam splitter as shown in Fig. \ref{fig:PACS_prep}. When vacuum is detected in one the top output mode we know that the single photon has been added to the coherent state in the other output port, and thus a SPACS has been heralded~\cite{bib:Dakna98, bib:Dakna98_2, bib:Zavatta04, bib:Zavatta05}.

\begin{figure}[!htb]
\centering
\includegraphics[width=0.45\columnwidth]{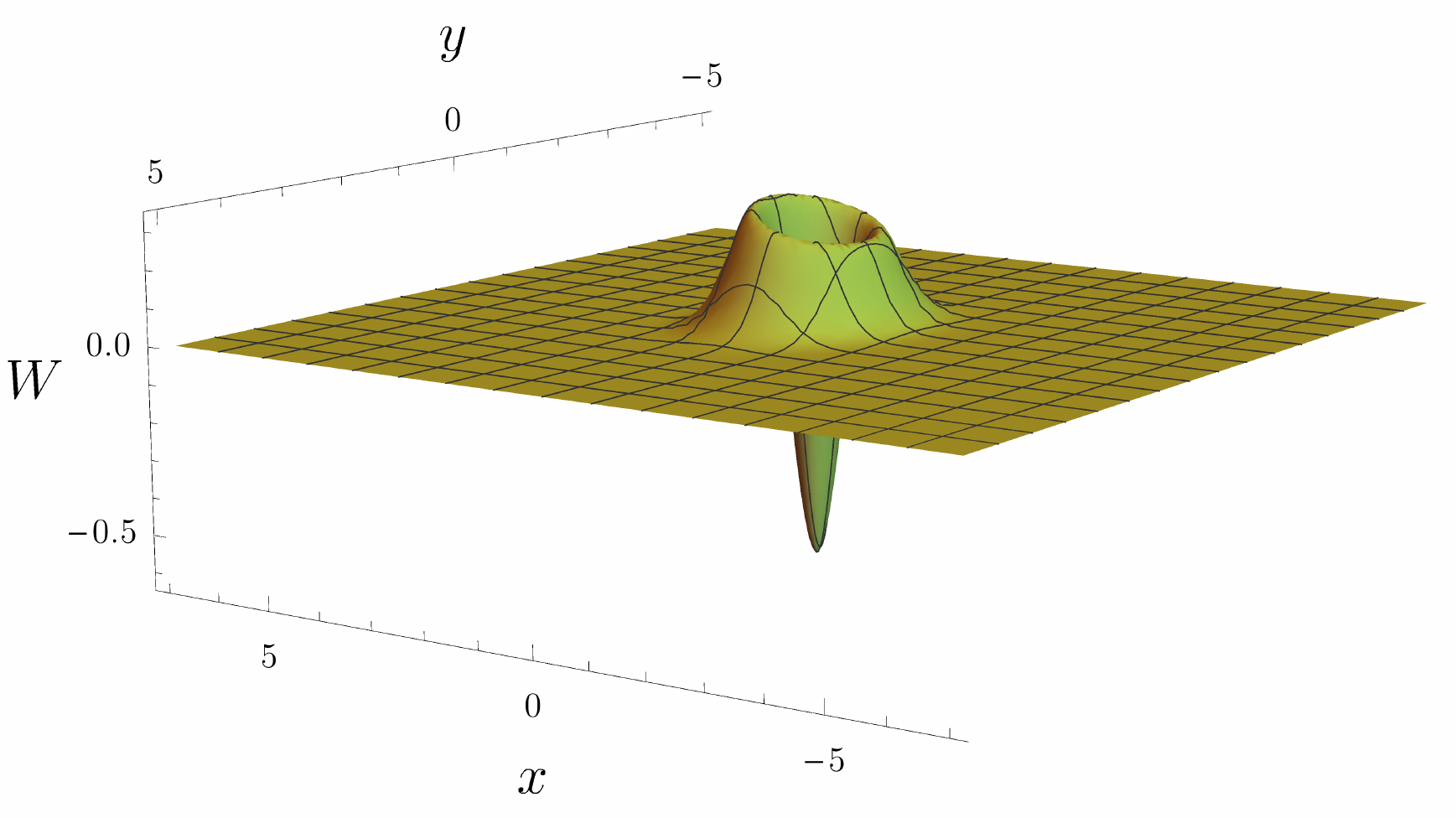} 
\includegraphics[width=0.45\columnwidth]{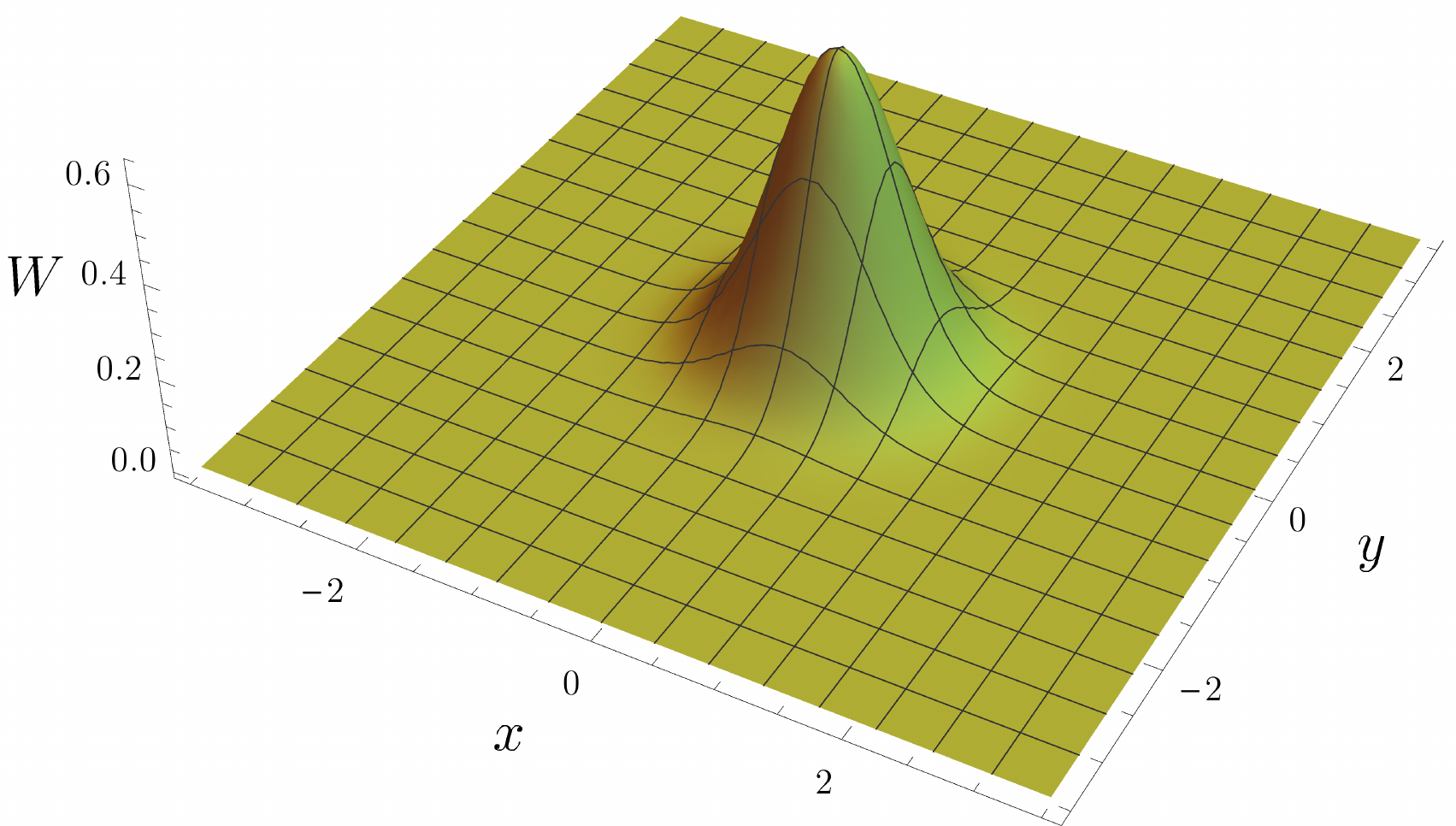}
\caption{Wigner function of a SPACS (left) and a coherent state (right), with amplitude \mbox{$|\alpha|^2=0.01$}. The former is seen to take negative values close to the phase-space origin, while that of the latter is strictly positive everywhere thus measuring $W(0)$ would be able to distinguish between a SPACS and a coherent state.} \label{wigner}
\end{figure}

\begin{figure}[!htb]
\centering
\includegraphics[width=\imageWidthThree\columnwidth]{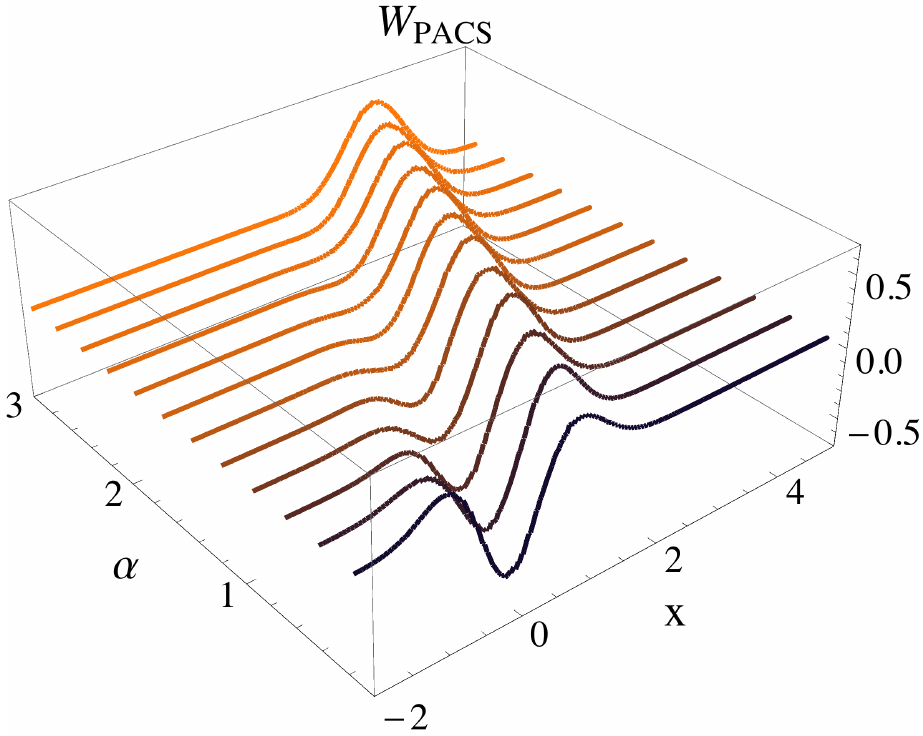} 
\caption{Two-dimensional slices along position of the Wigner function of a SPACS taken at a fixed momentum of zero as a function of the coherent amplitude $|\alpha|$. For increasing values of $|\alpha|$ we see that the negativity of the Wigner function vanishes and that it becomes Gaussian.} \label{wignerslices}
\end{figure}

The quantum-classical transition of SPACS have been studied since they allow for a seamless interpolation between the highly nonclassical Fock state $|1\rangle$ ($\alpha\rightarrow 0$) and a highly classical coherent state $|\alpha\rangle$ ($|\alpha|\gg1$)~\cite{bib:Zavatta04}. The Wigner function of a SPACS can be expressed as \cite{bib:Agarwal91}
\begin{equation}
W(z)=\frac{2(|2z-\alpha|^2-1)}{\pi(1+|\alpha|^2)}e^{-2|z-\alpha|^2},
\end{equation}
where $z=x+iy$ is the phase-space complex variable, and $\alpha$ the coherent amplitude in the state. Fig.~\ref{wigner} shows the Wigner functions of a SPACS and a coherent state. The former attains negative values at points close to the origin in phase space, which clearly demonstrates the nonclassical nature of the state. Fig.~\ref{wignerslices} shows 2-d slices along position of the Wigner function of a SPACS taken at a fixed momentum of zero as a function of the coherent amplitude $|\alpha|$. It can be seen that the Wigner function loses its negativity as $\alpha$ increases and tends towards being a Gaussian state. 

The SPACS-based input that we consider to a linear-optical sampling device can be written as
\begin{equation}
\label{input}
|\psi_{\rm in}\rangle^{\mathrm{SPACS}}=\mathcal{N}\prod_{i=1}^{n}\hat{a}_i^\dagger\hat{D}_i\left(\alpha^{(i)}\right)\ket{0_1,\dots,0_m},
\end{equation}
with
\begin{equation}
\label{inputNorm}
\mathcal{N}=\prod_{j=1}^{n}\frac{1}{\sqrt{1+|\alpha^{(j)}|^2}},
\end{equation}
where $\alpha^{(i)}$ represents the complex coherent amplitude in the \mbox{$i$th} mode and $\mathcal{N}$ is the overall normalization factor. In words this is saying that the input to the first $n$ modes are SPACS and the remaining $m-n$ modes are in the vacuum state.  A unitary operation $\hat{U}$ then transforms the state as
\begin{eqnarray}
\label{UxformPACS1}
|\psi_{\rm out}\rangle^{\mathrm{SPACS}}&=&\hat{U}|\psi_{\rm in}\rangle^{\mathrm{SPACS}}\nonumber\\
&=&\mathcal{N}\hat{U} \left(\prod_{i=1}^{n}\hat{a}_i^\dagger\hat{D}_i\left(\alpha^{(i)}\right)\right)\hat{U}^{\dagger}\hat{U}\ket{0_1,\dots,0_m}.
\end{eqnarray}
This state can also be written as
\begin{align}
\label{UxformPACS2}
|\psi_{\rm out}\rangle^{\mathrm{SPACS}}&=\mathcal{N}\hat{U} \left\{\prod_{i=1}^{n}\left(\hat{D}_i\left(\alpha^{(i)}\right)\hat{a}_i^\dagger+{\alpha^{(i)}}^*\hat{D}_i\left(\alpha^{(i)}\right)\right)\right\}\hat{U}^{\dagger}\nonumber\\
&\quad\times\ket{0_1,\dots,0_m},
\end{align}
where we have used the commutation relation between the displacement operator and the photon-creation operator
\begin{align}
\left[a^\dagger, \hat{D}(\alpha)\right]=\alpha^*\hat{D}(\alpha).
\end{align} 
We can further simplify the state
\begin{align}
|\psi_{\rm out}\rangle^{\mathrm{SPACS}}&=\mathcal{N}\hat{U} \prod_{i'=1}^{n}\hat{D}_{i'}\left(\alpha^{(i')}\right)\hat{U}^{\dagger}\hat{U}\prod_{i=1}^{n}\left(\hat{a}_i^\dagger+{\alpha^{(i)}}^*\right)\hat{U}^{\dagger}\nonumber\\
&\quad\times\ket{0_1,\dots,0_m},\nonumber\\
&=\mathcal{N}\prod_{i'=1}^{n}\left(\hat{U} \hat{D}_{i'}\left(\alpha^{(i')}\right)\hat{U}^\dagger\right)\prod_{i=1}^{n}\left(\hat{U}\hat{a}_i^\dagger\hat{U}^{\dagger}+{\alpha^{(i)}}^*\right)\nonumber\\
&\quad\times\ket{0_1,\dots,0_m}\nonumber\\
&=\mathcal{N}\prod_{j=1}^{m}\hat{D}_{j}\left(\beta^{(j)}\right) \prod_{i=1}^{n}\left(\hat{U}\hat{a}_i^\dagger\hat{U}^{\dagger}+{\alpha^{(i)}}^*\right)\ket{0_1,\dots,0_m},
\end{align}
where $\beta^{(j)} = \sum_{i'} U_{i',j} \alpha^{(i')}$ is the new displacement amplitude in the $j$th mode. As in the case of DSPFS sampling from subsection \ref{sec:SPACS_DSPFS}, we can now apply a counter-displacement operation of amplitude $\prod_{j=1}^{m}\hat{D}_j\left(-\beta^{(j)}\right)$, which is efficiently computed, so that the output state reduces to
\begin{align}
\label{pacsundis_state}
|\psi_{\rm out}\rangle^{\mathrm{SPACS}}&=\mathcal{N}\prod_{i=1}^{n}\left(\hat{U}\hat{a}_i^\dagger\hat{U}^{\dagger}+{\alpha^{(i)}}^*\right)\ket{0_1,\dots,0_m}.
\end{align}

We will now denote the state $\prod_{i=1}^{n}\left(\hat{U}\hat{a}_i^\dagger\hat{U}^{\dagger}\right)\ket{0_1,\dots,0_m}$, which corresponds to AA-type \BS as $|AA\rangle$. Further, for simplicity, we choose all the input coherent amplitudes to be equal to $\alpha$. Then, the output state in Eq. (\ref{pacsundis_state}) can be written as
\begin{align} \label{eq:spacsAAterm}
|\psi_{\rm out}\rangle^{\mathrm{SPACS}}&=\mathcal{N'}\left(\sum_{i=0}^{n-1}{\alpha^*}^{n-i}\left(\hat{U}\hat{\mathcal{A}}^{(i)}\hat{U}^\dagger\right)\ket{0_1,\dots,0_m}+|AA\rangle\right),
\end{align}
where $\hat{\mathcal{A}}^{(i)}$ is defined for $i\in\{0,1,\cdots,n\}$ as
\begin{equation}
\hat{\mathcal{A}}^{(i)}\equiv
\left\{
	\begin{array}{cl}
		\frac{1}{i!(n-i)!}\sum_{\sigma\in S_n}\prod_{k=1}^{i}\hat{a}^\dagger_{\sigma(k)},  & \mbox{if } i \geq 1 \\
		{\hat{\mathbb{I}}}, & \mbox{if } i =0,
	\end{array}
\right.
\end{equation}
with $S_n$ being the symmetric group of degree $n$, $\hat{\mathbb{I}}$ being the identity operator, and $\mathcal{N'}=1/(\sqrt{1+|\alpha|^2})^n$. Now we are left with performing photon number detection at the output. Detection events consisting of detecting all photons $n$ at the output correspond to sampling of the $|AA\rangle$ term from the superposition. The probability of detecting a total of $i$ photons at the output can be written as
\begin{align}
P_{i}=\mathcal{N'}^{2} {n\choose{i}}\left(|\alpha|^2\right)^{n-i},
\end{align}
since there are ${n\choose{i}}$ terms in $\hat{\mathcal{A}}^{(i)}$, each with a weight of $\mathcal{N'}^{2}\left(|\alpha|^2\right)^{n-i}$.

\subsection{The Quantum-Classical Divide and Computational-Complexity Transitions} \label{sec:SPACStransition}

Now we would like to analyse the computational complexity of this scheme. We know that the $|AA\rangle$ term in Eq. (\ref{eq:spacsAAterm}) is computationally complex to sample from and is obtained when detecting $n$ photons at the output. To show computational hardness we ask how should $|\alpha|$ scale in terms of $n$ (i.e. the total number of SPACS in the input) so that the probability of detecting $n$ photons at the output may be obtained in a polynomial number of measurements. In other words the post-selection probability of the interferometer scales inverse polynomially in $n$ since this scaling would guarantee that a polynomial number of measurements would obtain samples from the desired $|AA\rangle$ term at the output. 

To show this we will let ${\rm poly} (n)=n^k$, where $k\in \mathbb{Z}^+$ (the set of positive integers). Using the state of a SPACS from Eq. (\ref{input}) and solving for $|\alpha|$ that satisfies the above scaling requirement in the limit of a large $n$, we have
\begin{align}
\frac{1}{(1+|\alpha|^2)^n}&\geq \frac{1}{{\rm poly} (n)}\nonumber\\
\Rightarrow 1+|\alpha|^2&\leq ({\rm poly} (n))^{1/n}\nonumber\\
&\leq 1+\epsilon(n),
\label{psp1}
\end{align}
where the third inequality is due to the fact that for all $k\in \mathbb{Z}^+$,
\begin{align}
\lim_{n\rightarrow\infty}(n^k)^{1/n}&=\lim_{n\rightarrow\infty}e^{\frac{k}{n}\log n}\nonumber\\
&=\lim_{n\rightarrow\infty}e^{\frac{k}{n}}=e^{0^+}=1+\epsilon(n).
\label{psp2}
\end{align}
From Eq. (\ref{psp1}), we have
\begin{equation}
|\alpha|^2\leq \epsilon(n),
\label{psp3}
\end{equation}
and the large-$n$ expansion
\begin{equation}
e^{\frac{k}{n}\log n}=1+\frac{k}{n}\log n+O(\frac{1}{n^2}),
\end{equation}
tells us that $\epsilon(n)\geq (k/n)\log n$. The chain of inequalities
\begin{align}
\epsilon(n)\geq \frac{k\log n}{n}\geq\frac{1}{n}
\end{align}
implies $|\alpha|^2\leq 1/n$ is a sufficient condition on $|\alpha|$ to ensure that the post-selection probability of the $|AA\rangle$ term scales inverse polynomially in $n$. For $|\alpha|^2=1/n$, in the limit of large $n$, the probability of the term $|AA\rangle$ being detected at the output is
\begin{align}
P_n=\lim_{n\rightarrow\infty}\frac{1}{(1+\frac{1}{n})^n}=\frac{1}{e}\approx 36\%.
\end{align}
 
Further, the probability $P_n$ converges to one when $|\alpha|^2=1/n^2$. In other words the sampling problem with SPACS inputs reduces to AA \BS without the need for post selection. This result is consistent with AA's original result that \BS is robust against small amounts of noise \cite{bib:AaronsonArkhipov}. 

Another way analyse the computational complexity is by asking how should $|\alpha|$ scale so that the photon number sampling almost always yields the $m$-mode vacuum? For $|\alpha|^2=n^2$, we find that the probability of the $m$-mode vacuum term being detected at the output is
\begin{align}
P_0&=\lim_{n\rightarrow\infty}\frac{{(n^2)}^n}{(1+n^2)^n}\nonumber\\
&=\lim_{n\rightarrow\infty}\frac{1}{(1+\frac{1}{n^2})^n}=1.
\end{align}
This shows that the sampling problem with SPACS inputs becomes classically easy when $|\alpha|^2$ scales as $n^2$, or larger because it always results in the detection of the $m$-mode vacuum at the output. 

W found that that the computational complexity of sampling the SPACS goes from being just as hard as AA's \BS for coherent amplitudes when
\begin{equation}
|\alpha|^2\leq 1/n,
\end{equation}
to being classically simulatable when 
\begin{equation}
|\alpha|^2\geq n^2,
\end{equation}
where $n$ is the total number of SPACS inputs. There is an intermediate regime between $1/n$ and $n^2$ where we were not able to prove the computational complexity and so regime is left open as in interesting phenomena to investigate. Interestingly, this problem becomes classically easy to simulate when the number of photons at the output exceeds the fluctuation of photon number in the coherent states. 

\section{Photon-Added or -Subtracted Squeezed Vacuum State Sampling} \label{sec:SampOtherPASSV}

In this section we demonstrate that the \BS problem using photon-added or -subtracted squeezed vacuum (PASSV) states at the input and parity measurements at the output is of equal computational complexity to Fock state \BS for an arbitrary amount of squeezing. To do this we prove that this problem implements the same logical problem as \BS whereby the output statistics of the device is given by the same matrix permanent sampling problem. This is advantageous because we avoid doing the full complexity proof as done by AA, which was about one hundred pages long, yet we can still use their results. To do this we are careful to show that this problem is equivalent to the \BS problem. Also, since we are able to show an exact mapping to \BStwo, then the error threshold for approximate \BS holds. For consistency and simplicity, we consider in this analysis for the case of photon-added squeezed vacuum (PASV) states throughout this work and show that the subtracted (PSSV) case arbitrarily follows.

In section \ref{sec:PASSVinput} we discuss the PASSV input state and describe a non-deterministic method for preparing this input state. In section \ref{sec:PASSVevolution} we discuss the evolution of this state, which is similar to the evolution that is used in standard \BS except that we choose a Haar-random unitary with all real elements. In section \ref{sec:PASSVoutput} the output state is calculated and in section \ref{sec:PASSVmeasurement} describe how parity measurement is used to obtain the same computationally complex output statistics as \BStwo. Next we mention some complexity concerns in section \ref{sec:PASSVcomplexity} and wrap this section up with some discussion about this work in \ref{sec:PASSVdiscussion}.   

\subsection{Input} \label{sec:PASSVinput}

\begin{figure}[!htb]
\centering
\includegraphics[width=0.3\columnwidth]{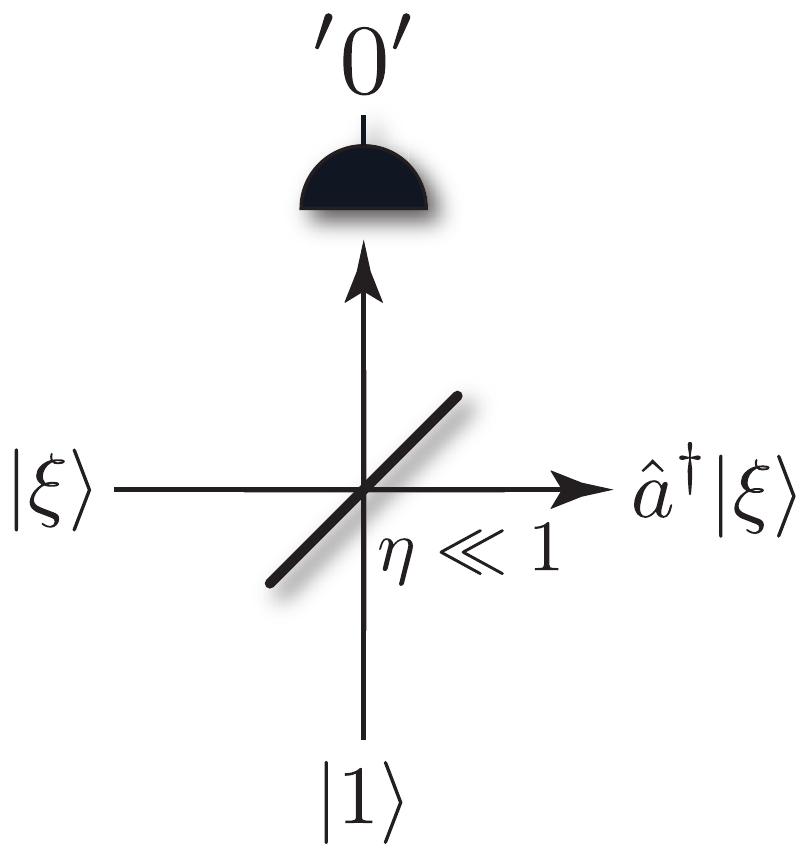}
\caption{Preparation of a PASV state. A SV state is mixed with a single-photon state on a low reflectivity beamsplitter. The top output mode is detected. Upon successful post-selection of the vacuum state the PASV state is prepared in the right output mode. The process is highly non-deterministic but can be performed before the sampling protocol.} \label{fig:prep}
\end{figure}

PASV states may be prepared by mixing a SV state (obtained from a degenerate parametric down-converter) with a single-photon state on a low reflectivity beamsplitter and post-selecting upon detecting the vacuum state in one of the output modes. The preparation scheme is shown in Fig. \ref{fig:prep}. When the post-selection is successful the PASV state is heralded in the other output mode. This scheme is non-deterministic, as are most quantum optical shemes, but may be performed in advance to the sampling protocol. PSSV states may be prepared similarly by sending in a squeezed state and a vacuum state to the inputs and post-selecting on one photon in one of the modes. 

For PASV \BS we prepare the first $n$ modes with PASV states and the remaining \mbox{$m-n$} modes with squeezed vacuum (SV) states. We let each mode have the same amount of squeezing given by the squeezing parameter $\xi$, which is of arbitrary value. The input state is
\begin{eqnarray} \label{eq:input_state}
\ket{\psi}_\mathrm{in}^\mathrm{SV} &=& \hat{a}_1^\dag\hat{S}_1(\xi)\dots\hat{a}^\dag_n\hat{S}_n(\xi)\hat{S}_{n+1}(\xi)\dots \hat{S}_m(\xi)\ket{0_1,\dots,0_m} \nonumber \\
&=& \hat{a}^\dag_1\dots \hat{a}^\dag_n\ket{\xi_1,\dots,\xi_m},
\end{eqnarray}
where we have abbreviated $\hat{S}_i(\xi)\ket{0_i}=\ket{\xi_i}$ and again the subscript indicates mode number. The input state in Eq. (\ref{eq:input_state}) is not normalized but can be normalized with the state $\mathcal{N}\ket{\psi}_\mathrm{in}^\mathrm{SV}$, where
\begin{equation} \label{eq:normalization}
\mathcal{N}= \Big[\sqrt{1+\sinh^2(\xi)}\;\Big]^{-n}.
\end{equation}
This normalization does not affect our result so we leave it out for simplicity. The squeezing operator is
\begin{equation}
\hat{S}(\xi)=\exp\left[\frac{1}{2}(\xi^*\hat{a}^2-\xi\hat{a}^\dag{}^2)\right],
\end{equation}
where $\hat{a}^\dag$ and $\hat{a}$ are the photon creation and annihilation operators respectively. In the Fock basis, if \mbox{$\xi=re^{i\theta}$}, then \mbox{$\hat{S}(\xi)\ket{0}=\ket{\xi}$} may be expressed as \cite{bib:GerryKnight05}
\begin{equation} \label{eq:sv}
\ket{\xi}=\frac{1}{\sqrt{\cosh(r)}}\sum_{m=0}^\infty (-1)^m \frac{\sqrt{(2m)!}}{{2^m m!}}e^{im\theta}\tanh^m(r)\ket{2m}.
\end{equation}
It can be seen that the SV state contains only even photon-number terms. When a creation or annihilation operator acts on a PASSV state then the resulting state contains only odd photon-number terms. In the limit of zero squeezing the SV state approaches the vacuum state
\begin{equation}
\lim_{\xi\to 0}\ket{\xi} = \ket{0},
\end{equation}
and the PASV state approaches the single-photon state
\begin{equation}
\lim_{\xi\to 0}\hat{a}^\dag\ket{\xi} = \ket{1}.
\end{equation}
Thus, we see that in the limit of vanishing squeezing, PASV \BS reduces to ideal Fock state \BStwo.

\subsection{Evolution} \label{sec:PASSVevolution}

The input state is fed into a passive linear optics interferometer  consisting of beamsplitters and phaseshifters, like in the original \BS protocol, which transforms the creation operators according to a linear map
\begin{equation} \label{eq:unitary_map}
\hat{U}\hat{a}_i^\dag\hat{U}^\dag \to \sum_j U_{i,j} \hat{a}_j^\dag,
\end{equation}
where $\hat{U}$ is an {$m\times m$} matrix. For PASSV \BS we consider an interferometer consisting of \emph{real} beamsplitters that implements an orthogonal matrix, which is also chosen to be Haar-random. So the difference is that for Fock state \BS \mbox{$\hat{U}_\mathrm{AA}\in SU(m)$}, whereas for PASSV \BS \mbox{$\hat{U}_\mathrm{SV} \in SO(m)$}. Reck \emph{et al.} showed that any \mbox{$m\times m$} unitary or orthogonal matrix can be implemented with at most $O(m^2)$ optical elements and that there is an efficient algorithm for finding the decomposition \cite{bib:Reck94}.

The complexity of choosing an orthogonal matrix instead of a unitary one is a concern because there is the possibility of choosing a subset of matrices from $SU(m)$, whose permanent is efficiently simulatable by a classical computer. If we were to sample from an efficiently simulatable distribution, then the result would not be interesting since the whole interesting aspect of \BS is that it simulates a classically intractable system. Later we prove that the associated complexities are equivalent and so the problem remains interesting.

\subsection{Output} \label{sec:PASSVoutput}

\begin{figure}[t]
\centering
\includegraphics[width=\imageWidth\columnwidth]{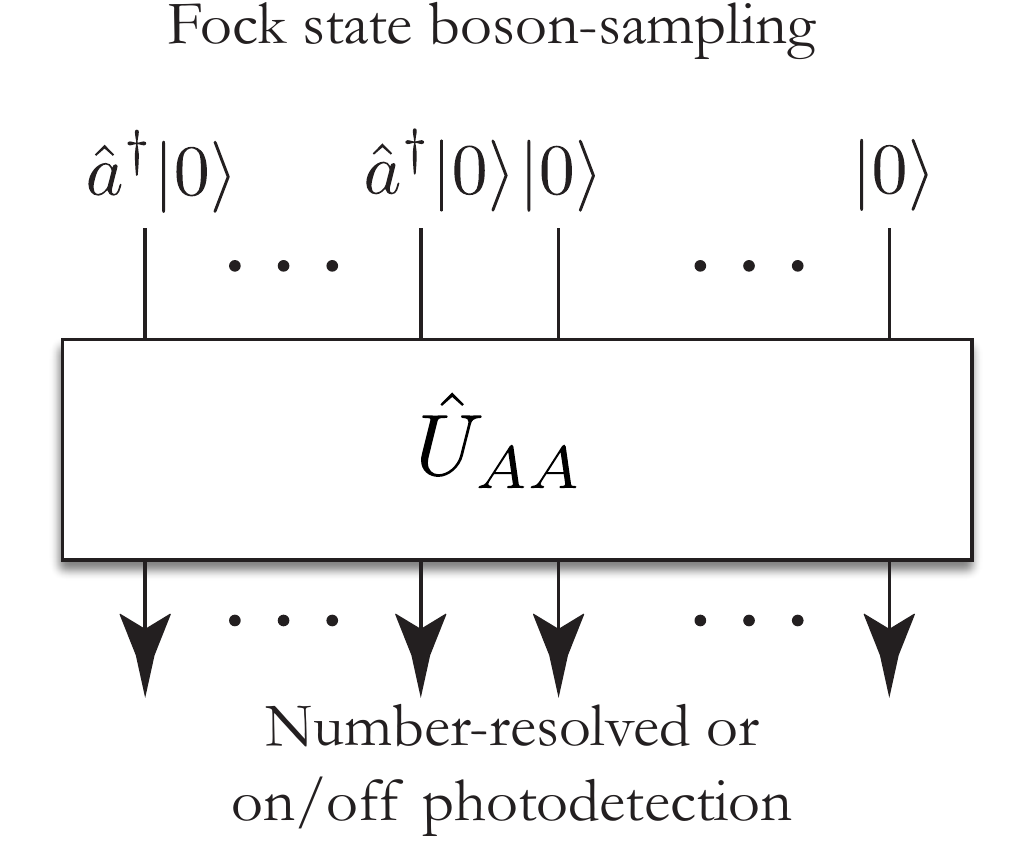}
\includegraphics[width=\imageWidth\columnwidth]{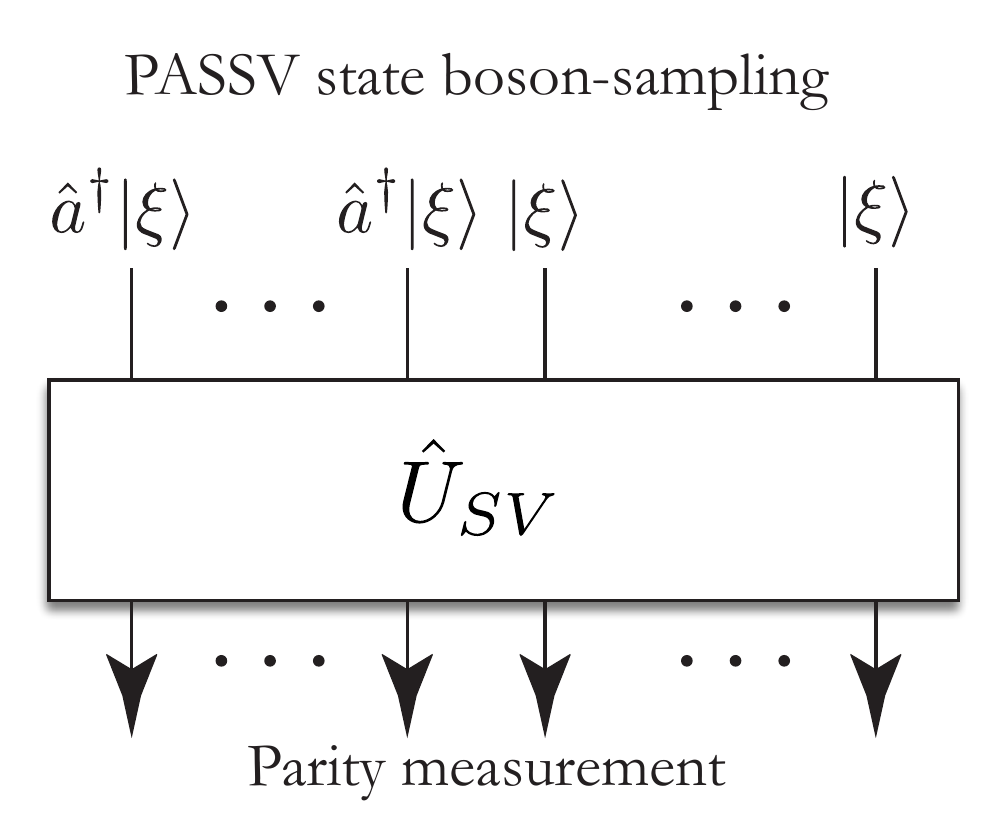}
\caption{(Left) The original Fock state model of \BS whereby we feed an $m$-mode linear optical interferometer with $n$-single photons and \mbox{$m-n$} vacuum states and sample via coincidence number-resolved photodetection. (Right) PASSV \BS whereby we prepare $n$ PASV states instead of single photons and \mbox{$m-n$} SV states instead of vacuum states and sample via coincidence parity measurement.}
\end{figure}

The output state for the original Fock state model of \BS after passing through the interferometer can be expressed as
\begin{eqnarray} \label{eq:aaoutput1}
\ket{\psi}_\mathrm{out}^\mathrm{AA} &=& \hat{U}_\mathrm{AA}\ket{\psi}_\mathrm{in}^\mathrm{AA} \nonumber \\
&=& \hat{U}_\mathrm{AA}\left[\hat{a}_1^\dag\dots\hat{a}_n^\dag\ket{0_1,\dots,0_m}\right] \nonumber \\
&=& \left[\hat{U}_\mathrm{AA}(\hat{a}_1^\dag\dots\hat{a}_n^\dag)\hat{U}_\mathrm{AA}^\dag\right]\hat{U}_\mathrm{AA}\ket{0_1,\dots,0_m} \nonumber \\
&=& \left[\hat{U}_\mathrm{AA}(\hat{a}_1^\dag\dots\hat{a}_n^\dag)\hat{U}_\mathrm{AA}^\dag\right]\ket{0_1,\dots,0_m},
\end{eqnarray}
where the last equality holds because $U_{AA}\ket{0}=\ket{0}$  (i.e. $U_{AA}$ represents passive optical elements which cannot generate new photons). 

With PASSV \BS we can use the same technique as in Eq. (\ref{eq:aaoutput1})
\begin{eqnarray} \label{eq:svoutput1}
\ket{\psi}_\mathrm{out}^\mathrm{SV} &=& \hat{U}_\mathrm{SV}^\mathrm{}\ket{\psi}_\mathrm{in}^\mathrm{SV} \nonumber \\
&=& \left[\hat{U}_\mathrm{SV}(\hat{a}^\dag_1\dots\hat{a}^\dag_n)\hat{U}_\mathrm{SV}^\dag\right]\hat{U}_\mathrm{SV}\ket{\xi_1,\dots,\xi_m}. 
\end{eqnarray}
It was shown by Jiang \emph{et al.} \cite{bib:PhysRevA.88.044301} that a pure product state input to a linear optical network is entangled at the output unless the input is either a tensor product of coherent states or a tensor product of squeezed states (with the same squeezing) provided that the network does not mix the squeezed and anti-squeezed quadratures. Not mixing the quadratures can be achieved by using a network comprised of real beamsplitters. This condition is satisfied since we are using \mbox{$\hat{U}_{SV}\in SO(m)$} and so the output state becomes
\begin{equation}
\label{eq:svoutput1b}
\ket{\psi}_\mathrm{out}^\mathrm{SV}= \left[\hat{U}_\mathrm{SV}(\hat{a}^\dag_1\dots\hat{a}^\dag_n)\hat{U}_\mathrm{SV}^\dag\right]\ket{\xi_1,\dots,\xi_m}.
\end{equation}
The leading operator corresponds to a configuration of $n$ creation operators as in Eq. (\ref{eq:aaoutput1}). The output can therefore be represented in a form like \BS where we distinguish all of the possible output distributions
\begin{equation} \label{eq:svoutput2}
\ket{\psi}_\mathrm{out}^\mathrm{SV} =\sum_{S}\gamma_{S}' \left[(\hat{a}_1^\dag)^{S_1} \dots (\hat{a}_m^\dag)^{S_m}\right]\ket{\xi_1,\dots,\xi_m},
\end{equation}
where,
\begin{equation}
\gamma_S' = \frac{\gamma_S}{\sqrt{S_1!\dots S_m!}}=\frac{\textrm{Per}(U_S)}{\sqrt{S_1!\dots S_m!}}.
\end{equation}
In the binary regime \mbox{$\gamma_S'=\gamma_S$}. For PSSV states the output is of the same form. This can be seen by replacing $\hat{a}^\dag_i$ with $\hat{a}_i$, but $\gamma_S$ now relates to $\hat{U}^\dag_{SV}$ instead of $\hat{U}_{SV}$, which is also Haar-random. We exclude the case of the PSSV states when $\xi=0$ since $\hat{a}\ket{0}=0$.

We know that from Eq. (\ref{eq:sv}) that squeezed states represented in the Fock basis have only even photon-number terms and so for a particular configuration $S$ where mode $i$ does not have a creation/annihilation operator acting on it, mode $i$ is a superposition of only even photon number states, whereas if mode $i$ has a creation/annihilation operator applied it contains only odd photon-number terms.

\subsection{Measurement} \label{sec:PASSVmeasurement}

The last step in \BS is to measure the output distribution. For PASSV \BS we perform a parity measurement that distinguishes  between odd and even photon number. These measurements may be characterised by the measurement operators
\begin{eqnarray}
\hat{\Pi}_+ &=& \ket{0}\bra{0} + \ket{2}\bra{2} + \ket{4}\bra{4} + \dots \\ \nonumber
\hat{\Pi}_- &=& \ket{1}\bra{1} + \ket{3}\bra{3} + \ket{5}\bra{5} + \dots.
\end{eqnarray}
A photon-number-resolving detector could easily implement this measurement scheme. Importantly this measurement scheme implies that measuring an even photon-number at output mode $i$ means that there was no creation/annihilation operator associated with that mode, whereas measuring an odd photon-number implies that there was. This measurement perfectly recovers the configuration $S$ since we are sampling the same creation/annihilation operators as in \BStwo. An interesting observation is that the squeezing parameter $\xi$ has no effect on the parity of the state and so the sampling statistics are  independent of the squeezing. In other words an arbitrary amount of squeezing can be done and a computationally complex sampling problem is still implemented.


\subsection{Complexity Concerns} \label{sec:PASSVcomplexity}

We have shown that the PASSV model samples permanents of submatrices in the same way as Fock state sampling. The only barrier to completing our proof is that \BS and PASSV sampling is in the same complexity class is to show whether choosing an orthogonal matrix has any implications for the complexity of PASSV sampling. 

The first thing to consider is whether or not a Haar-random matrix in $SO(m)$ might have an efficiently computable exact or approximate permanent, where $SO(m)$ is the $m$-dimensional rotation group. It is known that the exact permanent case is \#\textbf{P}-hard even for binary entries of the matrix, \mbox{$U_{i,j}\in\{0,1\}$} \cite{bib:Valiant79}. If the matrix has entries consisting of only non-negative real numbers then there is also a known algorithm for efficiently approximating a permanent. Also, in the same work it was shown that for a matrix with a single negative entry that an efficient approximation algorithm would allow one to compute an \textit{exact} \mbox{$\{0,1\}$}-permanent efficiently \cite{bib:Jerrum04}. Although computing a difficult permanent is a necessary but not sufficient condition for computational hardness since $SO(m)$ is considered to be universal for linear optics \cite{bib:Bouland}, there is no such complexity condition between unitary and orthogonal matrices.

More concretely, it is known that $SU(m)\subset SO(2m)$, where $SU(m)$ is the Lie group of \mbox{$m\times m$} unitary matrices \cite{bib:Georgi99}. This statement means that for a $2m$-mode interferometer, the set of all orthogonal transformations includes all unitary $m$-mode transformations as a subgroup. This means that the complexity of sampling the output of a \BS device implementing an arbitrary matrix from $SO(2m)$ is at least as hard as sampling matrices from $SU(m)$ with only a linear increase of mode number. The same complexity extends to an odd number of modes since $SO(2m)\subset SO(2m+1)$. This also implies that Fock state \BS itself remains hard under orthogonal transformations.

Now it is clear that PASSV \BS is in the same complexity class as Fock state AA \BStwo. If we let $A$ be some complexity class containing Fock state \BStwo, then we have shown that $A$ also contains PASSV \BStwo. The output of PASSV \BS is completely independent of the squeezing parameter $\xi$ and so we may assume without loss of generality that $\xi=0$. In this limit $\ket{\xi_i}=\ket{0_i}$ and thus this case of PASSV \BS reduces to an instance of Fock state \BS since $SO(m)\subset SU(m)$. Now suppose $B$ is some complexity class containing PASSV \BStwo.  Again choosing $\xi=0$, the inclusion $SU(m)\subset SO(2m)$ similarly implies that $B$ also contains Fock state \BStwo.

\subsection{Discussion of PASSV Sampling} \label{sec:PASSVdiscussion}

This PASSV result can be thought of in terms of letting the ket in Eq. (\ref{eq:aaoutput1}) act as a `background' signal which is invariant during the evolution of $\hat{U}_{SV}$. Since the leading operator in Eq. (\ref{eq:svoutput1b}) takes the same form as Eq. (\ref{eq:aaoutput1}), we would like the ket to also be independent of the choice of $\hat{U}_{SV}$ under some measurement, while still being distinguishable from a state which has an added or subtracted photon. The technique used in this work may be able to be used to characterize other states which implement a logically equivalent classically intractable sampling problem. A goal of ours is to prove an even more experimentally friendly set of states and measurements that implements the same computationally complex sampling problem.

PASSV \BS is harder to implement than ordinary \BStwo, which is already quite challenging. One particularly strong criticism of PASSV \BS is that it may require the use of photon-number resolving detectors to implement parity measurements instead of bucket detectors. Whilst this is true, one only needs to distinguish the parity of the even and odd photon-number Fock states and not the value of the Fock state. 

For any given $\xi$ and error rate the maximum number of necessarily distinguishable Fock states may be reduced. PASSV \BS may be regarded as a generalization of Fock state \BStwo, since in the limit of small squeezing (\mbox{$\xi\rightarrow 0$}), the SV reduces to a vacuum state and an on/off detector suffices. Additional experimental hurdles arise because squeezed states become more susceptible to noise. We do not address experimental errors such as this in our work. Our goal is to theoretically demonstrate the non-uniqueness of Fock states for computationally hard sampling problems and develop new techniques for understanding the computational complexity of other sampling problems. 

In this work we have shown that orthogonal matrices are sufficiently hard for PASSV sampling. A natural question is whether or not choosing another matrix, such as a unitary matrix, could change the complexity of the PASSV sampling problem.  This question is not so easily solved since Eq. (\ref{eq:svoutput1b}) no longer holds. Intuitively we expect that the problem would not become easier. In the limit of zero squeezing, we know there is no special computational complexity transition since PASSV sampling reduces to Fock state sampling. If such a complexity transition did exist, such as in SPACS sampling of section \ref{sec:SampOtherSPACS}, then we would expect a complexity phase transition at $\xi=0$. The same sampling probabilities however may be constructed with a different measurement scheme but this remains an open question.

\section{Generalized Cat State Sampling} \label{sec:SampOtherCATS}

In this section we consider a \BS device where the input states are arbitrary superpositions of coherent states. This constitutes a  broad class of continuous-variable optical states. In this work we focus on analysing cat sampling in three separate limits:
\begin{enumerate}
\item First, in section \ref{sec:CATzeroAmplitude} we review the coherent state, analyse even and odd cat states, and show that their Taylor expansions reduce to the vacuum and single-photon Fock state respectively as $\alpha\to0$. Thus, in the zero amplitude limit, cat sampling exactly reduces to \BS and therefore yields a computationally hard problem. 
\item Second, in section \ref{sec:CATsmallAmplitude} we analyse small, but non-zero amplitude odd cat states. This is equivalent to Fock state sampling with some components that are treated as an error. This error is related to the AA proof for approximate \BStwo, where it is required that the error rate satisfies a 1/poly(n) bound. Thus small, but non-zero, amplitude odd cat states are also computationally hard.
\item Third, in section \ref{sec:CATarbitraryAmplitude} we analyse general cat states which are arbitrary superpositions of two or more coherent states. We demonstrate that the output state is a highly entangled superposition of an exponential number of multi-mode coherent states \cite{bib:PhysRevA.46.2966, bib:PhysRevA.55.2478, bib:gerry2002nonlinear, bib:gerry2007nonlocal, bib:gerry2009maximally}, where the amplitude of each term is related to a permanent-like combinatoric problem, which would require exponential resources to compute via a brute-force approach. This provides strong evidence that such generalized optical sampling problems might be implementing classically hard problems. Determining a complete characterization of the computational complexity of such problems is a notoriously difficult open problem, but based on the evidence we present here, it likely resides in a classically hard class comparable to ideal \BStwo.
\end{enumerate}

Next, in section \ref{sec:CATcomplexity} we present a complexity theoretic argument for the hardness of cat state sampling to further support our evidence. We show that unless the polynomial hierarchy collapses to the third level there must not exist an efficient randomized classical algorithm which can produce an output distribution approximating that of an arbitrary interferometer with multiplicative error of $\sqrt{2}$ or less.

While such states may be more challenging to prepare than Fock states, addressing this question sheds light on what makes a quantum optical system classically hard to simulate, and may provide motivation for developing technologies for preparing quantum states of light beyond Fock states. We end this work with section \ref{sec:CATpreparing} by discussing the prospects for experimentally preparing general cat states.

\subsection{Zero Amplitude Cat analysis} \label{sec:CATzeroAmplitude}

\emph{Cat state} is a generic term for an arbitrary superposition of macroscopic states and may be used for quantum information processing \cite{bib:gilchrist04}. In quantum optics, this is generally understood to mean a superposition of two coherent states, potentially with large amplitudes. This is the definition we will use in this work. 

Firstly, a coherent state is state of the quantum harmonic oscillator \cite{bib:GerryKnight05} that most closely resembles a state of a classical harmonic oscillator. It may be defined in the Fock basis as,
\begin{equation}
\ket{\alpha} = e^{-\frac{1}{2}|\alpha|^2}\sum_{n=0}^{\infty}\frac{\alpha^n}{\sqrt{n!}}\ket{n},
\end{equation}
where $\alpha=|\alpha|e^{i\theta}$ is a complex number, $|\alpha|$ represents the amplitude of the state, and $\theta$ represents the phase of the state.   

Two illustrative examples of superpositions of these coherent states are the \emph{even} ($+$) and \emph{odd} ($-$) cat states, so-called because they contain only even or odd photon-number terms respectively,
\begin{equation}
\ket{\mathrm{cat}_\pm} = \frac{(\ket\alpha \pm \ket{-\alpha})}{\sqrt{2(1\pm e^{-2|\alpha|^2})}}.
\end{equation}

The odd cat state has the property that all of the even photon number terms vanish. In the limit of $\alpha\to0$ its amplitude identically approaches the single-photon state as shown here,
\begin{eqnarray}
\lim_{\alpha\to 0} \ket{\mathrm{cat}_-} &=& \lim_{\alpha\to0} \frac{\sqrt{2}e^{\frac{-|\alpha|^2}{2}}}{\sqrt{1-e^{-2|\alpha|^2}}}\left(\alpha\ket{1}+\frac{\alpha^3\ket{3}}{\sqrt{3!}}+\dots\right) \nonumber \\
&\approx& \ket{1}+O(\alpha^2)\ket{3} \nonumber \\
&\to& \ket{1}.
\end{eqnarray}
In the limit as $\alpha\to0$ we ignore all higher order $\alpha$ terms. 

Furthermore, the vacuum state is given by a trivial cat state containing only a single term in the superposition with a respective amplitude of \mbox{$\alpha=0$}. Alternately, the vacuum state can be regarded as the zero amplitude limit of the even cat state,
\begin{eqnarray}
\lim_{\alpha\to 0} \ket{\mathrm{cat}_+} &=& \lim_{\alpha\to0} \frac{\sqrt{2}e^{\frac{-|\alpha|^2}{2}}}{\sqrt{1+e^{-2|\alpha|^2}}}\left(\alpha\ket{0}+\frac{\alpha^2\ket{2}}{\sqrt{2!}}+\dots\right) \nonumber \\
&\approx& \ket{0}+O(\alpha^2)\ket{2} \nonumber \\
&\to& \ket{0}.
\end{eqnarray}

Thus, it is immediately clear that in the $\alpha\to0$ amplitude limit, cat state sampling reduces to ideal \BStwo, using an appropriate configuration of odd and even cat states, which is a provably hard problem. We use the term `provably hard' to mean computationally hard, assuming that ideal and approximate \BS are computationally hard. Specifically, to implement exact \BS with cat states, we choose our input state to be,
\begin{eqnarray}
\ket{\psi_\mathrm{in}} &=& \lim_{\alpha\to 0} (\ket{\mathrm{cat}_-}_1\dots \ket{\mathrm{cat}_-}_n \ket{\mathrm{cat}_+}_{n+1}\dots \ket{\mathrm{cat}_+}_m) \nonumber \\
&=& \ket{1_1,\dots,1_n,0_{n+1},\dots,0_m},
\end{eqnarray}
which is exactly the form of Eq. (\ref{eq:BSintroInputState}). This example is trivial but the point is to show a simple example of cat states leading to a computationally hard problem in a particular limit, which raises the question as to whether it remains hard as we transition out of that limit. In App.~\ref{app:odd_cat} we present an example of this reduction in the case of Hong-Ou-Mandel interference to explicitly demonstrate that small amplitude cats behave as single photons. This demonstrates that in certain regimes, cat state sampling reproduces single-photon statistics.

\subsection{Small Amplitude Cat Analysis} \label{sec:CATsmallAmplitude}
Having established that cat sampling reduces to \BS in the zero amplitude limit, the obvious next question is `what if the amplitude is small but non-zero?'. It was shown by AA that \BStwo, when corrupted by erroneous samples, remains computationally hard provided that the error rate scales as $1/\mathrm{poly}(n)$. If we consider a small, but non-zero, amplitude odd cat state, we can treat the non-single-photon terms, which scale as a function of $\alpha$, as erroneous terms. The error that these erroneous terms induce must be kept below the $1/\mathrm{poly}(n)$ bound. Specifically,
\begin{equation}
\ket{\mathrm{cat}_-} = \underbrace{\gamma_1(\alpha)\ket{1}}_{\mathrm{single\,photon}} + \underbrace{\gamma_3(\alpha) \ket{3} + \dots}_{\mathrm{error\,terms}},
\end{equation}
where $\gamma_i(\alpha)$ defines the odd photon-number distribution and follows from Eq. (\ref{eq:fna}). The left underbraced component represents the desired single-photon term and the right underbraced component represents the remaining photon-number terms, which are treated as errors. 

In App.~\ref{app:poly_bound} we show that the bound on the amplitude of the cat states for a provably hard sampling problem to take place is,
\begin{equation} \label{eq:bound_alpha_hard}
\alpha^{2n}\mathrm{csch}^n(\alpha^2) > 1/\mathrm{poly}(n),
\end{equation}
where we input odd cat states in every mode requiring a $\ket{1}$ and vacuum in the remaining modes. Although this function is exponential in $n$ the probability of successfully sampling from the correct distribution will satisfy this bound for sufficiently small values of $n$ and $\alpha$. The value of $n$ may still be large enough however to implement a post-classical \BS device. Thus, it follows that for non-zero, but sufficiently small $\alpha$, cat sampling remains computationally hard.

We have established that cat state \BS is a provably computationally hard problem in two regimes: (1) in the $\alpha\to0$ amplitude limit, in which case we reproduce ideal \BStwo, and (2) for non-zero but sufficiently small amplitudes, in which case the non-single-photon-number terms may be regarded as errors, which remains a computationally hard problem, subject to the bound given in Eq. (\ref{eq:bound_alpha_hard}). Having established this, the remainder of this work is dedicated to the completely general case where the terms in the cat states may have arbitrary amplitude, potentially at a macroscopic scale.

%
%

\subsection{Arbitrary Amplitude Cat Analysis} \label{sec:CATarbitraryAmplitude}

In this section we will consider generalized cat state sampling which are arbitrary superpositions of an arbitrary number of coherent states of the form
\begin{equation}
\ket{\mathrm{cat}} = \sum_{j=1}^t \lambda_j \ket{\alpha_j}.
\end{equation}

We let the input state to our  more generalized \BS model comprise $m$ arbitrary superpositions of $t$ coherent states
\begin{equation}
\ket{\psi_\mathrm{in}} = \bigotimes_{i=1}^m \sum_{j=1}^t \lambda_j^{(i)}\ket{\alpha_{j}^{(i)}},
\end{equation}
where $\ket{\alpha_{j}^{(i)}}$ is the coherent state of amplitude \mbox{$\alpha\in\mathbb{C}$} of the $j$th superposition term in the $i$th mode, and \mbox{$\lambda_j^{(i)}\in\mathbb{C}$} is the amplitude of the $j$th term of the superposition in the $i$th mode\footnote{Continuous superpositions are a simple generalization of our formalism, and with this generalization arbitrary states could be expressed as continuous superpositions of coherent states.}. In line with traditional \BStwo, we can choose a number of the modes to be the vacuum. This is achieved by setting \mbox{$\lambda_1^{(i)}=1$} and \mbox{$\alpha_1^{(i)}=0$}.

\begin{figure}[!htb]
\centering
\includegraphics[width=0.45\columnwidth]{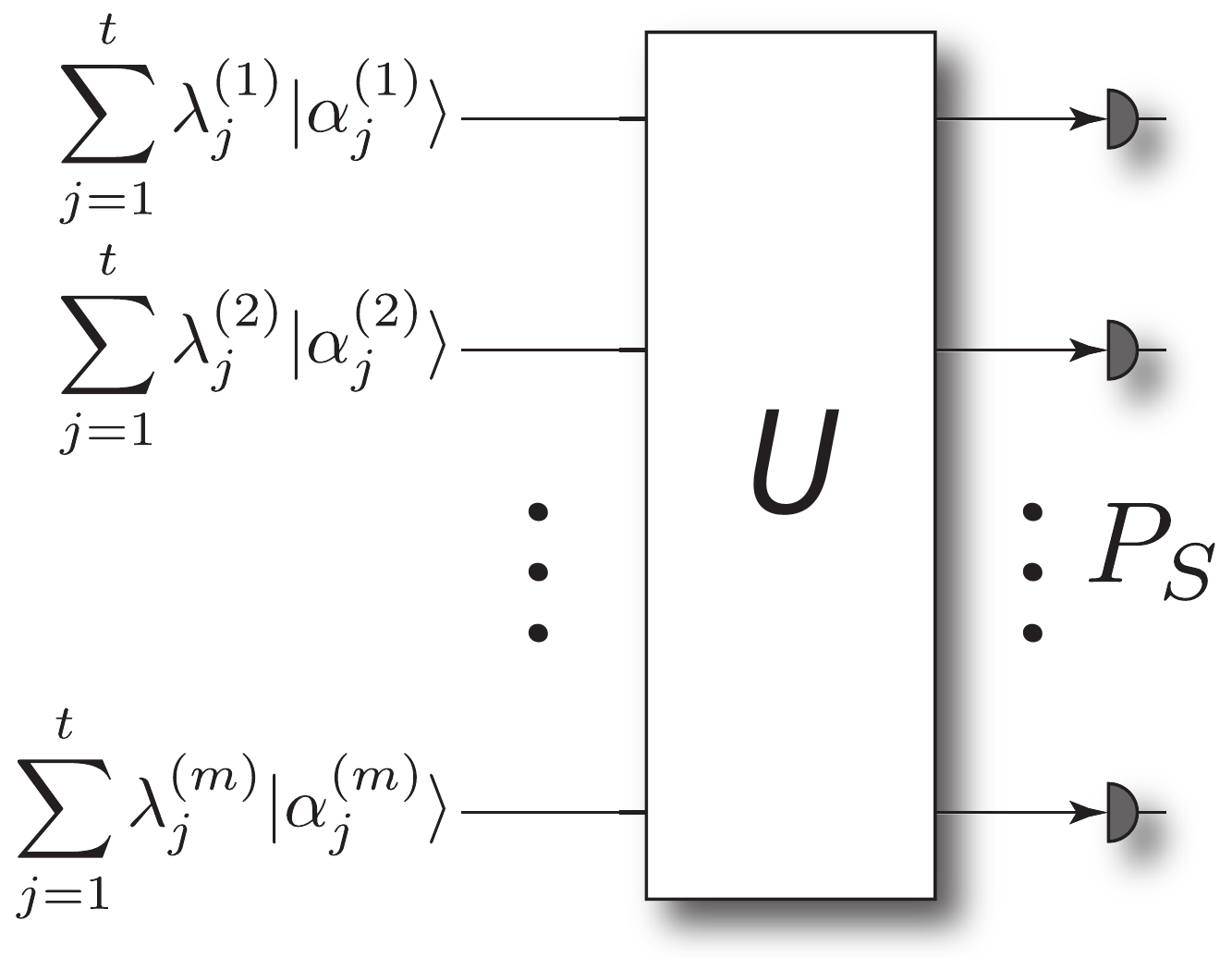}
\caption{The model for generalized \BS with generalized cat states. The input state to each mode is a tensor product of an arbitrary superposition of coherent states with a unique superposition in each mode. Some of these may be set to the vacuum. Following the application of a linear optics network, the distribution is sampled via number-resolved photo-detection.} \label{fig:cat_model}
\end{figure}

Expanding this expression yields a superposition of multi-mode coherent states of the form
\begin{equation} \label{eq:psi_in}
\ket{\psi_\mathrm{in}} = \sum_{\vec{t}=1}^{t} \lambda_{t_1}^{(1)}\dots \lambda_{t_m}^{(m)} \ket{\alpha_{t_{1}}^{(1)}, \dots, \alpha_{t_{m}}^{(m)}},
\end{equation}
where $\vec{t}$ is shorthand for $\{t_1,...,t_m\}$. We propagate this state through the passive linear optics network $\hat{U}$ as illustrated in Fig.~\ref{fig:cat_model}. Such a unitary network has the property that a multi-mode coherent state is mapped to another multi-mode coherent state,
\begin{equation} \label{eq:coherent_map}
\hat{U} \ket{\alpha^{(1)},\dots,\alpha^{(m)}} \to \ket{\beta^{(1)},\dots,\beta^{(m)}},
\end{equation}
where the relationship between the input and output amplitudes is given by
\begin{equation} \label{eq:coherent_map_relation}
\beta^{(j)} = \sum_{k=1}^m U_{j,k} \alpha^{(k)},
\end{equation}
as shown in App. \ref{app:coherent_map}. $\hat{U}$ acts on each term in the superposition of Eq. (\ref{eq:psi_in}) independently yielding an output state of the form,
\begin{eqnarray} \label{eq:psi_out}
\ket{\psi_\mathrm{out}} &=& \hat{U} \ket{\psi_\mathrm{in}} \nonumber \\
&=& \sum_{\vec{t}=1}^{t} \lambda_{t_1}^{(1)}\dots \lambda_{t_m}^{(m)} \ket{\beta_{\vec{t}}^{(1)}, \dots, \beta_{\vec{t}}^{(m)}}.
\end{eqnarray}
The number of terms in the output superposition is $t^m$, scaling exponentially with the number of modes given that \mbox{$t>1$}.

Our goal is to sample this distribution using number-resolved photodetectors, which are described by the measurement projectors,
\begin{equation}
\hat\Pi_i(n) = \ket{n}_i\bra{n}_i,
\end{equation}
where $n$ is the photon-number measurement outcome on the $i$th mode. Multi-mode measurements are described by the projectors,
\begin{equation} \label{eq:projector_S}
\hat\Pi(S) = \hat\Pi_1(S_1) \otimes \dots \otimes \hat\Pi_m(S_m),
\end{equation}
where \mbox{$S=\{S_1,\dots,S_m\}$} is the multi-mode measurement signature, with $S_i$ photons measured in the $i$th mode. The sample probabilities are given by,
\begin{equation} \label{eq:sample_prob}
P_S = \bra{\psi_\mathrm{out}} \hat\Pi(S) \ket{\psi_\mathrm{out}}.
\end{equation}
In the case of continuous-variable states, the number of measurement signatures, $|S|$, is unbounded as the photon-number is undefined, unlike Fock states where the total photon-number is conserved. 

Now we argue, without presenting a rigorous complexity argument, that this sampling problem is likely computationally hard if three criteria are satisfied:
\begin{enumerate}
\item There must be an exponential number of terms in the output distribution to rule out brute-force simulation.
\item The terms in the superposition are entangled so that the distribution cannot be trivially obtained by sampling each mode independently.
\item To ensure that the individual output amplitudes are not easy to simulate the output distribution must be related to a computationally hard problem.
\end{enumerate}

These criteria are general properties that classically hard problems are known to exhibit, but there is no proof that these criteria are sufficient to establish whether a problem is classically hard. For example, ideal \BS is known to be computationally hard, satisfying all 3 criteria but fermionic-sampling is known to be classically efficient since it violates criteria (3) because its output amplitudes relate to matrix determinants rather than permanents, which reside in the computational complexity class \textbf{P}.

Criteria (1) is achieved due to our choice of input state --- there are $t^m$ terms in the output distribution. It is easily seen that criteria (2) holds in general. As a simple example, consider the input state,
\begin{equation}
\ket{\psi_\mathrm{in}} = \mathcal{N}^2 (\ket\alpha + \ket{-\alpha})\otimes (\ket\alpha + \ket{-\alpha}) = \ket{\mathrm{cat_+},\mathrm{cat_+}},
\end{equation}
a tensor product of two even cat states. Passing this separable two-mode state through a 50/50 beamsplitter gives rise to the output state,
\begin{eqnarray}
\ket{\psi_\mathrm{out}} = \hat{H}\ket{\psi_\mathrm{in}} = \ket{\mathrm{cat}',0} + \ket{0,\mathrm{cat}'},
\end{eqnarray}
where \mbox{$\ket{\mathrm{cat}'}=\mathcal{N}^{2}_{+}(\ket{\sqrt{2}\alpha} + \ket{-\sqrt{2}\alpha})$} is a cat state. This is a path-entangled superposition of a cat state across two modes. Thus, while Eq. (\ref{eq:coherent_map}) demonstrates that a unitary network maps a tensor product of coherent states to a tensor product of coherent states, such a network will generate path-entanglement when the input state is a tensor product of superpositions of coherent states. Note the structural similarity between cat state interference and two-photon Hong-Ou-Mandel (HOM) \cite{bib:HOM87} type interference. In the case of HOM interference we have \mbox{$\hat{H}\ket{1,1}= (\ket{2,0}+\ket{0,2})/\sqrt{2}$}, whereas for cat states we have \mbox{$\hat{H}\ket{\mathrm{cat},\mathrm{cat}}= \ket{\mathrm{cat}',0} + \ket{0,\mathrm{cat}'}$}.

It was recently and independently reported by Jiang \emph{et al.} \cite{bib:PhysRevA.88.044301} that linear optics networks fed with nonclassical pure states of light almost always generates modal entanglement, consistent with our observation here. This ensures that the output state to our generalized \BS device is highly entangled, thus satisfying criteria (2). However, Jiang \textit{et al.} present no discussion about our hardness criteria (3); they do not connect their states to a computationally hard problem. Thus their work provides a necessary but not sufficient proof of computational hardness. It is important, as in our work here, to examine such non-classical input states individually and make the case for the importance of criteria (3). For example it is well known from the Gottesman-Knill theorem that some systems with exponentially large Hilbert spaces that satisfy our criteria (1) and (2) can nevertheless be efficiently simulated. An example is the circuit model for quantum computation that deploys only gates from the Clifford algebra.

Finally let us consider criteria (3). Let the expansion for a coherent state be,
\begin{equation}
\ket\alpha = \sum_{n=0}^\infty f_n(\alpha)\ket{n},
\end{equation}
in the photon-number basis, where,
\begin{equation} \label{eq:fna}
f_n(\alpha) = e^{-\frac{|\alpha|^2}{2}} \frac{\alpha^n}{\sqrt{n!}},
\end{equation}
is the amplitude of the $n$-photon term. Then,
\begin{equation}
\langle n|\alpha\rangle = f_n(\alpha).
\end{equation}
Thus, acting the measurement projector for configuration $S$, Eq. (\ref{eq:projector_S}), on the output state, Eq. (\ref{eq:psi_out}), we obtain, 
\begin{equation}
\hat\Pi(S) \ket{\psi_\mathrm{out}} = \gamma_S \ket{S_1, S_2, \dots,S_m},
\end{equation}
where,
\begin{eqnarray} \label{eq:gamma}
\gamma_S = \sum_{\vec{t}=1}^{t} \left(\prod_{j=1}^m \lambda_{t_j}^{(j)} f_{S_j}\!\!\left(\sum_{k=1}^m U_{j,k} \alpha_{t_j}^{(k)}\right) \right),
\end{eqnarray}
and the sampling probability takes the form \mbox{$P_S = |\gamma_S|^2$}. We can group the terms under the product and label them $A_{j,\vec{t}}^{(S)}$. Then the amplitudes are given by,
\begin{equation} \label{eq:S_perm}
\gamma_S = \sum_{\vec{t}=1}^t \prod_{j=1}^m A_{j,\vec{t}}^{(S)},
\end{equation}
which has the same analytic structure as the permanent when $t=m$ but sums over additional terms that are not present in the permanent. Evaluating this combinatoric problem requires exponential resources using brute-force. Via brute force, evaluating this expression requires summing $t^m$ terms. Given that Eq. (\ref{eq:S_perm}) has the same analytic form as the matrix permanent, which is known to be classically hard, this implies a striking similarity between cat state sampling and Fock state sampling, with the constraint that $A$ is of a form whose permanent is not trivial. In fact, in the $\alpha\to0$ limit, evaluating this combinatoric expression \emph{must} be as hard as calculating an $n\times n$ matrix permanent, since we know that in this limit the problem reduces to ideal \BStwo. In the original proof by Aaronson \& Arkhipov, it is required that $U$ is Haar-random. It is an open question as to whether $A$ can be made Haar-random in the presented generalized \BS model.

In the trivial case of \mbox{$t=1$} this expression simplifies to,
\begin{equation}
\gamma_S = \prod_{j=1}^m f_{S_j}\!\!\left(\sum_{k=1}^m U_{j,k} \alpha_{1}^{(k)}\right),
\end{equation}
which evaluates in polynomial time. In this case the input state is simply a tensor product of coherent states, and the runtime is consistent with the known result that simulating coherent states is trivial as the tensor product structure allows sampling to proceed by independently sampling each mode, each of which is an efficient sampling problem. However, when \mbox{$t>1$} the complexity of directly arithmetically evaluating Eq. (\ref{eq:gamma}) grows exponentially.

%
%

\subsection{A Computational Complexity Argument} \label{sec:CATcomplexity}


We found that unless the polynomial hierarchy collapses to the third level (i.e. $\textbf{P}^{\#}\textbf{P}=\textbf{BPP}^{N}\textbf{P}$ \cite{bib:AaronsonArkhipov}) there must not exist an efficient randomized classical algorithm which can approximate generalized cat state sampling with a multiplicative error of $\sqrt{2}$ or less. The existence of this algorithm would imply the collapse of computational complexity classes believed to be distinct. The polynomial hierarchy \textbf{PH} is composed of an infinite number of levels $k$, which are themselves composed of complexity classes \mbox{$\Sigma_k \mathbf{P} = \mathbf{NP}^{\Sigma_{k-1} \mathbf{P}}$}, \mbox{$\Delta_k \mathbf{P} = \mathbf{P}^{\Sigma_{k-1} \mathbf{P}}$}, and \mbox{$\Pi_k \mathbf{P} = \mathbf{coNP}^{\Sigma_{k-1} \mathbf{P}}$}, where \mbox{$\Sigma_0 \mathbf{P} = \Delta_0 \mathbf{P} = \Pi_0 \mathbf{P} = \mathbf{P}$}. 


Bremner, Jozsa \& Shepherd \cite{bib:bremner2011classical} developed a method for showing the intractability of classical simulation of certain circuits composed of commuting gates that we use. This technique has since been used in other work \cite{bib:morimae2014hardness, bib:morimae2014classical}. Post-selection is important in the method as the computational cost for certain computation is used only if that computation results in a particular post-selection measurement.

We provide informal definitions used in our proof of two computational complexity classes based on post-selection also used in \cite{bib:morimae2014hardness}. For formal definitions please see \cite{bib:bremner2011classical}. These classes are defined in terms of either classical or quantum circuits with output and post-selection registers $O_x$ and $P_x$ respectively. A language $L$ is in the class $\mathbf{PostBPP}$ if and only if there exists a uniform family of classical circuits and a \mbox{$0<\delta<1/2$} such that:
\begin{enumerate}
\item if $x \in L$ then $\text{Prob}(O_x=1|P_x = 0\dots 0) \geq \frac{1}{2}+\delta$,
\item if $x \notin L$ then $\text{Prob}(O_x=1|P_x = 0\dots 0) \leq \frac{1}{2}-\delta$.
\end{enumerate}
Similarly, a language $L$ is in the class \textbf{PostBQP} if the above criteria is satisfied for a uniform family of quantum circuits. We define a third complexity class \textbf{PostCAT} to understand post-selection applied to the interferometry experiments. We let a language $L$ be in \textbf{PostCAT} if a uniform family of $n$-port linear interferometers acting on cat state inputs satisfying the preceding criteria exist.

It is known that \textbf{PostBPP} corresponds to $\mathbf{BPP}_\text{path}$, which is contained within \mbox{$\Delta_3 \mathbf{P}$} \cite{bib:han1997threshold}. It was shown by Aaronson that \mbox{$\mathbf{PostBQP} = \mathbf{PP}$} \cite{bib:aaronson2005quantum}. This is surprising, since \mbox{$\mathbf{PH} \subseteq \mathbf{P}^\mathbf{PP}$}, implying a difference between the power of post-selected classical and quantum computation unless \mbox{$\mathbf{P}^\mathbf{PP} = \mathbf{PH} = \Delta_3 \mathbf{P}$}. Following from the definitions of \textbf{PostBPP} and \textbf{PostBQP} given above, this strengthens the argument that an efficient classical randomized algorithm cannot produce the output of any quantum circuit. This is because it would yield \mbox{$\mathbf{PostBQP}=\mathbf{PostBPP}$} and collapse the polynomial hierarchy, which is believed to be highly unlikely. More specifically: The existence of a polynomial time randomized classical algorithm which approximates the output distribution to an arbitrary quantum circuit to within multiplicative error of $\sqrt{2}$ would yield \mbox{$\mathbf{PH} = \Delta_3 \mathbf{P}$} \cite{bib:bremner2011classical,bib:morimae2014hardness}. 

To finish the proof we need to show that $\mathbf{PostBQP} \subseteq \mathbf{PostCAT}$, from which it would follow that the existence of an efficient randomized classical algorithm which can approximate the output distribution of an arbitrary linear interferometer applied to cat state inputs to within a multiplicative error of $\sqrt{2}$ would imply that \mbox{$\mathbf{PostCAT} \subseteq \mathbf{PostBPP}$} and hence \mbox{$\mathbf{PH} = \Delta_3 \mathbf{P}$}. It has already been shown that \mbox{$\mathbf{PostBQP} \subseteq \mathbf{PostCAT}$} in the work of Ralph \emph{et al.} \cite{bib:ralph}. They proved that arbitrary quantum circuits could be probabilistically implemented exactly on qubits encoded as a superposition of even and odd parity cat states and there is always some probability of obtaining a measurement result which correctly implements the desired gate with unit fidelity. This means that by post-selecting on these outcomes, the system can be made to implement an arbitrary quantum circuit. Now because post-selection could be applied to the output of the circuit and remaining in \textbf{PostCAT}, it means that any computation in \textbf{PostBQP} is also in \textbf{PostCAT}. Therefore, unless \mbox{$\mathbf{PH} = \Delta_3 \mathbf{P}$} there does not exist an efficient randomized classical algorithm which can produce an output distribution approximating that of an arbitrary interferometer with multiplicative error of $\sqrt{2}$ or less.

%
%

\subsection{Preparing Cat States} \label{sec:CATpreparing}

Finally, we will discuss the prospects for experimentally preparing cat states of the form used in our derivation. There exists a significant number of schemes for generating a finite number of superpositions of coherent states all of which are extremely difficult to scale to higher order cat states. For example, superpositions of coherent states with equal amplitudes but different phases can be produced with quantum nondemolition (QND) measurements \cite{bib:PhysRevA.37.2970} via the interaction of a strong Kerr nonlinearity \cite{bib:Haroche06, bib:GerryKnight05}. Another approach is to use strong Kerr nonlinearities together with coupled Mach-Zehnder interferometers \cite{bib:GerryKnight05} but this is impractical as outside the cavity a strong Kerr would require a coherent Electromagnetically Induced Transparency \cite{bib:harris2008, bib:PhysRevLett.64.1107} effect in an atomic gas cloud and even there in practice the nonlinearities are too weak for our purposes.

In a similar way that measurements of photon number can produce discrete coherent state superpositions in phase. Measurements of the phase can produce discrete coherent state superpositions in amplitude. This can be understood via the number-phase uncertainty relation. Any improved knowledge of the phase of a state induces kicks in the number and vice versa. In this way, by combining such different measurements, one can produce discrete superpositions in both phase and amplitude, which approaches the arbitrary superpositions of coherent states we require. Exactly such a scheme was proposed by Jeong \emph{et al.} in 2005 \cite{bib:Lund04, bib:JeongLundRalph05}. By combining both types of detection schemes, even with detectors of non-unit efficiency, they show that a large number of propagating superpositions of coherent states may be produced. These states then could be used in proof-of-principle experiments for our protocol outlined here.

\section{Summary}

An open question is `what classes of quantum states of light yield hard sampling problems using linear optics?' In this chapter we have studied this question in detail. We began by motivating this problem in section \ref{sec:SampOtherMotivation}. Specifically, we studied single-photon-added coherent states (SPACS) in section \ref{sec:SampOtherSPACS}, photon-added or -subtracted squeezed vacuum (PASSV) states in section \ref{sec:SampOtherPASSV}, and generalized superpositions of coherent states called cat states in section \ref{sec:SampOtherCATS}. Below we present a summary of the conclusions we were able to draw from sampling these three states:

\begin{itemize}
\item SPACS: We first showed that displaced single-photon Fock state sampling remains hard to efficiently simulate for all values of the displacement with a coincidence photon number detection. We then considered a more interesting problem using SPACS sampling and found that this problem transitions from computationally hard to simulate too computationally easy as the amplitude of the coherent states increases from near zero into the limit of large coherent amplitudes.  

\item PASSV: We have shown a direct mapping between Fock state \BS and PASSV \BS and that it operates in \emph{all} squeezing parameter regimes. In other words there are no bounds on the amount of squeezing and no approximations need to be made. This is unique as compared to the case of SPACS sampling. In this protocol we use a Haar random matrix with all real elements for the evolution which is different than the original \BS protocol where the Haar random matrix has complex values as well. This was so the squeezed and anti-squeezed quadratures of the squeezed vacuum do not mix with each other. 

\item Generalized Cat State Sampling: We have presented evidence that a linear optics network, fed with arbitrary superpositions of coherent states, and sampled via number-resolved photodetection, is likely to be a classically hard problem. Our argument relies on three realistic criteria for computational hardness of a sampling problem of which these satisfy. We led the reader into the more difficult generalized case by first analysing what happens in the zero amplitude case and then the small amplitude limit case. Also, because coherent states form an over-complete basis, any pure optical state can be expressed in terms of coherent states, suggesting that most quantum states of light may yield hard sampling problems.

\end{itemize}

These results demonstrate that there are a large class of non-Fock states that have associated sampling problems of equal computational complexity to \BS which means that there is nothing unique about the computational complexity of single-photon Fock state \BStwo. In fact, it seems that there is a plethora of other quantum states of light that exhibit similar sampling computational complexity. These results support a conjecture presented in \cite{bib:GardCrossDowling} that computational complexity relates to the negativity of the Wigner function as all of these analysed states have negative Wigner functions. This work helps us to understand what constitutes a computationally hard sampling problem and what it is about quantum states that are computationally complex to sample from. The states presented here are more experimentally challenging to create than single-photon Fock states. This work will further motivate the need to develop quantum state sources than can produce states other than Fock states. 

\begin{savequote}[45mm]
If you fail, never give up because F.A.I.L means ``First Attempt In Learning.'' End is not the end, in fact E.N.D. means ``Effort Never Dies.'' If you get no as an answer remember N.O. means ``Next Opportunity.'' So Let us be positive.
\qauthor{A.P.J. Abdul Kalam}
\end{savequote}

\chapter{\BS inspired Linear Optical Quantum Metrology --- An Application} \label{Ch:MORDOR}


\section{Synopsis}
It is known that quantum number-path entanglement is a resource for super-sensitive quantum metrology as it allows for sub-shotnoise or even Heisenberg-limited sensitivity. All known methods for generating such number-path entanglement are extremely challenging because it requires either very strong nonlinearities, or nondeterministic preparation schemes with feed-forward, which are difficult to implement. We know from studying quantum random walks with multi-photon walkers as well as \BS that passive linear optical devices generate a superexponentially large amount of number-path entanglement.

In this work we show a method to use this resource of entanglement for quantum metrology, which is motivated in section \ref{sec:MORDOR_Motivation}. We show in section \ref{sec:MORDOR_Device} that a simple, passive, linear-optical interferometer --- fed with only uncorrelated, single-photon inputs, coupled with simple, single-mode, disjoint photodetection --- is capable of beating the shotnoise limit. It is important to note that this protocol is an alteration to the original \BS protcol as presented by AA and so it is an application inspired by \BS and not exactly an aplication of \BStwo. Nonetheless, our result allows for practical quantum metrology with readily available technology. Due to the uniqueness of our architecture we use a new resource counting method that we coined ordinal resource counting (ORC) as discussed in section \ref{sec:MORDOR_ORC}.\footnote{It is worth noting that this metrology work has three Lord of the Rings \cite{bib:tolkien} references in it. See if you can find them.}

\section{Motivation} \label{sec:MORDOR_Motivation}
Quantum number-path entanglement is a resource for super-sensitive quantum metrology, as shown by Yurke \& Yuen, allowing for sensors that beat the shotnoise limit \cite{bib:yurke1986input, bib:yuen1986generation}. These sensors have applications in super-sensitive gyroscopy \cite{bib:dowling1998correlated}, gravimetry \cite{bib:yurtsever2003interferometry}, optical coherence tomography \cite{bib:nasr2003demonstration}, ellipsometry \cite{bib:toussaint2004}, magnetometry \cite{bib:jones2009magnetic}, protein concentration measurements \cite{bib:crespi2012measuring}, and microscopy \cite{bib:rozema2014scalable, bib:israel2014supersensitive}. This line of work culminated in the analysis of the bosonic NOON state (\mbox{$(\ket{N,0}+\ket{0,N})/\sqrt{2}$}, where $N$ is the total number of photons). This was shown to be optimal for local phase estimation with a fixed, finite number of photons, and in fact allows one to hit the Heisenberg limit and the Quantum Cram{\'e}r-Rao Bound \cite{bib:holland1993interferometric, bib:lee2002quantum, bib:durkin2007local, bib:dowling2008quantum}.

The NOON state is used in a two-mode interferometer where all $N$ particles are in a superposition of being in the first mode (with zero in the second mode) or in the second mode (with zero in the first mode). This state is known to be optimal as it reaches the Heisenberg limit but its generation is known to be quite difficult. There are two main methods for preparing NOON states: the first is to deploy very strong optical nonlinearities \cite{bib:gerry2001generation, bib:kapale2007bootstrapping}, and the second is to prepare them using measurement and feed-forward \cite{bib:lee2002linear, bib:vanmeter2007general, bib:cable2007efficient}. These are similar requirements to building a universal optical quantum computer and thus NOON-states are just as difficult to build \cite{bib:Kok07}. In addition to being complicated to prepare the detection scheme is quite challenging as parity measurements at each output port also likely need to be performed \cite{bib:seshadreesan2013phase}. 

Recently two independent lines of research, the study of quantum random walks with multi-photon walkers in passive linear-optical interferometers \cite{bib:mayer2011counting, bib:gard2013quantum, bib:gard2014inefficiency}, as well as the computational complexity analysis of the sampling problem using such devices \cite{bib:AaronsonArkhipov, bib:introChapter}, has led to a startling conclusion --- passive, multi-mode, linear-optical interferometers, fed with only uncorrelated single photons in each mode, produce quantum mechanical states with path-number entanglement that grows superexponentially fast in the two resources of mode and photon-number. For another practical application inspired by \BS see \cite{bib:vibSpec}. It is remarkable is that such a large degree of number-path entanglement is generated without the use of strong optical nonlinearities or with complicated measurement and feed-forward schemes. It is generated using the evolution of the single photons in a passive linear optical device. It is commonly misunderstood that such passive devices have `non-interacting' photons in them. There is however a type of photon-photon interaction due to the demand of bosonic state symmetrization from multiple applications of the Hong-Ou-Mandel effect \cite{bib:gard2014inefficiency}, which yields a superexponentially large amount of number-path entanglement. It is known that the evolution of single photons in a linear optical device, followed by projective measurements, can give rise to `effective' strong optical nonlinearities. We conjecture that there is a hidden Kerr-like nonlinearity in these interferometers \cite{bib:lapaire2003conditional}. Like \BS \cite{bib:AaronsonArkhipov}, and unlike universal quantum computing schemes such as that by Knill, Laflamme, and Milburn \cite{bib:LOQC}, this protocol is deterministic and does not require any ancillary photons.

The advantage of our \BS inspired method for quantum metrology is that generating and detecting single photons is quite standardized and relatively straightforward to implement in the lab \cite{bib:matthews2011heralding, bib:Spring2, bib:Broome2012, bib:crespi2013integrated, bib:ralph2013quantum, bib:motes13, bib:Spagnolo13}. The linear optical community is moving towards single photons, linear interferometers, and single-photon detectors all on a single, integrated, photonic chip, which allows for the scalability of linear optical devices to large numbers of modes and photons. This implies that scalable quantum metrology using our technique is feasible in the near future. We now show a method for using passive linear optics for quantum metrology. 

\section{Metrological Device} \label{sec:MORDOR_Device}
The phase-sensitivity, $\Delta\varphi$, of a metrology device can be defined in terms of the standard error propagation formula as, 
\begin{eqnarray} \label{eq:phaseSensitivity}
\Delta\varphi = \frac{\sqrt{\expec{\hat{O}^2}-\expec{\hat{O}}^2}}{\left|\frac{\partial\expec{\hat{O}}}{\partial\varphi}\right|}, 
\end{eqnarray}
where $\expec{\hat{O}}$ is the expectation of the observable being measured and $\varphi$ is the unknown phase we seek to estimate.

The photons evolve through a unitary network according to $U a_i^{\dag} U^{\dag} = \sum_j U_{ij} a_j^{\dag}$. In our protocol, we construct the $n$-mode interferometer $\hat{U}$ to be,
\begin{equation} \label{eq:MORDORU}
\hat{U} = \hat{V} \cdot \hat{\Phi} \cdot \hat{\Theta} \cdot \hat{V}^{\dag},
\end{equation}
which we call the quantum fourier transform interferometer (QuFTI) because $\hat{V}$ is the $n$-mode quantum Fourier transform matrix, with matrix elements given by,
\begin{equation}
\mathrm{V}_{j,k}^{(n)} = \frac{1}{\sqrt{n}}\mathrm{exp}\left[\frac{- 2 i j k \pi}{n}\right].
\end{equation} 
$\hat\Phi$ and $\hat\Theta$ are both diagonal matrices with linearly increasing phases along the diagonal represented by,
\begin{eqnarray} \label{eq:PhiTheta}
\Phi_{j,k} = \delta_{j,k} \exp{\Big[i(j-1)\varphi\Big]} \nonumber \\
\Theta_{j,k} = \delta_{j,k} \exp{\Big[i(j-1)\theta\Big]},
\end{eqnarray}
where $\varphi$ is the unknown phase one would like to measure and $\theta$ is the control phase. $\hat\Theta$ is introduced as a reference, which can calibrate the device by tuning $\theta$ appropriately. To see this tuning we combine $\hat\Phi$ and $\hat\Theta$ into a single diagonal matrix with a gradient given by,
\begin{equation}
\Phi_{j,k}\cdot\Theta_{j,k} = \delta_{j,k} \exp{\bigg[i(j-1)(\varphi+\theta)\bigg]}.
\end{equation}
The control phase $\theta$ can shift this gradient to the optimal measurement regime, which can be found by minimizing $\Delta\varphi$ with respect to $n$ and $\varphi$. Since this is a shift according to a known phase, we can for simplicity assume (and without loss of generality) that $\varphi$ is in the optimal regime for measurements and $\theta=0$. Thus, $\hat\Theta=\hat{I}$ and is left out of our analysis for simplicity.

\begin{figure}[!htb]
\centering
\includegraphics[width=0.6\columnwidth]{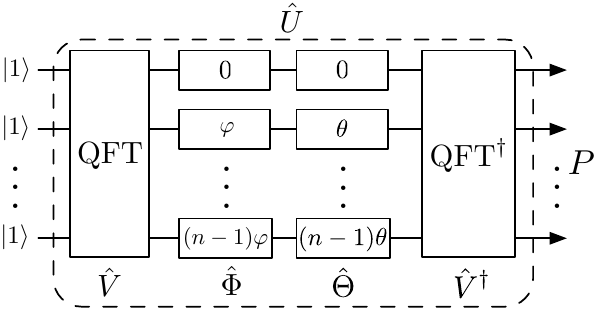}
\caption{Architecture of the quantum Fourier transform interferometer (QuFTI) for metrology using single-photon states. The input state comprises $n$ single photons, \mbox{$\ket{1}^{\otimes n}$}. The state evolves via the passive linear optics unitary \mbox{$\hat{U} = \hat{V} \cdot \hat{\Phi} \cdot \hat{\Theta} \cdot \hat{V}^\dag$}, where $\hat{V}$ is the quantum Fourier transform, $\hat\Phi$ is an unknown, linear phase gradient, and $\hat\Theta$ is a reference phase gradient used for calibration. At the output we perform a coincidence photodetection projecting on exactly one photon per output mode, measuring the observable $\hat{O}=(\ket{1}\bra{1})^{\otimes n}$, which, over many measurements, yields the probability distribution $P(\varphi)$ that acts as a witness for the unknown phase $\varphi$.} \label{fig:arch}
\end{figure}

In order to understand how such a linearly increasing array of unknown phase shifts may be arranged in a practical device, it is useful to consider a specific example. Let us suppose that we are to use the QuFTI as an optical magnetometer. We consider an interferometric magnetometer of the type discussed in \cite{bib:scully1992high} where each of the sensing modes of the QuFTI contains a gas cell of Rubidium prepared in a state of electromagnetically induced transparency whereby a photon passing through the cell at the point of zero absorption in the electromagnetically induced transparency spectrum acquires a phase shift   that is proportional to the product of an applied uniform (but unknown) magnetic field   and the length of the cell. We assume that the field is uniform across the QuFTI, as would be the case if the entire interferometer was constructed on an all optical chip and the field gradient across the chip were negligible. Since we are carrying out local phase measurements (not global) we are not interested in the magnitude of the magnetic field but wish to know if the field changes and if so by how much. (Often we are interested in if the field is oscillating and with what frequency.) Neglecting other sources of noise then in an ordinary Mach-Zehnder interferometer this limit would be set by the photon shotnoise limit. To construct a QuFTI with the linear cascade of phase shifters, as shown in Fig. \ref{fig:arch}, we simply increase the length of the cell by integer amounts in each mode. The first cell has length $L$, the second length $2L$, and so forth. This will then give us the linearly increasing configuration of unknown phase shifts required for the QuFTI to beat the SNL. 

One might question why one would employ a phase gradient rather than just a single phase. Investigation into using a single phase in $\hat\Phi$ indicates that this yields no benefit. We conjecture that this is because the number of paths interrogating a phase in a single mode is not superexponential as is the case when a phase gradient is employed.

The interferometer may always be constructed efficiently following the protocol of Reck \emph{et al.} \cite{bib:Reck94}, who showed that an \mbox{$n\times n$} linear optics interferometer may be constructed from $O(n^2)$ linear optical elements (beamsplitters and phase-shifters), and the algorithm for determining the circuit has runtime polynomial in $n$. Thus, an experimental implementation of our protocol may always be efficiently realized.

The input state to the device is \mbox{$\ket{1}^{\otimes n}$}, i.e. single photons inputed in each mode. If \mbox{$\varphi=0$} then $\hat\Phi=\hat{I}$ and thus $\hat{U}=\hat{V}\cdot\hat{I}\cdot\hat{V}^{\dag}=\hat{I}$. In this instance, the output state is exactly equal to the input state, \mbox{$\ket{1}^{\otimes n}$}. Thus, if we define $P$ as the coincidence probability of measuring one photon in each mode at the output, then \mbox{$P=1$} when \mbox{$\varphi=0$}. When \mbox{$\varphi\neq 0$}, in general \mbox{$P<1$}. Thus, intuitively, we anticipate that \mbox{$P(\varphi)$} will act as a witness for $\varphi$. 

In the protocol, assuming a lossless device, no measurement events are discarded. Upon repeating the protocol many times, let $x$ be the number of measurement outcomes with exactly one photon per mode, and $y$ be the number of measurement outcomes without exactly one photon per mode. Then $P$ is calculated as $P=x/(x+y)$. Thus, all measurement outcomes contribute to the signal and none are discarded. Note that, due to preservation of photon-number and the fact that we are considering the anti-bunched outcome, $P(\varphi)$ may be experimentally determined using non-number-resolving detectors if the device is lossless.  If the device is assumed to be lossy, then number-resolving detectors would be necessary to distinguish between an error outcome and one in which more than one photon exits the same mode.  The circuit for the architecture is shown in Fig. \ref{fig:arch}.

The state at the output to the device is a highly path-entangled superposition of $\binom{2n-1}{n}$ terms, which grows superexponentially with $n$.  This corresponds to the number of ways to add $n$ non-negative integers whose sum is $n$, or equivalently, the number of ways to put $n$ indistinguishable balls into $n$ distinguishable boxes. We conjecture that this superexponential path-entanglement yields improved phase-sensitivity as the paths query the phases a superexponential number of times.

The observable being measured is the projection onto the state with exactly one photon per output mode, \mbox{$\hat{O} = (\ket{1}\bra{1})^{\otimes n}$}. Thus, \mbox{$\langle\hat{O}\rangle = \langle\hat{O}^2\rangle = P$}. And, the phase-sensitivity estimator reduces to,
\begin{equation} \label{eq:phaseSenP}
\Delta\varphi = \frac{\sqrt{P - P^2}}{\left|\frac{\partial P}{\partial \varphi}\right|}.
\end{equation}

Following the result of \cite{bib:Scheel04perm}, $P$ is related to the permanent of $\hat U$ as,
\begin{equation}
P = \big|\mathrm{Per}(U)\big|^2.
\end{equation}
Here the permanent of the full \mbox{$n\times n$} matrix is computed, since exactly one photon is going into and out of every mode. This is unlike the \BS protocol \cite{bib:AaronsonArkhipov} where permanents of sub-matrices are computed.

We will now  examine the structure of this permanent. The matrix form for the $n$-mode unitary $\hat U^{(n)}$ is given by,
\begin{equation} \label{eq:Ujk}
U_{j,k}^{(n)} =\frac{1-e^{i n\varphi}}{n\left(e^{\frac{2 i \pi(j-k)}{n}}-e^{i \varphi}\right)},
\end{equation}
as derived in App. \ref{app:Ujk}. Taking the permanent of this matrix is challenging as calculating permanents are in general \mbox{\textbf{\#P}-hard}. However, based on calculating $\mathrm{Per}(\hat U^{(n)})$ for small $n$, we observe the empirical pattern,
\begin{equation} \label{eq:permU}
\mathrm{Per}(\hat{U}^{(n)})= \frac{1}{n^{n-1}}\prod_{j=1}^{n-1}\Big[je^{i n \varphi}+n-j\Big],
\end{equation}
as conjectured in App. \ref{app:series}. This analytic pattern we observe is not a proof of the permanent, but an empirical pattern --- a conjecture --- that has been verified by brute force to be correct up to $n=25$. Although we do not have a proof beyond that point, $n=25$ is well beyond what will be experimentally viable in the near future, and thus the pattern we observe is sufficient for experimentally enabling super-sensitive metrology with technology available in the foreseeable future.

\begin{figure}[!htb]
\centering
\includegraphics[width=\imageWidthThree\columnwidth]{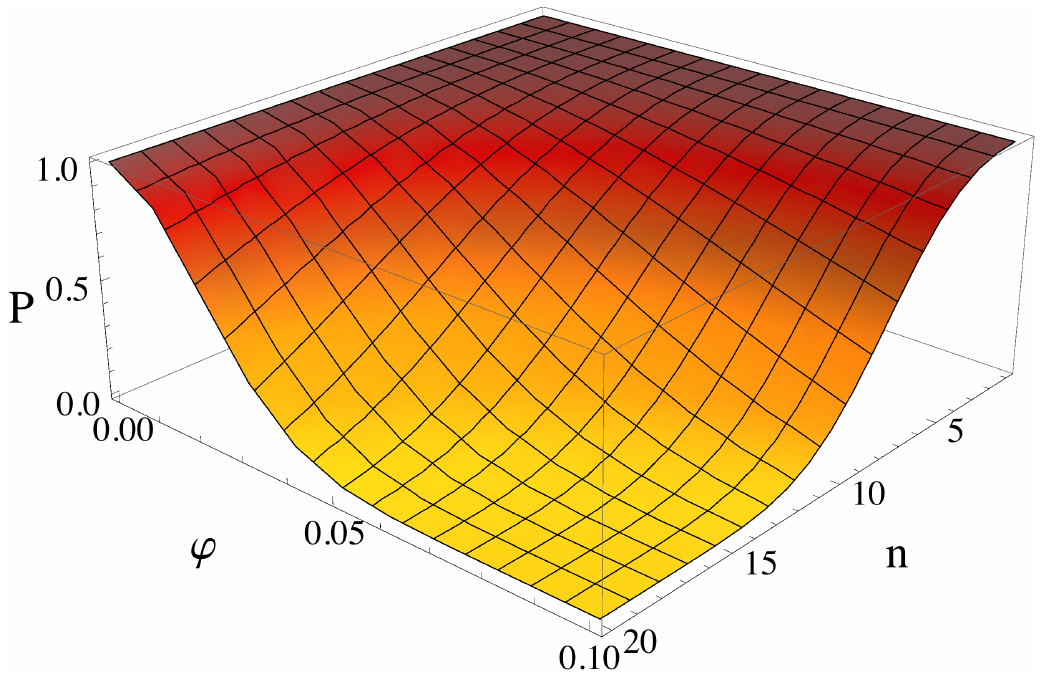}
\caption{Coincidence photodetection probability $P$ against the unknown phase $\varphi$ and the number of photons and modes $n$. As $n$ increases, the dependence of $P$ on $\varphi$ increases, resulting in improved phase-sensitivity.} \label{fig:P}
\end{figure}

Following as a corollary to the previous conjecture, the coincidence probability of measuring one photon in each mode is,
\begin{eqnarray} \label{eq:P_Result}
P &=& \Big|\mathrm{Per}(\hat{U}^{(n)})\Big|^2 \nonumber \\
&=& \frac{1}{n^{2n-2}}\prod_{j=1}^{n-1} \Big[a_n(j)\mathrm{cos}(n\varphi)+b_n(j) \Big],
\end{eqnarray} 
as shown in App. \ref{app:P}, where
\begin{eqnarray}
a_n(j) &=& 2j(n-j), \nonumber \\
b_n(j) &=& n^2-2jn+2j^2.
\end{eqnarray}
The dependence of $P$ on $n$ and $\varphi$ is shown in Fig. \ref{fig:P}.

It then follows that,
\begin{equation} \label{eq:dP}
\left|\frac{\partial P}{\partial \varphi}\right| = nP\big|\mathrm{sin}(n\varphi)\big|\sum_{j=1}^{n-1} \left|\frac{a_n(j)}{a_n(j)\mathrm{cos}(n\varphi)+b_n(j)}\right|,
\end{equation}
as shown in App. \ref{app:dP}.

\begin{figure}[!htb] 
\centering
\includegraphics[width=\imageWidthThree\columnwidth]{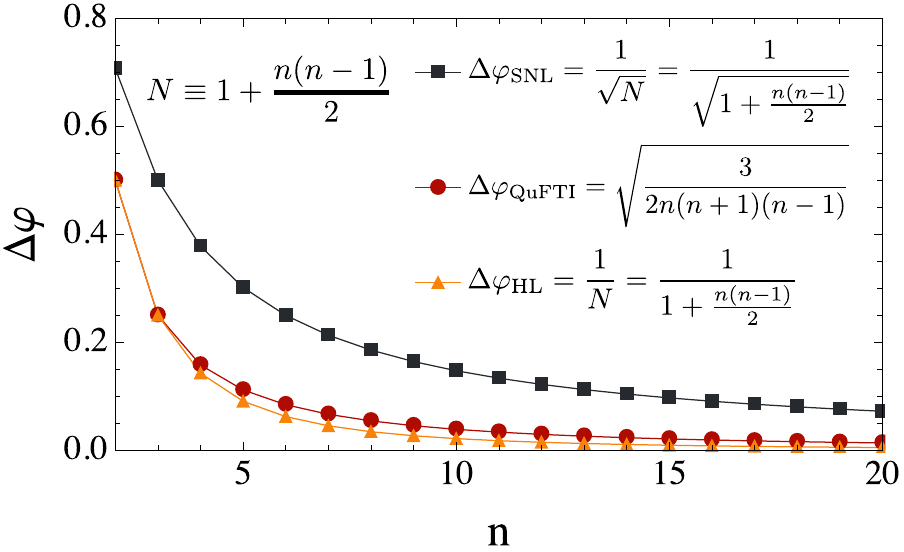}
\caption{Phase-sensitivity $\Delta\varphi$ against the number of photons $n$ (red circles). The shotnoise limit of $1/\sqrt{N}$ (black squares) and Heisenberg limit of $1/N$ (orange triangles) are shown for comparison. The QuFTI exhibits phase-sensitivity significantly better than the shotnoise limit, and only slightly worse than the Heisenberg limit. \label{fig:Delta}}  
\end{figure}

Finally, we wish to establish the scaling of $\Delta\varphi$. With a small $\varphi$ approximation (\mbox{$\mathrm{sin}(\varphi)\approx\varphi$}, \mbox{$\mathrm{cos}(\varphi)\approx 1-\frac{1}{2}\varphi^2$}) we find,
\begin{eqnarray} \label{eq:DeltaVarPhi}
\Delta\varphi &=& \sqrt{\frac{3}{2n(n+1)(n-1)}} \\ \nonumber
&=& \frac{1}{2\sqrt{{{n+1}\choose{3}}}},
\end{eqnarray}
as shown in App. \ref{app:dphi}. Thus, the phase sensitivity scales as \mbox{$\Delta\varphi = O(1/n^{3/2})$} as shown in Fig. \ref{fig:Delta}.

\section{Ordinal Resource Counting (ORC)} \label{sec:MORDOR_ORC}
We would like to compare the performance of our QuFTI to an equivalent multimode interferometer baseline for which we will construct the shotnoise limit (SNL) and Heisenberg limit (HL). This is a subtle comparison, due to the linearly increasing unknown phase-shifts, \{\mbox{$0,\varphi,\dots,(n-1)\varphi$}\}, that the QuFTI requires to operate. The mathematical relation is shown in Fig. \ref{fig:Delta}, where we have converted the number of resources, $N$, to the number of photons, $n$. There is disagreement on how such resources should be counted. This is the method, which we call Ordinal Resource Counting (ORC), that we feel most fairly counts our resources. A more detailed supporting discussion can be found in App. \ref{app:counting}.

While computing the sensitivity using the standard error propagation formula of Eq. (\ref{eq:phaseSensitivity}) provides clear evidence that our scheme does indeed beat the SNL, it would be instructive to carry out a calculation of the quantum Fisher information and thereby provide the quantum Cram{\'e}r-Rao bound, which would be a true measure of the best performance of this scheme possible, according to the laws of quantum theory. However, due to the need to compute the permanent of large matrices with complex entries, this calculation currently remains intractable. We are continuing to investigate such a computation for a future work. In general, analytic solutions to matrix permanents are not possible. In this instance, the analytic result is facilitated by the specific structure of the QuFTI unitary. Other phase gradients may yield analytic results, but we leave this for future work as well.

\section{Efficiency} \label{sec:MORDORefficiency}

We consider how in the presence of inefficient photon sources and photo-detectors the success probability of the protocol will drop exponentially with the number of photons. Specifically, if $\eta_s$ and $\eta_d$ are the source and detection efficiencies respectively, the success probability of the protocol is $\eta = (\eta_s \eta_d)^n$. Current cutting edge transition edge detectors operate at 98\% efficiency, with negligible dark count \cite{bib:fukuda2011titanium}. SPDC sources are the standard photon-source technology but they are non-deterministic. However, there are techniques that can greatly improve the heralding efficiency up to 42\% at 2.1 MHz \cite{bib:LPOR201400404}. Also, other source technologies, such as quantum dot sources are becoming viable with efficiencies also up to 42\% \cite{bib:Maier14}. For $n=10$, which is already well beyond current experiments, this yields $\eta= (0.98*0.42)^{10} \approx 0.00014$, which is about 300 successful experimental runs per second when operating with 2.1 MHz sources. 

In App. \ref{app:dephasing} we analyse dephasing, which is a source of decoherence, and find that the QuFTI protocol is far more robust against dephasing than the NOON state is.

\section{Summary}
We began this chapter with section \ref{sec:MORDOR_Motivation} by discussing some of the history of quantum metrology and why we need more simplistic implementations for quantum metrology. Then, in section \ref{sec:MORDOR_Device}, we showed that a passive linear optics network fed with single-photon Fock states may implement quantum metrology with phase-sensitivity that beats the shotnoise limit. This scheme was inspired by \BStwo. Unlike other schemes that employ exotic states such as NOON states which are notoriously difficult to prepare, single-photon states may be readily prepared in the laboratory using present-day technology. This new approach to metrology via easy-to-prepare single-photon states and disjoint photodetection provides a road towards improved quantum metrology with frugal physical resources. Importantly we use a new resource counting technique called ordinal resource counting as discussed in section \ref{sec:MORDOR_ORC}. There remains several open questions to be answered which are the subject of a soon to be published followup manuscript \cite{bib:MORDOR2}. Particularly, we will compare and contrast different interferometric schemes, discuss resource counting, calculate exact quantum Cram\'er-Rao bounds, and study details of experimental errors.

\begin{savequote}[45mm]
The creative principle resides in mathematics. In a certain sense, therefore, I hold it true that pure thought can grasp reality, as the ancients dreamed.
\qauthor{Albert Einstein}
\end{savequote}

\chapter{$\textbf{\#P}$-hardness of Certain Multidimensional Integrals} \label{ch:sharpPhard}

\section{Synopsis}
Here we show that multi-dimensional integrals of a certain form can be classified as \textbf{\#P}-hard like in \BStwo. In \BS the output statistics are given by matrix permanents. We map the \BS output statistics to an integral formalism using using quantum optical characteristic functions, which represent the state of the system in phase-space. The output statistics in the integral formalism mapping is equivalent to matrix permanents yielding a structure of integrals that can in general be categorised as \textbf{\#P}-hard. This yields new insight into the computational complexity of solving certain classes of multidimensional integrals. Our work provides a new approach for using methods from quantum physics to prove statements in computer science. 

In section \ref{sec:IntegralMotivation} we motivate this work and present the main result of this work. In section \ref{sec:IntegralProof} we show a detailed proof and in section \ref{sec:IntegralPermutation} we show an example using permutation matrices, which yield the expected result of unity.

\section{Motivation} \label{sec:IntegralMotivation}

The \BS work by Aaronson \& Arkhipov \cite{bib:AaronsonArkhipov} has led to much interest in the physics community because it is a simple approach to implementing a computationally hard problem. By mapping the \BS problem to an integral formalism we can obtain new insights into integrals of a certain form. We do this by using quantum optical characteristic functions to represent the output state of \BS as a multi-dimensional integral. This integral formalism directly maps to matrix permanents which are known to be \textbf{\#P}-hard in general. Thus we have shown a class of integrals that are also \textbf{\#P}-hard. Our work shows broad applications for utilizing quantum optics tools, in particular, and quantum physics paradigms, in general, to pose and to answer questions about the computational complexity of certain mathematical problems.

We show that integrals of the following form are \textbf{\#P}-hard in general:
\begin{eqnarray}
\mathcal{P} &=& \left(\frac{2}{\pi}\right)^m \idotsint e^{-2|\vec{\alpha}|^2} \prod_{j=1}^n \left(4\left|\sum_{k=1}^m \alpha_kU_{k,j}\right|^2-1\right) \prod_{j=1}^n \left(|\alpha_j|^2 - \frac{1}{2}\right) d^2\vec{\alpha}.
\end{eqnarray}
This construction may be used as a new tool for examining open problems regarding the complexity of \BS\ like problems. As examples of this formalism we show that the identity and permutation matrices are trivial to solve. Recently Rahimi-Keshari \emph{et al.} \cite{bib:rahimi2015efficient} showed the conditions necessary for the efficient classical simulation of quantum-optics experiments given particular input states in a quantum process with measurements at the output, which is relevant to this work because they use similar techniques. 

\section{Proof} \label{sec:IntegralProof}

To prove this we model the output photostatistics of a \BS system with characteristic functions \cite{bib:GerryKnight05}. This represents the state in phase-space and allows other representations (e.g. Wigner function) to be calculated. This  formalism is identical but expresses the problem in terms of multidimensional integrals. The integral equations that arise then must also be \mbox{\textbf{\#P}-hard}.

We begin with an $m$-mode separable input state of the form
\begin{align} \label{eq:sep_input}
\hat\rho = \hat\rho_1 \otimes \ldots \otimes \hat\rho_m.
\end{align}
The single-mode characteristic $W$ function \cite{bib:GerryKnight05} is defined as
\begin{align}
\chi_W(\alpha) = \mathrm{tr}[\hat\rho \cdot \hat{D}(\lambda)],
\end{align}
where \mbox{$\hat{D}(\alpha)$} is the displacement operator, given by
\begin{align}
\hat{D}(\alpha) = \mathrm{exp}(\lambda \hat{a}^\dag - \lambda^* \hat{a}),
\end{align}
and $\alpha$ is an arbitrary complex number representing the amplitude of the displacement in phase-space. This generalizes to the multi-mode case as
\begin{align}
\chi_W(\lambda_1,\ldots,\lambda_m) = \mathrm{tr}[\hat\rho \cdot \hat{D}_1(\lambda)\ldots \hat{D}_m(\lambda_m)],
\end{align}
where \mbox{$\hat{D}_j(\alpha_j)$} is the displacement operator on the $j^{th}$ mode. Then, the characteristic function for the state evolved via linear optics is
\begin{eqnarray}
\chi_W^U(\lambda_1,\ldots,\lambda_m) &=& \mathrm{tr}[\hat{U} \hat\rho \hat{U}^\dag \cdot \hat{D}_1(\lambda_1)\ldots \hat{D}_m(\lambda_m)] \nonumber \\
&=& \mathrm{tr}[\hat\rho \cdot \hat{U}^\dag \hat{D}_1(\lambda_1)\ldots \hat{D}_m(\lambda_m) \hat{U}] \nonumber \\
&=& \mathrm{tr}[\hat\rho \cdot \hat{D}_1(\mu_1)\ldots \hat{D}_m(\mu_m)],
\end{eqnarray}
where
\begin{align} \label{eq:beta_from_alpha}
\mu_j = \sum_{k=1}^m \lambda_k U_{j,k},
\end{align}
as shown in App. \ref{app:disp_ev}.
When the input state $\hat\rho$ is separable, as per Eq. (\ref{eq:sep_input}), 
with \mbox{$\hat\rho=(\ket{1}\bra{1})^{\otimes n}\otimes (\ket{0}\bra{0})^{\otimes (m-n)}$} so there are $n$ single photons in the first $n$ modes,
the multi-mode characteristic $\chi$ function reduces to
\begin{eqnarray} \label{eq:INTmmCW}
\chi_W^U(\lambda_1,\ldots,\lambda_m) &=& \mathrm{tr}[\hat\rho_1\cdot \hat{D}_1(\mu_1)] \ldots \mathrm{tr}[\hat\rho_m\cdot \hat{D}_m(\mu_m)] \nonumber \\
&=& \prod_{j=1}^n \bra{1} \hat{D}(\mu_j) \ket{1} \prod_{j=n+1}^m \bra{0}\hat{D}(\mu_j)\ket{0}
\nonumber \\
&=& \prod_{j=1}^n e^{-\frac{1}{2}|\mu_j|^2}\left(1-|\mu_j|^2\right) \prod_{j=n+1}^m e^{-\frac{1}{2}|\mu_j|^2} \nonumber \\
&=& \prod_{j=1}^m e^{-\frac{1}{2}|\mu_j|^2} \prod_{j=1}^n \left(1-|\mu_j|^2\right) \nonumber \\
&=& e^{-\frac{1}{2}\sum_{j=1}^m|\mu_j|^2} \prod_{j=1}^n \left(1-|\mu_j|^2\right) \nonumber \\
&=& e^{-\frac{1}{2}\mathcal{E}(\vec\lambda)} \prod_{j=1}^n \left(1-|\mu_j|^2\right),
\end{eqnarray}
We have used the identity shown in App. \ref{app:disp_ov}. Here
\begin{align}
\mathcal{E}(\vec\lambda) = \sum_{j=1}^m|\mu_j|^2 = \sum_{j=1}^m|\lambda_j|^2
\end{align}
is the total energy of the system with amplitudes $\vec\lambda$ (or equivalently $\vec\mu$ due to energy conservation).

At this stage the characteristic $W$ function, $\chi_W^U$, can always be efficiently calculated with any separable input state, since it is factorizable and there is no exponential growth in the number of terms. 
The complexity arises when we wish to determine properties of the state, such as reconstructing the Wigner function or determining individual amplitudes within the state.

Next we consider the Wigner function, which may be computed as a type of Fourier transform of $\chi_W$ \cite{bib:GerryKnight05},
\begin{align} \label{eq:SingleModeWigner}
W(\alpha) = \frac{1}{\pi^2} \int e^{\lambda^*\alpha - \lambda\alpha^*}  \chi_W(\lambda) d^2\lambda,
\end{align}
in the single-mode case, which again logically generalizes to the multi-mode case as
\begin{align}
W(\vec\alpha) = \frac{1}{\pi^{2m}} \idotsint e^{\vec{\lambda}^* \cdot \vec\alpha - \vec{\lambda}\cdot \vec\alpha^*} \chi_W(\vec{\lambda}) d^2\vec{\lambda},
\end{align}
where all our integrals implicitly run over the range \mbox{$\{-\infty,\infty\}$}.

Let us denote
\begin{align}
\beta_j = \sum_{k=1}^m \alpha_k U_{j,k},
\end{align}
We can then evaluate the Wigner function as, for the $n$-photon input,
\begin{eqnarray}
W(\vec\alpha) &=& \frac{1}{\pi^{2m}}
\idotsint e^{\vec{\lambda}^*\cdot\vec{\alpha}-\vec{\lambda}\cdot\vec\alpha^*} e^{-\frac{1}{2}\mathcal{E}(\vec\lambda)} \prod_{j=1}^n \left(1-|\mu_j|^2\right) d^2\vec{\lambda} \nonumber \\
&=&\frac{1}{\pi^{2m}}\idotsint e^{\vec{\mu}^*\cdot\vec{\beta}-\vec{\mu}\cdot\vec\beta^*} \prod_{j=1}^m \mathrm{exp}\left(-\frac{1}{2}|\mu_j|^2\right) \prod_{j=1}^n (1-|\mu_j|^2)  d^2\vec{\mu} \nonumber \\
&=& \left(\frac{2}{\pi}\right)^m e^{-2|\vec{\beta}|^2} \prod_{j=1}^n (4|\beta_j|^2-1)  \nonumber \\
&=& \left(\frac{2}{\pi}\right)^m e^{-2|\vec{\alpha}|^2} \prod_{j=1}^n \left(4\left|\sum_{k=1}^m \alpha_k U_{k,j}\right|^2-1\right).
\end{eqnarray}
We have focused on the case where the input state is \mbox{$\hat\rho=(\ket{1}\bra{1})^{\otimes n}\otimes (\ket{0}\bra{0})^{\otimes (m-n)}$}. Now we consider a particular output amplitude where a single photon is measured in the first $n$ output modes. This is determined by calculating the expectation value of the projector \mbox{$\hat\Pi = (\ket{1}\bra{1})^{\otimes n}\otimes (\ket{0}\bra{0})^{\otimes (m-n)}$}, which is equal to the expectation value of the $n$-dimensional number operator, \mbox{$\langle\hat{n}_1\ldots \hat{n}_n\rangle$}, where \mbox{$\hat{n}_j=\hat{a}_j^\dag\hat{a}_j$}. In the usual permanent-based approach, this amplitude corresponds to the permanent of a $n\times n$ submatrix of $U$.

Now we will consider a phase-space approach. For a single-mode state, the expectation value of the number operator may be obtained from the Wigner function as \cite{bib:GerryKnight05},
\begin{align}
\langle \hat{n} \rangle =& \int W(\alpha) \left(|\alpha|^2-\frac{1}{2} \right) d^2\alpha,
\end{align}
where the \mbox{$|\alpha|^2-\frac{1}{2}$} term is obtained by expressing $\hat{n}$ in symmetrically ordered form, \mbox{$\frac{1}{2}(\hat{a}^\dag\hat{a}+\hat{a}\hat{a}^\dag-1)$}, and making the substitution \mbox{$\hat{a}^\dag \to \alpha^*$}, \mbox{$\hat{a}\to \alpha$}.

In the multimode case this generalizes to
\begin{eqnarray} \label{eq:int_perm}
\mathcal{P} &=& \langle \hat{n}_1 \ldots \hat{n}_n \rangle \nonumber \\
&=& \idotsint W(\vec\alpha) \prod_{j=1}^n \left(|\alpha_j|^2-\frac{1}{2}\right) d^2\vec\alpha \nonumber \\
&=& \left(\frac{2}{\pi}\right)^m \idotsint e^{-2|\vec{\alpha}|^2} \prod_{j=1}^n \left(4\left|\sum_{k=1}^m \alpha_kU_{k,j}\right|^2-1\right) \prod_{j=1}^n \left(|\alpha_j|^2 - \frac{1}{2}\right) d^2\vec{\alpha}, \nonumber \\
\end{eqnarray}
where $\mathcal{P}$ is the probability of a single-photon output in each of the first $n$ input modes \mbox{($\ket{1}^{\otimes n}$)}.

To further simplify this expression, we use the identity
\begin{align}
\label{idzer}
\int e^{-2|{\alpha}|^2}\left(|\alpha|^2 - \frac{1}{2}\right)d^2{\alpha} = 0.
\end{align}
This is true for all components of the vector $\vec{\alpha}$.
This leads us to simplify Eq. (\ref{eq:int_perm}) as such.
We expand out the first product in the last line to give a polynomial in $\alpha_k$ and $\alpha_k^*$. We then eliminate terms by multipling \mbox{$\prod_{j=1}^n \left(|\alpha_j|^2 - \frac{1}{2}\right)$}.

We note that if the terms contain $\alpha_k$ and $\alpha_k^*$ with a total odd power, then the function is odd, and so the integral must yield zero.
Also, for any $k$ with \mbox{$1\le k \le n$} the term does not contain $\alpha_k$ or $\alpha_k^*$ and so the integral for that term must yield zero due to the identity in Eq. (\ref{idzer}).
Using these two properties, we can see that the remaining terms must contain each $\alpha_k$ or $\alpha_k^*$ a nonzero even number of times for each \mbox{$1\le k \le n$}.
Since the first product in Eq. (\ref{eq:int_perm}) goes to $n$, the maximum total power of $\alpha_k$ or $\alpha_k^*$ in any term is \mbox{$2n$} and so they must appear exactly twice.

With these properties all of the terms in the first product of the sum in Eq. (\ref{eq:int_perm}) for $m>n$ must yield zero.
This means that the sum can be truncated to $n$.
The expression then takes the form
\begin{align} \label{eq:int_perm2}
\mathcal{P} = \left(\frac{2}{\pi}\right)^m \idotsint e^{-2|\vec{\alpha}|^2} \left(4\left|\sum_{k=1}^n \alpha_kU_{k,j}\right|^2-1\right) \prod_{j=1}^n \left(|\alpha_j|^2 - \frac{1}{2}\right) d^2\vec{\alpha}.
\end{align}
It is now clear that $\mathcal{P}$ only depends on the \mbox{$n\times n$} submatrix of $U$ as expected because our knowledge from \BS tells us that the probability is given from the permanent of this submatrix.
Also, the $-1$ in the first product of Eq. (\ref{eq:int_perm}) only yields terms that integrate to zero.
We can then further simplify $\mathcal{P}$ to
\begin{align} \label{eq:int_perm3}
\mathcal{P} = 4^n \left(\frac{2}{\pi}\right)^m \idotsint e^{-2|\vec{\alpha}|^2} \prod_{j=1}^n \left|\sum_{k=1}^n \alpha_kU_{k,j}\right|^2 \prod_{j=1}^n \left(|\alpha_j|^2 - \frac{1}{2}\right) d^2\vec{\alpha}.
\end{align}
Next, using the identity
\begin{align}
\label{idpi}
\int e^{-2|{\alpha}|^2} d^2{\alpha} = \frac {\pi} 2,
\end{align}
and integrating over all $\alpha_k$ for \mbox{$k>n$} gives
\begin{align} \label{eq:int_perm4}
\mathcal{P} = \left(\frac{8}{\pi}\right)^n \idotsint e^{-2\sum_{j=1}^n|{\alpha_j}|^2} \prod_{j=1}^n \left|\sum_{k=1}^n \alpha_kU_{k,j}\right|^2 \prod_{j=1}^n \left(|\alpha_j|^2 - \frac{1}{2}\right) d^2\vec{\alpha}.
\end{align}
We can simplify further by expanding the first product.
It is easy to check that
\begin{align}
\label{idzer2}
\int e^{-2|{\alpha}|^2}\alpha^2\left(|\alpha|^2 - \frac{1}{2}\right)d^2{\alpha} = 0,
\end{align}
and similarly for $(\alpha^*)^2$.
That means that any terms with $\alpha_k^2$ or $(\alpha_k^*)^2$ will integrate to zero.
The terms that do not integrate to zero are those with the product \mbox{$|\alpha_1|^2|\alpha_2|^2\ldots |\alpha_n|^2$}.
We can also check that
\begin{align}
\label{idpi2}
\int e^{-2|{\alpha}|^2}|\alpha|^2\left(|\alpha|^2 - \frac{1}{2}\right)d^2{\alpha} = \frac{\pi} 8,
\end{align}
and therefore
\begin{align}
\left(\frac{8}{\pi}\right)^n \idotsint e^{-2\sum_{j=1}^n|{\alpha_j}|^2} \prod_{j=1}^n |\alpha_j|^2 \prod_{j=1}^n \left(|\alpha_j|^2 - \frac{1}{2}\right) d^2\alpha_1 \ldots d^2 \alpha_n = 1.
\end{align}
Therefore, the value of $\mathcal{P}$ corresponds to the coefficient of \mbox{$|\alpha_1|^2|\alpha_2|^2\ldots |\alpha_n|^2$} in the product \mbox{$\prod_{j=1}^n \left|\sum_{k=1}^n U_{k,j}\alpha_k\right|^2$}.
It is convenient to express this as
\begin{align}
\prod_{j=1}^n \left|\sum_{k=1}^n U_{k,j}\alpha_k\right|^2 = \left(\prod_{j=1}^n \sum_{k=1}^n U_{k,j}\alpha_k \right)\left(\prod_{j=1}^n \sum_{k=1}^n U_{k,j}^*\alpha_k^*\right).
\end{align}
We can find the coefficient of \mbox{$|\alpha_1|^2|\alpha_2|^2\ldots |\alpha_n|^2$} by using the MacMahon Master theorem on both expressions in brackets on the right-hand side \cite{bib:macmahon60}.
The coefficient of \mbox{$\alpha_1\alpha_2\ldots\alpha_n$} in the first brackets is ${\rm perm}(U^{n\times n})$, where \mbox{$U^{n\times n}$} denotes the \mbox{$n\times n$} submatrix of $U$.
Similarly, the coefficient of \mbox{$\alpha_1^*\alpha_2^*\ldots\alpha_n^*$} in the second brackets is \mbox{${\rm perm}((U^{n\times n})^*)$}.
The coefficient of \mbox{$|\alpha_1|^2|\alpha_2|^2\ldots |\alpha_n|^2$} in the product \mbox{$\prod_{j=1}^n \left|\sum_{k=1}^n U_{k,j}\alpha_k\right|^2$} is then \mbox{$|{\rm perm}(U^{n\times n})|^2$} and so that means that we can evaluate $\mathcal{P}$ as \mbox{$\mathcal{P} =|{\rm perm}(U^{n\times n})|^2$}.

We have found several forms of the integral for the probability. As expected this integral can be evaluated to the square of the permanent. Any of these forms of the integral will suffice but for simplicity we will focus on Eq. (\ref{eq:int_perm}). This equation gives an equivalence of two different forms, that is a matrix permanent and a multidimensional integral, for a particular output amplitude of a linear optics network. It follows that since matrix permanents are known to be \mbox{\textbf{\#P}-hard} in general then integrals of this form are also \mbox{\textbf{\#P}-hard} in general.

The integral formalism may be approximated with Monte Carlo sampling. Since the quantity being sampled takes positive and negative values it is difficult to accurately approximate. Nevertheless, there may be some examples where Monte Carlo sampling is more efficient in the integral formalism than for the permanent formalism but we have not found an example.

\section{Permutation Matrix Example} \label{sec:IntegralPermutation}

It is known that certain examples of matrix permanents can be calculated efficiently due to certain symmetries or sparsities in the matrix. One example is permutation matrices \mbox{$\sigma\in S_m$}, where $S_m$ are elements of the symmetric group. We show that example using our integral formalism is as expected using matrix permanents.

When \mbox{$U=\sigma$}, we have the property
\begin{align}
\sum_{k=1}^m  \alpha_kU_{j,k}^* = \alpha_{\sigma_j},
\end{align}
and we see that $\mathcal{P}$ is separable across $\vec{\alpha}$. In this case, the $n$-dimensional integral from Eq. (\ref{eq:int_perm}) also becomes separable, forming an $n$-dimensional product of integrals,
\begin{eqnarray} \label{eq:permutationResult}
\mathcal{P} &=& \left(\frac{2}{\pi}\right)^m \idotsint e^{-2|\vec{\alpha}|^2} \prod_{k=1}^n \left(4\left| \alpha_{\sigma_k}\right|^2-1\right) \prod_{j=1}^n \left(|\alpha_j|^2 - \frac{1}{2}\right) d^2\vec{\alpha} \nonumber \\
&=& \left(\frac{2}{\pi}\right)^m\left\{ \prod_{j=1}^n \int e^{-2|\alpha_j|^2} \left(4\left| \alpha_{j}\right|^2-1\right) \left(|\alpha_j|^2 - \frac{1}{2}\right) d^2\alpha_j\right\} \left\{\prod_{j=n+1}^{m} \int e^{-2|\alpha_j|^2}  d^2\alpha_j \right\}\nonumber \\
&=& \left(\frac{2}{\pi}\right)^m \left(\frac{\pi}{2}\right)^n \left(\frac{\pi}{2}\right)^{m-n} \nonumber \\
&=& 1,
\end{eqnarray}
where the integrals can easily be shown. Given the separable structure this is clearly computationally efficient to evaluate. Thus, we have confirmed that \mbox{$\mathrm{Per}(\sigma)=1$} for all $\sigma$ and $m$, as expected.


\section{Summary}

We have shown that a particular class of integrals is \mbox{\textbf{\#P}-hard} in general. This is motivated in section \ref{sec:IntegralMotivation}. We did this by representing the known state of the output amplitudes of a \BS device to a integral formalism as shown in section \ref{sec:IntegralProof}. We have shown that by employing two alternate but equivalent approaches to expressing the output amplitudes of linear optics networks fed with single-photon inputs, we are able to provide a quantum optical equivalence between matrix permanents and a particular class of multidimensional integrals. This implies that this class of integrals is \mbox{\textbf{\#P}-hard} in the worst case. The equivalence provides two important insights with broad impact. Firstly, it demonstrates the \mbox{\textbf{\#P}-hardness} of these multi-dimensional integrals. Secondly, by expressing the permanent in integral form, existing knowledge of the structure of integrals provides further insight into the computational complexity of permanents. Finally, we have shown that tools from quantum optics can be used to prove results in computational complexity theory. In addition we showed in section \ref{sec:IntegralPermutation} that this integral-based formalism is consistent with our understanding of permutation matrices. 

\begin{savequote}[45mm]
Any sufficiently advanced technology is indistinguishable from magic.
\qauthor{Arthur C. Clark}
\end{savequote}

\chapter{Preparation Strategies of Large Fock States from Single Photons} \label{ch:MotherFocker}

\section{Synopsis}

The photonic Fock state plays an important role in quantum technologies such as quantum information processing and quantum metrology. While single-photon Fock states are relatively easy to create and the single-photon source technology is standard in labs, it remains challenging to create large-photon Fock states. In this work we show how to efficiently prepare large-photon Fock states using specific strategies that require single-photon sources, beamsplitters, number-resolved photo-detectors, fast-feedforward, and an optical quantum memory. 

In section \ref{sec:MFmotivation} we motivate this work explaining why it is important. In section \ref{sec:MFSPDC} we show the most obvious way to prepare large-photon Fock states using spontaneous parametric down conversion (SPDC) sources and see that it is inefficient. In section \ref{sec:MFsingleshot} we show a method inspired by \BS that uses post-selected linear optics and find that it is also inefficient. In section \ref{sec:MFbootstrapped} we describe and derive the formalism for our bootstrapped approach that leads us to efficient large-photon Fock state preparation. In section \ref{sec:MFfusion} we describe how we fuse stored Fock states so that we can grow them into larger Fock states, we give some analytic approximations to some fusion strategies, we discuss more complex fusion strategies, and we discuss a hybrid scheme. In section \ref{sec:MFreduction} we discuss how to reduce the photon Fock state in case the state is made too large. In section \ref{sec:MFresults} we summarise the results of this work and numerically simulate the different Fock state preparation algorithms.

\section{Motivation} \label{sec:MFmotivation}

Fock states are an essential resource for many quantum technologies \cite{bib:NielsenChuang00, bib:KokLovett11} including quantum communication, quantum cryptography, quantum metrology \cite{bib:kapale2007bootstrapping, bib:yurke1986input, bib:yuen1986generation, bib:dowling1998correlated, bib:gerry2001generation}, interferometric protocols \cite{bib:MagdaLeap}, and quantum information processing \cite{bib:LOQC}. There have been a lot of advances made in single-photon source technology but technologies that prepare large photon-number Fock states efficiently do not exist. Na\"ive approaches for generating large photon-number Fock states may be made from single-photon Fock states using non-deterministic linear optics or by heralded spontaneous parametric down-conversion (SPDC). Both of these methods are inefficient because the success probability is exponentially low as we show in section \ref{sec:MFSPDC} and \ref{sec:MFsingleshot} respectively. 

It is known that quantum enhanced metrology is optimal (i.e. it reaches the Heisenberg limit of phase sensitivity) when NOON states are used \cite{bib:dowling2008quantum}. Creating NOON states with large photon number can be as hard as building a universal optical quantum computer, as it requires many of the same technologies, such as a quantum memory, and feedforward. Nonetheless, one needs Fock states with large photon number to first generate NOON states with large photon number for quantum enhanced metrology \cite{bib:kapale2007bootstrapping}. Thus, efficient schemes for generating Fock states with large photon number, as presented in this work, are an important stepping stone for realizing optimal quantum enhanced metrology.

In this work we show an iterative protocol for building up large photon-number Fock states from readily available single photons. We show that our technique exponentially improves scalability compared to na\"ive methods giving efficient methods for preparing large photon-number Fock states. The requirements for implementing our protocol are mostly the same as for universal linear optical quantum computing (LOQC) \cite{bib:LOQC}, which further motivates building LOQC related technologies.

\section{Spontaneous Parametric Down-Conversion with Post-Selection} \label{sec:MFSPDC}

A very trivial approach to preparing large photon-number Fock states is to use SPDC sources with post-selection. For a review of SPDC sources see section \ref{sec:SPDCSPDC}. We know that the photon number between the signal and idler modes is correlated so we can prepare an arbitrarily large $n$ photon-number Fock state by increasing the pump power. Experimentally demonstration of this approach has been done for small Fock states up to three photons \cite{bib:MerlinCooper13}.

\begin{figure}[!htb]
\centering
\includegraphics[width=0.55\columnwidth]{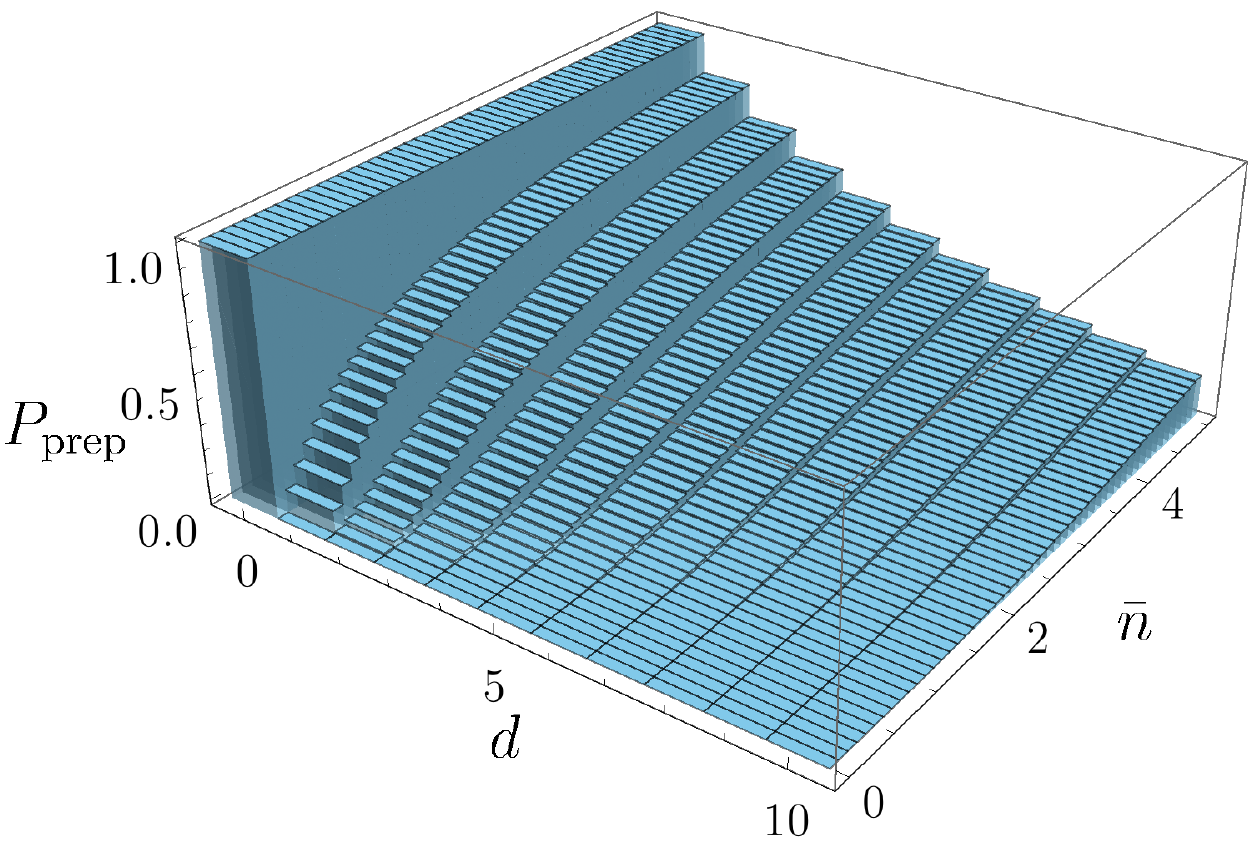}
\caption{The probability of preparing at least $d$ photons in the first mode of an SPDC source with mean photon number $\bar{n}$, by post-selecting upon measuring at least $d$ photons in the other mode.} \label{fig:SPDC_prep}
\end{figure}

To prepare large photon-number Fock states from SPDC sources with at least $d$ photons we add all possible probabilities as shown in Eq. (\ref{eq:SPDCprobabilitydistribution}). The probability is then given by
\begin{align}
P_\mathrm{prep}(d) = \sum_{n=d}^\infty |\lambda_n|^2 = \left(\frac{\bar{n}}{\bar{n}+1}\right)^d.
\end{align}
We see that this decreases exponentially with $d$. This means that to obtain a desired Fock state one would have to wait a time that depends exponentially with $d$. In addition currently available mean photon numbers are \mbox{$\bar{n} \ll 1$} \cite{bib:finger15}, which means that this is unviable for large $d$. In Fig. \ref{fig:SPDC_prep} we show this preparation probability as a function of $d$ and $\bar{n}$.

\section{Single-Shot Post-Selected Linear Optics} \label{sec:MFsingleshot}

\begin{figure}[!htb]
\centering
\includegraphics[width=0.3\columnwidth]{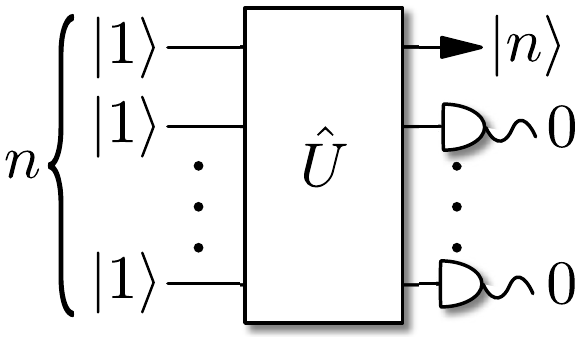}
\caption{Single-shot preparation on an $n$-photon Fock state from $n$ single photons. A photon is incident upon each input mode, and we post-select upon detecting vacuum in all but one of the output modes, where all $n$ photons must have exited the remaining output mode.} \label{fig:single_shot}
\end{figure}

One method to prepare a large photon Fock state would be to use a multimode linear optical interferometer as illustrated in Fig. \ref{fig:single_shot}. In this method single photons are input into every input mode where the number of input modes is equivalent to the photon-number Fock state desired. These photons evolve through a linear optical interferometer and then at the output all modes except one are post-selected upon to have the vacuum state. The remaining non-measured mode would then obtain the desired photon Fock state since photon number is conserved. 

Now we want to determine the efficiency of doing this. First we consider an $n$-mode interferometer with a single photon at each input mode. The input state is
\begin{align}
\ket{\psi_\mathrm{in}} = \ket{1}^{\otimes n} = \hat{a}^\dag_1 \dots \hat{a}^\dag_n \ket{0}^{\otimes{n}} =\left[\prod_{i=1}^n \hat{a}^\dag_i \right] \ket{0}^{\otimes n}.
\end{align}
The state then evolves through a linear optical network
\begin{eqnarray}
\ket{\psi_\mathrm{out}} &=& \hat{U} \ket{\psi_\mathrm{in}} \nonumber \\
&=& \left[ \prod_{i=1}^n \sum_{j=1}^n U_{i,j} \hat{a}^\dag_j \right] \ket{0}^{\otimes n}.
\end{eqnarray}
Now we post-select upon all photons exiting the first mode
\begin{eqnarray}
\ket{\psi_\mathrm{proj}} &=& \left[ \prod_{i=1}^n U_{i,1} \hat{a}^\dag_1 \right] \ket{0}^{\otimes{n}} \nonumber \\
&=& \sqrt{n!} \left[\prod_{i=1}^n U_{i,1} \right] \ket{n}\ket{0}^{\otimes{n-1}},
\end{eqnarray}
where the probability of this event occurring is
\begin{align}
P_\mathrm{bunch} = n! \left| \prod_{i=1}^n U_{i,1} \right|^2.
\end{align}
It turns out that a balanced interferometer maximizes $P_\mathrm{bunch}$ due to the triangle inequality. A balanced interferometer has weightings 
\begin{align}
|U_{i,1}| = \frac{1}{\sqrt{n}} \,\forall\, i.
\end{align}
For these weightings $P_\mathrm{bunch}$ becomes
\begin{align} \label{eq:single_shot_P}
P_\mathrm{bunch} = n! \left| \prod_{i=1}^n \frac{1}{\sqrt{n}} \right|^2 = \frac{n!}{n^n} \sim \frac{\sqrt{n}}{e^n},
\end{align}
which was obtained using Stirling's approximation. 

The probability of preparing an $n$-photon Fock state using this method decreases exponentially with $n$ and thus implementation is limited. As an example let us consider two-photon Hong-Ou-Mandel (HOM) \cite{bib:HOM87} interference. In this instance \mbox{$n=2$}, and
\begin{align}
P_\mathrm{bunch} = \frac{2!}{2^2} = \frac{1}{2},
\end{align}
as expected, since the output state of a HOM interferometer is \mbox{$\ket{\psi_\mathrm{out}} = (\ket{2,0}-\ket{0,2})/\sqrt{2}$}.

\section{Bootstrapped Preparation with Post-Selected Linear Optics} \label{sec:MFbootstrapped}

\begin{figure}[!htb]
\centering
\includegraphics[width=0.5\columnwidth]{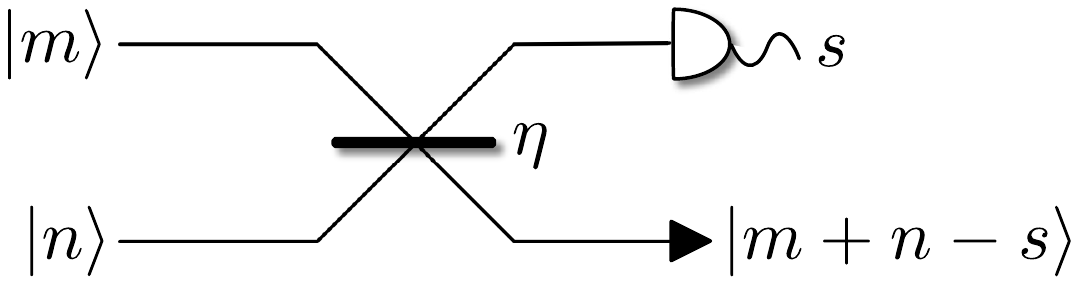}
\caption{The Fock state fusion operation. Two Fock states with $m$ and $n$ photons are mixed on a beamsplitter of reflectivity $\eta$. Upon detecting $s$ photons in the first output mode, an \mbox{$m+n-s$} photon state is prepared in the other mode.} \label{fig:fusion}
\end{figure}

We have devised a method to improve the inefficient scaling of previous methods. We call this a bootstrapped approach where we iteratively fuse two Fock states with photon numbers $m$ and $n$ attempting to create a larger Fock state. The fusion is done with a beamsplitter with some reflectivity $\eta$ and one of the output modes is post-selected upon. At the other output mode a Fock state with known photon number is created and is a usable resource. Specifically, when $s$ photons are detected we have prepared a new state of size \mbox{$m+n-s$}. This bootstrapped method is shown in Fig. \ref{fig:fusion}. This is the basis of how our efficient large photon-number Fock state generation protocol works. We expand on this in the next section describing specific fusion strategies or algorithms for improving efficiency further. Now we describe this protocol in detail.

The input state to the fusion operation is
\begin{eqnarray}
\ket{\psi_\mathrm{in}} &=& \ket{m,n} \nonumber \\
&=& \frac{1}{\sqrt{m!n!}} (\hat{a}^\dag)^m (\hat{b}^\dag)^n \ket{0,0},
\end{eqnarray}
which is incident upon a beamsplitter of reflectivity $\eta$, given by a biased beamsplitter unitary
\begin{align}
U_\mathrm{BS} = \left(\begin{array}{cc}
\eta & \sqrt{1-\eta^2} \\
\sqrt{1-\eta^2} & -\eta 
\end{array}\right),
\end{align}
where local phases are irrelevant. Then the output state is
\begin{align}
\ket{\psi_\mathrm{out}} &= \hat{U}_\mathrm{BS} \ket{\psi_\mathrm{in}} \nonumber \\
&= \frac{1}{\sqrt{m!n!}} \left(\eta\hat{a}^\dag + \left(1-\eta^2\right)^{1/2}\hat{b}^\dag\right)^m  \left(\left(1-\eta^2\right)^{1/2}\hat{a}^\dag - \eta \hat{b}^\dag\right)^n \ket{0,0} \nonumber \\
&= \frac{1}{\sqrt{m!n!}} \sum_{j=0}^m \binom{m}{j} \eta^j \left(1-\eta^2\right)^{(m-j)/2} (\hat{a}^\dag)^j (\hat{b}^\dag)^{m-j} \nonumber \\
&\quad\times \sum_{k=0}^n \binom{n}{k} \left(1-\eta^2\right)^{k/2} \eta^{n-k} (-1)^{n-k} (\hat{a}^\dag)^k (\hat{b}^\dag)^{n-k} \ket{0,0} \nonumber \\
&= \frac{1}{\sqrt{m!n!}} \sum_{j=0}^m \sum_{k=0}^n \binom{m}{j} \binom{n}{k} \eta^{n+j-k} \left(1-\eta^2\right)^{(m+k-j)/2} \nonumber \\
&\quad\times (-1)^{n-k} (\hat{a}^\dag)^{j+k}(\hat{b}^\dag)^{m+n-j-k}\ket{0,0} \nonumber \\
&= \frac{1}{\sqrt{m!n!}} \sum_{j=0}^m \sum_{k=0}^n \binom{m}{j} \binom{n}{k} \eta^{n+j-k} \left(1-\eta^2\right)^{(m+k-j)/2} \nonumber \\
&\quad\times (-1)^{n-k} \sqrt{(j+k)!(m+n-j-k)!} \ket{j+k,m+n-j-k}.
\end{align}
We are interested in the case where $s$ photons are measured in the first mode, thereby producing an $s$-photon-subtracted state in the second mode. Thus, we let \mbox{$s=j+k$}, and the unnormalized post-selected state reduces to
\begin{align}
\ket{\psi_\mathrm{ps}}&=\sqrt{\frac{s!(m+n-s)!}{m!n!}} \sum_{j=0}^s \binom{m}{j} \binom{n}{s-j} \eta^{n+2j-s} \left(1-\eta^2\right)^{(m+s-2j)/2} \nonumber \\
&\quad\times (-1)^{n-s+j} \ket{s,m+n-s}. \nonumber \\
\end{align}
The probability of detecting $s$ photons is, therefore,
\begin{align}
P_\mathrm{sub}(s|m,n) &= \eta^{2(n-s)} \left(1-\eta^2\right)^{m+s} \frac{s!(m+n-s)!}{m!n!} \left| \sum_{j=0}^s \binom{m}{j} \binom{n}{s-j} \left[\frac{\eta^2}{\eta^2-1}\right]^{j} \right|^2.
\end{align}

The state will have grown if the $s$-photon-subtracted state is larger than both $m$ and $n$ so we require \mbox{$s<m+n-\mathrm{max}(m,n)$}. The probability of preparing a state at least as large as both the input Fock states is
\begin{equation} \label{eq:P_grow}
P_\mathrm{grow}(m,n) = \sum_{s=0}^{\mathclap{m+n-\mathrm{max}(m,n)-1}} P_\mathrm{sub}(s|m,n).
\end{equation}
In the case of a protocol where we will not recycle resource states we only keep the \mbox{$s=0$} outcome.

For given configurations of input Fock states $m$ and $n$ we would like to maximize this probability so that state growth is maximized, so we optimize $\eta$ to maximize $P_\mathrm{grow}$ and obtain,
\begin{eqnarray}
P_\mathrm{opt}(m,n) &=& \mathrm{max}_\eta [P_\mathrm{grow}(m,n)], \nonumber \\
\eta_\mathrm{opt}(m,n) &=& \mathrm{argmax}_\eta [P_\mathrm{grow}(m,n)].
\end{eqnarray}
In Fig. \ref{fig:opt_eta_P} we show the optimized beamsplitter reflectivities and growth probabilities for \mbox{$1\leq m \leq 10$} and \mbox{$1 \leq n \leq 10$} for the both cases where we accept all possible outcomes and only the $s=0$ outcome.

\section{Fusion} \label{sec:MFfusion}

\begin{figure}[!htb]
\centering
\includegraphics[width=0.45\columnwidth]{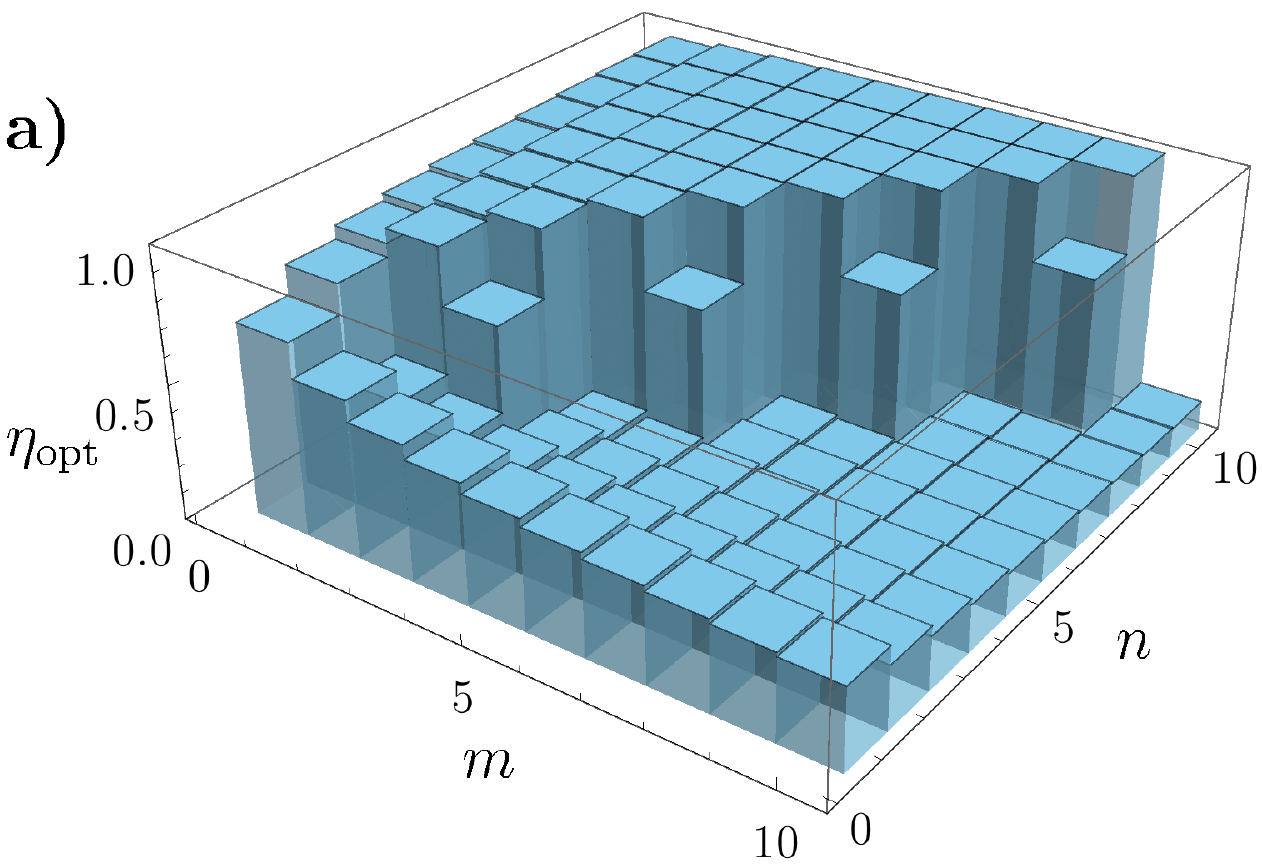}
\includegraphics[width=0.45\columnwidth]{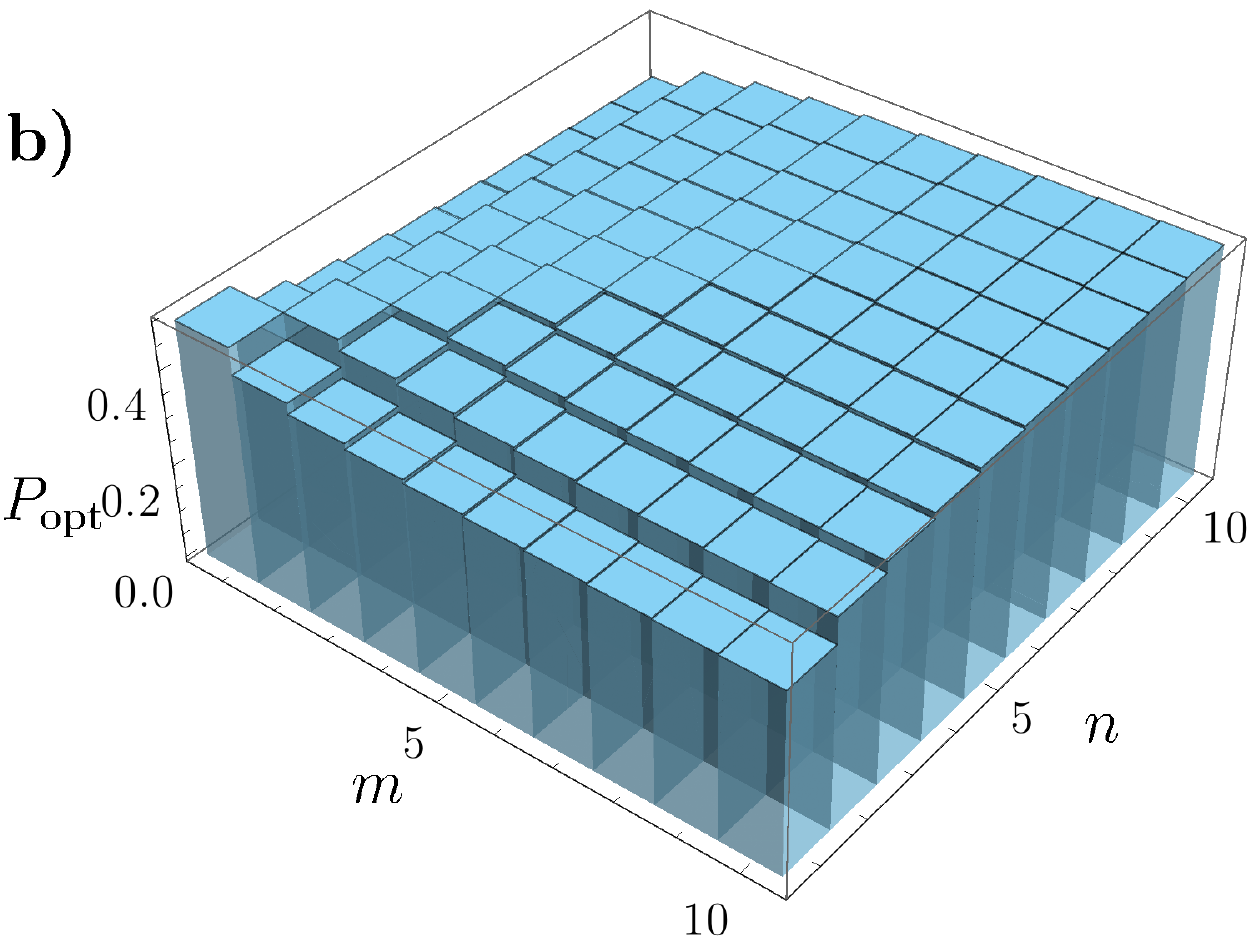} \\
\includegraphics[width=0.45\columnwidth]{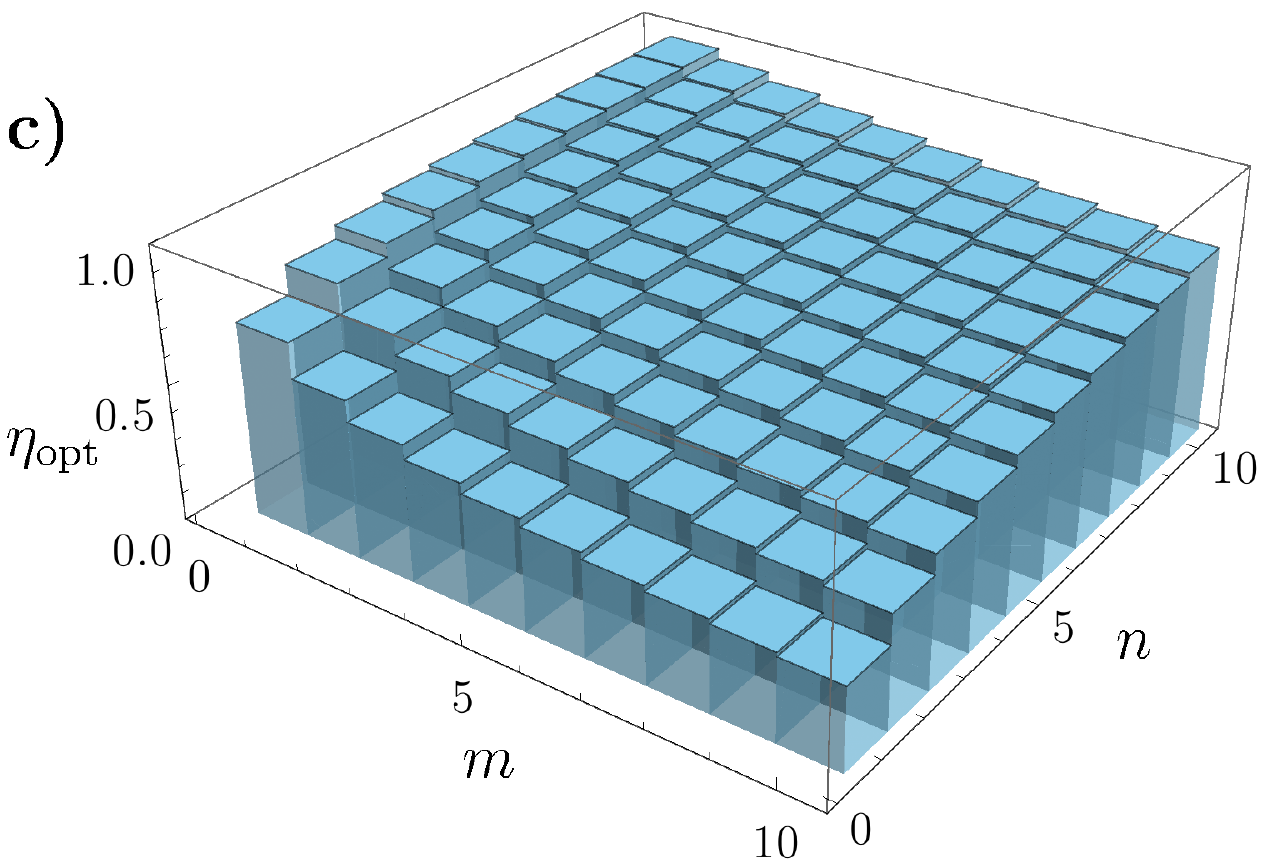}
\includegraphics[width=0.45\columnwidth]{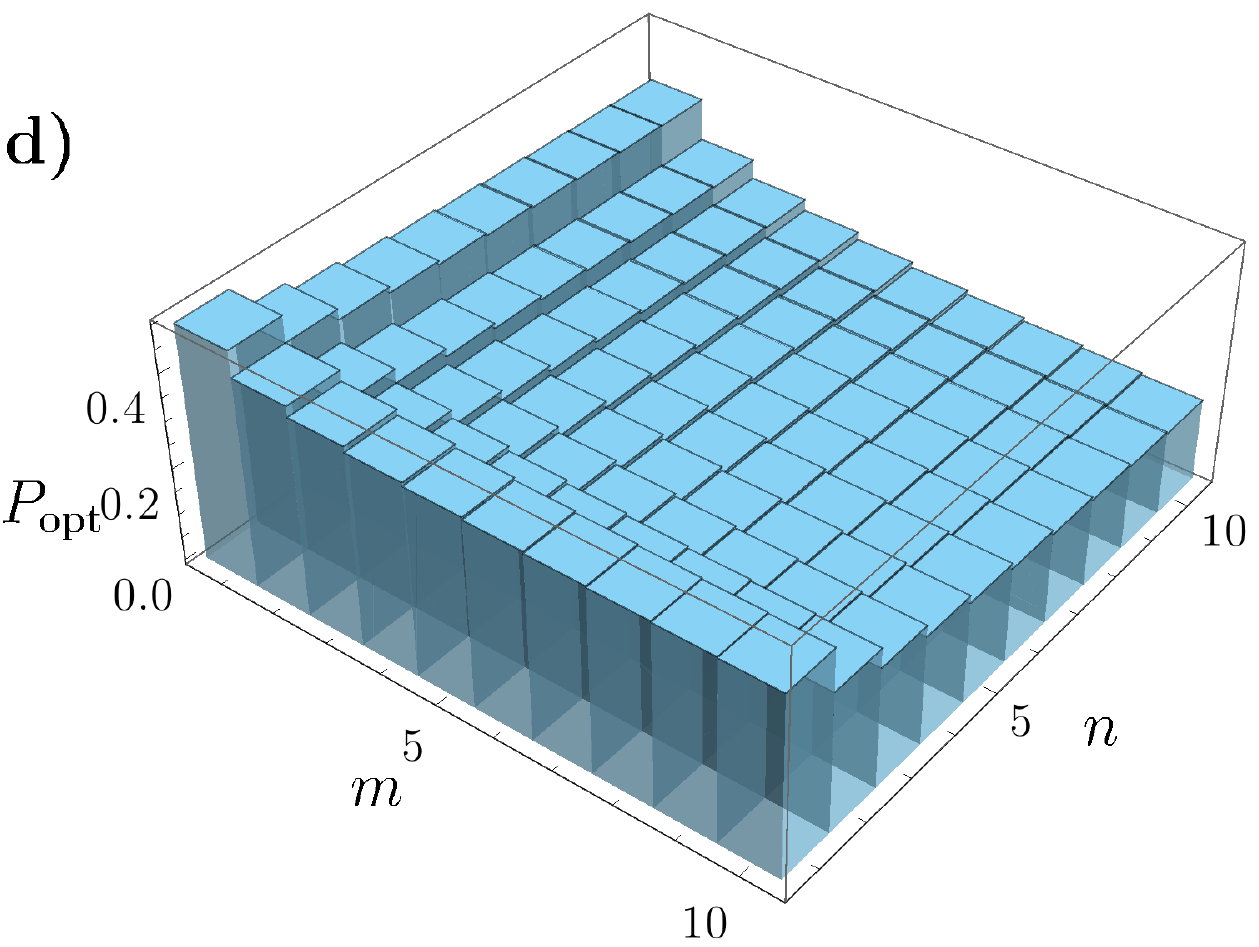}
\caption{(a) The optimal beamsplitter reflectivity ($\eta_\mathrm{opt}$) for maximizing growth probability when fusing two Fock states of photon number $m$ and $n$, accepting all outcomes $s$. (b) The associated growth probability ($P_\mathrm{opt}$) for the fusion operation. (c) The optimal beamsplitter reflectivity for maximizing only the probability of the \mbox{$s=0$} outcome, i.e. for the non-recycled protocol. (d) The associated success probability. Evidently the fusion operation with recycling has higher success probability than without recycling, since we are accepting all events $s$ as opposed to only the \mbox{$s=0$} cases.} \label{fig:opt_eta_P}
\end{figure}

We can see from Fig. \ref{fig:opt_eta_P} that when using recycling the growth probability is maximized when fusing Fock states of equal photon number, \mbox{$m=n$} and so we are led to believe that the best strategy is to always fuse states of equal size. This is called the balanced strategy. This is similar to cluster states \cite{bib:Raussendorf01, bib:Raussendorf03} since Rohde \& Barrett showed that a balanced strategy is optimal \cite{bib:RohdeBarrett07}. This balanced fusion strategy can be simplified because the only probabilities of interest are \mbox{$P_\mathrm{sub}(s|m,m)$} with an optimized growth probability of \mbox{$P_\mathrm{opt}(m,m)=1/2\,\,\forall\, m$}. This is very favorable when attempting to create large photon-number Fock states. When no recycling is used the growth probability is maximized when fusing a given Fock state with a single-photon state (i.e. \mbox{$m=1$} or \mbox{$n=1$}). Rohde \& Barrett call this the modesty strategy since we are attempting to fuse only single-photon Fock states. This strategy however decreases exponentially with the number of fusion operations since each time we require that $s=0$. We can overcome this inefficiency by employing ideas of recycling from cluster state protocols \cite{bib:Nielsen04, bib:Gross06, bib:RohdeBarrett07}, where we reuse any $s$ outcome as a resource to progressively build up a large-photon Fock state. 

\subsection{Generalized Fusion Protocol}
In our fusion protocol we assume that we have an unlimited resource of single-photon states. We also have quantum memories that store Fock states of varying photon numbers. We let $c_n(t)$ be the number of $n$-photon states that we have in the respective memory after the $t^{th}$ fusion operation. Because we assume that we have an unlimited supply of single-photon states, we let \mbox{$c_1(0)=\infty$}. All other buckets are initially empty, \mbox{$c_{i>1}(0)=0$}.

The next part of fusion is to retrieve two Fock states from the memory that depends on a specific fusion strategy, which we let be $m$ and $n$, and send them through the fusion as depicted in Fig. \ref{fig:fusion}. With probability \mbox{$P_\mathrm{sub}(s|m,n)$} the \mbox{$(m+n-s)$}-photon state is prepared. The corresponding memories are then updated according to
\begin{eqnarray} \label{eq:transitions}
c_m &\to& c_m-1, \nonumber \\
c_n &\to& c_n-1, \nonumber \\
c_{m+n-s} &\to& \left\{ \begin{array}{ll}
    c_{m+n-s} + 1 & \mathrm{with\,\,recycling;} \\
    c_{m+n-s} + \delta_{s,0} & \mathrm{without\,\,recycling}.
 \end{array}
\right.
\end{eqnarray}
A state from each of the $m$- and $n$-photon memory has been removed while the new state is added to the \mbox{$m+n-s$} photon memory. With fusion then a random walk of populations between the memories occurs. 

Now suppose that we would like to achieve a photon Fock state with photon number of at least $d$. Then the memories we are interested in are
\begin{align}
c_{\geq d} = \sum_{j=d}^\infty c_j.
\end{align}
The rate at which these states are prepared, per fusion operation, is then
\begin{align}
r(d) = \lim_{t\to\infty} \frac{c_{\geq d}(t)}{t},
\end{align}
where $t$ is the number of fusion operations. We establish the steady state flow dynamics of states by considering the \mbox{$t\to\infty$}.

\subsection{Analytic Approximations}
For some simplified schemes, we can establish analytic results that show that the rate is improved exponentially over the single-shot case discussed in Sec. \ref{sec:MFsingleshot}.
First we consider a non-recycled scheme, where we attempt to construct a Fock state with $d$ photons, where $d$ is a power of $2$.
For values of $d$ that are not a power of $2$, we can simply construct a Fock state with a number that is the next largest power of $2$.
The rate will be unchanged, and the scaling will not be significantly affected.

As this is a non-recycled scheme we only retain successes.
We therefore fuse single-photon states until we obtain $2$-photon states, fuse $2$-photon states until we obtain $4$-photon states, and so forth, which is why we consider powers of two.
As was found above (see Fig. \ref{fig:opt_eta_P}), the success probability is maximized for 50/50 beam splitters when using equal photon numbers, so we use 50/50 beam splitters.
To estimate the rate, we will estimate the average number of single-photon states needed to produce one $d$-photon state.
Then the preparation rate of $d$-photon Fock states per \emph{fusion operation} will have scaling equal to the inverse of this number.
That is because there can be no more than a factor of $2$ between the number of single-photon states used and the number of fusion operations.

To show that result, consider first the ideal case where every fusion operation is successful.
Then for $d$ single-photon states, there would be $d/2$ fusion operations, and $d/4$ fusion operations on the 2-photon states, and so forth.
Adding them together for $d$ a power of two gives a total number of fusion operations equal to $d-1$, or one less than the number of single-photon states.
As the success rate is reduced, the number of fusion operations can only be reduced for a given number of single photons.
Hence the number of fusion operations cannot be any larger than the number of single photons.
For this scheme we perform fusion operations on all pairs of single photons, so the number of fusion operations must be at least half the number of single-photon states.

Now we estimate the average number of single-photon states required to produce one $d$-photon state.
The expected number of attempts to fuse two $d/2$-photon states to produce the $d$-photon state will be $1/P_{\rm sub}(0|d/2,d/2)$.
This corresponds to an expected number of $d/2$-photon states of $2/P_{\rm sub}(0|d/2,d/2)$, as there are two in each attempt.
Then, the expected number of $d/4$-photon states required to produce each $d/2$-photon state is $2/P_{\rm sub}(0|d/4,d/4)$.
As a result, the expected number of $d/4$-photon states required to produce one $d$-photon state is $4/[P_{\rm sub}(0|d/2,d/2)P_{\rm sub}(0|d/4,d/4)]$.
Continuing this reasoning, the expected number of single photons required to produce one $d$-photon state is
\begin{align}
\label{expsin}
\prod_{j=1}^{\log_2 d} \frac{2^{\log_2 d}}{P_{\rm sub}(0|d/2^{j},d/2^{j})}.
\end{align}

To estimate this expected number of photons, we can use
\begin{equation}
P_{\rm sub}(0|n,n) = 2^{-2n} \frac{(2n)!}{(n!)^2} \sim \frac {1}{\sqrt{\pi n}},
\end{equation} 
where the approximation is via Stirling's formula.
Using this approximation with Eq. (\ref{expsin}) gives the expected number of single-photon states scaling as
\begin{eqnarray}
2^{\log_2 d} \prod_{j=1}^{\log_2 d} \sqrt{\pi d/2^j} &=& d (\pi d)^{(\log_2 d)/2} 2^{-\log_2 d (\log_2 d+1)/4)} \nonumber \\
&=& d^{3/4+(\log_2 \pi)/2+(\log_2 d)/4}.
\end{eqnarray}
Testing this expression numerically, we find that the expected number of single photons is about $1.2777$ times this value.

As discussed above, the preparation rate of $d$-photon Fock states per fusion operation will scale as the inverse of this expression, and is therefore
\begin{equation}
r(d) \propto \frac 1{d^{3/4+(\log_2 \pi)/2} d^{(\log_2 d)/4}} .
\end{equation}
There is an exponential improvement over the case with just a single interferometer and single-shot preparation.
The scaling is not exponential in $d$, but it is also not polynomial in $d$, because the power is logarithmic in $d$.

\begin{figure}[!htb]
\centering
\includegraphics[width=0.55\columnwidth]{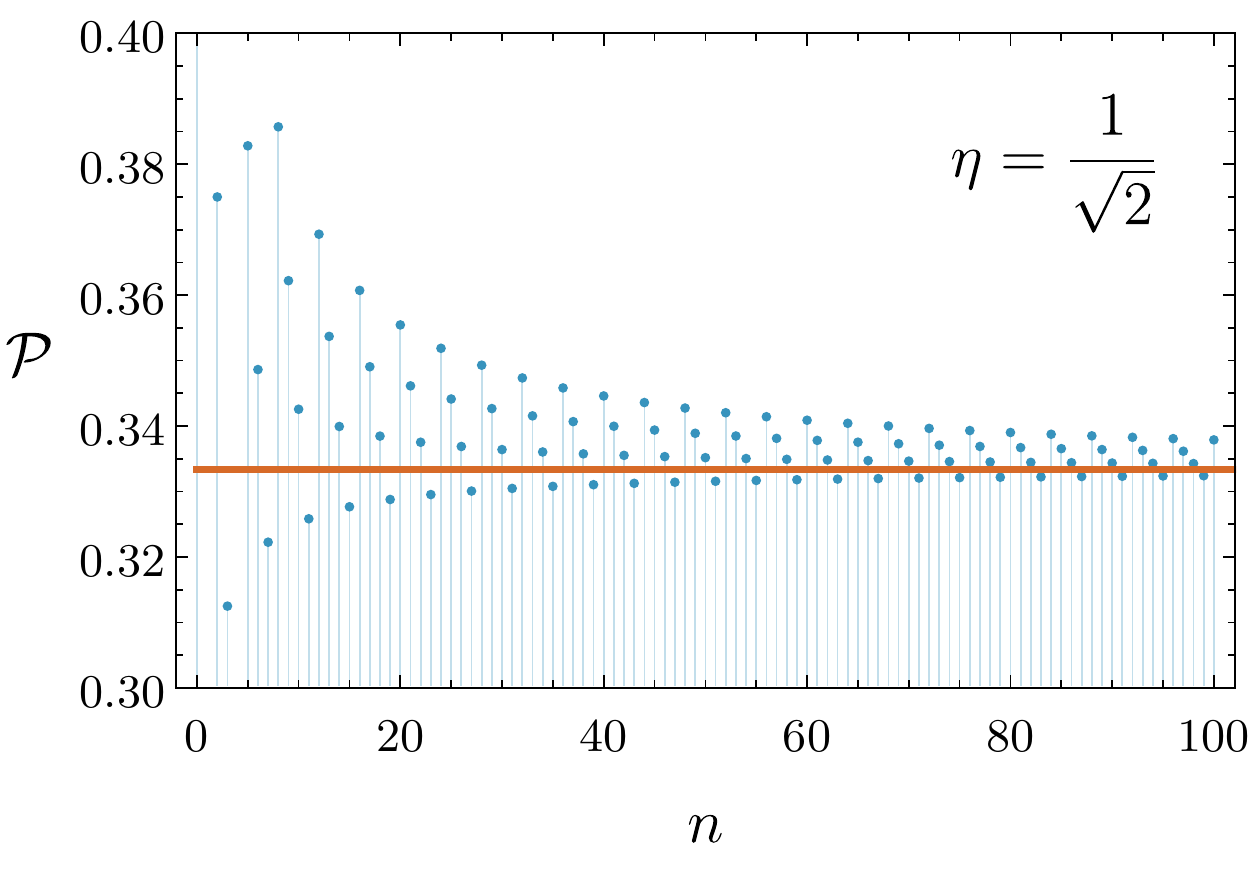}
\caption{The probability of an improved recycling scheme with $\eta=1/\sqrt{2}$, where we require that the number of photons lost is no more than $n/2$ when fusing two $n$-photon states. We see that this probability approaches $\sim1/3$ as $n\to\infty$.} \label{fig:limited_recycling}
\end{figure}

For a further improvement, we can add limited recycling.
Rather than just requiring zero photons to be lost at each stage, we require that the number of photons lost is no more than $n/2$ when fusing two $n$-photon states so that the probability of success is given by 
\begin{equation}
\mathcal{P}=\sum_{s=0}^{\mathrm{Floor}(n/2)}P_\mathrm{sub}(s|n,n),
\end{equation}
with $\eta=1/\sqrt{2}$.
Then we find numerically that the probability of success approaches $\sim1/3$ in the limit $n\to\infty$ as can be seen in Fig. \ref{fig:limited_recycling}. An analytic solution for this result currently eludes us so we have provided numerical evidence. 
It is smaller for smaller values of $n$, but because we are calculating the scaling for large $d$ we will take the probability of success to be $1/3$.
Without loss of generality we can require that on success the photon number is $\lceil 3n/2 \rceil$.
If the photon number is larger than that, we can reduce the photon number with the state reduction scheme described in Sec. \ref{sec:MFreduction}.
Now, in order to obtain photon number $d$, we need a number of levels scaling as $\log_{3/2} d$, rather than $\log_2 d$.
However, the multiplying factor for the number of repetitions at each stage is smaller.

Therefore, the number of single photons required to obtain a single $d$-photon Fock state scales as
\begin{equation}
6^{\log_{3/2}d} = d ^{\log_{3/2}6} = d^{4.419\ldots}.
\end{equation}
Testing this expression numerically, the number of single photons required is about $47$ times less.
The corresponding rate for this scheme then scales as
\begin{equation} \label{eq:MFscaling}
r(d) \propto d^{-4.419\ldots} .
\end{equation}
This scaling is again an exponential improvement, and is now strictly polynomial.



\subsection{Fusion Strategies} \label{sec:MFstrategies}
An analytic bound for more advanced recycling strategies is nontrivial and we instead simulate these protocols as classical Markov processes between the buckets, with transition probabilities given by the parameters $P_\mathrm{sub}(s|m,n)$, and transition rules from Eq. (\ref{eq:transitions}).

We see that Fig. \ref{fig:opt_eta_P} intuitively hints that the balanced strategy may be optimal we consider several other strategies as a comparison to provide further insight into the importance of fusion strategy. We will consider four recycling strategies in total:
\begin{itemize}
\item Balanced: Fuse the largest two available states of equal size, \mbox{$m=n$}. This strategy is based on the observation of Fig. \ref{fig:opt_eta_P} that fusing states of equal size maximizes the growth probability, $P_\mathrm{grow}$.
\item Modesty: Always attempt to grow our state by a single photon, by fusing the state $m$ with a single-photon state, \mbox{$n=1$}.
\item Random: Randomly choose any two available states, irrespective of their relative sizes.
\item Frugal: Same as balanced, except that it does not attempt to fuse two equally sized states if \mbox{$m=n>\lfloor d'/2 \rfloor$}, where \mbox{$d' \geq d$}. For states of size \mbox{$n > d'/2$}, it instead attempts to fuse available states such that \mbox{$d\leq m+n \leq d'$}. This is because larger number states are costly to prepare, so it is wasteful to fuse two states with a total photon number well in excess of the target $d$. 
\end{itemize}
The optimization technique for the frugal strategy is different than for the other strategies, which use $P_{\mathrm{grow}}$. If the total input photon number \mbox{$m+n \geq d$}, then for each configuration of input Fock states $m$ and $n$, we optimize $\eta$ to maximize the probability of getting at least $d$ photons. Specifically, we maximize
\begin{align}
	\sum_{s=0}^{\mathclap{m+n-d}}P(s|m,n).
\end{align}

If \mbox{$m+n<d$}, then for each configuration of input Fock states $m$ and $n$, we optimize $\eta$ to maximize the probability of increasing the maximum photon number with increasing weightings for growing the photon number larger. Specifically, we maximize
\begin{align}
	\sum_{s=0}^{\mathclap{m+n-\mathrm{max}(m,n)}}[m+n-s-\mathrm{max}(m,n)]P(s|m,n). \nonumber\\
\end{align}

\subsection{Hybrid Schemes} \label{sec:MFhybrid}

We have considered the scenario where single-photon states are freely available. This is an appropriate choice as single-photon sources are becoming mainstream. Improved photon-source technologies, such as quantum dot sources, may have the ability to prepare small-photon-number Fock states higher than $\ket{1}$ on demand. A simple modification then to our protocol is to begin with a resource of infinitely available resource states that are larger than one and we find that this further improves preparation rates.

Our framework can account for this by letting \mbox{$c_x=\infty$} rather than \mbox{$c_1=\infty$}, where $x$ is the photon number that can be prepared on demand. The rest of the protocol then proceeds as usual using a given fusion strategy.

\section{State Reduction} \label{sec:MFreduction}

In our analysis so far we calculate the rate of preparing at least $d$ photons but for some applications we may want exactly $d$ photons. For this we may need to then reduce the photon number of a Fock state if we were to prepare one too large. Luckily, this may also be done with post-selected linear optics by inputting the Fock state on a low-reflectivity beamsplitter with vacuum at the other input (\mbox{$n=0$}, \mbox{$\eta\ll 1$}). Most of the time no photons will be detected in the reflected mode since the beamsplitter reflectivity is low. With low probability a single photon will be detected and with even lower probability more than one photon will be detected. When photons are detected we have reduced the photon number by the number of photons we detected. We can do this repeatedly until we obtain the desired Fock state with photon number $d$. This protocol is efficient since it requires $O(s)$ beamsplitter operations on average when attempting to subtract $s$ photons. Also, this is experimentally easier to implement than state growth since the vacuum state is used in one of the modes meaning that there are no mode-matching requirements.

\section{Results} \label{sec:MFresults}

\begin{figure}[!htb]
\centering
\includegraphics[width=0.6\columnwidth]{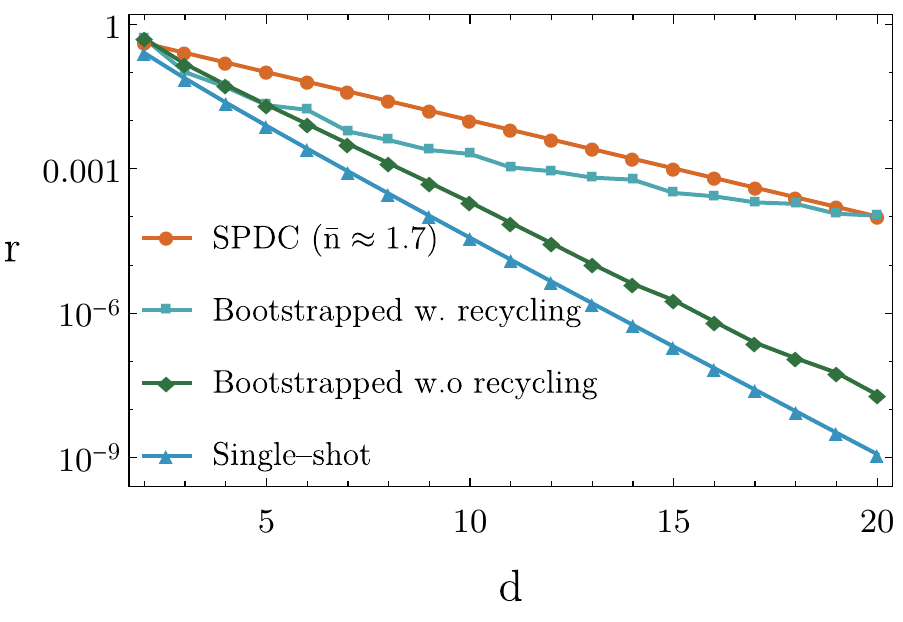}
\caption{Rate of preparation $r$ of Fock states with at least $d$ photons, for the recycled and non-recycled bootstrapped protocols (using the balanced strategy), the single-shot protocol, and the heralded SPDC protocol. In the cases of the two bootstrapped protocols and the SPDC protocol, we observe a strict exponential decrease in the preparation rate against the number of photons. It is evident that state recycling yields a far more favorable scaling than the non-recycled or single-shot protocols. For heralded SPDC, the exponential exhibits a dependence on the mean photon number ($\bar{n}$) of the source, which reflects the SPDC pump-power. Here we have chosen $\bar{n}$ such that the SPDC and recycled bootstrapped balanced protocols have the same 20-photon preparation rate, providing a baseline for the regime of SPDC operation such that it matches the preparation rate of the recycled protocol.} \label{fig:rate}
\end{figure}

First we show the rate of $d$-photon state preparation, $r(d)$, for both the recycled and non-recycled bootstrapped protocols (Sec. \ref{sec:MFbootstrapped}), the single-shot protocol (Sec. \ref{sec:MFsingleshot}), and heralded SPDC (Sec. \ref{sec:MFSPDC}) in Fig. \ref{fig:rate}. For the bootstrapped protocols we implemented the balanced fusion strategy and let the resource requirements be the number of beamsplitter operations. For the single-shot protocol, Eq. (\ref{eq:single_shot_P}), the number of trials of the entire interferometer is the resource that we consider and we convert this to the number of implemented beamsplitters by realising that the interferometer shown in Fig. \ref{fig:single_shot} can be built from $d$ beamsplitters in a linear array such that the photons are progressively routed to the top mode. And so the measured number of beamsplitter operations is given by Eq. (\ref{eq:single_shot_P}) but with an additional factor of \mbox{$1/d$}, yielding \mbox{$r(d) = d!/d^{d+1}$}. For heralded SPDC we cannot convert the resource requirements because the SPDC protocol does not use  beamsplitters or single photons as a resource. We decided to use as a resource the number of repetitions of the SPDC source. So that the SPDC case has the same 20-photon preparation rate as the balanced bootstrapped protocol with recycling, we chose the mean photon-number to be \mbox{$\bar{n}\approx 1.7$}. This is a threshold by which the SPDC is more efficient than the recycled bootstrapped protocol. Below it is less efficient. A mean photon number of \mbox{$\bar{n}\approx 1.7$} is well beyond the capabilities of typical experiments today. This plot clearly shows that all cases, other than the balanced recycled protocol, the rate of state preparation decreases exponentially with $d$. Also, recycling improves the preparation rate as can be seen in the example for 20-photon state preparation where the preparation rate improves by a factor of \mbox{$\approx 10^5$} over the single-shot protocol. 

\begin{figure}[!htb]
\centering
\includegraphics[width=0.6\columnwidth]{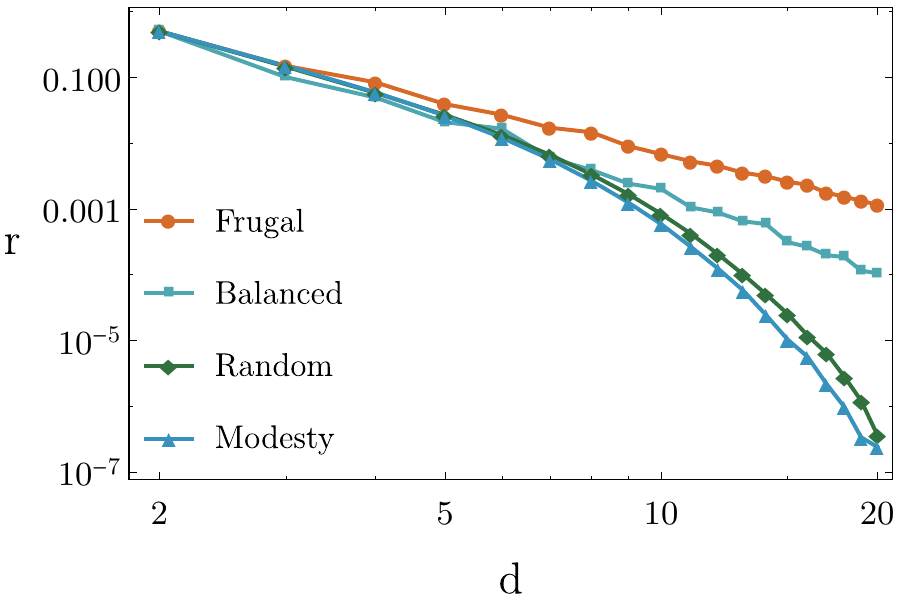}
\caption{Comparison of the strategies when employing state recycling, showing the strategies frugal, balanced, random, and modesty. The linear curves in the log-log plot are indicative of polynomial scaling of the preparation rate against the desired number of photons when employing the frugal and balanced strategies, whereas the random and modesty strategies exhibit exponential scaling.} \label{fig:rate_loglog}
\end{figure}

Now in Fig. \ref{fig:rate_loglog} we show the recycled protocol for some of the fusion strategies from Sec. \ref{sec:MFstrategies}. We show a log log plot and find that the preparation rate exhibits polynomial scaling against $d$ for the frugal and balanced recycled strategies and exponential scaling for the the random and modesty recycled strategies. This means that the frugal and balanced recycled protocols exhibit an exponential efficiency improvement. The scaling for the Frugal and Balanced strategies is $\sim1/d^{2.8}$ and $\sim1/d^{3.7}$ respectively, which is a significant improvement to the doubling approach that led to the rate in Eq. (\ref{eq:MFscaling}). 

\begin{figure}[!htb]
\centering
\includegraphics[width=0.6\columnwidth]{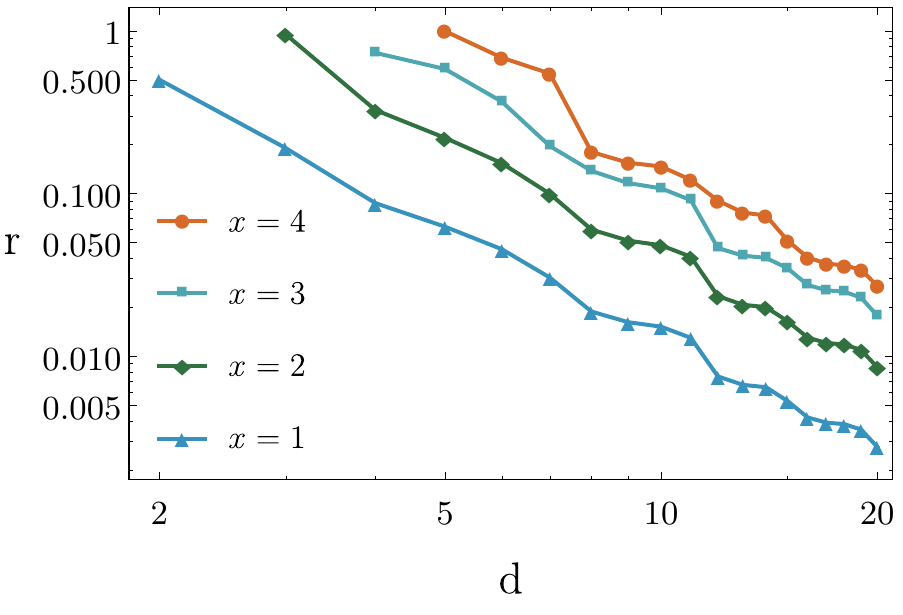}
\caption{Preparation rates for at least $d$ photons for hybrid schemes, where we begin with resource states of different photon numbers $x$ (i.e. \mbox{$c_x=\infty$}), and employ the frugal fusion strategy. Evidently, as we begin with larger starting resource states, the efficiency of the scheme improves, since now the steps that would ordinarily be required to prepare these resource states from single-photon states are mitigated.} \label{fig:hybrid}
\end{figure}

In Fig. \ref{fig:hybrid}, we show the hybrid schemes as discussed in section \ref{sec:MFhybrid}. Here we show the recycled frugal fusion strategy since this exhibits the best scaling.

\section{Summary} \label{sec:MFconclusion}

We have devised a protocol based on the idea of cluster states for efficiently preparing large photon-number Fock states as motivated in section \ref{sec:MFmotivation}. We began by describing some obvious ways to inefficiently prepare large photon-number Fock states in section \ref{sec:MFSPDC} and section \ref{sec:MFsingleshot}. Then in section \ref{sec:MFbootstrapped} we explain the basic formalism for our efficient approach. This protocol is non-deterministic, assumes that a resource of single-photon states is readily available, requires quantum memory, requires fast-feedforward, and uses linear optics with post-selection. Large photon-number Fock states are generated by iteratively fusing together smaller Fock states into larger ones that undergo a random walk in the quantum memory and shown in section \ref{sec:MFfusion}. The quantum memory allows for us to use state recycling where we can use all post-selected events as a resource. We found that our method was able to exponentially improve the state preparation rate by orders of magnitude compared to other na\"ive brute-force and single-shot methods. 
In section \ref{sec:MFreduction} we discuss how to reduce the photon number of a Fock state in case the resource state obtained is larger than desired. In section \ref{sec:MFresults} we show the results of this work and plot the various strategies showing the exponential improvement in scaling for certain fusion strategies. The requirements of this protocol like quantum memory and fast-feedforward are extremely challenging and are essentially the same requirements as LOQC. When LOQC is available so will the efficient preparation of large photon-number Fock states. Thus, the technology for generating these states will also help lead to realising LOQC. 

\begin{savequote}[45mm]
Son: Dad, why is cannabis illegal? \\
Dad: Well, son, legalizing the cannabis plant could free us from oil dependence, stop deforestation, become a saf alternative to many pharmaceuticals, and would cause prisons to shut down. \\
Now do you understand?
\qauthor{Anonymous}
\end{savequote}

\chapter{Preparing Continuous Variable Optical States for Quantum Error Correction by Coupling Atomic Ensembles to Squeezed States of Light} \label{ch:GKP}

\section{Synopsis}
In 2001 Gottesman, Kitaev, and Preskill (GKP) showed theoretically how to encode information fault-tolerantly on a continuous variable system using linear optical techniques; however, GKP did not show how to physically prepare these continuous variable states. In this work we present a non-deterministic scheme for generating optical continuous variable states by coupling an atomic ensemble with a squeezed state of light. The coupling creates a comb of Gaussian squeezed states of light in one channel. Upon particular measurement events of angular momentum in the other channel a desired resource state for encoding logical information is prepared. 

This work is first motivated in Sec. \ref{sec:GKPmotivation}. We then show how to prepare the resource state in Sec. \ref{sec:GKPpreparing}. Once we have the resource state we may encode logical states onto certain outcomes of the resource state as discussed in Sec. \ref{sec:GKPencoding}. In Sec. \ref{sec:GKPsymmetricencoding} we discuss the relationship between the total angular momentum required of the atomic ensemble for a given amount of squeezing in the squeezed state of light to prepare a resource state that is sufficient for resolving encoded states. In Sec. \ref{sec:GKPsuccessProb} we discuss the success probability of post-selecting upon a measurement that yields a useful resource state. Specifically, for a given total angular momentum $J$ of the atomic ensemble there are $2J+1$ possible outcomes. We find that the events that yield desirable resource states for encoding logical information also have the highest probability of being measured, which is a significant advantage of this scheme over others. In Sec. \ref{sec:PhysicalImplementation} we propose a physical implementation based on the Faraday interaction between an atomic ensemble and a squeezed optical field.

\section{Motivation} \label{sec:GKPmotivation}

Realising a quantum computer is laden with many difficult problems to overcome. Whatever architecture may be used for this realisation will require quantum error correction because the quantum state that is processed within the quantum computer is extremely susceptible to being destroyed by its environment and so it is a requirement that the state of the quantum system within the quantum computer remains coherent in the presence of such environmental noise. One way to overcome this noise is to encode logical information on either discrete or continuous variable (CV) systems. In general these systems use an ensemble of quantum particles to encode a logical state. There have been several proposals for encoding information in a larger system that is resilient against various types of errors such as the now famous toric code \cite{bib:kitaev03} among others \cite{bib:braunstein98, bib:lloyd98}.  

A promising CV proposal by Gottesman, Kitaev, and Preskill (GKP) in 2001 \cite{bib:GKP2001} showed how to encode logical states in an infinite-dimensional Hilbert space such that in the ideal encodings a universal set of fault-tolerant quantum gates can be implemented with linear optical operations, squeezing, homodyne detection, and photon counting. They described how a comb of delta functions can perfectly encode logical states; however, these delta functions are unphysical as they would require infinite energy, so they approximated the delta functions with Gaussians, which can encode the logical states with an associated error probability. These Gaussians approach the ideal delta functions as squeezing goes to infinity or equivalently their variance tends to zero. It remained unknown weather or not these error prone continuous variable encodings could enable fault-tolerant measurement-based quantum computation until Menicucci showed in 2014 that they can \cite{bib:menicucci2014}. 

GKP did not propose how to actually make these states and just presented them as a model for encoding. Since then proposals have been made to generate these GKP states optically \cite{bib:vasconcelos10} and by a variety of other methods \cite{bib:gottesman01, bib:Travaglione02, bib:pirandola04, bib:pirandola06, bib:pirandola06b}, but none are particularly easy to implement. In this work we propose a new way of generating optical GKP states by coupling an atomic ensemble to a squeezed state of light. 

\section{Preparing the Resource State} \label{sec:GKPpreparing}

\begin{figure}[!htb]
\centering
\includegraphics[width=.55\columnwidth]{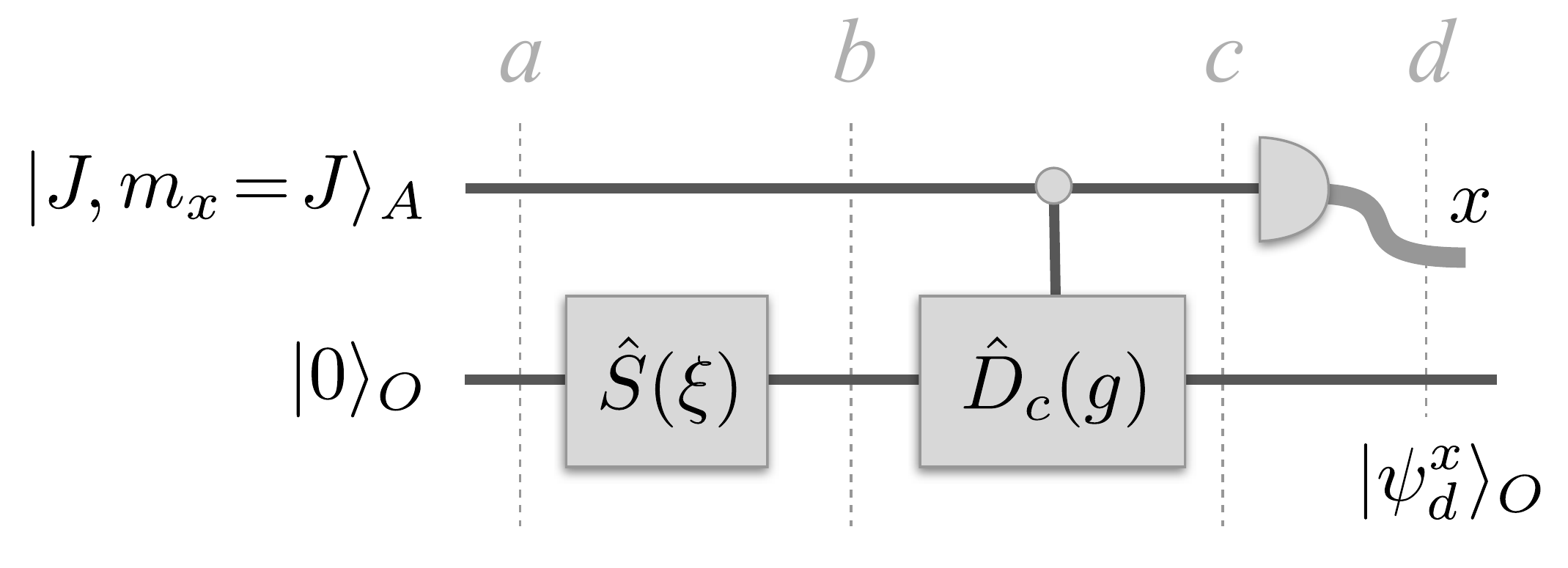}
\caption{The circuit model that creates the resource state $\ket{\psi_d^x}_O$. $\ket{J, m_x=J}_A$ is the orbital angular momentum eigenstate with total angular momentum $J$ prepared in the $x$-basis and $\ket{0}_O$ is the vacuum state of light which is squeezed via the standard squeezing operator $\hat{S}(\xi)$. A controlled position shift operator $\hat{D}_c(g)$ couples the atomic ensemble and the squeezed state of light in the $z$-basis. At the bottom output channel a resource state $\ket{\psi_d^x}_O$ is obtained based on post selecting particular measurements of angular momentum $x$ in the top channel.} \label{fig:CircuitModel}
\end{figure}

In this section we show how the optical resource state is conditionally prepared using a controlled interaction with an angular momentum. Where appropriate, the state of the angular momentum and of the optical field will be labelled with $A$ and $O$, respectively. The calculations follow the progress of the joint state through the circuit in Fig.~\ref{fig:CircuitModel}, which begins with an unentangled spin-optical state $\ket{\psi_a}_{AO}=\ket{J,m_x\!=\!J}_A\ket{0}_O$ at (a) and ends with a conditional optical state prepared upon a measurement of the spin at (d). A succinct derivation of the protocol for more general initial angular momentum and optical states is given in Appendix~\ref{sec:GKPKraus}. 

Initially the atomic ensemble with total collective angular momentum $J$ is prepared in a spin-polarized state along the $x$-axis which we will denote as $\ket{J, m_x\!=\!J}_A $. 
Expressed in terms of eigenstates along the $z$-axis, $\ket{J,m}_A$, the initial spin state is
\begin{align}
  \ket{J,m_x\!=\!J}_A 
  &= \sum_{m=-J}^J d_{m,J}^{(J)}(\pi/2) \ket{J,m}_A,
\end{align}
where the coefficients, $d_{m,m'}^{(J)}(\beta) \equiv \bra{m} \hat{R}_y(\beta) \ket{m'}$, are the matrix elements of the operator $\hat{R}_y(\beta)=\mathrm{exp}(-i \beta \hat{J}_y)$ that rotates the state around the $y$-axis by an angle $\beta$. 
For our purposes $J$ is fixed and $\beta=\pi/2$; henceforth, we drop unnecessary notation and define
\begin{equation}\label{eq:dmmd}
d_{m,m'} \equiv d_{m,m'}^{(J)}(\pi/2) = \frac{1}{2^J} \sum_{k=0}^{J+m'}(-1)^{k-m'+m}\frac{\sqrt{(J+m')!(J-m')!(J+m)!(J-m)!}}{(J+m'-k)!k!(J-k-m)!(k-m'+m)!},
\end{equation}
where we have used the explicit form for $d_{m,m'}^{(J)}(\beta)$ that appears, $e.g.$, in Sakurai \cite{bib:sakurai2011modern} Sec. 3.8. 

When $m'=\pm J$ only a single term in the summation in Eq. (\ref{eq:dmmd}) has non-negative factorials in the denominator, and the matrix element simplifies to
\begin{equation} \label{eq:GKPmatrixElements}
  d_{m,\pm J} = \frac{(\pm 1)^{J+m}}{2^J}\begin{pmatrix} 2J \\ J-m\end{pmatrix}^{1/2}.
\end{equation}
Specifically, the factorials in the denominator are non-negative only when $k=J-m$ for $m'=+J$; and for $k=0$ for $m'=-J$. The values of $m'=\pm J$ are significant as these yield desirable resource states, as we will see in Sec. \ref{sec:GKPencoding}.

The initial optical state is the bosonic vacuum $\ket{0}$ with creation and annihilation operators $\hat{a}$ and $\hat{a}^\dagger$ obeying the usual commutation relations $[\hat{a}, \hat{a}^\dagger]=1$. The squeezing operator \cite{bib:KokLovett11}
\begin{align}
\hat{S}(\xi) = e^{\frac{1}{2}(\xi^*\hat{a}^2-\xi \hat{a}^{\dagger 2})}
\end{align}
acts on this initial state to produce a squeezed vacuum,
\begin{equation} \label{eq:NickSucks}
\hat{S}(\xi)\ket{0}_O=\ket{\xi}_O.
\end{equation}
Hence the joint state at (b) in Fig.~\ref{fig:CircuitModel} is 
\begin{equation}
\ket{\psi_b}_{AO}= \sum_{m=-J}^J d_{m,J}\ket{J,m}_A\ket{\xi}_O.
\end{equation}

\begin{figure*}[t]
\centering
\includegraphics[width=\columnwidth]{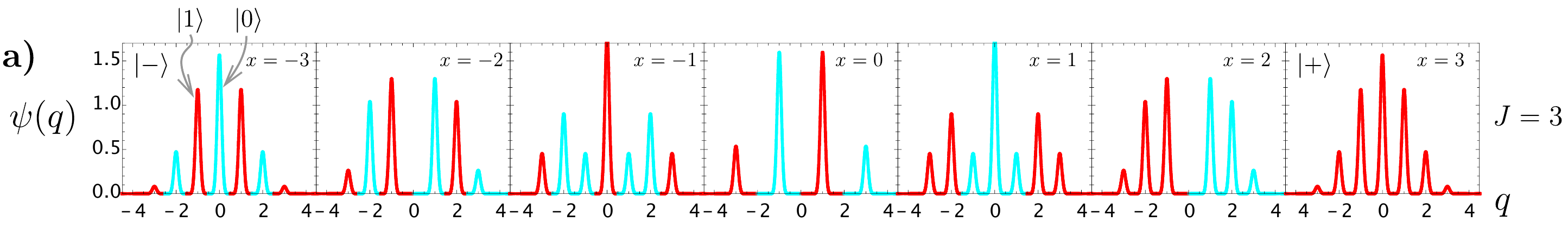}
\includegraphics[width=\columnwidth]{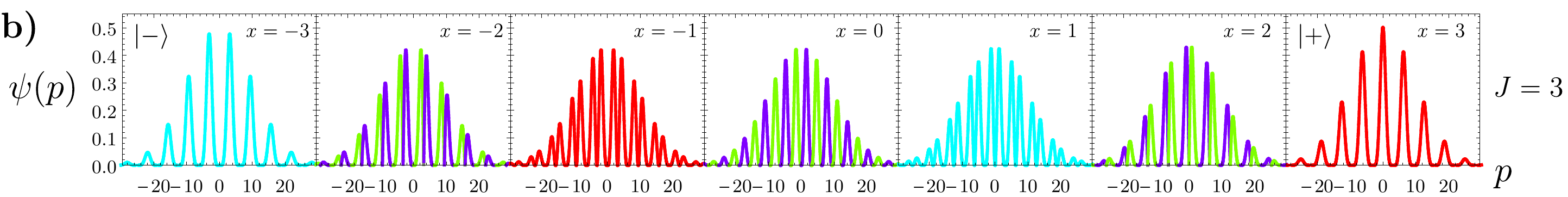}
\includegraphics[width=\columnwidth]{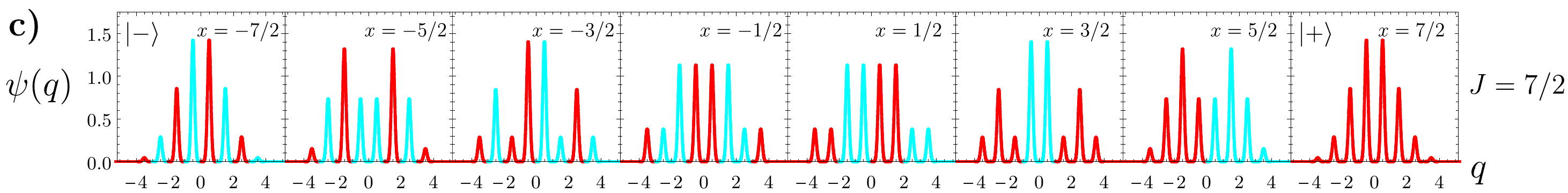}
\includegraphics[width=\columnwidth]{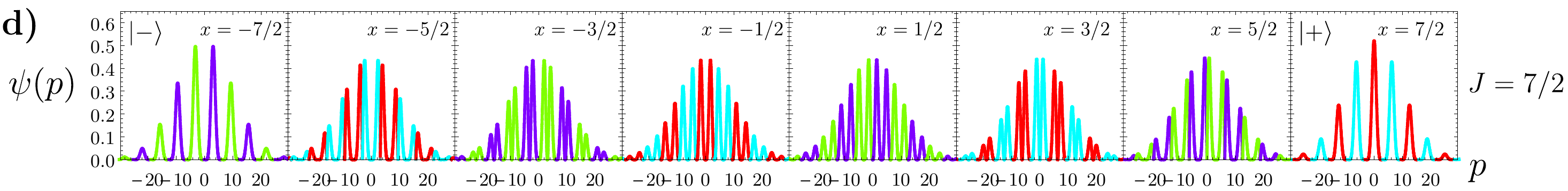}
\caption{Position- and momentum-space wavefunctions, Eqs. (\ref{eq:psicondJ}), for the unnormalised conditional optical state, $\ket{\psi_d^x}_O$, given total angular momentum $J=3$ (a,b) and $J=7/2$ (c,d), for each of the $2J+1$ measurement outcomes, $x$. When $x=\pm J$ or $x=0$ is measured the resulting state may be used to encode a logical state. In this work we focus on analyzing the $x=\pm J$ outcomes. The colors red, blue, purple, and green indicate relative phases of 0, $\pi$, $i$, and $-i$ respectively. Note $g=1$ and $\sigma_q=0.1$ in these plots so that the peaks have sufficient spread to show the features.} \label{fig:resourceState}
\end{figure*}

The angular momentum and optical field are entangled with a controlled interaction that applies a displacement to the field proportional to the angular momentum in the $z$-direction given by
\begin{align} \label{eq:BigD}
	\hat{D}_c(g) = e^{-i g \hat{p}\hat{J}_z},
\end{align}
where $\hat{p}=-i(\hat{a}-\hat{a}^\dagger)/\sqrt{2}$ is the momentum-quadrature operator. The key feature is that the controlled displacement has the following effect:
\begin{equation} \label{eq:displacedsqueezed}
\ket{J,m}_A\ket{\xi}_O \rightarrow \ket{J,m}_A\left|\gdisp{m},\xi\right\rangle_O,
\end{equation}
where $\ket{\alpha,\xi}_O$ is a displaced squeezed state \cite{bib:KokLovett11}. The scaling of the displacement has been chosen to match the work of GKP \cite{bib:GKP2001} for the encoded logical states. After the controlled displacement given by Eq. (\ref{eq:BigD}), the joint state at (c) in Fig.~\ref{fig:CircuitModel} is
\begin{align}
\ket{\psi_c}_{AO} = \sum_{m=-J}^J d_{m,J}\ket{J,m}_A \left|\gdisp{m}, \xi\right\rangle_O.
\end{align}

Finally, the angular momentum is projectively measured along the $x$-direction as shown in (d) in Fig.~\ref{fig:CircuitModel}.  We express $\ket{\psi_c}_{AO}$ in the basis of $x$-eigenstates, 
\begin{align}
\ket{\psi_c}_{AO} 
&= \sum_{m,m'=-J}^J  
d_{m,J}d_{m,m'}\ket{J, m_x=m'}_A \left|\gdisp{m}, \xi\right\rangle_O. \label{eq:psic}
\end{align}
where we used the fact that the matrix elements in Eq. (\ref{eq:dmmd}) satisfy $\bra{J,m'}\hat{R}_y (-\pi/2)\ket{J,m} = d_{m,m'}$. 
Then, measurement trivially collapses the $m'$ summation in Eq.~(\ref{eq:psic}) to a single term. That is, given measurement outcome $x$, the optical state is projected to
\begin{align} \label{eq:projectedOpticalState}
 \ket{\psi_d^x}_O = \frac{1}{\sqrt{\mathcal{P}(x)}}\sum_{m=-J}^J d_{m,J}d_{m,x}\left|\gdisp{m}, \xi\right\rangle_O.
\end{align}
The state is normalized by the probability of obtaining $x$, 
\begin{equation} \label{eq:GKPmeasProb}
\mathcal{P}(x) = \sum_{m,m'}d_{m,J}d_{m,x}d_{m',J}d_{m',x}e^{-\frac{1}{4}g^2e^{2\xi}(m-m')^2},
\end{equation}
as shown in Appendix~\ref{app:MeasurementProbability}, noting that the squeezing parameter $\xi$ is taken to be real. 

The conditional optical state can be written as a wavefunction in the position and momentum quadratures, $\hat{q}=(\hat{a}+\hat{a}^\dagger)/\sqrt{2}$ and  $\hat{p}=-i(\hat{a}-\hat{a}^\dagger)/\sqrt{2}$ respectively,
\begin{eqnarray} \label{eq:psicondJ}
    \psi_d(q|x) &=& \braket{q}{\psi_d^{x}}_O = \frac{e^{\xi/2}}{\mathcal{P}(x)^{1/2}\pi^{1/4}}\sum_{m=-J}^J 
    d_{m,J}d_{m,x}
     \exp\left(-\frac{(q-gm)^2}{2 e^{-2\xi}}\right), \label{eq:psicondJpos} \nonumber \\
    \psi_d(p|x) &=& \braket{p}{\psi_d^{x}}_O = \frac{e^{-\xi/2}}{\mathcal{P}(x)^{1/2}\pi^{1/4}}
    \sum_{m=-J}^J d_{m,J}d_{m,x}\exp\left( -igmp-\frac{p^2}{2e^{2\xi}}\right).
\end{eqnarray}
For spins of size $J = 3$ and $J=7/2$, the position- and momentum-space wavefunctions are plotted in Fig \ref{fig:resourceState}. Each measurement outcome $x$ prepares a conditional position-space wavefunction, Eq. (\ref{eq:psicondJpos}), composed of a superposition of displaced squeezed states separated by $mg$ (for integer $m$) with amplitudes governed by the product of matrix elements, $d_{m,J}d_{m,x}$. 

\section{Encoding} \label{sec:GKPencoding}

Logical information is encoded into the continuous variable degrees of freedom of the optical states using the scheme proposed by GKP. 
We seek maximally distinguishable encoded states with logical operations corresponding to translations in position and momentum.  
Of the conditional optical states after the angular momentum measurement, those corresponding to $x=\pm J$ or $x=0$ are suitable for GKP encoding. However, the state produced by $x=0$ has a much smaller probability than $x=\pm J$ (see Sec. \ref{sec:GKPsuccessProb}), is not available for half-integer $J$, and cannot be used together with the states generated by the other outcomes. Consequently, we concentrate exclusively on the resource states generated by the outcomes $x=\pm J$. The outcomes $x=\pm J$ respectively generate the logical states $\ket{+} = (\ket{0}+\ket{1})/\sqrt{2}$ and $\ket{-} = (\ket{0}-\ket{1})/\sqrt{2}$.
The corresponding wavefunctions, which appear in the far right and far left panels of Fig. \ref{fig:resourceState}, follow from Eqs. (\ref{eq:psicondJ}),
\begin{subequations} \label{eq:psiJ}
\begin{align}
    \psi_d(q|\pm J) & =  \frac{e^{\xi/2}}{\mathcal{P}(\pm J)^{1/2}\pi^{1/4}4^J}\sum_{m=-J}^J 
    \begin{pmatrix}2J \\ J-m\end{pmatrix} (\pm 1)^{J+m}\exp\left(-\frac{(q-gm)^2}{2 e^{-2\xi}}\right), \label{eq:psiqJ}\\
    \psi_d(p|\pm J) & = \frac{e^{-\xi/2}}{\mathcal{P}(\pm J)^{1/2}\pi^{1/4}}\exp\left(-\frac{p^2}{2e^{2\xi}}\right)
     \left\{ \begin{matrix}\cos^{2J}(g p/2) \\ e^{iJ\pi}\sin^{2J}(g p/2)\end{matrix} \right\}. \label{eq:psipJ}
\end{align}
\end{subequations}
where we have used the $d_{m,\pm J}$ matrix elements in Eq. (\ref{eq:GKPmatrixElements}).
The position-space wavefunction, $\psi(q|\pm J)$, can be described as a product of a comb of Gaussians each with variance
\begin{equation} \label{eq:GKPvarQ}
\sigma_{q}^2=e^{-2\xi},
\end{equation}
separated by $g$; and an envelope arising from the binomial distribution that appears in Eq. (\ref{eq:psiqJ}), namely
\begin{equation} \label{eq:GKPvarEnvQ}
\sigma_{q,\mathrm{env}}^2=g^2J/2.
\end{equation}
Similarly, the momentum-space wavefunction, $\psi(p| \pm J)$, can be described as a product of a Gaussian envelope with variance
\begin{equation} \label{eq:GKPvarEnvP}
\sigma_{p,\mathrm{env}}^2=e^{2\xi},
\end{equation}
and a comb of approximately Gaussian peaks generated by $\cos^{2J}(g p/2)$, hence separated by $2\pi/g$. The variance of the individual peaks is given by
\begin{eqnarray} \label{eq:GKPvarP}
    \sigma_{p}^2 &=& \frac{2(J^2\zeta(2,J)-1)}{g^2J^2} \nonumber \\
    &\approx& 1/\sigma_{q,\mathrm{env}}^2 +O(1/J^2), 
\end{eqnarray}
where $\zeta(s,a)$ is the Hurwitz zeta function, as shown in Appendix~\ref{app:cosvariance}.

\begin{figure*}[t]
\centering
\includegraphics[width=\columnwidth]{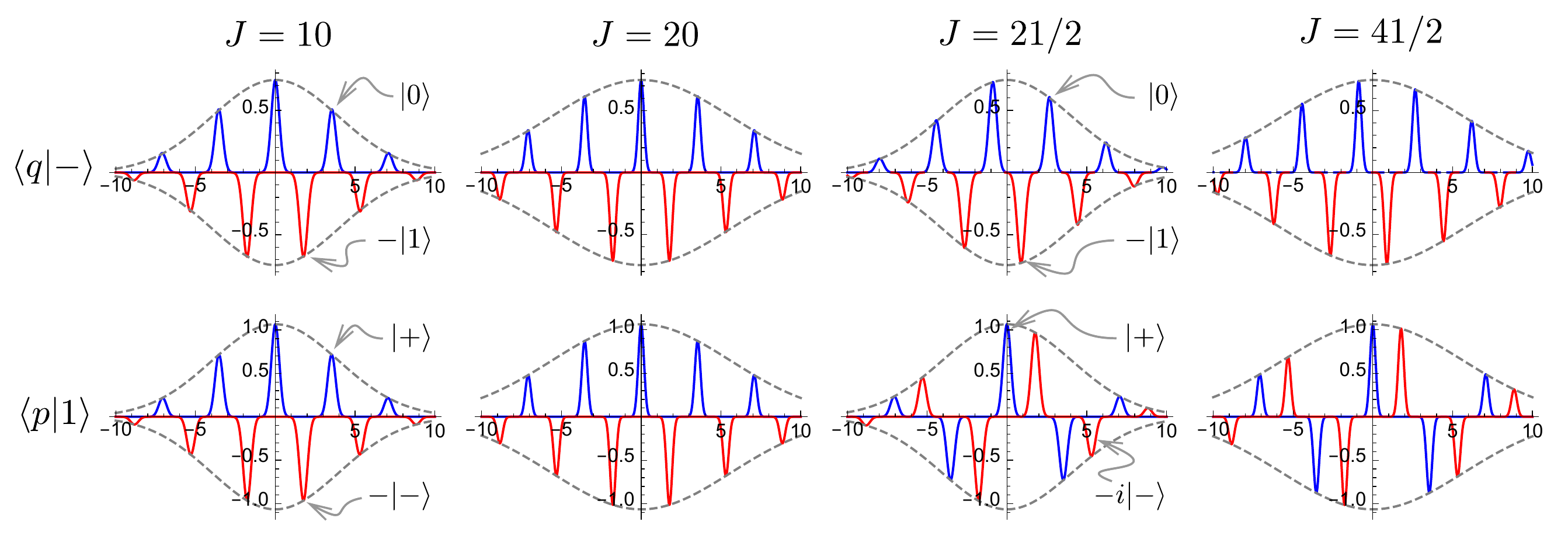}
\caption{Amplitudes of the logical $\ket{-}$ state in the $q$-representation and the logical $\ket{1}$ state in the $p$-representation for $J=10$, $20$, $21/2$ and $41/2$. For integer $J$, the logical $\ket{0}$ and $\ket{1}$ states can be read as the components above and below the line for $\braket{q}{-}$, similarly for $\ket{+}$ and $\ket{-}$ in $\braket{p}{1}$ plots. The half-integer $J$ values are displaced in the  $q$-representation so that they are centred at the origin. For half-integer $J$ the $\ket{-}$ component of $\braket{p}{1}$ is complex and has been rotated to the real plane for the figures. Parameters were chosen for a symmetric encoding; $g=\sqrt{\pi}$ and squeezing $\xi$ given by Eq.~(\ref{eq:xisymmetric}).}  \label{fig:SymmResourceState}
\end{figure*}

In the $q$-quadrature, the peaks in the logical $\ket{0}$ and $\ket{1}$ states are interleaved such that the spacing between peaks of one state and the other is a minimum of $g$. With this encoding $q$-shifts smaller than $g/2$ can be corrected, and Pauli-$X$ errors correspond to interchanging $\ket{0}$ and $\ket{1}$. These relationships are most clearly seen in the left-most panels of Fig.~\ref{fig:resourceState}(a,c) for the logical state $\ket{-}$. 
In the $q$-quadrature the peak locations are the same for the $\ket{+}$ and $\ket{-}$ logical states, so a $q$-measurement is unable to distinguish them. In the $p$-quadrature, however, the $\ket{+}$ and $\ket{-}$ states \emph{are} distinguishable while the $\ket{0}$ and $\ket{1}$ that are not. In this quadrature the minimum separation between peaks of $\ket{+}$ and of $\ket{-}$ is $\pi/g$, so $p$-shifts smaller than $\pi/(2g)$ can be corrected, and Pauli-$Z$ errors correspond to interchanging $\ket{+}$ and $\ket{-}$. Depending on the error model for a particular physical implementation, the interaction strength in the controlled displacement [Eq. (\ref{eq:BigD})], can be modified to produce encodings more robust to errors in one quadrature than the other.

\subsection{Symmetric encoding} \label{sec:GKPsymmetricencoding}

A specific encoding, laid out in the original GKP paper \cite{bib:GKP2001}, assumes that $q$- and $p$-quadrature errors are comparable in size. This \emph{symmetric encoding} is a resource for fault-tolerant quantum computing using continuous-variable cluster states. 
For the ideal (infinite squeezing) code, the size of a correctable shift error in the $q$-representation is half the minimum separation between a peak in the logical $\ket{0}$ state and a peak in the logical $\ket{1}$ state and similarly for $\ket{+}$ and $\ket{-}$ in the $p$-representation. Equating the two yields $g=\sqrt{\pi}$ in the code states, Eqs. (\ref{eq:psiJ}).


The prepared logical states are finite-squeezing approximations to the ideal GKP states. Thus, irrespective of shift errors, there is an \emph{embedded} probability of mistaking $\ket{0}$ and $\ket{1}$, or $\ket{+}$ and $\ket{-}$, due to finite overlap of neighboring Gaussians peaks. To balance the embedded errors, the symmetric encoding is completed by equating the peak variances in the $q$- and $p$-quadratures as follows.
For input squeezing $\xi$, the peak variance in $q$ is given by Eq. (\ref{eq:GKPvarQ}) and in $p$ follows from Eq. (\ref{eq:GKPvarP}) using $g = \sqrt{\pi}$. For fixed $\xi$ and $g$, the size of the angular momentum $J$ determines the peak variance in $p$. Setting $\sigma_{q}^2=\sigma_{p}^2$  gives
\begin{equation}\label{eq:xisymmetric}
    J(\xi) = \smallfrac{2}{\pi} e^{2 \xi}, 
\end{equation}
which carries the additional effect of symmetrizing the envelopes, $\sigma_{q,\mathrm{env}}^2=\sigma_{p,\mathrm{env}}^2=\pi J/2$. 
Table \ref{table:squeezing} gives the required $J$ for various values of squeezing (in dB). A symmetric encoding prepared with a total angular momentum $J\approx 127$ with $s\approx 20$ dB of squeezing is sufficient for fault-tolerant measurement-based quantum computing with continuous-variable cluster states \cite{bib:menicucci2014}.  
Examples of symmetric resource states prepared by the conditional procedure presented here are shown in Fig.~\ref{fig:SymmResourceState} along with the corresponding logical states. 

\begin{table}[!htb]
\centering
\begin{tabular}{ |c|c|c|c|c|c|c|c|c|c|c|c|  }
 \hline
 $s$ (dB) & 0 & 2 & 4 & 6 & 8 & 10 & 12 & 14 & 16 & 18 & 20 \\
 \hline
 $J$ & 1 & 2 & 3 & 5 & 8 & 13 & 20 & 32 & 51 & 80 & 127 \\
 \hline
\end{tabular}
\caption{Total angular momentum $J$ required to achieve a symmetric encoding for a given amount of squeezing. The squeezing expressed in dB, $s= -10 \, \mathrm{log}_{10}\left(\frac{e^{-2 \xi}}{1/2}\right)$, and $J$ has been rounded to the nearest integer.}
\label{table:squeezing}
\end{table}

\subsection{Success Probability} \label{sec:GKPsuccessProb}

\begin{figure}[h]
\centering
\includegraphics[width=0.55\columnwidth]{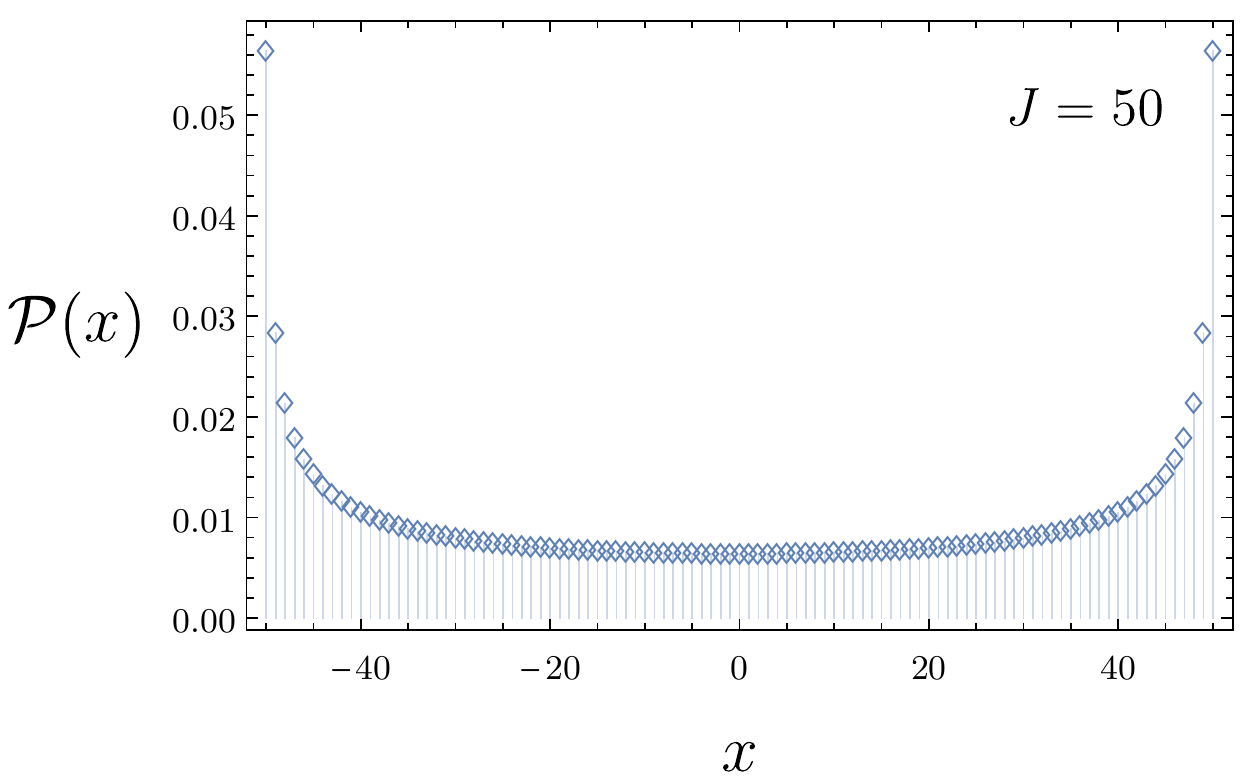}
\caption{The probability $\mathcal{P}(x)$ of measuring a particular angular momentum $x$ which corresponds to a particular resource state. The measurements of $x=\pm J$ are the most probable and are indeed the measurements the yield useful resource states which is an advantage of using this scheme.} \label{fig:hightails}
\end{figure}

The probabilities of obtaining the resource states for symmetric encoding follow from Eq. (\ref{eq:GKPmeasProb}). Given total angular momentum $J=50$, which arises from Eq. (\ref{eq:xisymmetric}), the probabilities for all spin measurement outcomes are shown in Fig. \ref{fig:hightails}. A significant advantage of the protocol proposed here is the relatively large success probabilities.

The outcomes we care most about (i.e. $x=\pm J$) are also the most probable ones to measure. This significantly boosts the success probability of obtaining a resource state that is usable for encoding logical states onto and is thus an advantage in our scheme. 
\begin{equation}\label{eq:JmeasProb}
     \mathcal{P}(\pm J) = \frac{1}{16^J} \sum_{m,m'} \begin{pmatrix}2J \\ J-m\end{pmatrix}
    \begin{pmatrix}2J \\ J-m'\end{pmatrix}e^{-\frac{\pi^2}{8} J (m-m')^2}.
\end{equation}
Since both resource states are compatible with the same encoding, and the probabilities are equal, the total success probability is $\mathcal{P}_s = 2\mathcal{P}(+J)$.

As shown in Appendix \ref{app:MeasurementProbability}, in the limit that the squeezing $\xi$ and associated $J$ are large, the embedded error vanishes and the probabilities become,
\begin{equation} \label{eq:PmeasLim}
\mathcal{P}(x) = \sum_{m}\left(d_{m,J}d_{m,x}\right)^2,
\end{equation}

\begin{figure}[h]
\centering
\includegraphics[width=0.55\columnwidth]{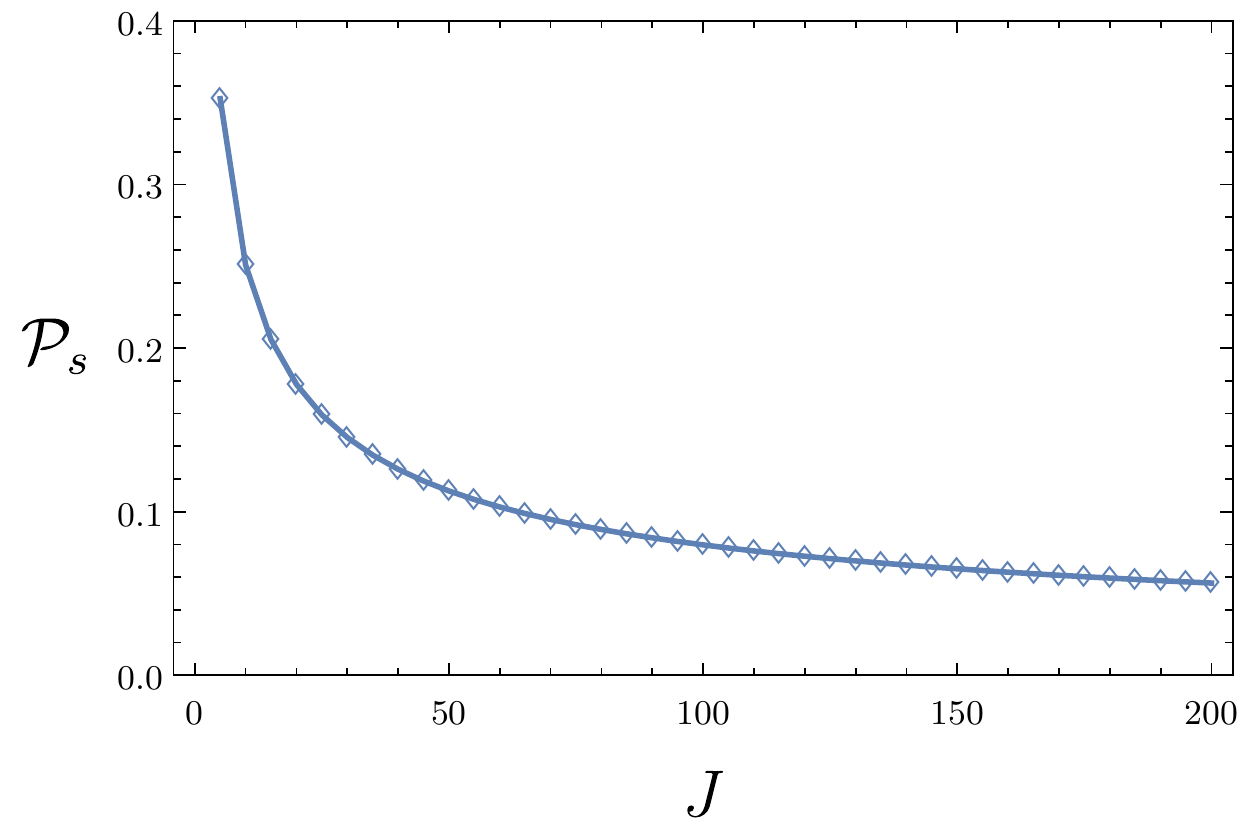}
\caption{The probability $\mathcal{P}_{s}$ of successfully obtaining a resource state that is useful to encode a logical state on in the large $J$ limit of Eq. (\ref{eq:PsuccessLargeJ}).} \label{fig:Psucc}
\end{figure}

Now the success probability of obtaining the desired resource state $\mathcal{P}_{s}$ is given by 
\begin{align} \label{eq:PsuccessLargeJ}
\mathcal{P}_{s} \approx \sqrt{\frac{2}{\pi J}},
\end{align}
which is shown in App.~\ref{app:MeasurementProbability} and the approximation holds for large $J$. Taking for instance $J=127$, which corresponds to a squeezing of 20 dB as shown in table \ref{table:squeezing}, we find $\mathcal{P}_{s}\approx 7\%$. This is a highly favourable success probability for such a large value of total angular momentum $J$. Fig. \ref{fig:Psucc} shows how the probability of successfully post-selecting on a desired resource state scales with $J$. Whereas the success probability of the iterated scheme scales exponentially poorly in the number of steps $N$ (equivalent to an angular momentum $N/2$ here), the probability here falls off only as $\sqrt{J}$. The reason comes from the nature of the measurements. Here, the measurement is collective and the outcomes arise solely from the highest irreducible representation of angular momentum for $N = 2J$ spins, where the $J=\smallfrac{1}{2}$ spins are measured individually, yielding outcomes spread over the full $2^N$ Hilbert space, the vast majority of them fail to produce the desired resource states.

\section{Physical Implementation} \label{sec:PhysicalImplementation}
In this section we show three important aspects for experimental implementation: (1) We present a method for experimental implementation using a Faraday-based quantum non-demolition (QND) interaction. (2) We show a method for measuring the angular momentum $x$ in the top channel of Fig. \ref{sec:GKPpreparing}.

\subsubsection{Faraday-based QND interaction} \label{sec:GKPinteraction}

To implement the protocol, we couple the squeezed field to the collective spin formed by an ensemble of polarizable neutral atoms. Consider $N$ such atoms, each with effective spin $\hat{j}_z = \frac{1}{2} \big(\ketbra{\uparrow}{\uparrow} - \ketbra{\downarrow}{\downarrow} \big)$ defined on metastable ground states $\{\uparrow, \downarrow \}$. The atoms couple to a common mode of light possessing two orthogonal linear polarizations, horizontal ($H$) and vertical ($V$), with respective annihilation operators $\hat{a}_H$ and $\hat{a}_V$. For an off-resonant field the atoms and light become entangled via the dispersive Faraday interaction, $\hat{U} = e^{-i\chi \hat{S}_3 \hat{J}_z}$, which describes a coupling of the collective atomic spin, $\hat{J}_z = \sum_{n=1}^N \hat{j}_z^{(n)}$, to the 3-component of the field's Stokes vector \cite{bib:deutsch10}, 
	\begin{align} \label{Eq::Stokes}
		\hat{S}_3 &= \smallfrac{1}{2i} \big( \hat{a}^\dagger_H \hat{a}_V - \hat{a}^\dagger_V \hat{a}_H \big).
	\end{align}
The Faraday interaction generates a rotation of the Stokes vector around the 3-axis proportional to the atomic spin projection along $\hat{J}_z$ with a strength characterized by the dimensionless, single-photon rotation angle $\chi$.

The controlled displacement required for the GKP resource-state protocol arises by preparing the $H$-mode of light in a coherent state with $N_L$ photons.
Making the linearization $\hat{a}_H \rightarrow \sqrt{N_L}$, the Stokes operator in Eq. ({\ref{Eq::Stokes}) becomes $\hat{S}_3 \approx  \sqrt{ N_L/2 } \hat{p},$ where $\hat{p} = -i(\hat{a}_V - \hat{a}_V^\dagger )/\sqrt{2}$ is the momentum quadrature of the $V$-polarized mode. This linearized Faraday interaction generates the requisite controlled translations of $V$-mode photons, Eq. (\ref{eq:BigD}), with effective coupling strength,
	\begin{align} \label{Eq::Effectiveg}
		g = \sqrt{ N_L /2  } \, \chi.
	\end{align}

\subsubsection{Projective spin measurement} \label{sec:GKPmeasurement}

Once the optical field and atoms have become entangled, the collective spin state is projectively measured.
A single atomic spin may be measured by driving a cycling transition and detecting the resulting fluorescence \cite{bib:Wineland11}. 
Concatenating with unitary transformations, a projective measurement can be realized. 
However, since our resource-state protocol benefits from spins larger than can be achieved using a single atom, we focus instead on a quantum nondemolition (QND) measurement of the collective spin of many atoms. 
The collective spin is coupled via the same Faraday interaction to a second field that serves as a \emph{meter}. 
The meter experiences a spin-dependent polarization rotation that is measured via homodyne polarimetry. 
When the spin-meter coupling is strong, relative rotations from different projective $m$-values become distinguishable over the meter's shot noise, and the collective spin measurement is projective. 
This is indeed the same strong-coupling requirement for the GKP-state peaks to be sufficiently separated. 
It become more challenging to distinguish angular momentums from each other as the total angular momentum increases (e.g. $J=126$ versus $J=127$ as opposed to $J=1$ versus $J=2$).

To implement the measurement, the collective spin is first rotated into the $x$-basis with a $\pi/2$-pulse. The meter is initialized with $N_L$ photons in the $H$-mode and squeezed vacuum in the quantum mechanical $V$-mode, $\ket{\xi}_M$. Polarimetry in the diagonal polarization basis implements an effective homodyne measurement of the position quadrature for $V$-mode photons, with the $H$-mode serving as the local oscillator \cite{bib:baragiola14}.  
The degree to which the measurement is projective is determined by the distinguishability of the meter states, $\ket{\gdisp{m}, \xi }_M = e^{ -i  g m \hat{p} } \ket{\xi}_M$, given by Eq. (\ref{eq:Overlap}),
	\begin{align} \label{Eq::Resolution}
		\left| \left\langle \gdisp{m}, \xi\bigg|\gdisp{m'},  \xi\right\rangle \right|^2 & = \exp \big[ - (g/ 2 \sigma_q)^2(m - m')^2 \big].
	\end{align}
In the limit that $(m - m')^2 \gg 4 \sigma_q^2/g^2 $, the meter states become orthogonal.  Thus, distinguishing neighboring eigenstates of $\hat{J}_z$ requires $ 8 \sigma_q^2/ N_L \chi^2 \ll 1 $, with limits set by the characteristic coupling strength $\chi$ and the squeezed shot noise in the polarimeter. Note that in contrast to the GKP encoding as discussed in Sec. \ref{sec:GKPencoding}, the spin-meter coupling is not constrained by a specific value of $g$ for a given $ \sigma_q^2$, since the goal is only to produce distinguishable meter states.

Practical limitations on $N_L$ in both the GKP mode and the meter arise for two related reasons.
First, the Faraday interaction is a valid description of the light-matter coupling when the quantum emitters remain far below saturation. 
Second, increased $N_L$ precipitates more spontaneous photon scattering that spoils the QND interaction and measurement. 
Indeed, this has restricted QND spin squeezing in free space to the Gaussian regime, far from a projective measurement. 
The requirement to overcome the effects of decoherence is that the coupling to the collective optical mode is large relative to all other modes.
This can be characterized by the optical density per atom, $\eta \equiv \sigma_0/A_m$, the ratio of the resonant atomic scattering cross section $\sigma_0$ to the transverse mode area $A_m$. 
While typical optical densities per atom in free space, $\eta \sim 10^{-5}$, are far too weak for our purposes \cite{bib:baragiola14}, those in engineered photonic environments, such as photonic crystal waveguides, can be much larger; $\eta \sim 1$. Future improvements are expected by operating near a band edge where ``slow light" can enhance the interaction by nearly two orders of magnitude \cite{bib:goban14}. 

A detailed study of optical pumping for atoms very near and strongly coupled to a waveguide is beyond the scope of our work; nevertheless, an estimate of the required coupling can be found from a free-space model for alkali atoms. 
Here, the Faraday rotation angle per photon per unit angular momentum given above is $\chi = \eta \Gamma / 2 \Delta $, for detuning $\Delta$ and spontaneous emission rate $\Gamma$ \cite{bib:deutsch10}.
To realize the coupling strength in Eq. (\ref{Eq::Effectiveg}) required for $qp$-symmetric codewords and an approximately projective measurement while limiting the number of free-space scattered photons, we find that for $N_L = 10^4$ and $\Delta = 1000 \Gamma$ the required optical density per atom is $\eta \sim 25$, within the reach of near-term technology.
It may be possible to augment the atom-light coupling with an optical cavity \cite{bib:mcconnell15} and to suppress the deleterious effects of optical pumping by judiciously selecting the effective spin within each atom \cite{bib:Bohnet14}. 
Alternative fruitful avenues have opened in other physical architectures, where demonstrated strong coupling of ``artificial atoms" to photonic environments could provide the necessary interaction strength \cite{bib:devoret13, bib:arnold15}. In such systems, Purcell enhancement of the total coupling rate has the potential to reduce the collective spin's susceptibility to other sources of noise.

\section{Summary}

In this work we show how to create a state that may be used for quantum error correction in quantum information processing. Quantum error correction is a critical component of realising any quantum computer. Specifically, we show a non-deterministic scheme for generating optical continuous variable states by coupling an atomic ensemble with a squeezed state of light, which can be used to encode and correct quantum information. This work is further motivated in Sec. \ref{sec:GKPmotivation}. The coupling creates a comb of Gaussian squeezed states of light, which upon particular measurement events of angular momentum, a desired resource state for encoding logical information is prepared. The details of preparing the resource state are presented in Sec. \ref{sec:GKPpreparing} and how to encode information on these states is discussed in Sec. \ref{sec:GKPencoding}. In Sec. \ref{sec:GKPsymmetricencoding} bounds between the total angular momentum required of the atomic ensemble for a given amount of squeezing in the squeezed state of light to sufficiently prepare a useable resource state is derived. In Sec. \ref{sec:GKPsuccessProb} the success probability of our scheme to yield desirable resource states for encoding logical information is shown. We find that the desired post-selection events also have the highest probability of occuring, which is a significant advantage of this scheme over others. Finally, we propose a physical implementation for realising our protocol based on the Faraday interaction in Sec. \ref{sec:PhysicalImplementation}.

\begin{savequote}[45mm]
You can tell more about a person by what they say about others than you can by what others say about them.
\qauthor{Leo Aikman}
\end{savequote}

\chapter{Summary}

In this thesis we developed research areas that will lead to novel quantum technologies such as a universal quantum computer. Such technologies promise significant breakthroughs in digital security, our ability to measure various phenomena in nature, condensed-matter physics, high-energy physics, atomic physics, quantum chemistry, cosmology, and medicine among others. Although the technologies required to build a universal quantum computer remain daunting, these breakthroughs are motivation enough for much of the world to focus resources into funding these futuristic technologies. 

To summarize this thesis we begin by introducing a specific type of quantum computing called linear optical quantum computing in Ch. \ref{ch:LOQC}. We then looked at quantum random walks in Ch. \ref{ch:QRW}, which are used in various quantum algorithms, and found that quantum walkers retain quantum advantages over classical walkers even in the presence of lattice congestion and dephasing. In Ch. \ref{Ch:BSIntro} we introduce \BStwo, which is a very simplified model of a linear optical quantum computer that simulates the interference of bosons and which contains many of the core technologies required to build a linear optical quantum computer. \BS is the topic of most of this thesis including this introduction and chapters \ref{Ch:SPDC}, \ref{Ch:FiberLoop}, \ref{Ch:sampOther}, \ref{Ch:MORDOR}, and \ref{ch:sharpPhard}. In Ch. \ref{Ch:SPDC} we found that \BS using the most readily available photon source, called a spontaneous parametric down conversion source, may be successfully implemented with a multiplexed device that can reroute photons. In Ch. \ref{Ch:FiberLoop} we invent how to implement \BS in a temporal implementation instead of spatial one and show that the temporal architecture simplifies the number of optical elements required from thousands or millions to just three. In addition we give a detailed error analysis of this new architecture. This was done in theory but has since been successfully experimentally demonstrated implementing the worlds largest ever \BS experiment. In Ch. \ref{Ch:sampOther} we show that the original protocol for \BS is not unique and that there is a large class of quantum states of light that may be used to implement a computationally hard to simulate instance of \BStwo. In Ch. \ref{Ch:MORDOR} we show another key discovery we made where we invented a world's first application inspired by \BS in the field of quantum metrology. This is one of only two \BS inspired applications that currently exist. In Ch. \ref{ch:sharpPhard} we use a quantum optics approach to map the \BS formalism into a multidimensional integral formalism using characteristic functions showing that they are equivalent and thus finding multidimensional integrals of a certain form that are also computationally complex. Finally, we switched away from \BS and focused on creating certain quantum states of light that are useful for quantum computing. In Ch. \ref{ch:MotherFocker} we give a protocol that can efficiently generate large-photon Fock states, which are a critical quantum state used in quantum information processing, and in Ch. \ref{ch:GKP} we show how to create a continuous variable optical state by coupling light to atoms that can be used to encode and correct quantum information in a quantum computer.

\bibliography{references}

\begin{thebibliography}{100}
\expandafter\ifx\csname url\endcsname\relax
  \def\url#1{\texttt{#1}}\fi
\expandafter\ifx\csname urlprefix\endcsname\relax\def\urlprefix{URL }\fi
\providecommand{\eprint}[2][]{\url{#2}}

\bibitem{bib:NielsenChuang00}
M.~A. Nielsen and I.~L. Chuang.
\newblock \emph{Quantum Computation and Quantum Information} (Cambridge
  University Press, Cambridge, 2000).

\bibitem{bib:cirac95}
J.~I. Cirac and P.~Zoller.
\newblock \emph{Quantum computations with cold trapped ions}.
\newblock Physical review letters \textbf{74}(20), 4091 (1995).

\bibitem{bib:kielpinski02}
D.~Kielpinski, C.~Monroe, and D.~J. Wineland.
\newblock \emph{Architecture for a large-scale ion-trap quantum computer}.
\newblock Nature \textbf{417}(6890), 709 (2002).

\bibitem{bib:wallraff04}
A.~Wallraff, D.~I. Schuster, A.~Blais, L.~Frunzio, R.-S. Huang, J.~Majer,
  S.~Kumar, S.~M. Girvin, and R.~J. Schoelkopf.
\newblock \emph{Strong coupling of a single photon to a superconducting qubit
  using circuit quantum electrodynamics}.
\newblock Nature \textbf{431}(7005), 162 (2004).

\bibitem{bib:cory98}
D.~G. Cory, M.~D. Price, and T.~F. Havel.
\newblock \emph{Nuclear magnetic resonance spectroscopy: An experimentally
  accessible paradigm for quantum computing}.
\newblock Physica D: Nonlinear Phenomena \textbf{120}(1), 82 (1998).

\bibitem{bib:Loss98}
D.~Loss and D.~P. DiVincenzo.
\newblock \emph{Quantum computation with quantum dots}.
\newblock Physical Review A \textbf{57}(1), 120 (1998).

\bibitem{bib:kane98}
B.~E. Kane.
\newblock \emph{A silicon-based nuclear spin quantum computer}.
\newblock nature \textbf{393}(6681), 133 (1998).

\bibitem{bib:LOQC}
E.~Knill, R.~Laflamme, and G.~J. Milburn.
\newblock \emph{A scheme for efficient quantum computation with linear optics}.
\newblock Nature \textbf{409}(6816), 46 (2001).

\bibitem{cerny}
V.~Cerny.
\newblock \emph{Quantum computers and intractable (np-complete) computing
  problems}.
\newblock Phys. Rev. A \textbf{48}, 116 (1993).

\bibitem{clauser}
J.~Clauser and J.~Dowling.
\newblock \emph{Factoring integers with young's n-slit interferometer}.
\newblock Phys. Rev. A \textbf{53}, 4587 (1996).

\bibitem{bib:Bartlett02b}
S.~D. Bartlett, B.~C. Sanders, S.~L. Braunstein, and K.~Nemoto.
\newblock \emph{Efficient classical simulation of continuous variable quantum
  information processes}.
\newblock Phys. Rev. Lett. \textbf{88}, 097904 (2002).

\bibitem{bib:Raussendorf01}
R.~Raussendorf and H.~J. Briegel.
\newblock \emph{A one-way quantum computer}.
\newblock Phys. Rev. Lett. \textbf{86}, 5188 (2001).

\bibitem{bib:Raussendorf03}
R.~Raussendorf, D.~E. Browne, and H.~J. Briegel.
\newblock \emph{Measurement-based quantum computation on cluster states}.
\newblock Phys. Rev. A \textbf{68}, 022312 (2003).

\bibitem{bib:freedman03}
M.~Freedman, A.~Kitaev, M.~Larsen, and Z.~Wang.
\newblock \emph{Topological quantum computation}.
\newblock Bulletin of the American Mathematical Society \textbf{40}(1), 31
  (2003).

\bibitem{bib:farhi}
E.~Farhi, J.~Goldstone, S.~Gutmann, J.~Lapan, A.~Lundgren, and D.~Preda.
\newblock \emph{A quantum adiabatic evolution algorithm applied to random
  instances of an np-complete problem}.
\newblock Science \textbf{292}, 472 (2001).

\bibitem{bib:ADZ}
Y.~Aharonov, L.~Davidovich, and N.~Zagury.
\newblock \emph{Quantum random walks}.
\newblock Phys. Rev. A \textbf{48}, 1687 (1993).

\bibitem{bib:bern}
E.~Bernstein and U.~Vazirani.
\newblock \emph{Quantum complexity theory}.
\newblock Usiam Journal on Computing \textbf{26}, 1411 (1997).

\bibitem{bib:jordan2}
S.~P. Jordan.
\newblock \emph{Permutational quantum computing}.
\newblock Quantum Information \& Computation \textbf{10}, 470 (2010).

\bibitem{bib:moussa}
O.~Moussa, C.~Ryan, D.~Cory, and R.~Laflamme.
\newblock \emph{Testing contextuality on quantum ensembles with one clean
  qubit}.
\newblock Phys. Rev. Lett. \textbf{104}, 160501 (2010).

\bibitem{bib:KokLovett11}
P.~Kok and B.~W. Lovett.
\newblock \emph{Introduction to Optical Quantum Information Processing}
  (Cambridge Press, 2010).

\bibitem{bib:Aharonov96}
D.~Aharonov and M.~Ben-Or.
\newblock \emph{Polynomial simulations of decohered quantum computers}.
\newblock 37th Annual Symposium on Foundations of Computer Science p.~46
  (1996).

\bibitem{bib:kitaev97}
A.~Y. Kitaev.
\newblock \emph{Quantum computations: algorithms and error correction}.
\newblock Russian Mathematical Surveys \textbf{52}(6), 1191 (1997).

\bibitem{bib:knill98}
E.~Knill, R.~Laflamme, and W.~H. Zurek.
\newblock \emph{Resilient quantum computation}.
\newblock Science \textbf{279}(5349), 342 (1998).

\bibitem{bib:preskill98}
J.~Preskill.
\newblock \emph{Reliable quantum computers}.
\newblock In \emph{Proceedings of the Royal Society of London A: Mathematical,
  Physical and Engineering Sciences}, vol. 454, pp. 385--410 (The Royal
  Society, 1998).

\bibitem{bib:lemr}
K.~Lemr, A.~\v{C}ernoch, J.~Soubusta, and M.~Du\v{s}ek.
\newblock \emph{Entangling efficiency of linear-optical quantum gates}.
\newblock Phys. Rev. A \textbf{86}, 032321 (2012).

\bibitem{bib:OuLu99}
Z.~Y. Ou and Y.~J. Lu.
\newblock \emph{Cavity enhanced spontaneous parametric down-conversion for the
  prolongation of correlation time between conjugate photons}.
\newblock Phys. Rev. Lett. \textbf{83}, 2556 (1999).

\bibitem{bib:Santori01}
C.~Santori, M.~Pelton, G.~Solomon, Y.~Dale, and Y.~Yamamoto.
\newblock \emph{Triggered single photons from a quantum dot}.
\newblock Phys. Rev. Lett. \textbf{86}, 1502 (2001).

\bibitem{bib:HOM87}
C.~K. Hong, Z.~Y. Ou, and L.~Mandel.
\newblock \emph{Measurement of sub-picosecond time intervals between two
  photons by interference}.
\newblock Phys. Rev. Lett. \textbf{59}, 2044 (1987).

\bibitem{bib:RohdeWebb07}
P.~P. Rohde, J.~G. Webb, E.~H. Huntington, and T.~C. Ralph.
\newblock \emph{Comparison of architectures for approximating number-resolving
  photo-detection using non-number-resolving detectors}.
\newblock New J. Phys. \textbf{9}, 233 (2007).

\bibitem{bib:Fitch03}
M.~J. Fitch, B.~C. Jacobs, T.~B. Pittman, and J.~D. Franson.
\newblock \emph{Photon number resolution using time-multiplexed single-photon
  detectors}.
\newblock Phys. Rev. A \textbf{68}, 043814 (2003).

\bibitem{bib:Achilles04}
D.~Achilles, C.~Silberhorn, C.~Sliwa, K.~Banaszek, I.~A. Walmsley, M.~J. Fitch,
  B.~C. Jacobs, T.~B. Pittman, and J.~D. Franson.
\newblock \emph{Photon number resolving detection using time-multiplexing}.
\newblock J. Mod. Opt. \textbf{51}, 1499 (2004).

\bibitem{bib:LPOR201400027}
T.~Meany, L.~A. Ngah, M.~J. Collins, A.~S. Clark, R.~J. Williams, B.~J.
  Eggleton, M.~J. Steel, M.~J. Withford, O.~Alibart, and S.~Tanzilli.
\newblock \emph{Hybrid photonic circuit for multiplexed heralded single
  photons}.
\newblock Laser \& Photonics Reviews  (2014).

\bibitem{bib:ma2011experimental}
X.-S. Ma, S.~Zotter, J.~Kofler, T.~Jennewein, and A.~Zeilinger.
\newblock \emph{Experimental generation of single photons via active
  multiplexing}.
\newblock Phys. Rev. A \textbf{83}, 043814 (2011).

\bibitem{bib:Nielsen04}
M.~A. Nielsen.
\newblock \emph{Optical quantum computation using cluster states} \textbf{93},
  040503 (2004).

\bibitem{bib:humphreys13}
P.~C. Humphreys, B.~J. Metcalf, J.~B. Spring, M.~Moore, X.-M. Jin, M.~Barbieri,
  W.~S. Kolthammer, and I.~A. Walmsley.
\newblock \emph{Linear optical quantum computing in a single spatial mode}.
\newblock Physical review letters \textbf{111}(15), 150501 (2013).

\bibitem{bib:GottesmanChuang99}
D.~Gottesman and I.~L. Chuang.
\newblock \emph{Demonstrating the viability of universal quantum computation
  using teleportation and single-qubit operations}.
\newblock Nature (London) \textbf{402}, 390 (1999).

\bibitem{bib:Shor97}
P.~W. Shor.
\newblock \emph{Polynomial-time algorithms for prime factorization and discrete
  logarithms on a quantum computer}.
\newblock SIAM J. Comput. \textbf{26}, 1484 (1997).

\bibitem{bib:BennetBrassard84}
C.~H. Bennett and G.~Brassard.
\newblock \emph{Quantum cryptography: public key distribution and coin
  tossing}.
\newblock Proceedings of IEEE International Conference on Computers, Systems
  and Signal Processing p. 175 (1984).

\bibitem{bib:grover}
L.~K. Grover.
\newblock Proc. 28th Annual ACM Symposium on the Theory of Computing (STOC) p.
  212 (1996).

\bibitem{bib:georgescu14}
I.~Georgescu, S.~Ashhab, and F.~Nori.
\newblock \emph{Quantum simulation}.
\newblock Reviews of Modern Physics \textbf{86}(1), 153 (2014).

\bibitem{bib:AAKV}
D.~Aharonov, A.~Ambainis, J.~Kempe, and U.~Vazirani.
\newblock STOC '01 Proceedings of the 33rd ACM symposium on Theory of computing
  \textbf{50} (2001).

\bibitem{bib:Kempe08}
J.~Kempe.
\newblock \emph{Quantum random walks - an introductory overview}.
\newblock Cont. Phys. \textbf{44}, 307 (2003).

\bibitem{bib:Salvador12}
S.~E. Venegas-Andraca.
\newblock \emph{Quantum walks: a comprehensive review}.
\newblock QIP \textbf{5}, 1015 (2012).

\bibitem{bib:increasing13}
P.~P. Rohde, A.~Schreiber, M.~Stefanak, I.~Jex, A.~Gilchrist, and
  C.~Silberhorn.
\newblock \emph{Increasing the dimensionality of quantum walks using multiple
  walkers}.
\newblock J. Comp. and Th. Nanosc. (in press)  (2013).

\bibitem{bib:childs03}
A.~M. Childs, R.~Cleve, E.~Deotto, E.~Farhi, S.~Gutmann, and D.~A. Spielman.
\newblock \emph{Exponential algorithmic speedup by a quantum walk}.
\newblock In \emph{Proceedings of the thirty-fifth annual ACM symposium on
  Theory of computing}, pp. 59--68 (ACM, 2003).

\bibitem{bib:ambainis07}
A.~Ambainis.
\newblock \emph{Quantum walk algorithm for element distinctness}.
\newblock SIAM Journal on Computing \textbf{37}(1), 210 (2007).

\bibitem{bib:Broome10}
M.~A. Broome, A.~Fedrizzi, B.~P. Lanyon, I.~Kassal, A.~Aspuru-Guzik, and A.~G.
  White.
\newblock \emph{Discrete single-photon quantum walks with tunable decoherence}.
\newblock Phys. Rev. Lett. \textbf{104}, 153602 (2010).

\bibitem{bib:lockhart13}
J.~Lockhart, C.~Di~Franco, and M.~Paternostro.
\newblock \emph{Performance of continuous time quantum walks under phase
  damping}.
\newblock Physics Letters A \textbf{378}, 338 (2014).

\bibitem{bib:Hagai08}
H.~B. Perets, Y.~Lahini, F.~Pozzi, M.~Sorel, R.~Morandotti, and Y.~Silberberg.
\newblock \emph{Realization of quantum walks with negligible decoherence in
  waveguide lattices}.
\newblock Phys. Rev. Lett. \textbf{100}, 170506 (2008).

\bibitem{bib:Schreiber10}
A.~Schreiber, K.~N. Cassemiro, V.~Poto{\u c}ek, A.~G{\' a}bris, P.~J. Mosley,
  E.~Andersson, I.~Jex, and C.~Silberhorn.
\newblock \emph{Photons walking the line: A quantum walk with adjustable coin
  operations}.
\newblock Phys. Rev. Lett. \textbf{104}, 050502 (2010).

\bibitem{bib:Peruzzo10}
A.~Peruzzo, M.~Lobino, J.~C.~F. Matthews, N.~Matsuda, A.~Politi, K.~Poulios,
  X.-Q. Zhou, Y.~Lahini, N.~Ismail, K.~W{\" o}rhoff, Y.~Bromberg,
  Y.~Silberberg, M.~G. Thompson, and J.~L. O'Brien.
\newblock \emph{Quantum walks of correlated particles}.
\newblock Science \textbf{329}, 1500 (2010).

\bibitem{bib:Schreiber11b}
A.~Schreiber, K.~N. Cassemiro, V.~Potocek, A.~Gabris, I.~Jex, and
  C.~Silberhorn.
\newblock \emph{Decoherence and disorder in quantum walks: From ballistic
  spread to localization}.
\newblock Phys. Rev. Lett. \textbf{106}, 180403 (2011).

\bibitem{bib:Matthews11}
J.~C.~F. Matthews, K.~Poulios, J.~D.~A. Meinecke, A.~Politi, A.~Peruzzo,
  N.~Ismail, K.~W{\" o}rhoff, M.~G. Thompson, and J.~L. O'Brien.
\newblock \emph{Simulating quantum statistics with entangled photons: a
  continuous transition from bosons to fermions}.
\newblock Sci. Rep. \textbf{3}, 1539 (2013).

\bibitem{bib:Owens11}
J.~O. Owens, M.~A. Broome, D.~N. Biggerstaff, M.~E. Goggin, A.~Fedrizzi,
  T.~Linjordet, M.~Ams, G.~D. Marshall, J.~Twamley, M.~J. Withford, and A.~G.
  White.
\newblock \emph{Two-photon quantum walks in an elliptical direct-write
  waveguide array}.
\newblock New. J. Phys. \textbf{13}, 075003 (2011).

\bibitem{bib:Schreiber12}
A.~Schreiber, A.~G{\'a}bris, P.~P. Rohde, K.~Laiho, M.~{\v S}tefa{\v n}{\' a}k,
  V.~Poto{\u c}ek, I.~Jex, and C.~Silberhorn.
\newblock \emph{A 2d quantum walk simulation of two-particle dynamics}.
\newblock Science \textbf{336}, 55 (2012).

\bibitem{bib:Sansoni12}
L.~Sansoni, F.~Sciarrino, G.~Vallone, P.~Mataloni, A.~Crespi, R.~Ramponi, and
  R.~Osellame.
\newblock \emph{Two-particle bosonic-fermionic quantum walk via integrated
  photonics}.
\newblock Phys. Rev. Lett. \textbf{108}, 010502 (2012).

\bibitem{bib:broadbent57}
S.~R. Broadbent and J.~M. Hammersley.
\newblock \emph{Percolation processes i. crystals and mazes}.
\newblock In \emph{Proc. Cambridge Philos. Soc}, vol.~53, p.~41 (1957).

\bibitem{bib:Shante71}
V.~K. Shante and S.~Kirkpatrick.
\newblock \emph{An introduction to percolation theory}.
\newblock Advances in Physics \textbf{20}(85), 325 (1971).

\bibitem{bib:blanc86}
R.~Blanc.
\newblock \emph{Introduction to percolation theory}.
\newblock In \emph{Contribution of Clusters Physics to Materials Science and
  Technology}, pp. 425--478 (Springer, 1986).

\bibitem{bib:grimmett99}
G.~R. Grimmett.
\newblock \emph{Percolation}, vol. 321 (Springer, 1999).

\bibitem{bib:yonezawa89}
F.~Yonezawa, S.~Sakamoto, and M.~Hori.
\newblock \emph{Percolation in two-dimensional lattices. i. a technique for the
  estimation of thresholds}.
\newblock Phys. Rev. B \textbf{40}, 636 (1989).

\bibitem{bib:sahimi94}
M.~Sahimi.
\newblock \emph{Applications of percolation theory} (CRC PressI Llc, 1994).

\bibitem{bib:kollar12}
B.~Koll\'ar, T.~Kiss, J.~Novotn\'y, and I.~Jex.
\newblock \emph{Asymptotic dynamics of coined quantum walks on percolation
  graphs}.
\newblock Phys. Rev. Lett. \textbf{108}, 230505 (2012).

\bibitem{bib:kendon10}
G.~Leung, P.~Knott, J.~Bailey, and V.~Kendon.
\newblock \emph{Coined quantum walks on percolation graphs}.
\newblock New Journal of Physics \textbf{12}(12), 123018 (2010).

\bibitem{bib:magnus1995matrix}
J.~R. Magnus, H.~Neudecker, \emph{et~al.}
\newblock \emph{Matrix differential calculus with applications in statistics
  and econometrics}  (1995).

\bibitem{bib:kempe2003quantum}
J.~Kempe.
\newblock \emph{Quantum random walks: an introductory overview}.
\newblock Contemporary Physics \textbf{44}(4), 307 (2003).

\bibitem{bib:AaronsonArkhipov}
S.~Aaronson and A.~Arkhipov.
\newblock \emph{The computational complexity of linear optics}.
\newblock In \emph{Proceedings of the 43rd annual ACM symposium on Theory of
  computing}, pp. 333--342 (ACM, 2011).

\bibitem{bib:ralph2013quantum}
T.~Ralph.
\newblock \emph{Quantum computation: Boson sampling on a chip}.
\newblock Nature Phot. \textbf{7}(7), 514 (2013).

\bibitem{bib:Broome2012}
M.~A. Broome, A.~Fedrizzi, S.~Rahimi-Keshari, J.~Dove, S.~Aaronson, T.~C.
  Ralph, and A.~G. White.
\newblock \emph{Photonic boson sampling in a tunable circuit}.
\newblock Science \textbf{339}, 6121 (2013).

\bibitem{bib:Spring2}
J.~B. Spring, B.~J. Metcalf, P.~C. Humphreys, W.~S. Kolthammer, X.-M. Jin,
  M.~Barbieri, A.~Datta, N.~Thomas-Peter, N.~K. Langford, D.~Kundys, J.~C.
  Gates, B.~J. Smith, P.~G.~R. Smith, and I.~A. Walmsley.
\newblock \emph{Boson sampling on a photonic chip}.
\newblock Science \textbf{339}, 798 (2013).

\bibitem{bib:anon}
Anonymous.
\newblock \emph{The rise of the boson-sampling computer}.
\newblock Photonics Spectra \textbf{47}, 33 (2013).

\bibitem{bib:Tillmann4}
M.~Tillmann, B.~Daki, R.~Heilmann, S.~Nolte, A.~Szameit, and P.~Walther.
\newblock \emph{Experimental boson sampling}.
\newblock Nature Photonics \textbf{7}, 540 (2013).

\bibitem{bib:crespi2013integrated}
A.~Crespi, R.~Osellame, R.~Ramponi, D.~J. Brod, E.~F. Galv{\~a}o, N.~Spagnolo,
  C.~Vitelli, E.~Maiorino, P.~Mataloni, and F.~Sciarrino.
\newblock \emph{Integrated multimode interferometers with arbitrary designs for
  photonic boson sampling}.
\newblock Nature Phot. \textbf{7}, 545 (2013).

\bibitem{bib:he16}
Y.~He, Z.-E. Su, H.-L. Huang, X.~Ding, J.~Qin, C.~Wang, S.~Unsleber, C.~Chen,
  H.~Wang, Y.-M. He, \emph{et~al.}
\newblock \emph{Scalable boson sampling with a single-photon device}.
\newblock arXiv:1603.04127  (2016).

\bibitem{bib:dowlingSchmampling}
J.~P. Dowling  (2013).
\newblock
  \urlprefix\url{http://quantumpundit.blogspot.com/2013/07/sampling-schmampling.html}.

\bibitem{bib:RohdeLowFid12}
P.~P. Rohde.
\newblock \emph{Optical quantum computing with photons of arbitrarily low
  fidelity and purity}.
\newblock Phys. Rev. A \textbf{86}, 052321 (2012).

\bibitem{bib:RohdeRalphErrBS}
P.~P. Rohde and T.~C. Ralph.
\newblock \emph{Error tolerance of the boson-sampling model for linear optics
  quantum computing}.
\newblock Phys. Rev. A \textbf{85}, 022332 (2012).

\bibitem{bib:PhysRevA.88.044301}
Z.~Jiang, M.~D. Lang, and C.~M. Caves.
\newblock \emph{Mixing nonclassical pure states in a linear-optical network
  almost always generates modal entanglement}.
\newblock Phys. Rev. A \textbf{88}, 044301 (2013).

\bibitem{bib:motes13}
K.~R. Motes, J.~P. Dowling, and P.~P. Rohde.
\newblock \emph{Spontaneous parametric down-conversion photon sources are
  scalable in the asymptotic limit for boson sampling}.
\newblock Phys. Rev. A \textbf{88}, 063822 (2013).

\bibitem{bib:catSampling}
P.~P. Rohde, K.~R. Motes, P.~A. Knott, J.~Fitzsimons, W.~J. Munro, and J.~P.
  Dowling.
\newblock \emph{Evidence for the conjecture that sampling generalized cat
  states with linear optics is hard}.
\newblock Phys. Rev. A \textbf{91}, 012342 (2015).

\bibitem{bib:Kaushik14}
J.~Olson, K.~Seshadreesan, K.~Motes, P.~Rohde, and J.~Dowling.
\newblock \emph{Boson sampling with photon-added coherent states}.
\newblock Bulletin of the American Physical Society \textbf{59} (2014).

\bibitem{bib:Pacs13}
K.~P., J.~P. Olson, K.~R. Motes, P.~P. Rohde, and J.~P. Dowling.
\newblock \emph{Boson sampling with displaced single-photon fock states versus
  single-photon-added coherent states: The quantum-classical divide and
  computational-complexity transitions in linear optics}.
\newblock Phys. Rev. A \textbf{91}, 022334 (2015).

\bibitem{bib:Shc14a}
V.~S. Shchesnovich.
\newblock \emph{Sufficient condition for the mode mismatch of single photons
  for scalability of the boson-sampling computer}.
\newblock Phys. Rev. A \textbf{89}, 022333 (2014).

\bibitem{bib:lou}
H.~Lou and D.~F.~V. James.
\newblock \emph{Proposal for a scalable universal bosonic simulator using
  individually trapped ions}.
\newblock Phys. Rev. A \textbf{85}, 062329 (2012).

\bibitem{bib:MORDOR}
K.~R. Motes, J.~P. Olson, E.~J. Rabeaux, J.~P. Dowling, S.~J. Olson, and P.~P.
  Rohde.
\newblock \emph{Linear optical quantum metrology with single photons:
  Exploiting spontaneously generated entanglement to beat the shot-noise
  limit}.
\newblock Phys. Rev. Lett. \textbf{114}, 170802 (2015).

\bibitem{bib:vibSpec}
J.~Huh, G.~G. Guerreschi, B.~Peropadre, J.~R. McClean, and A.~Aspuru-Guzik.
\newblock \emph{Boson sampling for molecular vibronic spectra}.
\newblock Nat Photon \textbf{9}(9), 615 (2015).

\bibitem{bib:feyn}
R.~P. Feynman.
\newblock \emph{Simulating physics with computers}.
\newblock International Journal Of Theoretical Physics \textbf{21}, 467 (1982).

\bibitem{bib:LloydSim}
S.~Lloyd.
\newblock \emph{Universal quantum simulators}.
\newblock Science \textbf{273}, 5278 (1996).

\bibitem{bib:DeutschJozsa92}
D.~Deutsch and R.~Jozsa.
\newblock \emph{Rapid solution of problems by quantum computation}.
\newblock Proc. R. Soc. Lond. A \textbf{439}, 553 (1992).

\bibitem{bib:gard2013quantum}
B.~T. Gard, R.~M. Cross, P.~M. Anisimov, H.~Lee, and J.~P. Dowling.
\newblock \emph{Quantum random walks with multiphoton interference and
  high-order correlation functions}.
\newblock JOSA B \textbf{30}(6), 1538 (2013).

\bibitem{bib:gard2014inefficiency}
B.~T. Gard, J.~P. Olson, R.~M. Cross, M.~B. Kim, H.~Lee, and J.~P. Dowling.
\newblock \emph{Inefficiency of classically simulating linear optical quantum
  computing with fock-state inputs}.
\newblock Phys. Rev. A \textbf{89}(2), 022328 (2014).

\bibitem{bib:Ryser63}
H.~J. Ryser.
\newblock Combinatorial Mathematics, Carus Mathematical Monograph No. 14
  (1963).

\bibitem{bib:lapaire2003conditional}
G.~Lapaire, P.~Kok, J.~P. Dowling, and J.~Sipe.
\newblock \emph{Conditional linear-optical measurement schemes generate
  effective photon nonlinearities}.
\newblock Physical Review A \textbf{68}(4), 042314 (2003).

\bibitem{bib:AA13response}
S.~Aaronson and A.~Arkhipov.
\newblock \emph{Bosonsampling is far from uniform}.
\newblock In \emph{Electronic Colloquium on Computational Complexity}, p.
  Report No. 135 (2013).

\bibitem{bib:Spagnolo13}
N.~Spagnolo, C.~Vitelli, M.~Bentivegna, D.~J. Brod, A.~Crespi, F.~Flamini,
  S.~Giacomini, G.~Milani, R.~Ramponi, P.~Mataloni, R.~Osellame, E.~F. Galvao,
  and F.~Sciarrino.
\newblock \emph{Efficient experimental validation of photonic boson sampling
  against the uniform distribution}.
\newblock Nature Photonics \textbf{8}, 610 (2014).

\bibitem{bib:Carolan13}
J.~Carolan, J.~D.~A. Meinecke, P.~Shadbolt, N.~J. Russell, N.~Ismail,
  K.~W�rhoff, T.~Rudolph, M.~G. Thompson, J.~L. O'Brien, J.~C.~F. Matthews,
  and A.~Laing.
\newblock \emph{On the experimental verification of quantum complexity in
  linear optics}.
\newblock Nature Photonics \textbf{8}, 621 (2014).

\bibitem{bib:Molmer13}
M.~C. Tichy, K.~Mayer, A.~Buchleitner, and K.~M{\o}lmer.
\newblock \emph{Stringent and efficient assessment of boson-sampling devices}.
\newblock Phys. Rev. Lett. \textbf{113}, 020502 (2014).

\bibitem{bib:QKUM14}
M.~Walschaers, J.~Kuipers, J.-D. Urbina, K.~Mayer, M.~C. Tichy, K.~Richter, and
  A.~Buchleitner.
\newblock \emph{Statistical benchmark for bosonsampling}.
\newblock New Journal of Physics \textbf{18}(3), 032001 (2016).

\bibitem{bib:Rohde15a}
P.~P. Rohde.
\newblock \emph{Boson sampling with photons of arbitrary spectral structure}.
\newblock Phys. Rev. A \textbf{91}, 012307 (2015).

\bibitem{bib:Shc14b}
V.~Shchesnovich.
\newblock \emph{Partial indistinguishability theory for multiphoton experiments
  in multiport devices}.
\newblock Physical Review A \textbf{91}(1), 013844 (2015).

\bibitem{bib:Tich14}
M.~C. Tichy.
\newblock \emph{Sampling of partially distinguishable bosons and the relation
  to the multidimensional permanent}.
\newblock Phys. Rev. A \textbf{91}, 022316 (2015).

\bibitem{bib:Lund13}
A.~P. Lund, A.~Laing, S.~Rahimi-Keshari, T.~Rudolph, J.~L. O'Brien, and T.~C.
  Ralph.
\newblock \emph{Boson sampling from a gaussian state}.
\newblock Phys. Rev. Lett. \textbf{113}, 100502 (2014).

\bibitem{bib:motes2014scalable}
K.~R. Motes, A.~Gilchrist, J.~P. Dowling, and P.~P. Rohde.
\newblock \emph{Scalable boson-sampling with time-bin encoding using a
  loop-based architecture}.
\newblock Phys. Rev. Lett. \textbf{113}, 120501 (2014).

\bibitem{bib:Shen14}
C.~Shen, Z.~Zhang, and L.-M. Duan.
\newblock \emph{Scalable implementation of boson sampling with trapped ions}.
\newblock Phys. Rev. Lett. \textbf{112}, 050504 (2014).

\bibitem{bib:peropadre15}
B.~Peropadre, G.~G. Guerreschi, J.~Huh, and A.~Aspuru-Guzik.
\newblock \emph{Microwave boson sampling}.
\newblock arXiv:1510.08064  (2015).

\bibitem{bib:gogolin2013}
C.~Gogolin, M.~Kliesch, L.~Aolita, and J.~Eisert.
\newblock \emph{Boson-sampling in the light of sample complexity}.
\newblock arXiv:1306.3995  (2013).

\bibitem{bib:introChapter}
S.~A. Malinovskaya and I.~Novikova.
\newblock \emph{From Atomic to Mesoscale: The Role of Quantum Coherence in
  Systems of Various Complexities} (World Scientific, 2015).

\bibitem{bib:birthday}
A.~Arkhipov and G.~Kuperberg.
\newblock \emph{The bosonic birthday paradox}.
\newblock Geometry \& Topology Monographs \textbf{18}, 1 (2012).

\bibitem{bib:Reck94}
M.~Reck, A.~Zeilinger, H.~J. Bernstein, and P.~Bertani.
\newblock \emph{Experimental realization of any discrete unitary operator}.
\newblock Phys. Rev. Lett. \textbf{73}, 58 (1994).

\bibitem{bib:kok2001multi}
P.~Kok and S.~L. Braunstein.
\newblock \emph{Multi-dimensional hermite polynomials in quantum optics}.
\newblock Journal of Physics A: Mathematical and General \textbf{34}(31), 6185
  (2001).

\bibitem{bib:RosenbergTES}
D.~Rosenberg, A.~E. Lita, A.~J. Miller, and S.~W. Nam.
\newblock \emph{Noise-free high-efficiency photon-number-resolving detectors}.
\newblock Phys. Rev. A \textbf{71}, 061803 (2005).

\bibitem{bib:Scheel04perm}
S.~Scheel.
\newblock \emph{Permanents in linear optical networks}.
\newblock Acta Physica Slovaca \textbf{58}, 675 (2008).

\bibitem{bib:GardCrossDowling}
B.~T. Gard, J.~P. Olson, R.~M. Cross, M.~B. Kim, H.~Lee, and J.~P. Dowling.
\newblock \emph{Inefficiency of classically simulating linear optical quantum
  computing with fock-state inputs}.
\newblock Physical Review A \textbf{89}(2), 022328 (2014).

\bibitem{bib:Valiant79}
L.~G. Valiant.
\newblock \emph{The complexity of computing the permanent}.
\newblock Theoretical Computer Science \textbf{8}, 189 (1979).

\bibitem{bib:Jerrum04}
M.~Jerrum, A.~Sinclair, and E.~Vigoda.
\newblock \emph{A polynomial-time approximation algorithm for the permanent of
  a matrix with nonnegative entries}.
\newblock Journal of the ACM \textbf{51}, 673 (2004).

\bibitem{bib:BSECT}
P.~P. Rohde, K.~R. Motes, P.~A. Knott, and W.~J. Munro.
\newblock \emph{Will boson-sampling ever disprove the extended church-turing
  thesis?}  (2014).
\newblock ArXiv:1401.2199.

\bibitem{bib:Shchesnovich13}
V.~S. Shchesnovich.
\newblock \emph{Sufficient bound on the mode mismatch of single photons for
  scalability of the boson sampling computer}.
\newblock Phys. Rev. A \textbf{89}, 022333 (2013).

\bibitem{bib:SourceAndDetectorReview}
M.~D. Eisaman, J.~Fan, A.~Migdall, and S.~V. Polyakov.
\newblock \emph{Invited review article: Single-photon sources and detectors}.
\newblock Review of Scientific Instruments \textbf{82}, 071101 (2011).

\bibitem{bib:Moreau01}
E.~Moreau, I.~Robert, J.~M. G\'erard, I.~Abram, L.~Manin, and V.~Thierry-Mieg.
\newblock \emph{Single-mode solid-state single photon source based on isolated
  quantum dots in pillar microcavities}.
\newblock App. Phys. Lett. \textbf{79}, 2865 (2001).

\bibitem{bib:PRLFiberLoop}
K.~R. Motes, A.~Gilchrist, J.~P. Dowling, and P.~P. Rohde.
\newblock \emph{Scalable boson sampling with time-bin encoding using a
  loop-based architecture}.
\newblock Phys. Rev. Lett. \textbf{113}, 120501 (2014).

\bibitem{bib:GerryKnight05}
C.~C. Gerry and P.~L. Knight.
\newblock \emph{Introductory quantum optics} (Cambridge University Press,
  2005).

\bibitem{bib:Peruzzo11b}
A.~Peruzzo, A.~Laing, A.~Politi, T.~Rudolph, and J.~L. O'Brien.
\newblock \emph{Multimode quantum interference of photons in multiport
  integrated devices}.
\newblock Nature Comm. \textbf{2}, 224 (2011).

\bibitem{bib:Politi02052008}
A.~Politi, M.~J. Cryan, J.~G. Rarity, S.~Yu, and J.~L. O'Brien.
\newblock \emph{Silica-on-silicon waveguide quantum circuits}.
\newblock Science \textbf{320}, 646 (2008).

\bibitem{bib:matthews2009}
J.~C. Matthews, A.~Politi, A.~Stefanov, and J.~L. O'Brien.
\newblock \emph{Manipulation of multiphoton entanglement in waveguide quantum
  circuits}.
\newblock Nature Photonics \textbf{3}, 346 (2009).

\bibitem{bib:Politi04092009}
A.~Politi, J.~C.~F. Matthews, and J.~L. O'Brien.
\newblock \emph{Shor's quantum factoring algorithm on a photonic chip}.
\newblock Science \textbf{325}, 1221 (2009).

\bibitem{bib:kok2001detection}
P.~Kok and S.~L. Braunstein.
\newblock \emph{Detection devices in entanglement-based optical state
  preparation}.
\newblock Physical Review A \textbf{63}(3), 033812 (2001).

\bibitem{bib:tillmann2014BS}
M.~Tillmann, S.-H. Tan, S.~E. Stoeckl, B.~C. Sanders, H.~de~Guise, R.~Heilmann,
  S.~Nolte, A.~Szameit, and P.~Walther.
\newblock \emph{Generalized multiphoton quantum interference}.
\newblock Physical Review X \textbf{5}(4), 041015 (2015).

\bibitem{bib:migdall2002tailoring}
A.~Migdall, D.~Branning, and S.~Castelletto.
\newblock \emph{Tailoring single-photon and multiphoton probabilities of a
  single-photon on-demand source}.
\newblock Phys. Rev. A \textbf{66}, 053805 (2002).

\bibitem{bib:bartlett2003requirement}
S.~D. Bartlett and B.~C. Sanders.
\newblock \emph{Requirement for quantum computation}.
\newblock Journal of Modern Optics \textbf{50}(15-17), 2331 (2003).

\bibitem{bib:beaujean2016oxazines}
P.~Beaujean, F.~Bondu, A.~Plaquet, J.~Garcia-Amor{\'o}s, J.~Cusido, F.~M.
  Raymo, F.~Castet, V.~Rodriguez, and B.~Champagne.
\newblock \emph{Oxazines: A new class of second-order nonlinear optical
  switches}.
\newblock Journal of the American Chemical Society \textbf{138}(15), 5052
  (2016).

\bibitem{bib:kawashima2016multi}
H.~Kawashima, K.~Suzuki, K.~Tanizawa, S.~Suda, G.~Cong, H.~Matsuura, S.~Namiki,
  and K.~Ikeda.
\newblock \emph{Multi-port optical switch based on silicon photonics}.
\newblock In \emph{Optical Fiber Communication Conference}, pp. W1E--6 (Optical
  Society of America, 2016).

\bibitem{bib:motes15}
K.~R. Motes, J.~P. Dowling, A.~Gilchrist, and P.~P. Rohde.
\newblock \emph{Implementing bosonsampling with time-bin encoding: Analysis of
  loss, mode mismatch, and time jitter}.
\newblock Physical Review A \textbf{92}(5), 052319 (2015).

\bibitem{bib:pant2015high}
M.~Pant and D.~Englund.
\newblock \emph{High-dimensional unitary transformations and boson sampling on
  temporal modes using dispersive optics}.
\newblock Phys. Rev. A \textbf{93}, 043803 (2016).

\bibitem{bib:RohdeUniversal}
P.~P. Rohde.
\newblock \emph{A simple scheme for universal linear optics quantum computing
  with constant experimental complexity using fiber-loops}.
\newblock Phys. Rev. A \textbf{91}, 012306 (2015).

\bibitem{bib:RohdeRalph05}
P.~P. Rohde and T.~C. Ralph.
\newblock \emph{Frequency and temporal effects in linear optical quantum
  computing}.
\newblock Phys. Rev. A \textbf{71}, 032320 (2005).

\bibitem{bib:spectralStructure}
P.~P. Rohde, W.~Mauerer, and C.~Silberhorn.
\newblock \emph{Spectral structure and decompositions of optical states, and
  their applications}.
\newblock New Journal of Physics \textbf{9}(4), 91 (2007).

\bibitem{bib:RohdeRalph06}
P.~P. Rohde and T.~C. Ralph.
\newblock \emph{Error models for mode-mismatch in linear optics quantum
  computing}.
\newblock Phys. Rev. A \textbf{73}, 062312 (2006).

\bibitem{bib:winzer2010spectrally}
P.~Winzer, A.~Gnauck, C.~Doerr, M.~Magarini, and L.~Buhl.
\newblock \emph{Spectrally efficient long-haul optical networking using
  112-gb/s polarization-multiplexed 16-qam}.
\newblock Lightwave Technology, Journal of \textbf{28}(4), 547 (2010).

\bibitem{bib:schindler2014monolithic}
P.~Schindler, D.~Korn, C.~Stamatiadis, M.~O'Keefe, L.~Stampoulidis,
  R.~Schmogrow, P.~Zakynthinos, R.~Palmer, N.~Cameron, Y.~Zhou, \emph{et~al.}
\newblock \emph{Monolithic gaas electro-optic iq modulator demonstrated at 150
  gbit/s with 64qam}.
\newblock Journal of Lightwave Technology \textbf{32}(4), 760 (2014).

\bibitem{bib:prosyk2012high}
K.~Prosyk, A.~Ait-Ouali, C.~Bornholdt, T.~Brast, M.~Gruner, M.~Hamacher,
  D.~Hoffmann, R.~Kaiser, R.~Millett, K.-O. Velthaus, \emph{et~al.}
\newblock \emph{High performance 40ghz inp mach-zehnder modulator}.
\newblock In \emph{Optical Fiber Communication Conference}, pp. OW4F--7
  (Optical Society of America, 2012).

\bibitem{bib:PCMichaelSteel}
M.~Steel.
\newblock private communication (2014).

\bibitem{bib:Resch}
J.~M. Donohue, M.~Agnew, J.~Lavoie, and K.~J. Resch.
\newblock \emph{Coherent ultrafast measurement of time-bin encoded photons}.
\newblock Phys. Rev. Lett. \textbf{111}, 153602 (2013).

\bibitem{bib:RohdeRalphAA12}
P.~P. Rohde and T.~C. Ralph.
\newblock \emph{Error tolerance of the bosonsampling model for linear optics
  quantum computing}.
\newblock Phys. Rev. A \textbf{85}, 022332 (2012).

\bibitem{bib:aaronson2016bosonsampling}
S.~Aaronson and D.~J. Brod.
\newblock \emph{Bosonsampling with lost photons}.
\newblock Physical Review A \textbf{93}(1), 012335 (2016).

\bibitem{bib:berazin2012}
F.~Berazin.
\newblock \emph{The method of second quantization}, vol.~24 (Elsevier, 2012).

\bibitem{bib:PCAndrewWhite}
A.~White.
\newblock private communication (2015).

\bibitem{bib:Han15}
Z.~Han, G.~Moille, X.~Checoury, J.~Bourderionnet, P.~Boucaud, A.~D. Rossi, and
  S.~Combri\'{e}.
\newblock \emph{High-performance and power-efficient 2$\times$2 optical switch
  on silicon-on-insulator}.
\newblock Opt. Express \textbf{23}(19), 24163 (2015).

\bibitem{bib:PhysRevB.89.161303}
M.~Davan\ifmmode~\mbox{\c{c}}\else \c{c}\fi{}o, C.~S. Hellberg, S.~Ates,
  A.~Badolato, and K.~Srinivasan.
\newblock \emph{Multiple time scale blinking in inas quantum dot single-photon
  sources}.
\newblock Phys. Rev. B \textbf{89}, 161303 (2014).

\bibitem{bib:scheidl2014crossed}
T.~Scheidl, F.~Tiefenbacher, R.~Prevedel, F.~Steinlechner, R.~Ursin, and
  A.~Zeilinger.
\newblock \emph{Crossed-crystal scheme for femtosecond-pulsed entangled photon
  generation in periodically poled potassium titanyl phosphate}.
\newblock Physical Review A \textbf{89}(4), 042324 (2014).

\bibitem{bib:kaltenbaek2006experimental}
R.~Kaltenbaek, B.~Blauensteiner, M.~{\.Z}ukowski, M.~Aspelmeyer, and
  A.~Zeilinger.
\newblock \emph{Experimental interference of independent photons}.
\newblock Physical review letters \textbf{96}(24), 240502 (2006).

\bibitem{bib:Olson14}
J.~P. Olson, K.~P. Seshadreesan, K.~R. Motes, P.~P. Rohde, and J.~P. Dowling.
\newblock \emph{Sampling arbitrary photon-added or photon-subtracted squeezed
  states is in the same complexity class as boson sampling}.
\newblock Physical Review A \textbf{91}(2), 022317 (2015).

\bibitem{bib:Bartlett02}
S.~D. Bartlett, E.~Diamanti, B.~C. Sanders, and Y.~Yamamoto.
\newblock \emph{Photon counting schemes and performance of non-deterministic
  nonlinear gates in linear optics}.
\newblock Proceedings of Free-Space Laser Communication and Laser Imaging II
  \textbf{4821} (2002).

\bibitem{bib:rahimi2014can}
S.~Rahimi-Keshari, A.~P. Lund, and T.~C. Ralph.
\newblock \emph{What can quantum optics say about computational complexity
  theory?}
\newblock Phys. Rev. Lett. \textbf{114}, 060501 (2015).

\bibitem{bib:BW99}
K.~Banaszek and K.~W\'odkiewicz.
\newblock \emph{Testing quantum nonlocality in phase space}.
\newblock Phys. Rev. Lett. \textbf{82}, 2009 (1999).

\bibitem{bib:WLD07}
C.~F. Wildfeuer, A.~P. Lund, and J.~P. Dowling.
\newblock \emph{Strong violations of bell-type inequalities for path-entangled
  number states}.
\newblock Phys. Rev. A \textbf{76}, 052101 (2007).

\bibitem{bib:Agarwal91}
G.~S. Agarwal and K.~Tara.
\newblock \emph{Nonclassical properties of states generated by the excitations
  on a coherent state}.
\newblock Phys. Rev. A \textbf{43}, 492 (1991).

\bibitem{bib:Dakna98}
M.~Dakna, L.~Knoll, and D.-G. Welsch.
\newblock \emph{Photon-added state preparation via conditional measurement on a
  beam splitter}.
\newblock Opt. Comm. \textbf{145}(1), 309  (1998).

\bibitem{bib:Dakna98_2}
M.~Dakna, L.~Knoll, and D.-G. Welsch.
\newblock \emph{Quantum state engineering using conditional measurement on a
  beam splitter}.
\newblock Eur. Phys. J. D \textbf{3}(3), 295 (1998).

\bibitem{bib:Zavatta04}
A.~Zavatta, S.~Viciani, and M.~Bellini.
\newblock \emph{Quantum-to-classical transition with single-photon-added
  coherent states of light}.
\newblock Science \textbf{306}(5696), 660 (2004).

\bibitem{bib:Zavatta05}
A.~Zavatta, S.~Viciani, and M.~Bellini.
\newblock \emph{Single-photon excitation of a coherent state: Catching the
  elementary step of stimulated light emission}.
\newblock Phys. Rev. A \textbf{72}, 023820 (2005).

\bibitem{bib:Bouland}
A.~Bouland and S.~Aaronson.
\newblock \emph{Generation of universal linear optics by any beam splitter}.
\newblock Phys. Rev. A \textbf{89}, 062316 (2014).

\bibitem{bib:Georgi99}
H.~M. Georgi.
\newblock \emph{Lie algebras in particle physics} (Perseus, 1999).

\bibitem{bib:PhysRevA.46.2966}
B.~C. Sanders.
\newblock \emph{Erratum: Entangled coherent states}.
\newblock Phys. Rev. A \textbf{46}, 2966 (1992).

\bibitem{bib:PhysRevA.55.2478}
C.~C. Gerry.
\newblock \emph{Generation of schr\"odinger cats and entangled coherent states
  in the motion of a trapped ion by a dispersive interaction}.
\newblock Phys. Rev. A \textbf{55}, 2478 (1997).

\bibitem{bib:gerry2002nonlinear}
C.~C. Gerry, A.~Benmoussa, and R.~Campos.
\newblock \emph{Nonlinear interferometer as a resource for maximally entangled
  photonic states: application to interferometry}.
\newblock Physical Review A \textbf{66}(1), 013804 (2002).

\bibitem{bib:gerry2007nonlocal}
C.~C. Gerry and R.~Grobe.
\newblock \emph{Nonlocal entanglement of coherent states, complementarity, and
  quantum erasure}.
\newblock Physical Review A \textbf{75}(3), 034303 (2007).

\bibitem{bib:gerry2009maximally}
C.~C. Gerry, J.~Mimih, and A.~Benmoussa.
\newblock \emph{Maximally entangled coherent states and strong violations of
  bell-type inequalities}.
\newblock Physical Review A \textbf{80}(2), 022111 (2009).

\bibitem{bib:gilchrist04}
A.~Gilchrist, K.~Nemoto, W.~J. Munro, T.~Ralph, S.~Glancy, S.~L. Braunstein,
  and G.~Milburn.
\newblock \emph{Schr{\"o}dinger cats and their power for quantum information
  processing}.
\newblock Journal of Optics B: Quantum and Semiclassical Optics \textbf{6}(8),
  S828 (2004).

\bibitem{bib:bremner2011classical}
M.~J. Bremner, R.~Jozsa, and D.~J. Shepherd.
\newblock \emph{Classical simulation of commuting quantum computations implies
  collapse of the polynomial hierarchy}.
\newblock Proc. Royal Soc. A: Math., Phys. and Eng. Sci. \textbf{467}, 459
  (2011).

\bibitem{bib:morimae2014hardness}
T.~Morimae, K.~Fujii, and J.~F. Fitzsimons.
\newblock \emph{Hardness of classically simulating the one-clean-qubit model}.
\newblock Phys. Rev. Lett. \textbf{112}, 130502 (2014).

\bibitem{bib:morimae2014classical}
K.~Fujii, H.~Kobayashi, T.~Morimae, H.~Nishimura, S.~Tamate, and S.~Tani.
\newblock \emph{Impossibility of classically simulating one-clean-qubit
  computation}.
\newblock arXiv:1409.6777  (2014).

\bibitem{bib:han1997threshold}
Y.~Han, L.~A. Hemaspaandra, and T.~Thierauf.
\newblock \emph{Threshold computation and cryptographic security}.
\newblock SIAM Journal on Computing \textbf{26}, 59 (1997).

\bibitem{bib:aaronson2005quantum}
S.~Aaronson.
\newblock \emph{Quantum computing, postselection, and probabilistic
  polynomial-time}.
\newblock Proc. Royal Soc. A: Math., Phys. and Eng. Sci. \textbf{461}, 3473
  (2005).

\bibitem{bib:ralph}
T.~C. Ralph, A.~Gilchrist, G.~J. Milburn, W.~J. Munro, and S.~Glancy.
\newblock \emph{Quantum computation with optical coherent states}.
\newblock Phys. Rev. A \textbf{68}, 042319 (2003).

\bibitem{bib:PhysRevA.37.2970}
P.~Alsing, G.~J. Milburn, and D.~F. Walls.
\newblock \emph{Quantum nondemolition measurements in optical cavities}.
\newblock Phys. Rev. A \textbf{37}, 2970 (1988).

\bibitem{bib:Haroche06}
S.~Haroche.
\newblock \emph{Exploring the quantum: atoms, cavities and photons} (Oxford
  Press, 2006).

\bibitem{bib:harris2008}
S.~E. Harris.
\newblock \emph{Electromagnetically induced transparency}.
\newblock Physics Today \textbf{50}(7), 36 (2008).

\bibitem{bib:PhysRevLett.64.1107}
S.~E. Harris, J.~E. Field, and A.~Imamo\ifmmode~\breve{g}\else \u{g}\fi{}lu.
\newblock \emph{Nonlinear optical processes using electromagnetically induced
  transparency}.
\newblock Phys. Rev. Lett. \textbf{64}, 1107 (1990).

\bibitem{bib:Lund04}
A.~P. Lund, H.~Jeong, T.~C. Ralph, and M.~S. Kim.
\newblock \emph{Conditional production of superpositions of coherent states
  with inefficient photon detection}.
\newblock Phys. Rev. A \textbf{70}, 020101(R) (2004).

\bibitem{bib:JeongLundRalph05}
H.~Jeong, A.~P. Lund, and T.~C. Ralph.
\newblock \emph{Production of superpositions of coherent states in traveling
  optical fields with inefficient photon detection}.
\newblock Phys. Rev. A \textbf{72}, 013801 (2005).

\bibitem{bib:tolkien}
J.~R.~R. Tolkien.
\newblock \emph{The Lord of the Rings: One Volume} (Houghton Mifflin Harcourt,
  2012).

\bibitem{bib:yurke1986input}
B.~Yurke.
\newblock \emph{Input states for enhancement of fermion interferometer
  sensitivity}.
\newblock Phys. Rev. Lett \textbf{56}, 1515 (1986).

\bibitem{bib:yuen1986generation}
H.~P. Yuen.
\newblock \emph{Generation, detection, and application of high-intensity
  photon-number-eigenstate fields}.
\newblock Phys. Rev. Lett. \textbf{56}, 2176 (1986).

\bibitem{bib:dowling1998correlated}
J.~P. Dowling.
\newblock \emph{Correlated input-port, matter-wave interferometer:
  Quantum-noise limits to the atom-laser gyroscope}.
\newblock Phys. Rev. A \textbf{57}, 4736 (1998).

\bibitem{bib:yurtsever2003interferometry}
U.~Yurtsever, D.~Strekalov, and J.~Dowling.
\newblock \emph{Interferometry with entangled atoms}.
\newblock The Euro. Phys. J. D-Atomic, Molecular, Optical and Plasma Physics
  \textbf{22}, 365 (2003).

\bibitem{bib:nasr2003demonstration}
M.~B. Nasr, B.~E. Saleh, A.~V. Sergienko, and M.~C. Teich.
\newblock \emph{Demonstration of dispersion-canceled quantum-optical coherence
  tomography}.
\newblock Phys. Rev. Lett. \textbf{91}, 083601 (2003).

\bibitem{bib:toussaint2004}
K.~Toussaint, Jr., G.~D. Giuseppe, K.~J. Bycenski, A.~V. Sergienko, B.~E.~A.
  Saleh, and M.~C. Teich.
\newblock \emph{Quantum ellipsometry using correlated-photon beams}.
\newblock Physical Review A \textbf{70}, 023801 (2004).

\bibitem{bib:jones2009magnetic}
J.~A. Jones, S.~D. Karlen, J.~Fitzsimons, A.~Ardavan, S.~C. Benjamin, G.~A.~D.
  Briggs, and J.~J. Morton.
\newblock \emph{Magnetic field sensing beyond the standard quantum limit using
  10-spin noon states}.
\newblock Science \textbf{324}, 1166 (2009).

\bibitem{bib:crespi2012measuring}
A.~Crespi, M.~Lobino, J.~C. Matthews, A.~Politi, C.~R. Neal, R.~Ramponi,
  R.~Osellame, and J.~L. O'Brien.
\newblock \emph{Measuring protein concentration with entangled photons}.
\newblock App. Phys. Lett. \textbf{100}, 233704 (2012).

\bibitem{bib:rozema2014scalable}
L.~A. Rozema, J.~D. Bateman, D.~H. Mahler, R.~Okamoto, A.~Feizpour, A.~Hayat,
  and A.~M. Steinberg.
\newblock \emph{Scalable spatial superresolution using entangled photons}.
\newblock Phys. Rev. Lett. \textbf{112}, 223602 (2014).

\bibitem{bib:israel2014supersensitive}
Y.~Israel, S.~Rosen, and Y.~Silberberg.
\newblock \emph{Supersensitive polarization microscopy using noon states of
  light}.
\newblock Phys. Rev. Lett. \textbf{112}, 103604 (2014).

\bibitem{bib:holland1993interferometric}
M.~Holland and K.~Burnett.
\newblock \emph{Interferometric detection of optical phase shifts at the
  heisenberg limit}.
\newblock Phys. Rev. Lett. \textbf{71}, 1355 (1993).

\bibitem{bib:lee2002quantum}
H.~Lee, P.~Kok, and J.~P. Dowling.
\newblock \emph{A quantum rosetta stone for interferometry}.
\newblock J. Mod. Opt. \textbf{49}(14-15), 2325 (2002).

\bibitem{bib:durkin2007local}
G.~A. Durkin and J.~P. Dowling.
\newblock \emph{Local and global distinguishability in quantum interferometry}.
\newblock Physical review letters \textbf{99}(7), 070801 (2007).

\bibitem{bib:dowling2008quantum}
J.~P. Dowling.
\newblock \emph{Quantum optical metrology--the lowdown on high-n00n states}.
\newblock Contemporary physics \textbf{49}(2), 125 (2008).

\bibitem{bib:gerry2001generation}
C.~C. Gerry and R.~Campos.
\newblock \emph{Generation of maximally entangled photonic states with a
  quantum-optical fredkin gate}.
\newblock Phys. Rev. A \textbf{64}, 063814 (2001).

\bibitem{bib:kapale2007bootstrapping}
K.~T. Kapale and J.~P. Dowling.
\newblock \emph{Bootstrapping approach for generating maximally path-entangled
  photon states}.
\newblock Phys. Rev. Lett. \textbf{99}, 053602 (2007).

\bibitem{bib:lee2002linear}
H.~Lee, P.~Kok, N.~J. Cerf, and J.~P. Dowling.
\newblock \emph{Linear optics and projective measurements alone suffice to
  create large-photon-number path entanglement}.
\newblock Phys. Rev. A \textbf{65}, 030101 (2002).

\bibitem{bib:vanmeter2007general}
N.~VanMeter, P.~Lougovski, D.~Uskov, K.~Kieling, J.~Eisert, and J.~P. Dowling.
\newblock \emph{General linear-optical quantum state generation scheme:
  Applications to maximally path-entangled states}.
\newblock Phys. Rev. A \textbf{76}, 063808 (2007).

\bibitem{bib:cable2007efficient}
H.~Cable and J.~P. Dowling.
\newblock \emph{Efficient generation of large number-path entanglement using
  only linear optics and feed-forward}.
\newblock Physical review letters \textbf{99}(16), 163604 (2007).

\bibitem{bib:Kok07}
P.~Kok, W.~J. Munro, K.~Nemoto, T.~C. Ralph, J.~P. Dowling, and G.~J. Milburn.
\newblock \emph{Linear optical quantum computing with photonic qubits}.
\newblock Rev. Mod. Phys. \textbf{79}, 135 (2007).

\bibitem{bib:seshadreesan2013phase}
K.~P. Seshadreesan, S.~Kim, J.~P. Dowling, and H.~Lee.
\newblock \emph{Phase estimation at the quantum cramer-rao bound via parity
  detection}.
\newblock Phys. Rev. A \textbf{87}, 043833 (2013).

\bibitem{bib:mayer2011counting}
K.~Mayer, M.~C. Tichy, F.~Mintert, T.~Konrad, and A.~Buchleitner.
\newblock \emph{Counting statistics of many-particle quantum walks}.
\newblock Phys. Rev. A \textbf{83}, 062307 (2011).

\bibitem{bib:matthews2011heralding}
J.~C. Matthews, A.~Politi, D.~Bonneau, and J.~L. O'Brien.
\newblock \emph{Heralding two-photon and four-photon path entanglement on a
  chip}.
\newblock Phys. Rev. Lett. \textbf{107}, 163602 (2011).

\bibitem{bib:scully1992high}
M.~O. Scully and M.~Fleischhauer.
\newblock \emph{High-sensitivity magnetometer based on index-enhanced media}.
\newblock Physical review letters \textbf{69}(9), 1360 (1992).

\bibitem{bib:fukuda2011titanium}
D.~Fukuda, G.~Fujii, T.~Numata, K.~Amemiya, A.~Yoshizawa, H.~Tsuchida,
  H.~Fujino, H.~Ishii, T.~Itatani, S.~Inoue, \emph{et~al.}
\newblock \emph{Titanium-based transition-edge photon number resolving detector
  with 98\% detection efficiency with index-matched small-gap fiber coupling}.
\newblock Opt. Express \textbf{19}(2), 870 (2011).

\bibitem{bib:LPOR201400404}
L.~A. Ngah, O.~Alibart, L.~Labont\'{e}, V.~D'Auria, and S.~Tanzilli.
\newblock \emph{Ultra-fast heralded single photon source based on telecom
  technology}.
\newblock Laser \& Photonics Reviews \textbf{9}, L1 (2015).

\bibitem{bib:Maier14}
S.~Maier, P.~Gold, A.~Forchel, N.~Gregersen, J.~M{\o}rk, S.~H\"{o}fling,
  C.~Schneider, and M.~Kamp.
\newblock \emph{Bright single photon source based on self-aligned quantum
  dot--cavity systems}.
\newblock Opt. Express \textbf{22}(7), 8136 (2014).

\bibitem{bib:MORDOR2}
J.~P. Olson, K.~R. Motes, P.~M. Birchall, N.~M. Studer, M.~LaBorde, T.~Moulder,
  P.~P. Rohde, and J.~P. Dowling.
\newblock \emph{Linear optical quantum metrology with single photons ---
  experimental errors, resource counting, and quantum cram\'er-rao bounds} .

\bibitem{bib:rahimi2015efficient}
S.~Rahimi-Keshari, T.~C. Ralph, and C.~M. Caves.
\newblock \emph{Efficient classical simulation of quantum optics}.
\newblock arXiv:1511.06526  (2015).

\bibitem{bib:macmahon60}
P.~MacMahon.
\newblock \emph{Combinatory analysis, 1915-16, two volumes (bound as one),
  chelsea pub} (1960).

\bibitem{bib:MagdaLeap}
M.~Stobinska, P.~P. Rohde, and P.~Kurzynski.
\newblock \emph{Quantum leap: how to complete a quantum walk in a single step}
  (2015).
\newblock \eprint{arXiv:1504.05480}.

\bibitem{bib:MerlinCooper13}
M.~Cooper, L.~J. Wright, C.~S{\" o}ller, and B.~J. Smith.
\newblock \emph{Experimental generation of multi-photon fock states}.
\newblock Opt. Ex. \textbf{21}, 5309 (2013).

\bibitem{bib:finger15}
M.~A. Finger, T.~S. Iskhakov, N.~Y. Joly, M.~V. Chekhova, and P.~S.~J. Russell.
\newblock \emph{Raman-free, noble-gas-filled photonic-crystal fiber source for
  ultrafast, very bright twin-beam squeezed vacuum}.
\newblock Physical review letters \textbf{115}(14), 143602 (2015).

\bibitem{bib:RohdeBarrett07}
P.~P. Rohde and S.~D. Barrett.
\newblock \emph{Strategies for the preparation of large cluster states using
  non-deterministic gates}.
\newblock New J. Phys. \textbf{9}, 198 (2007).

\bibitem{bib:Gross06}
D.~Gross, K.~Kieling, and J.~Eisert.
\newblock \emph{Potential and limits to cluster state quantum computing using
  probabilistic gates}.
\newblock Phys. Rev. A \textbf{74}, 042343 (2006).

\bibitem{bib:kitaev03}
A.~Y. Kitaev.
\newblock \emph{Fault-tolerant quantum computation by anyons}.
\newblock Annals of Physics \textbf{303}(1), 2 (2003).

\bibitem{bib:braunstein98}
S.~L. Braunstein.
\newblock \emph{Error correction for continuous quantum variables}.
\newblock In \emph{Quantum Information with Continuous Variables}, pp. 19--29
  (Springer, 1998).

\bibitem{bib:lloyd98}
S.~Lloyd and J.-J.~E. Slotine.
\newblock \emph{Analog quantum error correction}.
\newblock Physical Review Letters \textbf{80}(18), 4088 (1998).

\bibitem{bib:GKP2001}
D.~Gottesman, A.~Kitaev, and J.~Preskill.
\newblock \emph{Encoding a qubit in an oscillator}.
\newblock Physical Review A \textbf{64}(1), 012310 (2001).

\bibitem{bib:menicucci2014}
N.~C. Menicucci.
\newblock \emph{Fault-tolerant measurement-based quantum computing with
  continuous-variable cluster states}.
\newblock Physical review letters \textbf{112}(12), 120504 (2014).

\bibitem{bib:vasconcelos10}
H.~M. Vasconcelos, L.~Sanz, and S.~Glancy.
\newblock \emph{All-optical generation of states for Òencoding a qubit in an
  oscillatorÓ}.
\newblock Optics letters \textbf{35}(19), 3261 (2010).

\bibitem{bib:gottesman01}
D.~Gottesman, A.~Kitaev, and J.~Preskill.
\newblock \emph{Encoding a qubit in an oscillator}.
\newblock Physical Review A \textbf{64}(1), 012310 (2001).

\bibitem{bib:Travaglione02}
B.~C. Travaglione and G.~J. Milburn.
\newblock \emph{Preparing encoded states in an oscillator}.
\newblock Phys. Rev. A \textbf{66}, 052322 (2002).

\bibitem{bib:pirandola04}
S.~Pirandola, S.~Mancini, D.~Vitali, and P.~Tombesi.
\newblock \emph{Constructing finite-dimensional codes with optical continuous
  variables}.
\newblock EPL (Europhysics Letters) \textbf{68}(3), 323 (2004).

\bibitem{bib:pirandola06}
S.~Pirandola, S.~Mancini, D.~Vitali, and P.~Tombesi.
\newblock \emph{Continuous variable encoding by ponderomotive interaction}.
\newblock The European Physical Journal D-Atomic, Molecular, Optical and Plasma
  Physics \textbf{37}(2), 283 (2006).

\bibitem{bib:pirandola06b}
S.~Pirandola, S.~Mancini, D.~Vitali, and P.~Tombesi.
\newblock \emph{Generating continuous variable quantum codewords in the
  near-field atomic lithography}.
\newblock Journal of Physics B: Atomic, Molecular and Optical Physics
  \textbf{39}(4), 997 (2006).

\bibitem{bib:sakurai2011modern}
J.~J. Sakurai and J.~Napolitano.
\newblock \emph{Modern quantum mechanics} (Addison-Wesley, 2011).

\bibitem{bib:deutsch10}
I.~H. Deutsch and P.~S. Jessen.
\newblock \emph{Quantum control and measurement of atomic spins in polarization
  spectroscopy}.
\newblock Optics Communications \textbf{283}(5), 681 (2010).

\bibitem{bib:Wineland11}
D.~Wineland and D.~Leibfried.
\newblock \emph{Quantum information processing and metrology with trapped
  ions}.
\newblock Laser Phys. Lett. \textbf{8}(3), 175 (2011).

\bibitem{bib:baragiola14}
B.~Q. Baragiola, L.~M. Norris, E.~Montano, P.~G. Mickelson, P.~S. Jessen, and
  I.~H. Deutsch.
\newblock \emph{Three-dimensional light-matter interface for collective spin
  squeezing in atomic ensembles}.
\newblock Physical Review A \textbf{89}(3), 033850 (2014).

\bibitem{bib:goban14}
A.~Goban, C.-L. Hung, S.-P. Yu, J.~Hood, J.~Muniz, J.~Lee, M.~Martin,
  A.~McClung, K.~Choi, D.~Chang, \emph{et~al.}
\newblock \emph{Atom--light interactions in photonic crystals}.
\newblock Nature communications \textbf{5} (2014).

\bibitem{bib:mcconnell15}
R.~McConnell, H.~Zhang, J.~Hu, S.~{\'C}uk, and V.~Vuleti{\'c}.
\newblock \emph{Entanglement with negative wigner function of almost 3,000
  atoms heralded by one photon}.
\newblock Nature \textbf{519}(7544), 439 (2015).

\bibitem{bib:Bohnet14}
J.~G. Bohnet, K.~C. Cox, M.~A. Norcia, J.~M. Weiner, Z.~Chen, and J.~K.
  Thompson.
\newblock \emph{Reduced spin measurement back-action for a phase sensitivity
  ten times beyond the standard quantum limit}.
\newblock Nature Photonics \textbf{8}(9), 731 (2014).

\bibitem{bib:devoret13}
M.~H. Devoret and R.~J. Schoelkopf.
\newblock \emph{Superconducting circuits for quantum information: an outlook}.
\newblock Science \textbf{339}(6124), 1169 (2013).

\bibitem{bib:arnold15}
C.~Arnold, J.~Demory, V.~Loo, A.~Lema{\^\i}tre, I.~Sagnes, M.~Glazov, O.~Krebs,
  P.~Voisin, P.~Senellart, and L.~Lanco.
\newblock \emph{Macroscopic rotation of photon polarization induced by a single
  spin}.
\newblock Nature communications \textbf{6} (2015).

\bibitem{bib:scully1993quantum}
M.~O. Scully and J.~P. Dowling.
\newblock \emph{Quantum-noise limits to matter-wave interferometry}.
\newblock Physical Review A \textbf{48}(4), 3186 (1993).

\bibitem{bib:Bardhan2013}
J.~D. B.~R.~Bardan, K.~Jiang.
\newblock \emph{Effects of phase fluctuations on phase sensitivity and
  visibility of path-entangled photon fock states}.
\newblock Phys. Rev. A \textbf{88}(4), 023857 (2013).

\bibitem{bib:paris04}
R.~B. Paris.
\newblock \emph{The stokes phenomenon associated with the hurwitz zeta function
  $\zeta(s, a)$}.
\newblock Proc. R. Soc. A \textbf{461}, 297 (2004).

\bibitem{bib:QRWmotes16}
K.~R. Motes, A.~Gilchrist, and P.~P. Rohde.
\newblock \emph{Quantum random walks on congested lattices and the effect of
  dephasing}.
\newblock Scientific reports \textbf{6} (2016).

\bibitem{bib:rohde2016quantum}
P.~P. Rohde, D.~W. Berry, K.~R. Motes, and J.~P. Dowling.
\newblock \emph{A quantum optics argument for the \# p-hardness of a class of
  multidimensional integrals}.
\newblock arXiv:1607.04960  (2016).

\bibitem{bib:motes2016efficient}
K.~R. Motes, R.~L. Mann, J.~P. Olson, N.~M. Studer, E.~A. Bergeron,
  A.~Gilchrist, J.~P. Dowling, D.~W. Berry, and P.~P. Rohde.
\newblock \emph{Efficient recycling strategies for preparing large fock states
  from single-photon sources: Applications to quantum metrology}.
\newblock Phys. Rev. A \textbf{94}, 012344 (2016).

\bibitem{bib:motes2017encoding}
K.~R. Motes, B.~Q. Baragiola, A.~Gilchrist, and N.~C. Menicucci.
\newblock \emph{Encoding qubits into oscillators with atomic ensembles and
  squeezed light}.
\newblock arXiv:1703.02107  (2017).

\end{thebibliography}

\appendix
\chapter{Appendix}


\section{Intuitive Example of Loop Bias due to Loss} \label{app:MultipleLoopBiasExample}
For Sec. \ref{sec:OuterLoopLoss} an example of how $\hat{U}$ becomes biased is explained here. Let us consider two examples of a two-mode pulse-train --- a single inner loop and two inner loops. 

\subsection{One Loop}
The first mode can can exit the first output mode by traversing the inner loop once. Here it picks up loss due to the middle switch twice ${\eta_s}^2$, and loss due to the inner loop fiber once $\eta_f$, obtaining a net loss of \mbox{${\eta_s}^2\eta_f$}. The first mode can exit the second output mode by traversing the inner loop twice. In this case it obtains a net loss of \mbox{${\eta_s}^3{\eta_f}^2$}. A similar analysis can be performed for the other combinations. Then, we can write the loss amplitudes corresponding to the input (rows) and output (columns) modes in matrix form as, 
\begin{equation} 
\mathcal{\hat{L}} = \left(\begin{array}{cc}
{\eta_s}^2\eta_f & {\eta_s}^3{\eta_f}^2  \\
\eta_s &{\eta_s}^2\eta_f 
\end{array}\right) = \eta_s \left(\begin{array}{cc}
\eta & {\eta}^2  \\
1 &\eta 
\end{array}\right),
\end{equation}
where $\eta=\eta_s\eta_f$ and observe the bias accumulating in this input to output map. The net input-to-output mapping of amplitudes is given by taking the element-wise product of this loss matrix with the ideal unitary, \mbox{$\mathcal{\hat{L}}\circ \hat{U}$}, thereby leaving us with a biased map.

\subsection{Two Loops}
A similar analysis as above but following the paths for two consecutive applications of the inner loop (i.e one traversal of the outer loop), we find the input-to-output loss matrix to be,
\begin{equation} 
\mathcal{\hat{L}} = \left(\begin{array}{cc}
{\eta_s}^4{\eta_f}^2 & {\eta_s}^5{\eta_f}^3  \\
{\eta_s}^3\eta_f &{\eta_s}^4{\eta_f}^2 
\end{array}\right) = {\eta_s}^2 \eta \left(\begin{array}{cc}
\eta & {\eta}^2  \\
1 &\eta 
\end{array}\right). 
\end{equation}
where we have ignored the losses due to the outer loop as it yields an overall normalisation factor that does not bias $\hat{U}$. As we can see, for each iteration of the inner loop $\hat{U}$ accumulates more loss, with a decreasing overall success probability, but the amount of skew in the matrix remains the same.

\section{Wave-packet Simplifications}\label{app:WavePacketSimp}

For Sec. \ref{sec:ErrorFidelityMetric} we derive the overlap of two photons with temporal creation operator \mbox{$\hat{\mathcal{A}}^{\dag}(t,\Delta)$} as per Eq. (\ref{eq:TPCO}) and temporal density function $\psi(t,\Delta)$. For the purpose of this work we assume that the temporal spacing $\tau$ between each mode is much larger than the width of the wave packet $\omega$ such that the overlap of our temporal photons in different time-bins is negligible,
\begin{equation}
\bra{0}\hat{\mathcal{A}}(t,\Delta)\hat{\mathcal{A}}^{\dag}(t',\Delta')\ket{0}=0,
\end{equation}
for $t \neq t'$. 

For photons in the same time-bin and allowing for arbitrary temporal-shifts, the overlap is,
\begin{eqnarray}
&\ &\bra{0}\hat{\mathcal{A}}(t,\Delta')\hat{\mathcal{A}}^{\dag}(t,\Delta)\ket{0} \nonumber \\
&=&\bigg(\bra{0}\int_{-\infty}^{\infty}\psi^{*}(x'-t-\Delta')\hat{a}(x')dx'\bigg)\nonumber\\
&\times& \bigg(\int_{-\infty}^{\infty}\psi(x-t-\Delta)\hat{a}^{\dag}(x)dx \ket{0} \bigg) \nonumber \\
&=& \int_{-\infty}^{\infty}\int_{-\infty}^{\infty}\psi^{*}(x'-t-\Delta')\psi(x-t-\Delta) \nonumber \\
&\times& \underbrace{\bra{0}\hat{a}(x')\hat{a}^{\dag}(x)\ket{0}}_{\delta_{x',x}}dx'dx \nonumber \\
&=& \int_{-\infty}^{\infty} \psi^{*}(x-\Delta')\psi(x-\Delta)dx \nonumber \\
&=& e^{-\frac{(\Delta' -\Delta )^2}{4 \omega^2}}.
\end{eqnarray}

For ideal states where \mbox{$\Delta'=\Delta=0$}, we notice that \mbox{$\mathcal{F}=1$}, as expected when there is no mode-mismatch. We use these results for simplifying Eq. (\ref{eq:Fidai3}) in our analysis of mode-mismatch.


%
%

\section{Reproducing Hong-Ou-Mandel Interference Using Small Amplitude Odd Cat States} \label{app:odd_cat}

As for Sec. \ref{sec:CATzeroAmplitude} we begin with our generalized cat state result from Eq. (\ref{eq:gamma}). 
\begin{equation} \label{eq:mainresult}
\gamma_s = \sum_{\vec{t}=1}^{t} \left(\prod_{j=1}^m \lambda_{t_j}^{(j)} f_{S_j}(\beta_{\vec{t}}^{(j)})\right),
\end{equation}
and input the odd cat state which has the form 
\begin{eqnarray} \label{eq:app_odd_cate_state}
\ket{\mathrm{cat}_-} &=& \frac{\ket{\alpha}-\ket{-\alpha}}{\sqrt{2(1-\mathrm{exp}[-2\alpha^2])}}.
\end{eqnarray}

When considering the specific example of $\ket{\mathrm{cat}_-}$ the $\lambda_{t_j}^{(j)}$ of Eq.~(\ref{eq:mainresult}) goes to $(-1)^{t_j}$. Eq.~(\ref{eq:mainresult}) then becomes,
\begin{eqnarray} \label{eq:mainresult2}
\gamma_s &=& \sum_{\vec{t}=1}^{t} \left(\prod_{j=1}^m (-1)^{t_j} \frac{f_{S_j}(\beta_{\vec{t}}^{(j)})}{\sqrt{2(1-\mathrm{exp}[-2\alpha^2])}} \right).
\end{eqnarray}
Since the $\beta_{\vec{t}}^{(j)}$'s in Eq.~(\ref{eq:mainresult2}) depend on $\alpha$, we substitute the argument of $f_{S_j}$ using Eq.~ (\ref{eq:fna}),
\begin{eqnarray} 
\gamma_s &=& \frac{1}{\left(\sqrt{2(1-\mathrm{exp}[-2\alpha^2])}\right)^m} \\
&\times& \sum_{\vec{t}=1}^{t} \left(\prod_{j=1}^m (-1)^{t_j}  \mathrm{exp}\left[-\frac{|\beta_{\vec{t}}^{(j)}|^2}{2}\right] \frac{(\beta_{\vec{t}}^{(j)})^{S_j}}{\sqrt{S_j!}}\right)
\end{eqnarray}
Next we take a first order approximation. Since $\alpha$ is small, the exponential in the numerator goes to one while the exponential in the denominator goes to $\mathrm{exp}(x)\approx 1+x$ because otherwise this would diverge. This yields,
\begin{eqnarray} \label{eq:sub}
\gamma_s &\approx& \frac{1}{\left(\sqrt{2(1-(1-2\alpha^2)}\right)^m} \sum_{\vec{t}=1}^{t} \left(\prod_{j=1}^m (-1)^{t_j} \left(1\right) \frac{(\beta_{\vec{t}}^{(j)})^{S_j}}{\sqrt{S_j!}}\right) \nonumber \\
&=& \frac{1}{(2\alpha)^m\sqrt{S_1!S_2!\dots S_m!}} \sum_{\vec{t}=1}^{t} \left(\prod_{j=1}^m (-1)^{t_j} (\beta_{\vec{t}}^{(j)})^{S_j}\right) \nonumber \\
&=& \frac{1}{(2\alpha)^m\sqrt{S_1!S_2!\dots S_m!}} \sum_{\vec{t}=1}^{t} (-1)^{\sigma(\vec{t})}\prod_{j=1}^m (\beta_{\vec{t}}^{(j)})^{S_j}.
\end{eqnarray}

In the limit of small $\alpha$ we know that the odd cat state reduces to a single photon Fock state. Here we consider the case of a cat state being inputted into the first two modes and let the unitary be the Hadamard gate. In small $\alpha$ this corresponds to inputting a single photon Fock state into the first two modes and interfering them in a single 50/50 beamsplitter. Therefore, the corresponding bunching in the output modes would to be expected. In this section we show that our expression of Eq.~(\ref{eq:sub}) does show the expected bunching. 

We begin by putting an odd cat state $\ket{\mathrm{cat}_-}$ with $t=2$ terms into the first $m=2$ modes. Then Eq.~(\ref{eq:sub}) becomes,
\begin{eqnarray} \label{eq:exp}
\gamma_s &\approx& \frac{1}{(2\alpha)^2\sqrt{S_1!S_2!}} \sum_{t_1,t_2=1}^{2} (-1)^{\sigma(\vec{t})} \prod_{j=1}^2 (\beta_{t_1,t_2}^{(j)})^{S_j} \nonumber \\
&=& \frac{1}{(2\alpha)^2\sqrt{S_1!S_2!}} \sum_{t_1,t_2=1}^{2} (-1)^{\sigma(t_1+t_2)} (\beta_{t_1,t_2}^{(1)})^{S_1}(\beta_{t_1,t_2}^{(2)})^{S_2} \nonumber \\
&=&  \frac{1}{(2\alpha)^2\sqrt{S_1!S_2!}} \left[
(\beta_{1,1}^{(1)})^{S_1}(\beta_{1,1}^{(2)})^{S_2}
-(\beta_{1,2}^{(1)})^{S_1}(\beta_{1,2}^{(2)})^{S_2} \right.  \nonumber \\
&-&\left. (\beta_{2,1}^{(1)})^{S_1}(\beta_{2,1}^{(2)})^{S_2}
+(\beta_{2,2}^{(1)})^{S_1}(\beta_{2,2}^{(2)})^{S_2}
\right]
\end{eqnarray}

Now to calculate the $\beta_{\vec{t}}^{(j)}$'s for this case we first take the tensor product between the first two modes. Ignoring the normalization factor this yields, 
\begin{eqnarray} 
\ket{\mathrm{cat}\_}&=&(\ket{\alpha}-\ket{-\alpha})\otimes(\ket{\alpha}-\ket{-\alpha}) \nonumber \\
&=&\ket{\alpha,\alpha}-\ket{\alpha,-\alpha}-\ket{-\alpha,\alpha}+\ket{-\alpha,-\alpha}. \nonumber \\
\end{eqnarray}

Next we pass them through a 50/50 beamsplitter,
\begin{equation}
U\ket{\mathrm{cat}\_}=\ket{\sqrt{2}\alpha,0}-\ket{0,\sqrt{2}\alpha}-\ket{0,-\sqrt{2}\alpha}+\ket{-\sqrt{2}\alpha,0}.
\end{equation}
Now we read off the $\beta_{\vec{t}}^{(j)}$'s to be
\begin{eqnarray}
\beta_{1,1}^{(1)}&=&\beta_{1,2}^{(2)}=\sqrt{2}\alpha \nonumber \\
\beta_{2,2}^{(1)}&=& \beta_{2,1}^{(2)}=-\sqrt{2}\alpha \nonumber \\
\beta_{1,2}^{(1)}&=&\beta_{2,1}^{(1)}=\beta_{1,1}^{(2)}=\beta_{2,2}^{(2)}=0.
\end{eqnarray}

Now Eq.~(\ref{eq:exp}) becomes,
\begin{eqnarray} \label{eq:b}
\gamma_s &=&  \frac{1}{(2\alpha)^2\sqrt{S_1!S_2!}} \left[
(\sqrt{2\alpha})^{S_1}(0)^{S_2}
-(0)^{S_1}(\sqrt{2}\alpha)^{S_2} \right. \nonumber \\
&-& \left. (0)^{S_1}(-\sqrt{2}\alpha)^{S_2}
+(-\sqrt{2}\alpha)^{S_1}(0)^{S_2}
\right].
\end{eqnarray}

Because we are dealing in the limit of small $\alpha$, a non-zero number arbitrarily close to zero raised to a zero power is one, so the terms $0^{S_j}=\delta_{S_j,0}$. Now Eq.~(\ref{eq:b}) becomes,
\begin{eqnarray} \label{eq:Z}
\gamma_s &=&  \frac{1}{(2\alpha)^2\sqrt{S_1!S_2!}} \left[
(\sqrt{2\alpha})^{S_1}\delta_{S_2,0}
-(\sqrt{2}\alpha)^{S_2}\delta_{S_1,0} \right. \nonumber \\
&-& \left. (-\sqrt{2}\alpha)^{S_2}\delta_{S_1,0}
+(-\sqrt{2}\alpha)^{S_1}\delta_{S_2,0}
\right].
\end{eqnarray}

For this example we know that there are three possible signature outcomes. We expect that the configuration $S_1=S_2=1$ is not possible due to HOM photon bunching and thus in this case $\gamma_s=0$. For configurations $S_1=0$ and $S_2=2$ or $S_1=2$ and $S_2=0$ we would expect a non-zero configuration amplitude of $\gamma_s=1/2$ in each case. Next, we will show that this is indeed the case. 
\subsection[Hong Ou Mandel Example of Measuring One Photon at Each Output Port]{Configuration $S_1=S_2=1$}
With configuration $S_1=S_2=1$ Eq.~(\ref{eq:Z}) becomes,
\begin{eqnarray}
\gamma_s &\approx& \frac{1}{4\alpha^2} \left[ 
(\sqrt{2}\alpha)\delta_{1,0}
-(\sqrt{2}\alpha)\delta_{1,0} \right. \nonumber \\
&-&\left.(-\sqrt{2}\alpha)\delta_{1,0}
+(-\sqrt{2}\alpha)\delta_{1,0} \right] \nonumber \\
&=& 0,
\end{eqnarray}
which vanishes as expected. 

\subsection[Hong Ou Mandel Example of Measuring Zero and Two Photons at  Output Ports One and Two respectively]{Configuration $S_1=0$ and $S_2=2$}
With configuration $S_1=0$ and $S_2=2$ Eq.~(\ref{eq:Z}) becomes,
\begin{eqnarray} \label{}
\gamma_s &=&  \frac{1}{(2\alpha)^2\sqrt{0!2!}} \left[
(\sqrt{2\alpha})^{0}\delta_{2,0}
-(\sqrt{2}\alpha)^{2}\delta_{0,0} \right. \nonumber \\
&-& \left. (-\sqrt{2}\alpha)^{2}\delta_{0,0}
+(-\sqrt{2}\alpha)^{0}\delta_{2,0}
\right] \nonumber \\
&=& \frac{1}{4\alpha^2\sqrt{2}} \left[
-2\alpha^{2}
-2\alpha^{2} 
\right] \nonumber \\
&=& -\frac{1}{\sqrt{2}},
\end{eqnarray}
and the corresponding classical probability is $1/2$ as expected.

\subsection[Hong Ou Mandel Example of Measuring Two and Zero Photons at  Output Ports One and Two respectively]{Configuration $S_1=2$ and $S_2=0$}
With configuration $S_1=2$ and $S_2=0$ Eq.~(\ref{eq:Z}) becomes,
\begin{eqnarray} \label{}
\gamma_s &=&  \frac{1}{(2\alpha)^2\sqrt{2!0!}} \left[
(\sqrt{2\alpha})^{2}\delta_{0,0}
-(\sqrt{2}\alpha)^{0}\delta_{2,0} \right. \nonumber \\
&-& \left.(-\sqrt{2}\alpha)^{0}\delta_{2,0}
+(-\sqrt{2}\alpha)^{2}\delta_{0,0}
\right] \nonumber \\
&=& \frac{1}{4\alpha^2\sqrt{2}} \left[
2\alpha^{2}
+2\alpha^{2}
\right] \nonumber \\
&=& \frac{1}{\sqrt{2}},
\end{eqnarray}
again with classical probability $1/2$ as expected.

Thus, our result generalizes to the expected results for passing a single photon Fock state inputted in modes one and two through a Hadamard gate. This shows that our cat state generalization works for the odd cat state in the limit of small $\alpha$, which is equivalent to Aaronson \& Arkhipov's \BStwo. 

%
%
\section{Non-Zero Amplitude Odd Cat States as an Error Model} \label{app:poly_bound}

As for Sec. \ref{sec:CATsmallAmplitude} according to the error bound derived by Aaronson \& Arkhipov, the probability of sampling from the correct distribution must not exceed the bound of $1/\mathrm{poly}(n)$ in order for it to implement  classically hard \BStwo. The correct input distribution is $\ket{1,\dots,1,0,\dots,0}$ and the probability of successfully sampling from it depends on the odd cat states since we input odd cat states in every mode requiring a $\ket{1}$ and vacuum in the remaining modes. Thus, the single-photon component of the odd cat state must be successfully sampled $n$ times. 

Consider the odd cat state from Eq.~(\ref{eq:app_odd_cate_state}). The amplitude of the single photon term of an odd cat state is given by,
\begin{equation}
\gamma_1 = \frac{f_1(\alpha) - f_1(-\alpha)}{\sqrt{2(1-e^{-2|\alpha|^2})}},
\end{equation}
where $f_n(\alpha)$ is defined in Eq.~(\ref{eq:fna}). In order to get the classical probability we calculate $\gamma_1^2.$ The amplitude of the $n=1$ coherent state photon term is then,
\begin{eqnarray} \label{eq:fna1}
f_1(\alpha) &=& \alpha e^{-\frac{|\alpha|^2}{2}}, 
\end{eqnarray}
thus, the probability of having sampled from the correct term is,
\begin{eqnarray}
P &=& {\gamma_1}^{2n} \nonumber \\
&=& \left(\frac{\alpha e^{\frac{-|\alpha|^2}{2}} - (-\alpha)e^{\frac{-|\alpha|^2}{2}}}{\sqrt{2(1-e^{-2|\alpha|^2})}}\right)^{2n} \nonumber \\
&=& \left(\frac{2\alpha e^{\frac{-|\alpha|^2}{2}}}{\sqrt{2(1-e^{-2|\alpha|^2})}}\right)^{2n}\nonumber \\
&=&  \left(\frac{4\alpha^2 e^{-|\alpha|^2}}{2(1-e^{-2|\alpha|^2})}\right)^{n}\nonumber \\
&=& \left(\frac{2\alpha^2}{e^{|\alpha|^2}(1-e^{-2|\alpha|^2})}\right)^{n}\nonumber \\
&=& \left(\frac{2\alpha^2}{e^{|\alpha|^2}-e^{-|\alpha|^2}}\right)^{n}\nonumber \\
&=& \left(\alpha^2\mathrm{csch}(|\alpha|^2)\right)^{n}\nonumber \\
&=&\alpha^{2n} \mathrm{csch}^n(|\alpha|^2),
\end{eqnarray}
where the hyperbolic trigonometric identity $\mathrm{csch}(x)=2/(e^x-e^{-x})$ was used.
Following AA's given bound this requires that,
\begin{equation}
\alpha^{2n} \mathrm{csch}^n(|\alpha|^2) > 1/\mathrm{poly}(n),
\end{equation}
in order for the sampling problem to be in a regime which is provably computationally hard.

%

\section{Propagating Multi-Mode Coherent States Through Passive Linear Optics Networks} \label{app:coherent_map}

As for Sec. \ref{sec:CATarbitraryAmplitude} the unitary map describing a passive linear optics network is given by,
\begin{equation} \label{eq:app_unitary_map}
\hat{U}: \,\, \hat{a}_i^\dag \to \sum_{j=1}^m U_{i,j} \hat{a}_j^\dag,
\end{equation}
and taking the Hermitian conjugate yields,
\begin{equation}
\label{eq:app_unitary_map_conj}
\hat{U}: \,\, \hat{a}_i \to \sum_{j=1}^m U_{i,j}^* \hat{a}_j.
\end{equation}
A coherent state can be expressed in terms of a displacement operator acting on the vacuum state,
\begin{equation}
\ket{\alpha^{(i)}}_i = \hat{D}_i(\alpha^{(i)}) \ket{0}_i,
\end{equation}
where the displacement operator may be expressed in terms of creation and annihilation operators as,
\begin{equation}
\hat{D}_i(\alpha^{(i)}) = \mathrm{exp}(\alpha^{(i)} \hat{a}_i^\dag - {\alpha^{(i)}}^* \hat{a}_i).
\end{equation}
Applying the unitary map Eqs.~(\ref{eq:app_unitary_map}) and (\ref{eq:app_unitary_map_conj}), we obtain,
\begin{equation}
\hat{U}\hat{D}_i(\alpha_i) = \mathrm{exp}\!\left(\alpha^{(i)} \sum_{j=1}^m U_{i,j} \hat{a}_j^\dag - {\alpha^{(i)}}^* \sum_{j=1}^m U_{i,j}^* \hat{a}_j\right).
\end{equation}
Let,
\begin{equation}
\hat{U} \ket{\alpha^{(1)},\dots,\alpha^{(m)}} = \ket{\beta^{(1)},\dots,\beta^{(m)}}.
\end{equation}
Then,
\begin{equation}
\hat{U} \ket{\alpha^{(1)},\dots,\alpha^{(m)}} = \hat{U} \hat{D}_1(\alpha^{(1)}) \dots \hat{D}_m(\alpha^{(m)}) \ket{0_1,\dots,0_m}.
\end{equation}
For each term,
\begin{eqnarray}
\hat{U} \hat{D}_i(\alpha^{(i)}) &=& \prod_{j=1}^m \mathrm{exp}\!\left(\alpha^{(i)} U_{i,j} \hat{a}_j^\dag - {\alpha^{(i)}}^* U_{i,j}^* \hat{a}_j\right) \nonumber \\
&=& \prod_{j=1}^m \hat{D}_j(U_{i,j} \alpha^{(i)}).
\end{eqnarray}
Thus,
\begin{eqnarray}
\hat{U} \prod_{i=1}^m \ket{\alpha^{(i)}}_i &=& \hat{U} \prod_{i=1}^m \hat{D}(\alpha^{(i)}) \ket{0}_i \nonumber \\
&=& \prod_{i=1}^m \prod_{j=1}^m \hat{D}_j(U_{i,j} \alpha^{(i)}) \ket{0} \nonumber \\
&=& \bigotimes_{j=1}^m \left|\sum_{i=1}^m U_{i,j} \alpha^{(i)}\right\rangle_j \nonumber \\
&=& \bigotimes_{j=1}^m \ket{\beta^{(j)}}_j.
\end{eqnarray}
And,
\begin{equation}
\label{eq:final_beta}
\beta^{(j)} = \sum_{i=1}^m U_{i,j} \alpha^{(i)},
\end{equation}
as per Eq.~(\ref{eq:coherent_map_relation}). 

\section[Proof of the Algebraic Form of the MORDOR Unitary]{Proof of $U_{j,k}^{(n)}$} \label{app:Ujk}

For this derivation in Sec. \ref{sec:MORDOR_Device} we begin from Eq. (\ref{eq:MORDORU}) and setting $\hat{\Theta}=\hat{I}$,
\begin{eqnarray}
U_{j,k}^{(n)} &=& (\hat{V}\hat{\Phi}\hat{V^{\dag}})_{j,k} \nonumber \\ 
&=& \sum_{l,m=1}^{n}V_{j,l}\Phi_{l,m}V_{m,k}^{\dag} \nonumber \\
&=& \sum_{l,m=1}^{n} \underbrace{\frac{e^{- 2 i j l \pi/n}}{\sqrt{n}}}_{V_{j,l}}\underbrace{\delta_{l,m}e^{i(l-1)\varphi}}_{\Phi_{l,m}}\underbrace{\frac{e^{2 i m k \pi/n}}{\sqrt{n}}}_{V_{m,k}^{\dag}} \nonumber \\
&=& \frac{1}{n} \sum_{l=1}^{n} e^{\frac{- 2 i j l \pi}{n}} e^{i (l-1) \varphi} e^{\frac{2 i l k \pi}{n}}\nonumber \\
&=& \frac{1}{n} \sum_{l=1}^{n} e^{\frac{2 i l(k-j) \pi}{n} + i(l-1)\varphi} \nonumber \\
&=& e^{\frac{2 i (k-j) \pi}{n}}\frac{1}{n} \sum_{l=0}^{n-1} (e^{\frac{2 i (k-j) \pi}{n} + i\varphi})^l. \nonumber 
\end{eqnarray}
From the geometric series, it follows,
\begin{eqnarray}
U_{j,k}^{(n)}&=& \frac{1}{n(e^{\frac{2 i (j-k) \pi}{n}})}\frac{1-e^{i n\varphi}}{\left(1-e^{\frac{2 i (k-j) \pi}{n} +i \varphi}\right)}, \label{eq:U} \nonumber \\
&=& \frac{1-e^{i n\varphi}}{n\left(e^{\frac{2 i \pi(j-k)}{n}}-e^{i \varphi}\right)}
\end{eqnarray}
which is what we set out to prove.
which is Eq. (\ref{eq:Ujk}) that we set out to prove, where the last line follows from the geometric series.

\section[Conjecture for the Analytic Form of the MORDOR Permanent]{Conjecture for the Analytic Form of Per($\hat U^{(n)}$)} \label{app:series}

For this derivation in Sec. \ref{sec:MORDOR_Device} our goal is to find the analytic form for Per($\hat U^{(n)}$) where $U_{j,k}^{(n)}$ is as in Eq. (\ref{eq:U}).  We can perform a brute force calculation to obtain the analytic form for small $n$. Doing so up to $n=6$ is presented in the following table.
\begin{table}[!htb] \label{tab:MORDORtable}
\centering
\begin{tabular}{|c|c|} 
\hline
 $n$ & Per$(\hat U^{(n)})$ \nonumber \\
 \hline
 1 & 1 \nonumber \\
 2 & $e^{i \phi } \cos (\phi )$ \\
 3 & $\ \frac{1}{9} \left(2+e^{3 i \phi   }\right) \left(1+2 e^{3 i \phi   }\right)$ \\
 4 & $\frac{1}{32} \left(1+e^{4 i \phi   }\right) \left(3+e^{4 i \phi }\right)    \left(1+3 e^{4 i \phi }\right)$ \\
 5 & $\frac{1}{625} \left(4+e^{5 i \phi
   }\right) \left(3+2 e^{5 i \phi   }\right) \left(2+3 e^{5 i \phi   }\right) \left(1+4 e^{5 i \phi   }\right)$ \\
 6 & $\frac{1}{648} \left(1+e^{6 i \phi    }\right) \left(2+e^{6 i \phi }\right)    \left(5+e^{6 i \phi }\right)    \left(1+2 e^{6 i \phi }\right)    \left(1+5 e^{6 i \phi }\right)$ 
 \\
 \hline
\end{tabular}
\end{table} 
One can see the pattern that emerges is of the following form:
\begin{equation} \label{eq:permU2} 
\mathrm{Per}(\hat{U}^{(n)})= \frac{1}{n^{n-1}}\prod_{j=1}^{n-1}\Big[je^{i n \varphi}+n-j\Big],
\end{equation}
which is Eq. (\ref{eq:permU}) that we set out to show. This equation has been verified analytically up to $n=16$ and up to $n=25$ numerically..

\section[Calculation of the MORDOR Signal]{Calculation of $P$} \label{app:P}
For this derivation in Sec. \ref{sec:MORDOR_Device}, assuming our conjecture in Eq. (\ref{eq:permU}) holds, we can compute the coincidence probability of measuring one photon in each mode at the output,
\begin{eqnarray}
P &=& \big|\mathrm{Perm}(U^{(n)})\big|^2 \nonumber \\
&=& \left|\frac{1}{n^{n-1}} \prod_{j=1}^{n-1}\left(je^{i n \varphi}+n-j\right)\right|^2 \nonumber \\
&=& \frac{1}{n^{2n-2}} \prod_{j=1}^{n-1}\Big|\left(je^{i n \varphi}+n-j\right)\Big|^2 \nonumber \\
&=& \frac{1}{n^{2n-2}} \prod_{j=1}^{n-1}\Big|j\mathrm{cos}(n\varphi)+ij\mathrm{sin}(n\varphi)+n-j\Big|^2 \nonumber \\
&=& \frac{1}{n^{2n-2}} \prod_{j=1}^{n-1}\Big|\underbrace{j\mathrm{cos}(n\varphi)+(n-j)}_\mathrm{Re}+i\underbrace{j\mathrm{sin}(n\varphi)}_\mathrm{Im}\Big|^2. \nonumber \\
\end{eqnarray} 
Invoking the property that \mbox{$|z|^2= \mathrm{Re}(z)^2+\mathrm{Im}(z)^2$}, where \mbox{$z\in \mathbb{C}$}, 
\begin{eqnarray} \label{eq:Pproof}
P &=& \frac{1}{n^{2n-2}} \prod_{j=1}^{n-1} \Big[\big(j\mathrm{cos}(n\varphi)+(n-j)\big)^2+j^2\mathrm{sin}^2(n\varphi)\Big] \nonumber \\
&=& \frac{1}{n^{2n-2}} \prod_{j=1}^{n-1} \Big[\underbrace{j^2\mathrm{cos}^2(n\varphi)+j^2\mathrm{sin}^2(n\varphi)}_{=j^2} \nonumber \\
&+& 2j(n-j)\mathrm{cos}(n\varphi)+(n-j)^2 \Big] \nonumber \\
&=& \frac{1}{n^{2n-2}} \prod_{j=1}^{n-1} \Big[j^2 + 2j(n-j)\mathrm{cos}(n\varphi)+(n-j)^2 \Big]\nonumber \\
&=& \frac{1}{n^{2n-2}} \prod_{j=1}^{n-1} \Big[\underbrace{2j(n-j)}_{a_n(j)}\mathrm{cos}(n\varphi)+\underbrace{n^2-2jn+2j^2}_{b_n(j)} \Big] \nonumber \\
&=& \frac{1}{n^{2n-2}} \prod_{j=1}^{n-1} \Big[a_n(j)\mathrm{cos}(n\varphi)+b_n(j) \Big], \nonumber \\
\end{eqnarray} 
which is Eq. (\ref{eq:P_Result}) that we set out to show.

\section[Calculation of the Signal Rate of Change with Respect to the Unknown Phase]{Calculation of $\left|\frac{\partial P}{\partial \varphi}\right|$} \label{app:dP}

For this derivation in Sec. \ref{sec:MORDOR_Device} we begin with Eq. (\ref{eq:Pproof}), exploiting the logarithm product rule,
\begin{eqnarray}
\mathrm{ln}(P) &=& \underbrace{\mathrm{ln}\left(\frac{1}{n^{2n-2}}\right)}_{C} + \mathrm{ln}\left(\prod_{j=1}^{n-1} \Big[a_n(j)\mathrm{cos}(n\varphi) + b_n(j) \Big] \right) \nonumber \\
&=& C + \sum_{j=1}^{n-1} \mathrm{ln}\Big[a_n(j)\mathrm{cos}(n\varphi)+b_n(j) \Big],
\end{eqnarray} 
where $C$ is a constant. Now the derivative becomes, 
\begin{eqnarray}
\frac{1}{P}\frac{\partial P}{\partial \varphi} &=& -\sum_{j=1}^{n-1} \frac{na_n(j)\mathrm{sin}(n\varphi)}{a_n(j)\mathrm{cos}(n\varphi)+b_n(j)} \nonumber \\
\frac{\partial P}{\partial \varphi} &=& -nP\mathrm{sin}(n\varphi)\sum_{j=1}^{n-1} \frac{a_n(j)}{a_n(j)\mathrm{cos}(n\varphi)+b_n(j)}.\nonumber \\
\end{eqnarray} 
Thus,
\begin{equation}
\left|\frac{\partial P}{\partial\varphi}\right|=nP\big|\mathrm{sin}(n\varphi)\big|\sum_{j=1}^{n-1} \left|\frac{a_n(j)}{a_n(j)\mathrm{cos}(n\varphi)+b_n(j)}\right|,
\end{equation}
which is Eq. (\ref{eq:dP}) that we set out to show.

\section[Calculation of Phase Sensitivity in the Small Angle Approximation]{Calculation of $\Delta\varphi$ in the Small Angle Approximation} \label{app:dphi}
For this derivation in Sec. \ref{sec:MORDOR_Device} we wish to compute $\Delta\varphi$ in the limit that $n\varphi\ll1$. Then $P$ in the small angle regime of Eq. (\ref{eq:P_Result}) becomes,
\begin{eqnarray}
\label{eq:papprox}
P&\approx& \frac{1}{n^{2n-2}}\prod_{j=1}^{n-1} \bigg[a_n(j)\Big(1-\frac{1}{2}(n\varphi)^2\Big)+b_n(j)\bigg] \nonumber \\ 
&=& \frac{1}{n^{2n-2}}\prod_{j=1}^{n-1}\bigg[ n^2-(nj+j^2)n^2\varphi^2 \bigg]\nonumber \\ 
&=& \prod_{j=1}^{n-1}\Big[1-(nj+j^2)\varphi^2\Big],
\end{eqnarray}
where $\cos(n\varphi)$ is expanded to the first nonconstant term in its Taylor series. This product has the form of a binomial expansion. Dropping terms above order $\varphi^2$, $P$ reduces to,
\begin{eqnarray}
\label{eq:pfinal}
P&\approx& 1-\varphi^2\sum_{j=1}^{n-1}\Big[nj+j^2\Big] \nonumber \\
&=& 1-\varphi^2\Big[\frac{1}{6}(n-1)n(n+1)\Big] \nonumber \\
&=&1-k(n)\varphi^2,
\end{eqnarray}
where $k(n)=\frac{1}{6}n(n-1)(n+1)\geq0$ $\forall$ $n\geq1$.  From Eq. (\ref{eq:pfinal}) we can easily compute $P^2$ and $\big|\frac{\partial P}{\partial \varphi}\big|$ to be,
\begin{eqnarray}
P^2&\approx& 1-2k(n)\varphi^2 \\
\left|\frac{\partial P}{\partial \varphi}\right|&=&2k(n)|\varphi|,
\end{eqnarray}
where we have again dropped terms above order $\varphi^2$. Using Eq. (\ref{eq:phaseSenP}) the phase sensitivity $\Delta\varphi$ in the small angle regime is,
\begin{eqnarray}
\Delta\varphi &=& \frac{\sqrt{P-P^2}}{\left|\frac{\partial P}{\partial\varphi}\right|} \nonumber \\
&=&\frac{\sqrt{\Big(1-k(n)\varphi^2\Big)-\Big(1-2k(n)\varphi^2\Big)}}{2k(n)|\varphi|} \nonumber \\
&=&\frac{\sqrt{k(n)\varphi^2}}{2k(n)|\varphi|} \nonumber \\
&=& \frac{1}{2\sqrt{k(n)}} \nonumber \\
&=&\sqrt{\frac{3}{2{(n-1)n(n+1)}}},
\end{eqnarray}
which is Eq. (\ref{eq:DeltaVarPhi}) that we set out to show.

\section{Discussion of Ordinal Resource Counting (ORC)} \label{app:counting}

For Sec. \ref{sec:MORDOR_ORC} we would like to compare the performance of our QuFTI to an equivalent multimode interferometer baseline for which we will construct the shotnoise limit (SNL) and Heisenberg limit (HL). This is a subtle comparison, due to the linearly increasing unknown phase-shifts, \{\mbox{$0,\varphi,\dots,(n-1)\varphi$}\}, that the QuFTI requires to operate. There is a long and muddled history of increasing the interrogation time (or here length) of the probe particles with the unknown phase-shift followed by an incorrect reckoning of the true resources. Here we shall use a protocol we call Ordinal Resource Counting (ORC) whereby all resources, such as number of `calls' to the phase-shifter $\varphi$, are converted to the `currency' of the resource that is most precious to us, namely photon-number. We do this as follows. 

First we must construct a multimode interferometer with $n$ photon inputs that provides the baseline if the photons remain uncorrelated and the number-path entanglement remains minimal. Such a comparator is shown in Fig. \ref{fig:resources1}, and consists of $n$, two-mode Mach-Zehnder Interferometers (MZI) in a vertical cascade, fed with single-photon inputs, with the same linearly increasing unknown phase-shift sequence as the QuFTI. Since the MZIs are disconnected, the number-path entanglement remains constant and minimal, and of the form \mbox{$(\ket{1,0}+\ket{0,1})/\sqrt{2}$} inside each MZI. 

\begin{figure}[!htb]
\centering
\includegraphics[width=\imageWidth\columnwidth]{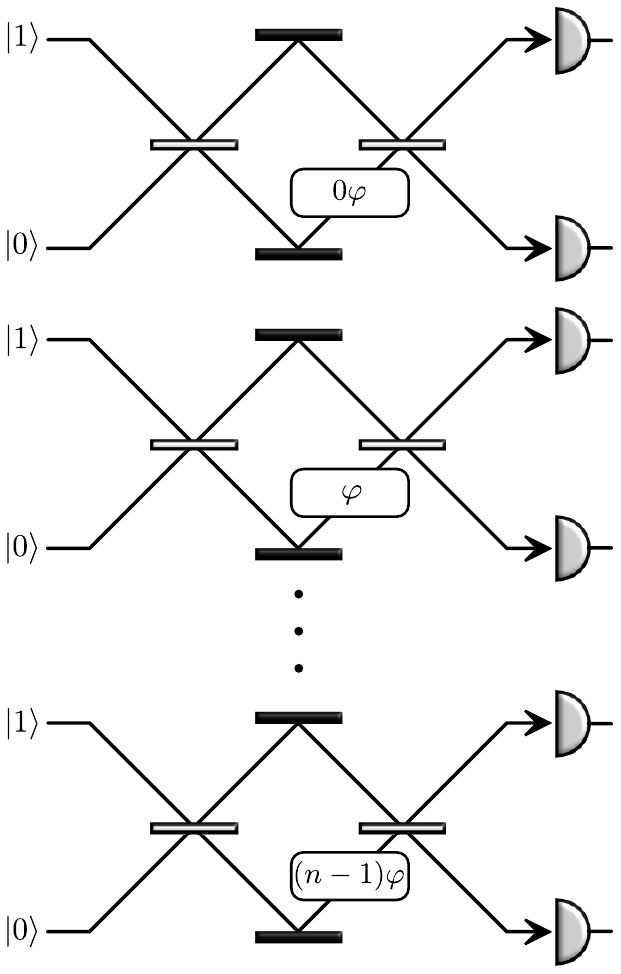}
\caption{$n$ instances of two-mode Mach-Zehnder interferometers, with a linearly increasing phase gradient. This system has the same configuration of phases as the QuFTI, but the photons are not allowed to interfere, and thus has minimal number-path entanglement.} \label{fig:resources1}
\end{figure}

Now to convert the linearly increasing interrogation lengths of the unknown phase-shifts, we note that a single photon interrogating a phase-shift of say $2\varphi$ is equivalent to a single photon interrogating a single phase-shift $\varphi$ twice, which is in turn equivalent to two uncorrelated photons entering the same port of the MZI containing a single phase-shift of $\varphi$. In this way we may convert `number of interrogations of the phase-shifter' into the currency of `number of photons' to carry out a fair reckoning of the resources. Following this logic we are led to Fig. \ref{fig:resources2} showing a cascade of MZIs where the linearly increasing phase-shifters are replaced with a single phase-shifter of $\varphi$ and the single photons at the MZI inputs are replaced with a linearly increasing number of photons. Then the `number of interrogations of the phase-shifter' becomes $n(n-1)/2$, but there is an additional photon that is part of the QuFTI resources so our total number of resources becomes,
\begin{equation} \label{eq:N}
N\equiv1+\frac{n(n-1)}{2}.
\end{equation}
\begin{figure}[!htb]
\centering
\includegraphics[width=0.5\columnwidth]{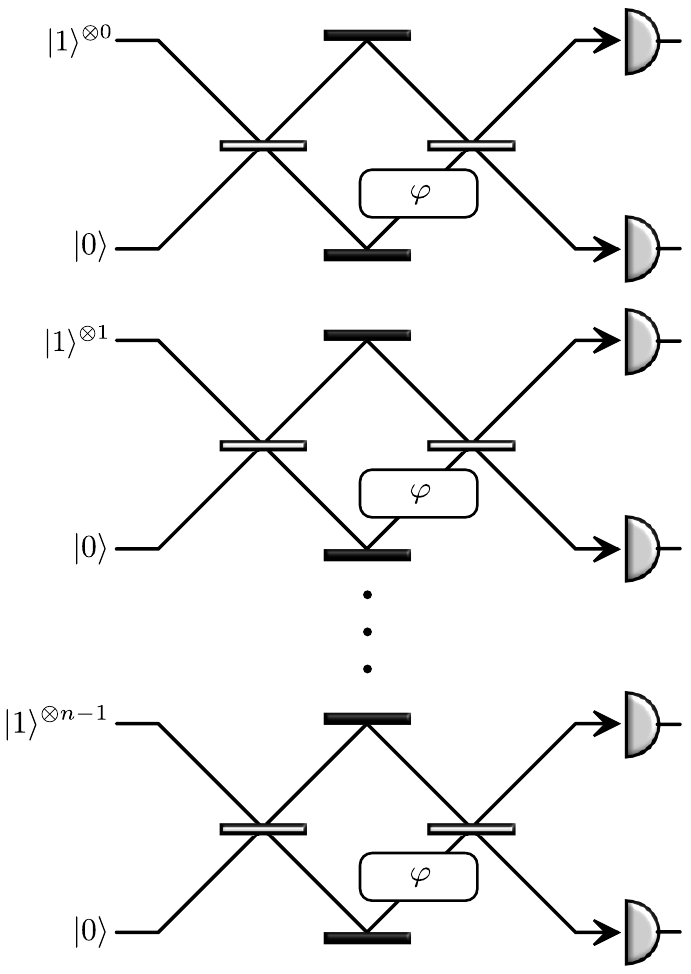}
\caption{Noting that a single photon interrogating a phase-shift of $n\varphi$ is equivalent to $n$ independent interrogations of $\varphi$, Fig. \ref{fig:resources1} can be represented in terms of the resource of photons as shown here. Here $\ket{1}^{\otimes j}$ means that $j$ independent (i.e distinguishable) photons have been prepared.} \label{fig:resources2}
\end{figure}

Next we note that this cascade of $n$ MZIs in Fig. \ref{fig:resources2} may be replaced with a single MZI, shown in Fig. \ref{fig:resources3}, where the input is now an ordinal grouped ranking of the uncorrelated photons following the same pattern as in Fig. \ref{fig:resources2}. Hence in the configuration in Fig. \ref{fig:resources3} we have a single MZI with vacuum entering the lower port, a stream of $N$ uncorrelated photons entering the upper port, and a single phase-shifter $\varphi$ between the beamsplitters. It is well-known that for this configuration the sensitivity of this system scales as the SNL \cite{bib:scully1993quantum, bib:dowling1998correlated}, namely,
\begin{equation}
\Delta\varphi_\mathrm{SNL} = \frac{1}{\sqrt{N}} = \frac{1}{\sqrt{1+\frac{n(n-1)}{2}}}.
\end{equation} 
This then provides us a fair reckoning of the SNL to be used gauging the performance of the QuFTI.

\begin{figure}[!htb]
\centering
\includegraphics[width=\imageWidthTwo\columnwidth]{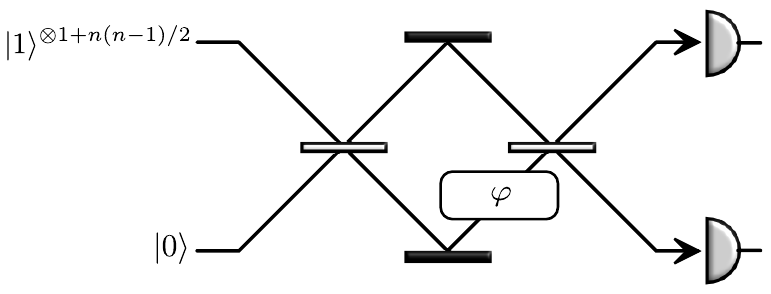}
\caption{Grouping all the independent interferometers in Fig. \ref{fig:resources2} together and including the extra photon from the QuFTI model, we obtain a single MZI with \mbox{$1+n(n-1)/2$} independent photons as input. This configuration achieves the shotnoise limit, and thus provides a benchmark for comparing our QuFTI protocol against the shotnoise and Heisenberg limits, with photons as the resource being counted.} \label{fig:resources3}
\end{figure}

Finally, if instead we were to maximally path-number entangle these resources into a NOON state of the form \mbox{$(\ket{N,0}+\ket{0,N})/\sqrt{2}$} (just to the right of the first beam splitter but before the phase-shifter) the sensitivity then becomes Heisenberg limited,
\begin{equation}
\Delta\varphi_\mathrm{HL} = \frac{1}{N} = \frac{1}{1+\frac{n(n-1)}{2}},
\end{equation}
which is a sensitivity known to saturate the Quantum Cram{\'e}r-Rao Bound (CRB) for sensitivity in local phase estimation with $N$ photons \cite{bib:lee2002quantum, bib:durkin2007local}. As the CRB is the best one may do, according to the laws of quantum mechanics, then in this case the HL is optimal. As discussed, the performance of the QuFTI falls between the SNL and the HL, but with the feature of not having to do anything resource intensive such as preparing a high-NOON state. 

Thus the SNL and the HL, computed via this Ordinal Resource Counting method, provides the fairest comparison of sensitivity performance of the QuFTI with such ambiguities such as how to handle `number of calls to the phase-shifter' removed by replacing such a notion with `number of photons' inputted into the interferometer. 

\section{Dephasing} \label{app:dephasing}

As for Sec. \ref{sec:MORDORefficiency} a form of decoherence to consider is dephasing. Dephasing in our work may be modelled with the result of Bardhan \emph{et al.} \cite{bib:Bardhan2013}, whereby dephasing occurs on each mode separately. When considering our example of a magnetometer, dephasing would occur in the magnetic field cells where atomic fluctuations may occur that differ between cells.  In the rest of the interferometer, dephasing can be made very close to zero, particularly on an all optical chip. 

\begin{figure}[t]
\centering
\includegraphics[width=\imageWidthThree\columnwidth]{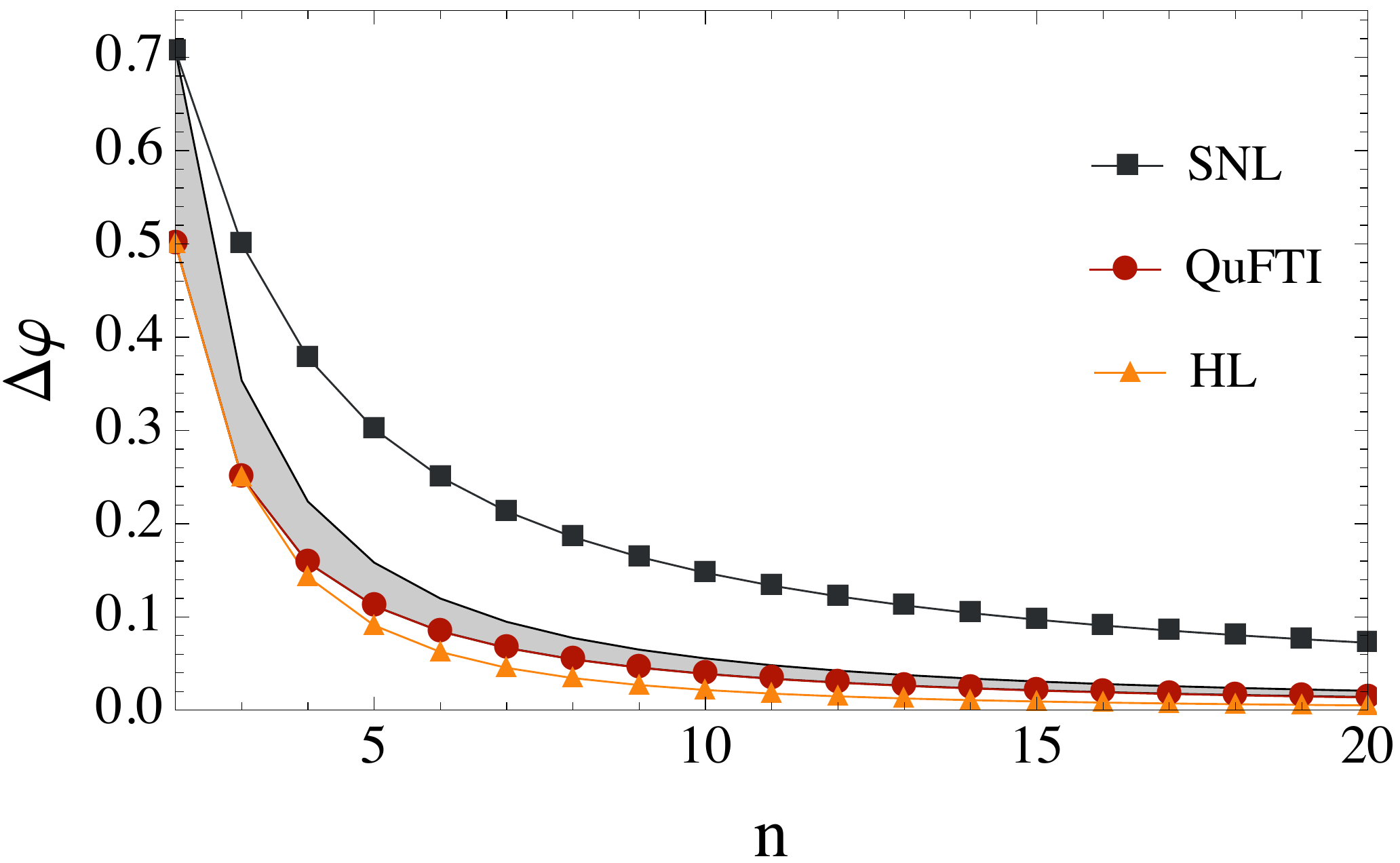}
\caption{Dephasing for $\varphi=0.01$.  The shaded region represents the phase sensitivity for the QuFTI where $0\leq\chi\leq 0.01$.} \label{fig:dephasing}
\end{figure}

\begin{figure}[b]
\centering
\includegraphics[width=\imageWidthThree\columnwidth]{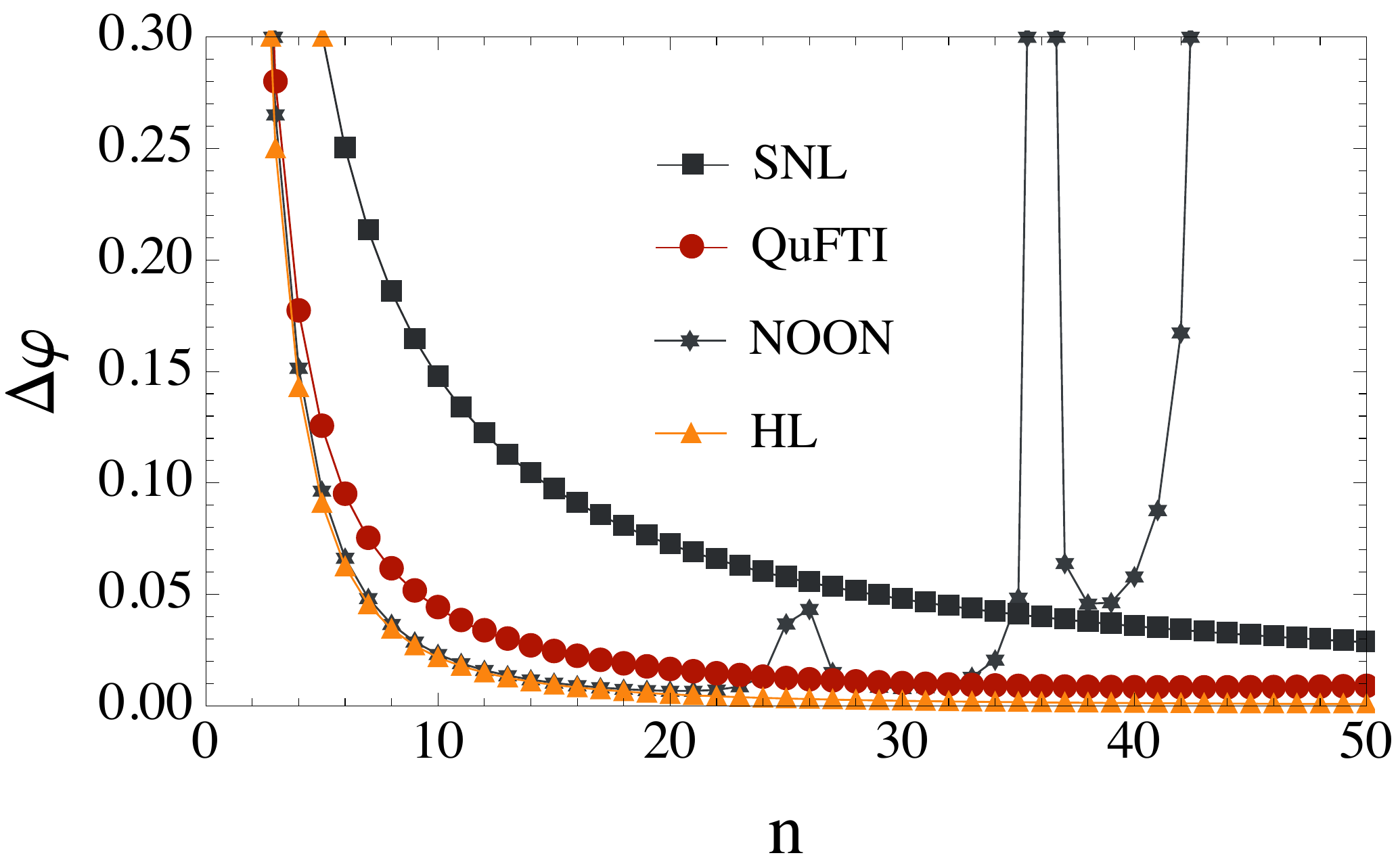}
\caption{The effect of dephasing on the NOON state and QuFTI where $\varphi=0.01, \chi=0.005$.  The NOON state is plotted with respect to $N$ for fair resource counting.} \label{fig:dephasingNOON}
\end{figure}

To model dephasing we investigate a random phase shift $\Delta\chi$ added to each mode separately. $\Delta\chi$ is a Gaussian random variable of zero mean but nonzero second order moment. The phase shift in the $j$th mode then becomes,
\begin{eqnarray}
e^{\pm i j \varphi}&\rightarrow& e^{\pm i j (\varphi + \Delta\chi)} \nonumber \\
&=& e^{\pm i j \varphi}e^{\pm i j \Delta\chi} \nonumber \\
&=& e^{\pm i j \varphi}\left(1 \pm i j\Delta\chi - \frac{1}{2}j\Delta\chi^2\pm\dots\right).
\end{eqnarray}
Using $\expec{\Delta\chi}=0$, $\expec{\Delta\chi^2}\neq0$, and that $\Delta\chi \ll \phi$ we simplify this to be, 
\begin{eqnarray}
e^{\pm i j \varphi}&\rightarrow& e^{\pm i j \varphi}\left(1- \frac{1}{2}j\Delta\chi^2 \pm\dots\right) \nonumber \\
&\approx& e^{\pm i j \varphi}e^{-\frac{1}{2} j^2 \Delta\chi^2}.
\end{eqnarray}

The signal $P$ in Eq. 10 from our work then changes in the presence of dephasing. The dependence that $P$ has on the unknown phase $\varphi$ does not depend on the mode number $j$. Then the term that depends on $\varphi$ becomes,
\begin{eqnarray}
\mathrm{cos}(n\phi) &=& \frac{1}{2}\left(e^{i n \varphi}+e^{-i n \varphi}\right) \nonumber \\
&\rightarrow& \frac{1}{2}\left(e^{i n \phi}+e^{- i n \phi}\right)e^{-\frac{�1}{2}n^2\Delta\chi^2} \nonumber \\
&=& \mathrm{cos}(n\phi)e^{-\frac{�1}{2}n^2\Delta\chi^2}
\end{eqnarray}
Using this substitution $P$ becomes,
\begin{eqnarray} \label{}
P &=& \Big|\mathrm{Per}(\hat{U}^{(n)})\Big|^2 \nonumber \\
&=& \frac{1}{n^{2n-2}}\prod_{j=1}^{n-1} \Big[a_n(j)\mathrm{cos}(n\phi)e^{-\frac{�1}{2}n^2\Delta\chi^2}+b_n(j) \Big].
\end{eqnarray} 
The factor $e^{-\frac{1}{2}n^2\Delta\chi^2}$ can be absorbed into $a_n(j)$ so that the derivation of $|\frac{\partial P}{\partial\phi}|$ in Eq. (\ref{eq:dP}) is identical.
Using this result we numerically plot the phase sensitivity with dephasing in Fig. \ref{fig:dephasing}.

In order to meaningfully analyze the dephased sensitivity, we would like to compare with other well known metrological schemes.  In Fig. \ref{fig:dephasingNOON}, we compare the QuFTI to the NOON state (with $N$ input photons for a fair resource comparison) and see that the QuFTI is far more robust against dephasing.

\section{Evolution of Displacement Operators Through a Linear Optics Network} \label{app:disp_ev}

In Sec. \ref{sec:IntegralProof} we use the following identity. A linear optics network maps a product of displacement operators over $m$ modes, with amplitudes $\lambda_j$, to another product of displacement operators with amplitudes given by \mbox{$\mu_j = \sum_{k=1}^m U_{j,k} \lambda_k$}.
\begin{align}
\hat{U} \left(\prod_{j=1}^m \hat{D}(\lambda_j)\right) \hat{U}^\dag =&\ \mathrm{exp}\left(\sum_{j=1}^m \hat{U}( \lambda_j \hat{a}^\dag_j - \lambda^*_j \hat{a}_j) \hat{U}^\dag\right) \nonumber \\
=&\ \mathrm{exp}\left(\sum_{j=1}^m \sum_{k=1}^m \lambda_j U_{j,k} \hat{a}^\dag_k - \lambda_j^* U_{j,k}^* \hat{a}_k\right) \nonumber \\
=&\ \prod_{j=1}^m \mathrm{exp}\left(\sum_{k=1}^m \lambda_j U_{j,k} \hat{a}^\dag_k - \lambda_j^* U_{j,k}^* \hat{a}_k\right) \nonumber \\
=&\ \prod_{j=1}^m \hat{D}(\mu_j),
\end{align}
where
\begin{eqnarray}
\mu_j &=& \sum_{k=1}^m \lambda_k U_{j,k},
\end{eqnarray}
which is Eq. (\ref{eq:beta_from_alpha}) that we set out to prove.

\section{Overlap of the Displacement Operator with the Single-Photon State} \label{app:disp_ov}

In Sec. \ref{sec:IntegralProof} we also use the following identity. 
The overlap between the single-photon state with the displacement operator \mbox{$\bra{1} \hat{D}(\lambda) \ket{1}$} is given by \mbox{$e^{-\frac{1}{2}|\lambda |^2}(1-|\lambda |^2)$}.
\begin{align} \label{eq:charProof}
\bra{1}\hat{D}(\lambda)\ket{1} =&\ \bra{0}\hat{a} \hat{D}(\lambda) \hat{a}^\dag\ket{0} \nonumber \\
=&\ \bra{0}\hat{a} (\hat{a}^\dag -\lambda ^*) \hat{D}(\lambda)\ket{0} \nonumber \\
=&\ \bra{0}\hat{a}\hat{a}^\dag \hat{D}(\lambda) \ket{0} - \bra{0}\lambda^*\hat{a} \hat{D}(\lambda)\ket{0} \nonumber \\
=&\ \langle 0|\lambda\rangle - \lambda ^*\langle 1|\lambda\rangle \nonumber \\
=&\ e^{-\frac{1}{2}|\lambda|^2} - |\lambda |^2 e^{-\frac{1}{2}|\lambda|^2} \nonumber \\
=&\ e^{-\frac{1}{2}|\lambda|^2}(1 - |\lambda |^2),
\end{align}
which is the identity that we used to obtain Eq. (\ref{eq:INTmmCW}). We have used the commutation relation between the displacement operator and the photon-creation operator \cite{bib:GerryKnight05},
\begin{align}
[\hat{a}^\dagger, \hat{D}(\lambda)]=\lambda^*\hat{D}(\lambda),
\end{align}
as well as the coherent state represented in the Fock basis
\begin{align}
\ket{\lambda}=e^{-\frac{|\lambda|^2}{2}}\sum_{n=0}^{\infty}\frac{\lambda^n}{\sqrt{n!}}\ket{n},
\end{align}
to read off the overlaps between the Fock states and the coherent state $\ket{\lambda}$.

%
%
%

\section{Kraus Operator Formalism} \label{sec:GKPKraus}

For Sec. \ref{sec:GKPpreparing} we note that the theory of generalized quantum measurements provides a convenient formalism to describe the state-preparation protocol described above.  
The conditional optical state is expressed using a set of Kraus operators corresponding to measurement outcomes on the spin. 

We begin with an unentangled state between a spin of total angular momentum $J$ and an optical field, $\ket{\psi_b} = \ket{\phi}_A \otimes \ket{\psi}_O$, where the spin and field states are arbitrary. 
For measurement outcome $x$ the normalized, conditional field state is given by 
	\begin{align} \label{eq:CondState}
		\ket{\psi_d^x}_O = \frac{\hat{A}_{x} \ket{\psi}_O}{ \sqrt{ \mathcal{P}(x) } },  
	\end{align}
where $\hat{A}_{x}$ is the Kraus operator and $\mathcal{P}(x)$ is the probability of outcome $x$. 
Expressing the spin state in the $z$-basis, $\ket{\phi}_A = \sum_m c_m \ket{m}$, the Kraus operator associated with the controlled-displacement interaction in Eq. (\ref{eq:BigD}) is
	\begin{align}
		\hat{A}_{x}  &\equiv \bra{x} \hat{D}_c(g) \ket{\phi}_A 
			 = \sum_{m=-J}^J c_m d_{m,x}  \,  e^{ -i g m \hat{p} } . \label{eq:KrausOp} 
	\end{align}
This describes the conditional operation that implements a set of displacements on the field proportional to the initial spin distribution, $c_m$, and the measurement outcome $x$. The probability of measurement outcome $x$ is obtained by tracing over the initial field state,
	\begin{align} \label{eq:ProbKraus}
		\mathcal{P}(x) & = \mbox{Tr} \big[ \hat{A}^{\dagger}_{x} \hat{A}_{x}  \ketbra{\psi}{\psi} \big]  \nonumber \\
		& = \sum_{m, m'= -J}^J c_m c_{m'}^* d_{m,x}  d_{m', x} \bra{\psi} e^{-i g(m-m')\hat{p}}\ket{\psi}.  
	\end{align}	

For a more general input state of light, $\hat{\rho}_O$, which may be mixed due to errors in the squeezing procedure and losses, the conditional field state is given by
 	\begin{align} \label{Eq:CondRho}
		\hat{\rho}^x_{d} = \frac{\hat{A}_{x} \hat{\rho}_O \hat{A}^\dagger_{x}}{ \mathcal{P}(x)  } 
	\end{align}
with probability $\mathcal{P}(x) = \mbox{Tr}[ \hat{A}_{x}^\dagger \hat{A}_{x} \hat{\rho}_O ]$.

\section{Measurement Probability} \label{app:MeasurementProbability}

In Sec. \ref{sec:GKPpreparing} we consider the case of a position-squeezed input field, Eq. (\ref{eq:NickSucks}), and an initial spin-coherent state corresponding to $c_m = d_{m,J}$ in Eq. (\ref{eq:KrausOp}). The Kraus operator description, Eq. (\ref{eq:CondState}), then gives the conditional field state, Eq. (\ref{eq:projectedOpticalState}). 
The probability of outcome $x$ follows directly from Eq. (\ref{eq:ProbKraus}),
\begin{align}
    \mathcal{P}(x) = 
     \sum_{m,m'} d_{m,J}d_{m,x}d_{m',J}d_{m',x}\braket{ \gdisp{m}  , \xi}{ \gdisp{m'} , \xi}.
\end{align}
To evaluate the expression, we note that for real $\alpha$ and $\xi$ a displaced squeezed state can be written as 
\begin{eqnarray}
\ket{\alpha,\xi} = \hat{D}(\alpha)\hat{S}(\xi)\ket{0} = \hat{S}(\xi)\hat{D}\big(\alpha e^{\xi} \big)\ket{0}.
\end{eqnarray}
Then, the overlap between states of different displacements is calculated simply,
\begin{align}
\overlap{\alpha,\xi}{\beta,\xi} 
&= \mathrm{exp}\left[ -\smallfrac{1}{2}e^{2\xi}(\alpha-\beta)^2\right], \label{eq:Overlap}
\end{align}
and the probability becomes 
\begin{equation} \label{eq:Pxappendix}
\mathcal{P}(x) = \sum_{m,m'}d_{m,J}d_{m,x}d_{m',J}d_{m',x}e^{-\frac{1}{4}g^2e^{2\xi}(m-m')^2}.
\end{equation}
The resource states for encoding arise for outcomes $x=\pm J$, which occur with probabilities 
\begin{equation}
    \mathcal{P}(\pm J) = \frac{1}{16^J} \sum_{m,m'} \begin{pmatrix}2J \\ J-m\end{pmatrix}
    \begin{pmatrix}2J \\ J-m'\end{pmatrix}e^{-\frac{1}{4}g^2e^{2\xi}(m-m')^2}.
\end{equation}

In the limit of large $g^2e^{2\xi}$, only the $m=m'$ term contributes significantly in the above expressions, and the probability is approximately 
\begin{equation} \label{eq:PmeasLimapp}
\mathcal{P}(x) \approx \sum_m \left(d_{m,J}d_{m,x}\right)^2.
\end{equation}
In the same limit, the probability $x=\pm J$ simplifies to
\begin{align}
  \mathcal{P}(\pm J) &\approx \frac{1}{16^J} \sum_{m=-J}^{J} \binom{2 J}{J-m}^2 \nonumber \\
&= \frac{1}{16^J} \sum_{k=0}^{2J} \binom{2 J}{k}^2 
= \frac{1}{16^J} \binom{4 J}{2J},
\end{align}
where the last step uses $\sum_{k=0}^{m}\binom{m}{k}^2=\binom{2m}{m}$. For large $J$, Taylor expanding $ \mathcal{P}(\pm J)$ to first order in $1/J$ yields 
\begin{align}
\mathcal{P}(\pm J)  &\approx \sqrt{\frac{1}{2 \pi J}}.
\end{align}
Both $+J$ and $-J$ yield a useful resource state, so the total resource preparation probability is 
\begin{equation}
    \mathcal{P}_{s} = 2\mathcal{P}(\pm J) \approx \sqrt{\frac{2}{\pi J}}.
\end{equation}


\section{Variance of Peaks in Momentum Representation} \label{app:cosvariance}

For section \ref{sec:GKPencoding} we note that the momentum representation of the resource states, Eq. (\ref{eq:psipJ}), is a Gaussian envelope multiplying a comb generated by $\cos^{2J}(pg/2)$. We want to approximate each peak in the comb by a Gaussian by matching the peak's variance. Treating a single peak as a probability distribution we have
\begin{equation}
    P(p) = \frac{g\Gamma(J+1)}{2\sqrt{\pi}\Gamma(J+1/2)}\cos^{2J}(pg/2)
\end{equation}
where the prefactors ensure the normalisation
\begin{equation}
    \int_{-\pi/g}^{\pi/g}P(p)dp = 1.
\end{equation}
The variance is then calculated in the usual way
\begin{equation}
    \sigma_p^2 = \langle p^2 \rangle - \langle p \rangle^2 =  \int_{-\pi/g}^{\pi/g}P(p)p^2dp.
\end{equation}
Performing the integral we find
\begin{equation}
    \sigma_p^2 = \frac{2(J^2\zeta(2,J)-1)}{g^2J^2},
\end{equation}
where $\zeta(s,J)$ is the Hurwitz zeta function
\begin{equation}
    \zeta(s,J) = \sum_{k=0}^\infty \frac{1}{(k+J)^{s}}.
\end{equation}
For large $J$, $\zeta(2,J)\approx \frac{1}{J}+\frac{1}{2J^2}+O(\frac{1}{J^3})$ (see for example \cite{bib:paris04}), hence
\begin{equation}
    \sigma_p^2 \approx \frac{2}{g^2 J} + O(1/J^2) = 1/\sigma_{q,\mathrm{env}}^2 + O(1/J^2).
\end{equation}



\chapter{Contributions}

Below is a by chapter summary of the contributions I have made to the results in this thesis. The authors are listed in the same order as they appear in the corresponding publications. 

\begin{itemize}
\item Chapter \ref{ch:LOQC}: Some of the material in this chapter has been published \cite{bib:introChapter} and is based on the work by Bryan T. Gard, Keith R. Motes, Jonathan P. Olson, Peter P. Rohde, and Jonathan P. Dowling. My contributions to this work include various sections of the writing. For this thesis I have rewritten much of the text and added a significant amount of new content. Some of this work may have been used as material in Jonathan P. Olson's and Bryan T. Gard's PhD theses at Louisiana State University.
\item Chapter \ref{ch:QRW}: The material in this chapter has been published \cite{bib:QRWmotes16} and is based on the work by Keith R. Motes, Alexei Gilchrist, and Peter P. Rohde. My contributions to this work include the majority of the writing, writing code for simulations, and producing the graphics. Much of this text is  verbatim to the article as I did the majority of the writing.
\item Chapter \ref{Ch:BSIntro}: Some of the material in this chapter has been published \cite{bib:introChapter} and is based on the work by Bryan T. Gard, Keith R. Motes, Jonathan P. Olson, Peter P. Rohde, and Jonathan P. Dowling. My contributions to this work include various sections of the writing. For this thesis I have rewritten much of the text and added a significant amount of new content. Some of this work may have been used as material in Jonathan P. Olson's and Bryan T. Gard's PhD theses at Louisiana State University.
\item Chapter \ref{Ch:SPDC}: The material in this chapter is based on the work by Keith R. Motes, Jonathan P. Dowling, and Peter P. Rohde \cite{bib:motes13}. My contributions to this work are writing much of the text, performing simulations, carrying out mathematical calculations and generating figures. Some of this text is verbatim to the article.
\item Chapter \ref{Ch:FiberLoop}: The material in this chapter has been published \cite{bib:motes2014scalable} and is based on the work by Keith R. Motes, Alexei, Gilchrist, Jonathan P. Dowling and Peter P. Rohde. My contributions to this work were working out the mathematics, writing the manuscript, coding simulations, and producing the similarity graphic. Some of this text is verbatim to the article.
\item Chapter \ref{Ch:sampOther}: The material in this chapter has been published \cite{bib:Olson14, bib:Kaushik14, bib:catSampling} and is based on the works by Jonthan P. Olson, Kaushik P. Shshadreesan, Keith R. Motes, Peter P. Rohde, Jonthan P. Dowling, Paul Knott, Joseph Fitzsimons, and William Munro. For \cite{bib:Olson14, bib:Kaushik14} I carried through calculations and edited the text and have therefore rewritten much of this text. For \cite{bib:catSampling} I carried through the calculations and did much of the writing and so some of this text is verbatim to the article. Certain parts including the computational complexity section that I did not do the original writing on has been rewritten. Some of this work may have been used as material in Jonathan P. Olson's and Kaushik P. Shshadreesan's PhD theses at Louisiana State University.
\item Chapter \ref{Ch:MORDOR}: The material in this chapter has been published \cite{bib:MORDOR} and is based on the work by Keith R. Motes, Jonathan P. Olson, Evan J. Rabeaux, Jonathan P. Dowling, S. Jay Olson, and Peter P. Rohde. My contributions to this work are writing much of the article, performing calculations, and writing code for simulations. Much of this text is  verbatim to the article as I did a majority of the writing. Some of this work may have been used as material in Jonathan P. Olson's PhD thesis at Louisiana State University. 
\item Chapter \ref{ch:sharpPhard}: The material in this chapter has been published \cite{bib:rohde2016quantum} and is based on the work by Peter P. Rohde, Dominic W. Berry, Ryan L. Mann, Keith R. Motes, and Jonathan P. Dowling. My contributions to this work are working through and writing several of the calculations and writing some parts of the main text. Much of the text here is written by me.
\item Chapter \ref{ch:MotherFocker}: The material in this chapter has been published \cite{bib:motes2016efficient} and is based on the work by Keith R. Motes, Ryan L. Mann, Jonathan P. Olson, Nicholas M. Studer, E. Annelise Bergeron, Alexei Gilchrist, and Jonathan P. Dowling. My contributions to this work are deriving some of the mathematics, writing the simulations for the fusion strategies, and producing most of the graphics. The majority of the text has been rewritten for this thesis.
\item Chapter \ref{ch:GKP}: The material in this chapter has been published \cite{bib:motes2017encoding} and is based on the work by Keith R Motes, Ben Q. Baragiola, Alexei Gilchrist, and Nicolas C. Menicucci. My contributions to this work are much of the writing, derivations, simulations, and producing most of the graphics. Much of this text is verbatim to the article as I did the majority of the writing.
\end{itemize}

{
\hypersetup{
colorlinks = true, 
linkcolor = [rgb]{0.00392156862745098, 0.24705882352941178, 
 0.8666666666666667} 
}
\listoffigures
\listoftables
}

\end{document}